\documentclass[a4paper,11pt,twoside,openright]{book}
\usepackage{graphicx}
\usepackage{epstopdf}

\usepackage{longtable}
\usepackage{amsmath}
\usepackage{amssymb}
\usepackage{cite}
\usepackage{booktabs} 
\usepackage{multirow}
\usepackage{subcaption}

\usepackage{ulem}   
\usepackage{color}
\definecolor{orange}{RGB}{255,127,0}
\definecolor{brown}{RGB}{102,51,0}
\definecolor{myred}{RGB}{192,0,0}
\definecolor{Darkgreen}{RGB}{30,120,30}
\definecolor{Darkblue}{RGB}{0,0,200}
\newcommand{\comment}[1]{}

\def\Comment#1{}
\newcommand{\bean}{\begin{eqnarray*}}
\newcommand{\eean}{\end{eqnarray*}}

\newcommand{\gapproxeq}{\lower
.7ex\hbox{$\;\stackrel{\textstyle >}{\sim}\;$}}
\newcommand{\lapproxeq}{\lower
.7ex\hbox{$\;\stackrel{\textstyle <}{\sim}\;$}}

\newcommand{\ie}{ {\it i.e.}, }

\newcommand\lsim{\mathrel{\rlap{\lower4pt\hbox{\hskip1pt$\sim$}}
    \raise1pt\hbox{$<$}}}
\newcommand\gsim{\mathrel{\rlap{\lower4pt\hbox{\hskip1pt$\sim$}}
    \raise1pt\hbox{$>$}}}
\newcommand{\ba}{\begin{array}}
\newcommand{\ea}{\end{array}}
\newcommand{\nn}{\nonumber}

\newcommand{\be}{\begin{equation}}
\newcommand{\ee}{\end{equation}}
\newcommand{\bear}{\begin{eqnarray}}
\newcommand{\eear}{\end{eqnarray}}

\newcommand{\ket}{\,\rangle}
\newcommand{\bra}{\langle \,}

\newcommand{\cO}{{\cal O}}
\newcommand{\bel}[1]{\be\label{#1}}
\newcommand{\chpt}{$\chi$PT}
\newcommand{\mL}{\mathcal{L}}
\newcommand{\mA}{\mathcal{A}}

\newcommand{\mF}{\mathcal{F}}
\newcommand{\mmF}{\mathbb{F}}
\newcommand{\mG}{\mathcal{G}}
\newcommand{\mI}{\mathcal{I}}
\newcommand{\mJ}{\mathcal{J}}
\newcommand{\mmJ}{\mathbb{J}}
\newcommand{\mM}{\mathcal{M}}
\newcommand{\mN}{\mathcal{N}}
\newcommand{\mO}{\mathcal{O}}
\newcommand{\mP}{\mathcal{P}}

\newcommand{\mT}{\mathcal{T}}
\newcommand{\mV}{\mathcal{V}}
\newcommand{\mX}{\mathcal{X}}
\newcommand{\mY}{\mathcal{Y}}
\newcommand{\Frac}[2]{\frac{\displaystyle #1}{\displaystyle #2}}
\newcommand{\Int}{\displaystyle{\int}}

\def\bat{\begin{array}{cc}}
	
\usepackage[toc,page]{appendix}

\newenvironment{dedication}
{\clearpage           
	\thispagestyle{empty}
	\vspace*{\stretch{1}}
	\itshape             
	\raggedleft          
}
{\par 
	\vspace{\stretch{3}} 
	\clearpage           
}

\title{A glimpse of heavy scales through\\ the electroweak resonance theory}
\author{Joaqu\'in Santos Blasco}
\date{(T\'itulo provisional \\ y pendiente de hacer portada)}

\begin{document}


\setlength{\unitlength}{1cm} 
\thispagestyle{empty}
\begin{center}

	\begin{picture}(3.5,2.4)
	\put(0,-3){\includegraphics[scale=0.06]{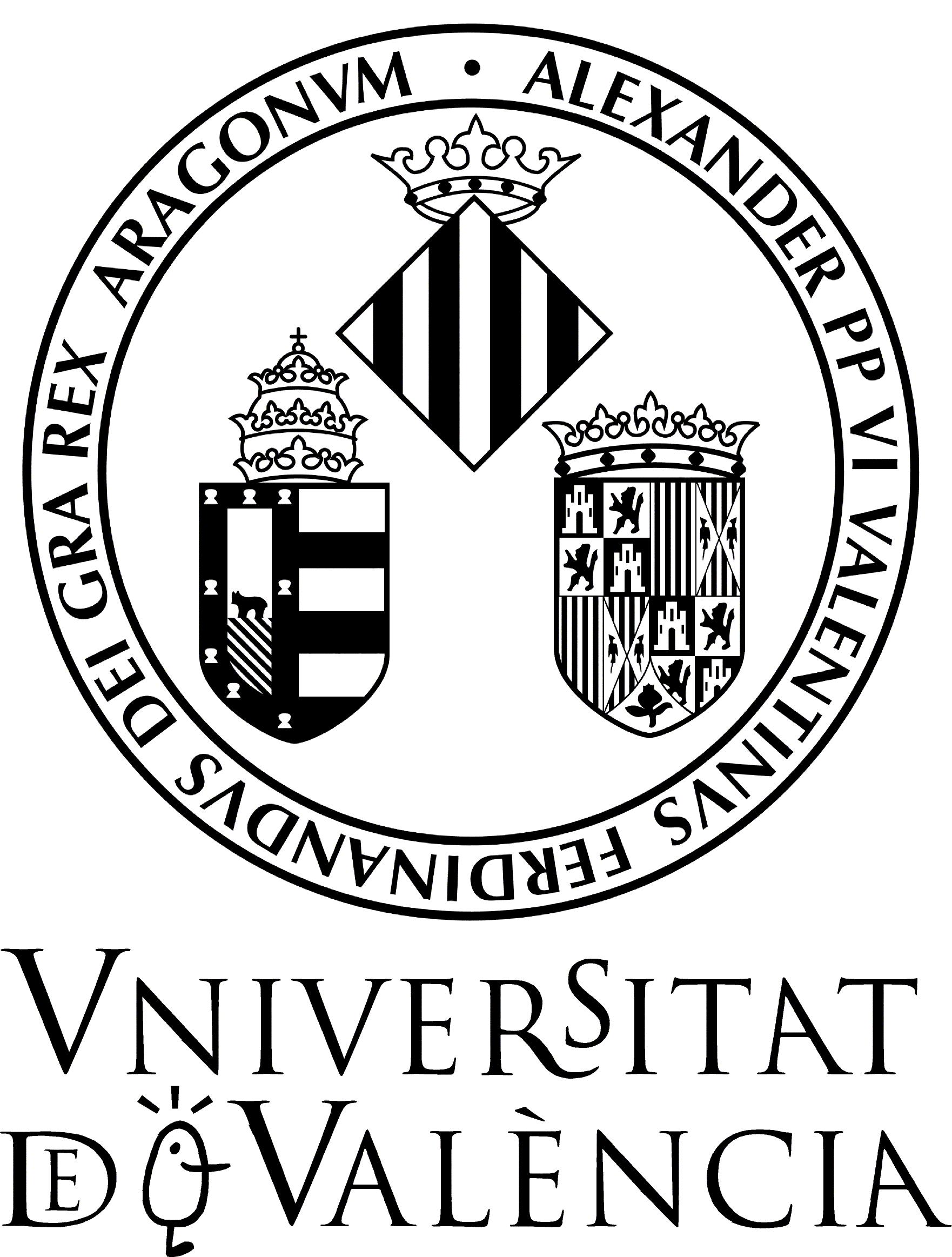}}
	\end{picture}
	\vspace*{4.35cm}
	
	\textbf{\huge{A glimpse of heavy scales \\ \vspace{2mm} through  the \\ \vspace{4mm} electroweak resonance theory}} \\[4ex]
	{\Large Tesis Doctoral}
	
	\vspace{\fill}

	{\LARGE \bf{Joaqu\'in Santos Blasco}} \\[2ex]
	{IFIC - Universitat de Val\`encia - CSIC}\\
	{Programa de Doctorado en F\'isica}\\
	{Departamento de F\'isica Te\'orica}\\[4ex]

	{\Large Director: Antonio Pich Zardoya}\\[2ex]
	{\Large Valencia, marzo de 2018}
	
\end{center}



\begin{dedication}
	A mis padres y a Blanca
\end{dedication}

\tableofcontents

\chapter{Introduction} \label{ch:intro}

From the positron discovery in 1932 to the Higgs announcement at LHC in 2012, there has been a ceaseless determination of new particles, revealing a new structure level of nature and configuring what nowadays we call modern particle physics. The personal and technological efforts required to take the research in this field up to the current status have been tremendous. Just the comparison between the first particle accelerators during the 1930s and the Large Hadron Collider (LHC) shows the magnitude of these changes: while the first cyclotron was a 4.8 MeV machine with 69 cm diameter, LHC external ring diameter is 27 km large and run-II collisions have reached 13 TeV in the  center-of-mass frame. Only with these two sample measures (6 orders of magnitude factor), it is enough to claim that they describe two different layers of reality.

Nowadays, the perspective is much distinct from those earlier days when particle physics was incomplete and incongruent. Nevertheless, let's go back for a moment to the discovery  of the pion in 1947, predicted some years before by Yukawa, which was determined by the analysis of cosmic rays. The strong interaction was constituted by nucleons (protons and neutrons) and this force was mediated by the exchange of pions. However, this discovery brought as well an unexpected partner: the kaon. The next decades, new particle detections 
and also resonance states continued endlessly. This chaos was restored with the Gell-Mann quark model and the consolidation of quantum chromodynamics (QCD) as the theory of the strong interactions. Eventually, pions were not considered to be elementary constituents of matter, but composite states from some more fundamental particles. 

The lesson here is that there is not a single path to reach a true theory in physics. Certainly, all the steps given with the pion and later with the anarchic hadron picture were indeed the primordial ingredients to even wonder about a more deeper level in the subatomic world.  

Coming back to more recent days, the Standard Model (SM) describes the most precise framework for particle physics and its success is unquestionable. Actually, its major prediction was already confirmed with the discovery of a Higgs-like scalar boson and, as a consequence, all the pieces are perfectly fastened with each other. Yet, the SM is definitely not the ultimate theory of nature, not even for particle physics, since it leaves some fundamental questions unanswered. Short and middle-term expectations in this field yield two possibilities: i) there is a large energy gap between the SM and new physics, so technology needs to be significantly improved to reach a more profound scale of reality. So far, this is the current status in particle physics, where LHC Run-II has not shown any significant deviation (five standard deviations) with respect to the SM predictions.\footnote{However, there are currently some 3$\sigma$ deviations which might indicate the presence of some physics beyond the SM eventually.} The other possibility ii) would lead us to the scenario lived by the generation of physicists from some years before the pion discovery. In this much more optimistic situation, new physics discoveries might come through direct observation, like the pion case, but they can also be produced indirectly, instead. It is indeed more plausible that they alter the SM theoretical value of some physical observable at an energy lower than the mass of the new particles. 

The last statement is reinforced with the example of chiral perturbation theory (\chpt). This effective field theory (EFT) describes the strong interaction in the low-momenta limit in terms of meson states, like the pion. Within a particular energy region, it provides a very useful framework to deal with these objects, which are conceived as the fundamental degrees of freedom of this theory, instead of the more elementary quarks.\footnote{Indeed, this much simpler description allow to perform many calculations that are unaffordable with QCD.} However, if one looks at the parameters that govern this model with the sufficient precision or at higher energies than before, they eventually do not match the expected values. In fact, heavier objects called resonances that are not included in the theory (like $\rho$ particles) alter remarkably the energy spectrum and their effects are even perceptible at the pion scale.

The main objective of this work is to build an efficient and useful theoretical framework to look for massive states in the TeV range. For this purpose, bottom-up effective field theories become a fundamental tool. In particular, they allow us to represent the SM and any possible extension in a model-independent picture. The main advantage of this approach to find new physics is that no basic assumption has to be made and, indeed, it is not required to know the underlying fundamental theory that explains it. Therefore, an optimal strategy consist in incorporating the most general set of high-energy states that can be generated by any UV theory to our SM-based effective theory and study their impact on the core parameters that characterize the SM. Following the analogy of chiral perturbation theory, we will denote these high-energy objects as resonances. Once compared with experimental data, these states are either disregarded or further constrained to higher energies attending to their discrepancies and compatibility with data. Otherwise, if experiments eventually present any significant deviation of some observable (or observables) with respect to the SM expectations, we would be able to know which kind of massive objects are generating it with no need to justify their origin nor the specific background model that explains it.

The thesis is based in 	\cite{Fingerprints, Colorful, Lowenergy, Rosell:2015bra, Santos:2015mqa, Rosell:2016zza, Rosell:2017kps} and it is organized in \ref{ch:conclusions} chapters (and 6 appendices). 
As it is suggested in this introduction, the Standard Model and effective field theories constitute the starting point for the development of the dissertation, which central concepts are gathered in chapter \ref{ch:foundations}. Right after that, we develop the electroweak effective theory in chapter \ref{ch:ewet} as a model-independent EFT, where the SM is included in the lowest order of the power expansion. The formalism employed for this task is the most appropriate for the incorporation of high-energy resonances, addressed in chapter \ref{ch:resonancetheory} through the resonance chiral theory. Therefore, once both the low-energy and the high-energy theories are well-established, we are able to match the theories in chapter \ref{ch:LECs} and make an analysis of the traces of the heavy objects in the low-energy couplings. Among all the possible types of resonances, we will analyze the spin-1 case more carefully in chapter \ref{ch:spin1}. There is a review on the current phenomenological status for the resonance bosonic sector in chapter \ref{ch:pheno} and, finally, the conclusions in chapter \ref{ch:conclusions}, both in English and Spanish, where all the thesis is summarized.



\chapter{Foundations: the SM and EFTs} \label{ch:foundations}

The Standard Model (SM) of particle physics is proved to be the best theoretical framework for describing the subatomic nature and it has allowed us to understand a whole new layer of the physical reality. Hence, its success has no precedents. Developed in the early 1970s, the SM has passed numerous precision tests being all of them consistent with the expected values from the model and showing no significant signal that points in a different direction. The predictive power of the SM is also remarkable. Indeed, the only remaining fundamental piece in the SM puzzle was finally filled with the discovery of a $m_H=125$ GeV Higgs-like scalar particle \cite{Pich:2015tqa}.\footnote{Natural units are implicit along the present work, $c=\hbar=1$.}

Nevertheless, despite the fact that there is no compromising experimental evidence in contradiction with this model,\footnote{Neutrino oscillations imply non-zero neutrino masses, not accounted for in the SM. However, they can be easily incorporated in the model.} the SM is certainly not the ultimate theory of nature. It is only sensitive to the electromagnetic, weak and strong interactions, leaving aside gravity and general relativity in its formulation. In addition, it does not explain phenomena from beyond the Standard Model (BSM) like dark matter and some open questions remain unanswered like why three is the number of families or what occasioned a slight asymmetry between matter and antimatter in the early universe after the Big Bang.

Direct experimental searches do not indicate the presence of new high-energy states below the TeV scale and there are no short-term expectations of any candidate to show up within this range. Therefore, the current situation motivates to modify the way new physics is pursued: instead of looking further into the barriers of the ever wider energy spectrum, it may be better to look deeper and more precise into the already known couplings that measure and intermediate the interactions among the quantum fields, expecting them to be sensitive to any high-energy physics somehow. 

The best theoretical framework for this strategy, also called indirect search, is the Effective field theory (EFT) picture. In the context of particle physics, EFTs are quantum field theories (QFTs) valid in a certain energy region and organized in such a way that i) only the degrees of freedom (dof) below a given energy scale $\Lambda$ are considered in the Lagrangian and ii) all the states above this cut-off scale are not directly included as interacting quantum fields in the EFT. Nonetheless, the presence of these high-energy objects is encoded somehow in the so-called low-energy theory. Our aim is to display a picture where we are able to collect SM physics in a low-energy frame, while the BSM physics, instead of being guessed in a model-dependent way, is inferred through the analysis of the parameters of the EFT, directly related to physical observables.

\section{A glance on the Standard Model}

The Standard Model of particle physics is the quantum field theory that governs the subatomic world. It describes with precision (up to the reach of the current experimental tests) the dynamics of the electromagnetic, weak and strong interactions, three of the four fundamental interactions of nature (known so far), where only gravity is not explained.\footnote{The formulation of a theory able to explain all the interactions of nature in a compact way, so-called theory of everything, would be undoubtedly one of the greatest achievement in science history.} Therefore, the extent of this theory is enormous: it explains from how nucleons are bound in an atom to why the light scatters and the sky is blue as a result. Even so, the SM only requires 19 independent (plus neutrino masses) input parameters that cannot be obtained from the model itself. Their explanations obey to a deeper underlying layer of nature, which nowadays is out of the reach of the physics understanding.

Firstly, and before the analysis of the core structure of the SM and its organization, it is convenient to introduce the fundamental particles conforming it. A fundamental particle is not a compound state of the other more primordial particles,\ie it is not a composite object. So far, there is no evidence that the following elements have a deeper structure. Particles (not necessarily fundamental) are divided into fermions and bosons, depending of their inner internal angular momentum, or spin, to be half-integer or integer, respectively. Hence, the first ones satisfy the Fermi-Dirac statistics whereas the second ones, the Bose-Einstein statistics. 

%
\begin{table}[h!] 		
	\begin{center}
		\renewcommand{\arraystretch}{1.54}
		\begin{tabular}{|c l|c|c|c}
			\cline{1-4}
			Quark, & $q^a$ & Charge ($e$) & Mass (MeV)  &
			\\ \cline{1-4}
			up, & $u$ & +2/3 & $2.2^{+0.6}_{-0.4}$& \multirow{2}{*}{$\begin{matrix} \text{1st} \vspace{-2mm}\\ \text{generation} \vspace{-1.5mm}\end{matrix}$}  
			\\ \cline{1-4}
			down, & $d$ & -1/3 & $4.7^{+0.5}_{-0.4}$ &
			\\ \hline
			charm, & $c$ & +2/3 & $(1.28 \pm 0.03)\times 10^3$& \multirow{2}{*}{$\begin{matrix} \text{2nd} \vspace{-2mm}\\ \text{generation} \vspace{-1.5mm} \end{matrix}$}    
			\\ \cline{1-4}
			strange, & $s$ & -1/3 & $96^{+8}_{-4}$ &
			\\ \hline
			top, & $t$ & +2/3 & $(173.1 \pm 0.6) \times 10^3\;\; (\dagger)$ 
			& \multirow{2}{*}{$\begin{matrix} \text{3rd} \vspace{-2mm}\\ \text{generation} \vspace{-1.5mm} \end{matrix}$}  
			\\ \cline{1-4}
			bottom, & $b$ & -1/3 & $4.18^{+0.04}_{-0.03} \times 10^3$ & 
			\\ \hline
			\multicolumn{5}{c}{} 
			\\[-2ex] \cline{1-4}
			Lepton, & $l$ & Charge ($e$) & Mass (MeV) &
			\\ \cline{1-4}
			electron, & $e$ & -1 & $0.5109989461 \pm 0.0000000031$& \multirow{2}{*}{$\begin{matrix} \\ \text{1st} \vspace{-2mm}\\ \text{generation} \end{matrix}$}  
			\\ \cline{1-4}
			$\begin{matrix} \text{electron} \vspace{-2mm}\\ \text{neutrino,} \end{matrix}$ & $\nu_e$  & 0 & $< 0.002 \quad (\dagger\dagger)$ &
			\\ \hline
			muon, & $\mu$ & -1 & $105.6583745 \pm 0.0000024$ & \multirow{2}{*}{$\begin{matrix} \\  \text{2nd} \vspace{-2mm}\\ \text{generation} \end{matrix}$}    
			\\ \cline{1-4}
			$\begin{matrix} \text{muon} \vspace{-2mm}\\ \text{neutrino,} \end{matrix}$ & $\nu_\mu$  & 0 &  $< 0.002 \quad (\dagger\dagger)$&
			\\ \hline
			tau, & $\tau$ & -1 & $ 1776.86 \pm  0.12$& \multirow{2}{*}{$\begin{matrix} \\ \text{3rd} \vspace{-2mm}\\ \text{generation} \end{matrix}$ }  
			\\ \cline{1-4} 
			$\begin{matrix} \text{muon} \vspace{-2mm}\\ \text{neutrino,} \end{matrix}$ & $\nu_\tau$  & 0 &  $< 0.002 \quad (\dagger\dagger)$
			\\ \hline
			\multicolumn{5}{c}{}
			\\[-3ex]
		\end{tabular}	
		\small{
			\caption{SM fermion elementary particles: electric charge and measured mass within current experimental bounds \cite{Patrignani:2016xqp}. Each particle in the table has its own anti-particle, with the same mass and opposite electric charge (and all other charges not explicit). Quarks (leptons) are placed in the upper (lower) block, and they are classified in three generations. $(\dagger)$ The top quark is specially sensitive to the way masses are defined. The listed value corresponds to a direct mass measurement, but if considered the pole mass instead, it turns to be $m_t^{\text{pole}} = 173.5 \pm 1.1$ GeV; and from cross section measurements it is $m_t^{\sigma}= 160^{+5}_{-4}$ GeV. The specific mass definitions for the rest of leptons are detailed in \cite{Patrignani:2016xqp}. 
			$(\dagger\dagger)$ At least two of the three neutrinos have a non-zero mass superior to 0.4 eV.}
		\label{tab:fermion}
		}		
		\vspace{-4mm}
	\end{center}	
\end{table}

Elementary fermions in the SM carry spin 1/2 and can be either quarks or leptons, depending on whether they are sensitive to the strong interaction or not. Each fermionic particle has its own anti-particle, with the same mass and opposite quantum numbers. In table \ref{tab:fermion}, the elementary fermions are listed with their respective electric charges and masses (anti-fermions are implicit). Both sets show 6 fundamental particles, grouped in 3 generations. The concept of generation, or family, makes reference to different subsets of particles leading to similar interactions, differing only in their masses. Indeed, fermions from distinct families bring the same quantum numbers (except for the flavor number, which is a generation number) and only differ (substantially) in their masses. For instance, the first lepton generation, compound by the electron and the electron neutrino --- also positron and electron anti-neutrino --- has the same properties as the second generation, made out of the muon and muon neutrino --- and anti-muon and muon anti-neutrino.

Quarks carry additionally color-charge. In contrast to the electric charge being a scalar, the charges associated to the strong interactions belong to three different types, conventionally called red, green and blue, like the RGB color pattern. Unlike leptons, quarks are always found in nature forming compound states, like pions or protons, and cannot exist freely in normal conditions.\footnote{In very high-energy and high-density enviroments, such as the early Big Bang, quarks are not able to hold united and they form what is called quark-gluon plasma. Actually, the higher the energy is, the weaker the strong interaction becomes, so that quarks are asymptotically free particles.} This property is called confinement. Pragmatically, it implies that physical states made out of quarks must compensate their global color charges so that the outcome is not red, green or blue. This can be accomplished\footnote{Recently, in 2015, LHCb detected a more exotic state made out of $u\bar c c u d$, being the first pentaquark ever detected \cite{Aaij:2015tga}.} 
by a combination of a quark and an antiquark of the same color, so-called meson, or with the bound of a red, a green and a blue quark, that would generate a `white' state, in analogy to the RGB pattern, referred as baryon. All the quark states (actually, all known particles) discovered so far and their main properties are listed in \cite{Patrignani:2016xqp}.

Fundamental bosons in the SM are typically in charge of mediating the physical interactions. This is the case of the spin-1 gauge bosons, being the photon, the force carrier of the electromagnetic interaction; the gluon, produced when quarks exchange color,  and the $W$ and $Z$ gauge bosons, which mediate the weak interaction. Otherwise, there is the Higgs boson, recently discovered at LHC and being the only spin-0 elementary particle in the SM, which is required in order to keep the theory unitary, well-behaved and renormalizable. Some of the SM bosons fundamental properties are collected in table \ref{tab:boson}.
%
\begin{table}[h!] 
	\begin{center}
		\renewcommand{\arraystretch}{1.55}
		\begin{tabular}{|c l|c|c|c|}
			\hline
			Boson, & $\phi$ & \ Spin\  & $\begin{matrix} \text{Electric} \\[-1ex]
				\text{charge}\; (e) \end{matrix} $ & Mass (GeV)  
			\\ \hline
			Photon, & $\gamma$ & 1 & 0 & $<1\times 10^{-24}$ 
			\\ \hline
			Charged weak, & $W^-$ & 1 & -1 & $80.385 \pm 0.015$ 
			\\ \hline
			Neutral weak, & $Z$ & 1 & 0 & $91.1876 \pm 0.0021$     
			\\ \hline
			Gluons, & $G^a$ & 1 & 0 & $0\;\;(\dagger)$
			\\ \hline
			Higgs, & $H$ & 0 & 0 & $125.09 \pm 0.24$ 
			\\ \hline
			\multicolumn{5}{c}{}
			\\[-3ex]
		\end{tabular}	
		\small{
			\caption{SM boson elementary particles: spin, electric charge and measured mass within current experimental bounds \cite{Patrignani:2016xqp}. Each particle in the table is its own antiparticle, with the exception of $W^-$ which antiparticle is $W^+$. $(\dagger)$ This value for the gluon mass is the theoretical value. A few MeVs mass, however, cannot be discarded yet \cite{Patrignani:2016xqp}.}
		\label{tab:boson}
		}		
	\end{center}		
\end{table}

The concepts of symmetry and gauge invariance become fundamental in order to understand how the SM and also the inner structure of nature are formulated. We call symmetry in QFT a transformation of the fields that does not alter the dynamics of the system or, in other words, keeps the action invariant \cite{Donoghue:1992dd}. The Noether theorem states that whenever a continuous and differentiable\footnote{Discrete symmetries, for instance, do not apply.} symmetry holds in a system, there is a conservation law underneath and, therefore, a conserved charge. For instance, if a system remains invariant under translations, (linear) momentum is a conserved quantity while if it remains unchanged under $U(1)_{{\text{e.m.}}}$ local phase transformations, electric charge is preserved. 

The last example, indeed, leads to the second essential concept for the SM construction: gauge symmetry. Transformations of the quantum fields can be either global or local. In the context of field theories, a global symmetry is associated to a given transformation acting identically throughout the entire space-time, whereas a local symmetry refers to transformations which depend on the space-time coordinates. The following field transformations show explicitly the difference between global (left-hand term) and local (right-hand term),
\be \label{eq:difgloballocal}
\psi(x)\,\longrightarrow\, {\rm e}^{i\beta}\,\psi(x)\,, \qquad\qquad \psi(x)\,\longrightarrow\, {\rm e}^{i\beta(x)}\,\psi(x)\,, 
\ee
where $\beta$ and $\beta(x)$ are a constant phase and an arbitrary scalar function over the space-time coordinates, respectively.\footnote{In general, we will denote these exponentials as $g$ and $g(x)$, correspondingly, with these elements belonging to a given Lie group (see appendix \ref{app:liealgebra}).} We call gauge transformations to the local mappings which leave the action invariant and local gauge symmetry (or simply gauge symmetry) to the underlying invariance under these transformations \cite{Donoghue:1992dd}.

The SM is mathematically formulated as a Yang-Mills theory,\ie it relies on reductive and compact Lie gauge symmetry groups which define its structure and interactions (see appendix \ref{app:liealgebra}). Electromagnetism responds to a $U(1)_{{\text{e.m.}}}$ gauge invariance of the SM Lagrangian. Therefore, every particle carries a definite electric charge and it is required the existence of an electromagnetic gauge boson, the photon, which is predicted to be massless. This interaction was actually the first discovered and its physical formulation, quantum electrodynamics (QED), is certainly the most precisely tested theory from all science subjects. Tables \ref{tab:fermion} and \ref{tab:boson} show the outstanding bounds in the determination of the electron and photon masses. Equivalently, quantum chromodynamics (QCD), the theory of the strong interactions, is accurately explained through the $SU(3)$ gauge symmetry group. It gives raise to a conserved color quantum number and eight massless gluon fields mediating this interaction. The non-abelian character of the Lie group also explains accurately the self-interactions of the gluon fields.

The weak interaction, however, cannot be explained from the action of a gauge symmetry group alone as the other fundamental interactions. Actually, the interaction has three significant and  contrastive features that make it differ from the electromagnetic and strong forces: i) it should explain quark flavor changing interactions, ii) charge conjugation, $C$, and parity, $P$, are not a good discrete symmetries of the theory\footnote{Actually, $CP$, the combination of charge conjugation and parity, is not a good symmetry of the weak interactions, neither.} and, finally, iii) the weak gauge bosons are massive (see table \ref{tab:boson}). The solution to this puzzle was introduced with the Glashow-Salam-Weinberg electroweak model during the 1960s \cite{Glashow:1961tr, Weinberg:1967tq, Salam:1968rm}. This theory collects as one single fundamental interaction the electromagnetic and weak forces, unified under the gauge symmetry group $SU(2)_L \otimes U(1)_Y$. The subindex $L$ means that only the left-handed fermions in the theory transform under the $SU(2)$ weak isospin subgroup, whereas the right-handed ones transform trivially; and the subscript $Y$ stand for weak hypercharge, the conserved charge of the weak interactions, related to the electric charge, $Q$, and the third component of the weak isospin,  
$T_3$, as
\be \label{eq:hypercharge}
Q \,=\, T_3 \,+\, Y\,.
\ee
In addition, this model configuration does not require the inclusion of right-handed neutrinos, which are not present in the minimal formulation.

\section{Spontaneous Symmetry Breaking in the SM} \label{sec:SSB}

The electroweak model, nonetheless, formulated as a Yang-Mills theory, is still not enough to explain the $W$ and $Z$ gauge bosons masses. 
 Therefore, it is necessary to `break' this symmetry, in the sense that, despite the SM Lagrangian being actually invariant under the electroweak group, the ground state of the theory is not. This phenomenon is known as spontaneous symmetry breaking (SSB).

There is, in fact, not only one way to break a symmetry. Given a specific model, it is possible to make some scalar field (or fields) acquire a vacuum expectation value (vev) in such a way that the field parametrization around this ground state does not accomplish the symmetry prescriptions of the model. Nevertheless, one can also induce the same effect without any scalar field. In this kind of models, the symmetry gets dynamically broken due to quantum corrections. The most paradigmatic example of this situations occurs with chiral symmetry at QCD, with a $SU(2)_L\otimes SU(2)_R$ global symmetry broken to $SU(2)_{(L+R)}$. Actually, QCD is a non-perturbative theory at low-energies and it is characterized by a quark condensate in the vacuum.

It is important not to confuse the concept of SSB, where actually the symmetry remains exact although it is somehow apparently hidden, with symmetries that are slightly but explicitly broken, like the isospin case, where the symmetry is broken by the different up and down quark masses, and by electromagnetic interactions. 

In the SM, the electroweak symmetry is broken like in the first case, through the so-called Higgs mechanism. It consists of the inclusion of a complex scalar doublet (i.e., four new degrees of freedom) in the SM, introduced as
\be \label{eq:scalardoublet}
\Phi(x) \,=\, \left( \begin{matrix} \Phi^+ \\ \Phi^0 \end{matrix} \right)\,,
\ee
and $\Phi^c\equiv i\sigma_2\Phi^*$ its charge-conjugate partner, being $\sigma_i$ with $i=1,2,3$ the Pauli matrices (see appendix \ref{app:liealgebra}). Besides, gauge invariance under the electroweak symmetry group fixes their covariant derivative as
\be
D_\mu\Phi \,\equiv\, \partial_\mu\Phi
\,+\, i g \,\frac{\vec{\sigma}}{2}\vec{W}_\mu \,\Phi \,+\, i g' \, y_\Phi\, B_\mu \,\Phi \, ,
\ee
where $g$ and $g'$ are defined as the weak couplings associated to the $\vec W_\mu$ and $B_\mu$ gauge fields\footnote{Recall that the photon, $A_\mu$, and the $Z_\mu$ fields are related to $W^3_\mu$ and $B_\mu$ through the Weingerg angle, $\theta_W$.} and $y_\Phi=1/2$ is the scalar doublet hypercharge.

\begin{figure}[h]
	\centering
	\includegraphics[width=0.9 \textwidth]{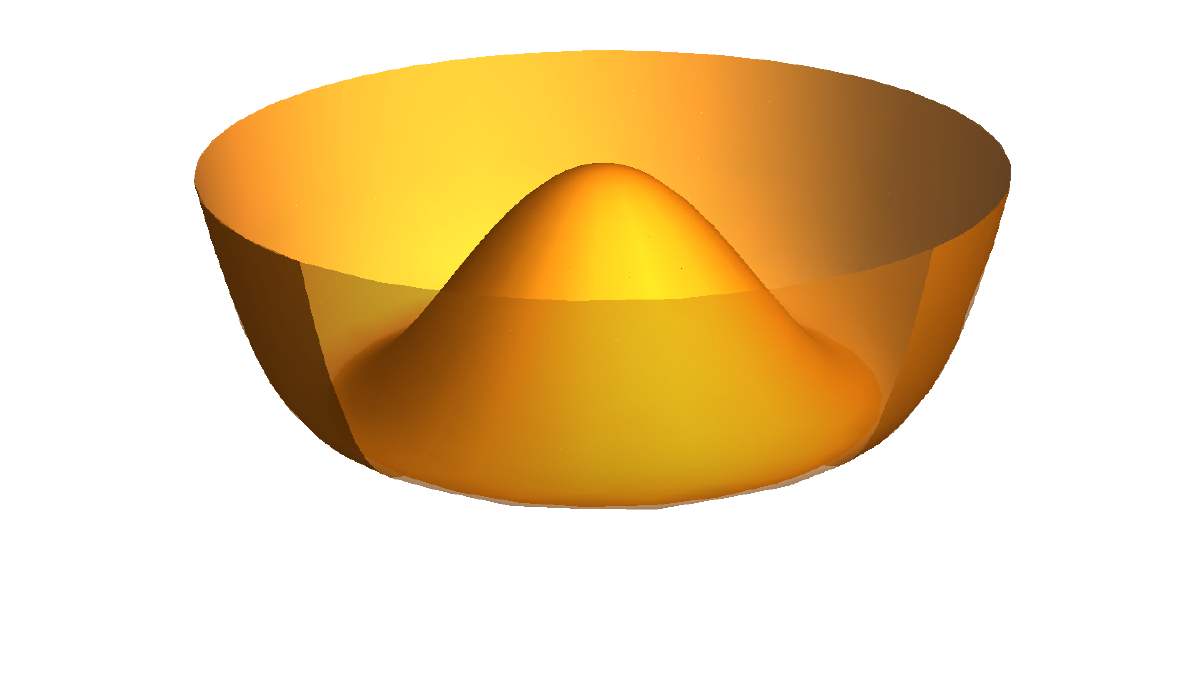}
	\vspace{-1cm}
	\caption{Potential function representation for two scalar real fields, analogous to eq.~(\ref{eq:scalarlagrangianPhi}), being $\mu^2<0$ and $\lambda>0$.}
	\label{fig:mexican}
\end{figure}

The additional demand of generating non-trivial ground states so that the symmetry is spontaneously broken leads to a gauged scalar Lagrangian 
\be \label{eq:scalarlagrangianPhi}
\mL(\Phi) \,=\, (D_\mu \Phi)^\dagger D^\mu \Phi \,-\, V(\Phi)\,, \;\qquad \text{with}\; \qquad V(\Phi)\,=\, \mu^2 \Phi^\dagger \Phi \,+\, \lambda\, (\Phi^\dagger \Phi)^2\,, \quad
\ee
being $V(\Phi)$ a generic scalar potential provided the quartic term satisfies $\lambda>0$, so that the theory is energetically bounded from below, and the quadratic one, $\mu^2<0$, in order to generate non-trivial minima. In figure \ref{fig:mexican}, we illustrate the analogous potential to eq.~(\ref{eq:scalarlagrangianPhi}), with real fields (and not complex) instead. It displays an infinite set of degenerate minimal energy states displayed along a circumference around the coordinates origin.
\vspace{.40cm}

The conservation of the electric charge yields that only the neutral scalar field is able to acquire a vev. It is defined in terms of the parameters of the scalar potential as
\be \label{eq:vev}
|\bra 0 | \Phi^0 |0 \ket| \,=\, \sqrt{\Frac{-\mu^2}{2\lambda}} \,\equiv\, \Frac{v}{\sqrt 2} \,,
\ee
once minimized the scalar function. The precise and arbitrary selection of one of the infinite minima as the ground state of the electroweak model breaks the symmetry in such a way that only electromagnetism, $U(1)_{\text{e.m.}}$, is a good symmetry of the vacuum \cite{Pich:2007vu}. Therefore, the electroweak gauge symmetry gets broken following the pattern:
\be \label{eq:SMssb}
SU(2)_L \,\otimes \, U(1)_Y \, \longrightarrow \, U(1)_{\text{e.m.}} \,.
\ee

According to the Goldstone theorem, whenever a continuous symmetry is broken spontaneously, as many massless Goldstone bosons as broken generators should appear as a result. In consideration of eq.~(\ref{eq:SMssb}), three Goldstone bosons should arise. We can actually acknowledge them just by properly re-expressing the doublet scalar field:
one can perform a polar decomposition in such a way that its four degrees of freedom  parametrize the physical excitations over the vacuum,
\be\label{eq:polarSM}
\Phi(x) \, = \, \frac{1}{\sqrt{2}}\; U(\varphi(x))\, \left( \begin{matrix} 0 \\ v + H(x) \end{matrix} \right)\, ,
\ee
where $H(x)$ is the Higgs field, a scalar under the chiral group $G$, and $U(\phi(x))$ a $2 \times 2$ unitary matrix\footnote{This parametrization allows to re-express the scalar potential, defined in eq.~(\ref{eq:scalarlagrangianPhi}), as a function of the Higgs field only, $V(H)$.} collecting the 3 massless Goldstone bosons, $\varphi^i(x)$, defined as
\be \label{eq:Goldstones}
U(\varphi) \, = \,  \exp{\left\{ i \vec{\sigma} \, \vec{\varphi} / v \right\} }\,,\qquad\quad \mbox{with} \qquad
D_\mu U \,=\, \partial_\mu U
+ i g \,\Frac{\vec{\sigma}}{2}\vec{W}_\mu \, U - i g'\, U \,\Frac{\sigma_3}{2} B_\mu\,.
\ee
In analogy to figure $\ref{fig:mexican}$, Goldstone bosons would represent the physical excitations along de valley of minima shown in the image. Therefore, the fields leading to this action must be massless. On the contrary, excitations in the radial direction, which are perpendicular to the minima, must be produced by a massive field, corresponding to the Higgs boson in this case.

Nonetheless, the implementation of the scalar doublet in the electroweak model (and thus the Goldstone bosons) is not only remarkable because it succeeds in unifying the electromagnetic and the weak interactions together with the SSB, but it also achieves to explain how the gauge bosons acquire mass. Actually, considering the scalar Lagrangian in eq.~(\ref{eq:scalarlagrangianPhi}) and taking the unitary gauge, where $U(\varphi)=1$, one obtains
\be \label{eq:unitarygauge}
\mL(\Phi) \,=\, \Frac{1}{2}\,\partial_\mu H \partial^\mu H \,+\, (v+H)^2 \left( \Frac{g^2}{4}\, W^\dagger_\mu W^\mu \,+\, \Frac{g^2}{8 \cos^2 \theta_W}\, Z_\mu^\dagger Z^\mu \right) \,,
\ee
being $\theta_W$ the Weinberg angle \cite{Pich:2007vu}.\footnote{It relates the Lagrangian gauge fields $W^\mu_3$ and $B^\mu$ with their mass eigenstates $Z^\mu$ and $A^\mu$.}  The electroweak SSB and the non-trivial vev are the very responsible of the generation of the $W^\pm$ and $Z$ masses, being
\be \label{eq:WZmass}
M_W \,=\, M_Z \cos \theta_W \,=\, \Frac{1}{2}\,v\,g
\ee
and, as a consequence, the three Goldstone boson have to be understood as the missing longitudinal polarizations of these gauge fields.

In addition, the Higgs mechanism implies the presence of a massive scalar singlet particle, which existence was proved some years ago at LHC with the discovery of a 125 GeV Higgs-like particle. Notwithstanding, this is not the actual genuine reason why the SM requires its presence. There are different Higgsless models that introduce three Goldstone bosons and explain the weak gauge bosons masses. The success of the Higgs particle lies in the ability to restore the unitarity and renormalizability of the theory, which is lacking in its absence. 

\section{The SM Lagrangian}

In consideration of the symmetries present in the subatomic world and the experimental inputs, the SM has consolidated as the paradigmatic model in particle physics. Its mathematical formulation can be summarized in the following Lagrangian:
\bear \label{eq:sm}
\mL_{\text{SM}} &= & \,-\;\,\frac{1}{2}\, \bra W_{\mu\nu} W^{\mu\nu} \ket_2 
\,-\,\frac{1}{4}\,  B_{\mu\nu}  B^{\mu\nu} 
\,-\,\frac{1}{2}\, \bra G_{\mu\nu} G^{\mu\nu} \ket_3  \nn\\ [1ex]
&& \,+\;\,  (D_\mu \Phi)^\dagger\, D^\mu \Phi \,-\, V(\Phi) \nn\\[1ex]
&& \,+\; \sum_{\psi_L}i \,\bar{\psi_L}\gamma^\mu D^L_\mu \psi_L \, +\, \sum_{\psi_R}i \,\bar{\psi_R}\gamma^\mu D^R_\mu \psi_R \nn\\
&& \,-\; \!\! \sum_{\psi_L,\psi_R} \bar{\psi_L}\, Y\, \Phi^c \psi_R \, + \, \text{h.c.}\quad .
\eear
where $\bra \cdot \ket_n$ stands for the $SU(n)$ trace and $\psi=l,\,q$.
The first line in eq.~(\ref{eq:sm}) collects the Yang-Mills Lagrangian associated to the electroweak and strong interactions, expressed in terms of the gauge field strength tensors. The scalar sector of the theory is compiled in the second line. As explained in the previous section, it is necessary in order to break the electroweak symmetry through  a non-zero vacuum expectation value.

In the next lines in the SM Lagrangian, we find the fermion kinetic and interaction terms. There is an explicit distinction, nonetheless, depending on the fermions being left or right-handed. Furthermore, the equation considers both leptons and quarks at the same time, where just the last ones are sensitive to the strong interactions. Therefore, the gauged covariant fermion derivative must be chiral and color-dependent, being defined as
\begin{align}
D^L_\mu\psi_L \,&=\, \Big(\partial_\mu \, + \, i\,g_S\, T^a\, G^a_\mu
\,+\, i g \,\frac{\vec{\sigma}}{2}\vec{W}_\mu \,+\, i g' \, y_\psi\, B_\mu  \Big)\, \psi_L \, , \nn\\
D^R_\mu\psi_R \,&=\, \Big(\partial_\mu \, + \, i\,g_S\, T^a\, G^a_\mu
\,+\, i g' \, y_\psi\, B_\mu \Big)\, \psi_R \, ,
\end{align}
where the gluon contribution, in terms of $G^a_\mu$, only applies for quark fields; being $g_S$ and $T^a$ the strong coupling and the $SU(3)$ Lie algebra generators (see appendix \ref{app:liealgebra}), respectively. 
Finally, the last contributions to the SM formulation in eq.~(\ref{eq:sm}) correspond to the Yukawa Lagrangian. Contrary to the previous fermion interactions, these terms mix the left and the right sectors of the theory, throughout the mediation of the doublet scalar field. Indeed, this is the SM mechanism to generate the fermion masses and the flavor mixing too.

\section{Effective field theories} \label{sec:EFTs}

Let's consider the following scenario: a solid object falling from some height. On the one hand, the dynamics of this process can be studied to a first approximation as a free falling object according the Newton's gravitational laws, corrected by fluid dynamics associated to the object being in a non-negligible atmosphere. On the other hand, it is well-known that Newtonian physics can be derived from general relativity, which is a much more complete and precise description of gravity. Nevertheless, it would be non-sense and certainly unaffordable to study the solid motion from this point of view. Therefore, the first model seems to be much more appropriate and more `effective' to describe it.

As reflected with the example, the selection of the variables in any problem is crucial. In QFT, this is, indeed, the justification of effective field theories (EFTs): the construction of the more convenient framework in order to describe the important degrees of freedom at a given energy scale. Yet, an EFT does not aim to be a general model nor a fundamental theory since its validity is restrained to a particular energy range. Considering some EFT valid for an energy scale $v$, the contributions of heavy fields from some higher scale $\Lambda\gg v$ are found to be suppressed with a factor of order $v/\Lambda$ with respect to the light fields.
This statement is, in fact, mathematically formulated as the Decoupling theorem, meaning that these heavy dofs are completely set apart when $\Lambda\to\infty$.

According to their purpose, there are two general approaches for EFTs \cite{Krause:2016uhw}.
First, in the top-down case, the EFT is built in order to describe the low-energy range of a more general and well-established theory. The validity of the EFT, also called low-energy theory, is limited up to a given energy threshold, whereas the original one, also high-energy theory or UV theory, still holds above it. For sure, both frameworks reproduce the same physics in the low-momenta region.

The second type of EFTs are denoted as bottom-up. In this scenario, the high-energy theory is unknown, as opposite to the previous approach, or not possible to apply at a given regime. Therefore, in any of these situations, the EFT cannot be derived directly from a more fundamental theory. The motivation for these effective models is diverse. They can be used either to search for new physics beyond the current energy barriers in a model-independent way or to determine the low-momenta regime of the theory which is otherwise not available computationally.

A well-known example of an EFT is the Fermi theory of weak interactions. In general, one would employ the SM Lagrangian in eq.~(\ref{eq:sm}) to analyze some weak transition between two fermion currents, which is mediated by the exchange of W gauge bosons, being the propagator,
\be \label{eq:propW}
i \Delta_W(q) \,=\, \Frac{1}{q^2\,-\, M_W^2} \left(-g_{\mu\nu} \, + \, \Frac{ q_\mu q_\nu}{M_W^2} \right) \,.
\ee
Nonetheless, if one is only interested in the low-energy limit of these interactions,\ie $p\ll M_W$, they can also be reproduced with a local effective Hamiltonian including only the lightest fermions \cite{Pich:1998xt}, 
\be 
\mathcal{H}_{\text{eff}} \,=\,\Frac{G_F}{\sqrt 2} \mJ_\mu^\dagger \mJ^\mu \,,  \qquad \text{with} \qquad \mJ_\mu \,=\, \sum_{i,j} \bar u_i \gamma_\mu (1-\gamma_5) V_{ij} d_j \,+\, \sum_l \bar \nu_l \gamma_\mu (1-\gamma_5) l\,,
\ee
being $G_F$ the Fermi constant and $V_{ij}$ the Calibbo-Kobayashi-Maskawa mixing matrix. 
Actually, both this EFT and the SM have the same behavior at low-energies. In the limit $p\ll M_W$, the propagator in eq.~(\ref{eq:propW}) becomes $-g_{\mu\nu}/M_W$ and thus a straightforward identification with the SM Lagrangian yields
\be
\Frac{G_F}{\sqrt 2} \,=\, \Frac{g^2}{8 M_W^2}\,.
\ee
In this scenario, the Fermi theory is interpreted as a top-down EFT and the main reason to employ this formalism is simplicity. However, back to the pre-SM era, this theory was understood as a bottom-up EFT since the high-energy theory (the SM) was still unknown.

Another well known example of the bottom-up approach and certainly the most paradigmatic one is chiral perturbation theory (\chpt), which is explained with more detail in the next section. It appears to be the more appropriate framework to reproduce the low-energy QCD dynamics, where hadrons are the actual dofs, instead of the more fundamental quarks and gluons.

The characterization of bottom-up EFTs lies on three primary aspects: the particle content, the symmetries and the power counting. The first requirement is definitely to determine which are the particles involved: the degrees of freedom present in the EFT. For this purpose, it is sometimes useful to set an energy threshold so that all the existing particles with masses from above are not considered explicitly in the theory. 
Next to this, it is necessary to establish the symmetries of the EFT Lagrangian in order to indicate the nature of the physical interactions to be analyzed. There are two possibilities for the symmetry implementation. On the one hand, the symmetry holds even at a fundamental level,\ie not only the EFT is invariant, but also the underlying theory. On the other hand, a given symmetry may be accidental, which means that it holds at the EFT level but there is no strong reason why it should. Hence, this symmetry can be either conserved at the UV theory or broken by an small amount. 
The last fundamental element, the power counting, describes how the effective field theory is organized and sets the relative dominance of one particular physical process with respect to the others. Therefore, it establishes a hierarchy among all the operators at the Lagrangian level.

Being the essence of bottom-up EFTs to be model-independent and not to assume any condition from the UV, one cannot afford to select which precise set of operators belong to it and which of them do not. In order to be consistent, one must include all the possible interactions that can be generated with a given field content and symmetries. This consideration yields an infinite number of possible operators, $\mO_i$. They can be collected in
\be \label{eq:EFT1}
\mL_{\text{EFT}} \,=\, \sum_i\, c_i\,\mO_i \,, 
\ee
where $c_i$ are the so-called Wilson coefficients, which respectively parametrize the operators in the EFT. The crucial fact here is that each of the operators in eq.~(\ref{eq:EFT1}) is assigned a different weight,\footnote{We will also refer to power counting dimension $d_i$ terms as order $\mO_{d}(p^{d_i})$ operators.} $d_i$, according to the power counting, which reveals information about the importance of each interaction within the EFT. For instance, a given operator with dimension $d=4$ is expected to be of the same size than any other interaction with the same power counting, but it should be subdominant with respect to some $d=2$ structure, whereas some $d=7$ operator must be negligible in comparison. 

The hierarchy introduced by the power counting allow us to re-express eq.~(\ref{eq:EFT1}) like
\be \label{eq:EFT2}
\mL_{\text{EFT}} \,=\, \sum_{d_i=0}^\infty\, \mL^{d_i} \,.
\ee
where $\mL^{d_i}$ contains all the operators with dimension $d_i$ or, in other words, it is the order $d_i$ effective Lagrangian. Despite the fact that the full Lagrangian contains infinite operators, the number of them at a given order is finite. This is actually the essential feature of bottom-up EFTs: they describe some non fundamental theory at a given scale with a finite number of operators, up to some controlled uncertainty level.

As a general consideration, it is not common to reach large order corrections in an EFT. Usually, it is enough to consider just the leading order (LO) Lagrangian, corresponding to the lowest order Lagrangian, and the next-to-leading order Lagrangian (NLO), which represents the first corrections to the previous one.\footnote{For some EFTs, the next-to-next-to-leading Lagrangian (NNLO) is also required, depending on the desired precision.}

\section{A theory of mesons} \label{sec:chpt}

There is no doubt about QCD being the right theory of the strong interactions. Based on gauge symmetry principles, it is valid at any energy scale (so far) and it has a remarkable predictive power. As already mentioned, it is formulated in terms of quarks and gluons, the fundamental elements of this interaction, which happen to be confined in hadronic states. However, its computation is often very hard or even unfeasible, particularly at low-energies, where perturbativity is eventually lost.

Instead, it seems much more convenient to describe the low-momenta limit of the strong interaction in terms of the physical composite degrees of freedom: mesons and baryons. Chiral perturbation theory is an EFT defined in terms of these particles and it has been proved to explain the low-energy phenomenology of the strong interaction. If we set the energy cut-off slightly above 1 GeV, the only physical states to be considered must be naively compound by quarks with masses from below this threshold\footnote{Depending on the purpose, it is also interesting to lower the cut-off to few hundred MeVs and analyze the two-flavor \chpt, where only $u$ and $d$ are involved.} at least:\footnote{Actually, the proton are the neutron are the only baryon particles which are within the established range. However, we are going to exclude them in this section, for the sake of clarity.} $u$, $d$ and $s$. 

Considering the three-flavor QCD Lagrangian, introduced in the third line in eq.~(\ref{eq:sm}), it is straightforward to prove that it is invariant under $SU(3)_L \otimes SU(3)_R$, also called (three-flavor) chiral symmetry. However, the physical vacuum of the theory (the \chpt\ ground state) is not invariant under this global group. Therefore, spontaneous chiral symmetry breaking (SCSB) occurs with the following pattern
\be
SU(3)_L \otimes SU(3)_R \, \longrightarrow SU(3)_{L+R}\,.
\ee
The acquired chiral vev, analogous to the SM vev ($v$) with the SSB, introduced in section \ref{sec:SSB}, is the pion decay constant, $f_\pi \equiv \sqrt 2 F_\pi= 130.41$ MeV, defined through
\be \label{eq:fpi}
\bra 0\, |\, \bar u \gamma_\mu \gamma_5 d\, |\, \pi^-(p)\ket \,=\, i \sqrt{2}\, F_\pi\, p_\mu \,.
\ee
Actually, this symmetry breaking justifies the absence of much lighter hadrons (with masses of the same order than their quark constituents) in the spectrum. Furthermore, it also explains the absence of degenerate hadron multiplets with different parity transformations \cite{Pich:1998xt}. 

The breaking of eight generators leads to the presence of eight Goldstone boson fields, $\phi^a$, which are precisely the scalar octet of mesons that we aimed to characterize with the EFT. We parametrize them in a compact way:
\be \label{eq:mesons}
\Phi(x) \,=\, \sqrt{2}\,\phi^a\, T^a \,=\, \left( \begin{matrix} \,\Frac{\pi^0}{\sqrt 2} + \Frac{\eta^8}{\sqrt 6} & \pi^+ & K^+ \\ \pi^- & -\Frac{\pi^0}{\sqrt 2} + \Frac{\eta^8}{\sqrt 6} & K^0 \\ K^- & K^0 & -2 \Frac{\eta^8}{\sqrt 6}\, \end{matrix} \right) \,,
\ee
where $T^a$ are the $SU(3)$ generators (see appendix \ref{app:liealgebra}). As in the SM case, the Goldstone bosons can be collected in an exponential matrix that accounts directly for the physical excitations over the chiral vacuum, defined as
\be \label{eq:chiralU}
U(\phi(x)) \,=\, \exp{ \left( i \Frac{\sqrt{2}\,\Phi(x)}{ F_\pi} \right)} \, \longrightarrow \, g_R\, U(\phi(x))\, g_L^\dagger \,,
\ee
and transforming linearly under the chiral group. 

Apart from the Goldstone bosons, \chpt\ includes a general set of local external sources, gauge invariant under the chiral group, that interact with the meson fields. We introduce the fields $l_\mu$ and $r_\mu$ as the left and right vector sources, which contribute through the Goldstone derivative, $D_\mu$, and their respective strength tensors
\begin{align}
D_\mu U \,&=\, \partial_\mu U \,-\, i r_\mu U \,+\, i U l_\mu\,, \nn\\
F_{L}^{\mu\nu} \,&\equiv\, \partial^\mu l^\nu \,-\, \partial^\nu l^\mu \,-\, i [l^\mu, l^\nu]\,, \nn\\
F_{R}^{\mu\nu} \,&\equiv\, \partial^\mu r^\nu \,-\, \partial^\nu r^\mu \,-\, i [r^\mu, r^\nu]\,,
\end{align}
and also the scalar, $s$, and pseudoscalar, $p$, gauge fields, which are usually grouped in $\chi\equiv 2 B_0 (s + i p )$, being $B_0$ a constant not only fixed by the symmetry prescriptions.

Once the symmetries and the field content have been introduced, it only remains to set the power counting, where the Lagrangian is organized in even\footnote{Indeed, parity conservation does not allow an odd number of derivatives.} powers of momentum, 
\be \label{eq:chptevenpowers}
\mL_{\text{\chpt}}\, =\, \sum_i \mL_{2i}\,.
\ee
Consistently, it is possible to assign to each of the \chpt\ fields a corresponding chiral dimension, $d$, yielding the following power counting
\bear\label{eq:chptpower_counting}
F_\pi\, ,\, \Frac{\Phi}{F_\pi}\,  , \, U(\phi) & \sim & \cO_d\left(p^0\right)\, ,
\nn\\
D_\mu U\,  & \sim & \cO_d\left( p^1\right)\, ,
\nn\\[1ex]
F^{\mu\nu}_L,\, F^{\mu\nu}_R,\, \chi & \sim & \cO_d\left( p^2\right)\, .
\eear
Therefore, we are able to construct a given order Lagrangian accordingly. For instance, the LO Lagrangian, $\mL_2$, is formed by all the operators with global dimension $d=2$ according to eq.~(\ref{eq:chptpower_counting}) and it reads
\be \label{eq:chiralL2}
\mL_2 \,=\, \Frac{F^2}{4}\bra (D_\mu U)^\dagger D^\mu U \,+\, U^\dagger \chi \,+\, \chi\, U \ket_2 \,.
\ee
The NLO corrections to the tree-level interactions from the previous Lagrangian come from three different sides: i) tree-level processes from local $\mO_d(p^4)$ operators, from the NLO chiral Lagrangian, ii) one-loop corrections with $\mL_2$ vertices and iii) the chiral anomaly from the  Wess-Zumino-Witten functional \cite{Wess:1971yu, Witten:1983tx}. More details can be found in \cite{Gasser:1983yg, Gasser:1984gg}.

Notwithstanding, the validity of \chpt\ is not as wide and it cannot be extended as much as suggested before in the energy spectrum, when we arbitrarily set a cut-off of order 1 GeV. However, it should be valid and consistent up to the appearance of any physical state not accounted in the \chpt\ Lagrangian. This situation actually occurs with the presence of resonance states like the $\rho$, a triplet vector multiplet, with a mass of 775 GeV and several other resonances within the 1 GeV to 2 GeV range. Indeed, the low-energy spectrum of strong interactions is dominated by this kind of resonance objects. 

While \chpt\ is the best framework to deal with strong interactions for energies $E\ll m_\rho$, it does not describe properly the phenomenology in the range $ m_\rho \lsim \,E\, <\, 2\,\text{GeV}$, where a theory including the lowest resonance objects is demanded. The best framework within these energies is the so-called resonance chiral theory (R$\chi$T) \cite{Ecker:1988te, Ecker:1989yg, Cirigliano:2006hb}. It includes both a $SU(3)$ chiral Lagrangian, which gives raise to an octet of scalar Goldstone bosons and $U(3)$ flavor multiplets, which account for the lightest sets of resonances.

Unlike \chpt\,, the resonance theory of mesons lacks a definite power counting because the resonance masses are of the same order as the chiral symmetry breaking scale.\footnote{Nevertheless, there is some guidance in the loop level expansion when a large number of colors, $N_C$, is taken.} As a consequence, operator building is not performed as in the previous case. One needs to include the more general resonance interaction Lagrangian involving light fields (from \chpt) and resonances. Furthermore, it is assumed that only the lightest set of resonances are relevant, since heavier ones are further suppressed. Then, resonances have to be integrated out from the Lagrangian and, therefore, we can evaluate how important their interactions are according to the \chpt\ power counting. 

This procedure actually matches R$\chi$T with \chpt\ \cite{Weinberg:1978kz,Gasser:1983yg,Gasser:1984gg,Pich:1995bw,Ecker:1994gg,Bijnens:1999sh,Bijnens:1999hw,Bijnens:2014lea}. Taking the resonances out of the action leads to the rise of non-trivial contributions to the Wilson coefficients that parametrize the different \chpt\ local operators from eq.~(\ref{eq:chptevenpowers}). They are the most significant signature of massive states when analyzing the low-energy limit of the theory. Indeed, one could predict their existence just by analyzing the precise pattern of these coefficients.

\begin{figure}
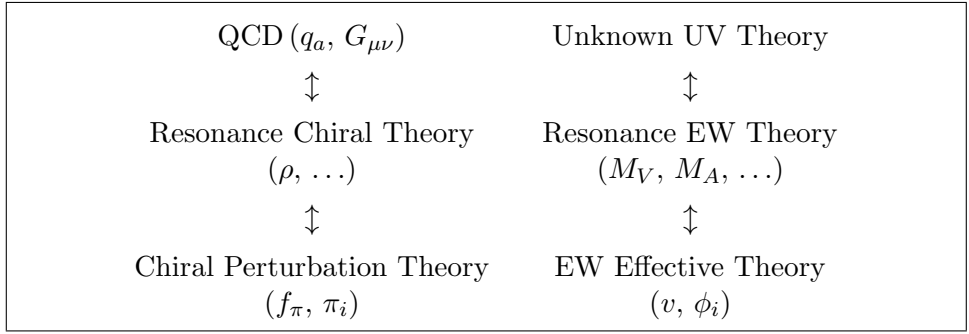

	\fbox{
		\begin{minipage}{0.9\textwidth}	
			\centering
			\vspace{-3mm}
			$$\begin{matrix}
			\mbox{QCD}\,(q_a,\, G_{\mu\nu}) & & \mbox{Unknown UV Theory} \\[1ex]		
			\updownarrow & & \updownarrow  \\[1ex]
			\mbox{Resonance Chiral Theory} & & \mbox{Resonance EW Theory} \\
			(\rho,\,\ldots) & & (M_V,\,M_A,\,\ldots) \\[1ex]
			\updownarrow & & \updownarrow \\[1ex]
			\mbox{Chiral Perturbation Theory} & & \mbox{EW Effective Theory} \\
			(f_\pi,\,\pi_i) & & (v,\,\phi_i)\\
			\end{matrix}
			$$
		\end{minipage}
	}
	\small{
		\caption{Comparative scheme of resonance frameworks for a meson theory (left) and for the electroweak theory (right). More general and fundamental theories are placed in the top of the diagram whereas low-energy EFTs are in the bottom. }
		\label{fig:motivationchpt}
	}
\end{figure}

The \chpt\ picture and the role of resonances constitute in fact one of the main motivations for this thesis. If nature has shown how resonance states can be present in a particle physics theory, why cannot this situation happen as well in the context of the electroweak theory? We will present an scenario where the SM, described in an EFT formalism, portraits the low-energy limit of an unknown fundamental theory. We will introduce the more general set of possible electroweak resonances, which will leave their imprints in the Wilson coefficients associated to the low-energy EFT. Considering the analogy with this section (figure \ref{fig:motivationchpt}), the SM would play the part of \chpt, the electroweak resonances would set a high-energy theory as R$\chi$T does in the meson theory, while the UV theory acts like QCD in this comparison.

	
\chapter{Electroweak Effective Theory } \label{ch:ewet}

The Electroweak Effective Theory\footnote{In other works also called electroweak chiral Lagrangian (EWChL) or Higgs effective field theory (HEFT).} (EWET) is the most general low-energy EFT containing the particles of the Standard Model, e.g. the gauge bosons, the SM fermions, the electroweak Goldstone bosons and a $m_h=125\, {\rm GeV}$ Higgs-like singlet scalar boson;\footnote{Despite the fact that it is not necessarily the SM Higgs, we will generally refer to this particle as the Higgs boson, denoted in small letters $h(x)$.} and including its symmetries. This effective field theory does not assume any particular structure embedding the Higgs field, in opposition to the SM with the complex scalar doublet field $\Phi(x)$ (see section \ref{sec:SSB}). Hence, the presence of three extra degrees of freedom, which are the Goldstone bosons, is required in order to allow the gauge bosons to be massive and to incorporate their missing longitudinal polarizations. As a consequence, one of the main advantages of the EWET is thus the absence of relation between the Higgs field and the Goldstone bosons, so that one can test their dynamics separately. This formulation contrasts with the Higgs mechanism in the SM, where the phenomenology of the scalar sector is fully dependent of the measurements of the Higgs mass, being the masses of the particles proportional to the Higgs coupling.

Nonetheless, this is not an exotic description of the electroweak nature but a more general picture, since one is able to recover the SM prescription provided a certain limit with a given set of known parameters.   
This non-SM Higgs framework aims the theory to be as model-independent as possible and allows to discern whether the $\Phi(x)$ doublet configuration is the truly responsible for the symmetry breaking or it happens with other implementation somehow. The only premise the EWET assumes is the pattern of electroweak symmetry breaking (EWSB).

\section{Symmetry breaking in the EWET}
The success of the Higgs mechanism in the SM and the spontaneous symmetry breaking of the electroweak symmetry is well acknowledged. They restore the unitarity of the SM at high energies once the doublet scalar is included in the theory and assures the SM Lagrangian renormalizability. So far, no significant deviations with respect to the SM predictions in the electroweak sector have been detected. 

In order to analyze how the symmetry is broken in the EWET, it is more convenient to re-express the SM scalar Lagrangian in a more appropriate way \cite{Pich:1998xt, Pich:1995bw} where the complex scalar doublet is accommodated in a in a $2\times 2$ matrix \cite{Appelquist:1980vg}, being 
\be
\Sigma \,\equiv\, \left( \Phi^c, \Phi\right) \, =\,
\left( \bat \Phi^{0*} & \Phi^+  \\ -\Phi^- &  \Phi^0 \ea\right)\,.
\label{eq:sigma_matrix}
\ee
Therefore, it reads
\be \label{eq:l_sm}
\mL(\Sigma)\, =\, \frac{1}{2}\, \bra \left(D^\mu\Sigma\right)^\dagger D_\mu\Sigma \ket_2
- \frac{\lambda}{16} \left( \bra \Sigma^\dagger\Sigma\,\ket_2
- v^2\right)^2 , 
\ee
where the gauge-covariant derivative, $D_\mu\Sigma$, is defined as 
\be
D_\mu\Sigma \,\equiv\, \partial_\mu\Sigma
\,+\, i g \,\frac{\vec{\sigma}}{2}\vec{W}_\mu \,\Sigma \,-\, i g' \,\Sigma \,\frac{\sigma_3}{2} B_\mu \, .
\ee

The expression of the scalar Lagrangian in terms of $\Sigma$ manifests its $G\equiv SU(2)_L\otimes SU(2)_R$ global symmetry, also called chiral symmetry, in a straightforward way, given the transformations 
\be
\{\Sigma,\, D_\mu \Sigma \} \quad \longrightarrow \quad g_L \,\{\Sigma,\, D_\mu \Sigma \}\, g_R^\dagger\, ,
\qquad\quad \mbox{with} \qquad
g_{L,R}  \in SU(2)_{L,R} \, .
\label{eq:sigma_transf}
\ee
One can recover the SM when this global symmetry group $G$ is gauged to a $SU(2)_L\otimes U(1)_Y$ local group. Recall that only the left component of $G$ can be promoted to local, since the $\sigma_3$ matrix in the covariant derivative does not allow the same thing to the right part. 

In the EWET, instead, the full group $G$ is formally gauged, both the left and the right sides, assuming that the symmetry is indeed not exactly preserved and thus the small breaking is treated as a perturbation. At the quantum field level, this can be done with the implementation of the gauge sources

\bel{eq:FakeTransform}
\hat{W}^\mu\,\longrightarrow\, g_L^{\phantom{\dagger}}\, \hat{W}^\mu g_L^\dagger + i\, g_L^{\phantom{\dagger}}\, \partial^\mu g_L^\dagger\, ,
\qquad\quad
\hat{B}^\mu\,\longrightarrow\, g_R^{\phantom{\dagger}}\, \hat{B}^\mu g_R^\dagger + i\, g_R^{\phantom{\dagger}}\, \partial^\mu g_R^\dagger\, ,
\ee
where there is no explicit $SU(2)_R$ breaking and $g_{L,R}\in SU(2)_{L,R}$.
With this configuration,
the vacuum is preserved under the custodial symmetry group $H\equiv SU(2)_{L+R}$, just in the case when $g_L=g_R$, and chiral symmetry is spontaneously broken into this group \cite{Sikivie:1980hm}. We will refer to this process as electroweak chiral symmetry breaking or simply EWSB, which summarizes as follows:
\be \label{eq:ChiralSymmetry}
G\equiv SU(2)_L\otimes SU(2)_R\quad \longrightarrow \quad H\equiv SU(2)_{L+R}\, .
\ee

Following the same technique as in eq.~(\ref{eq:polarSM}), one can perform a polar decomposition of the SM scalar fields, parametrizing the physical excitations over the vacuum,
\be\label{eq:polar}
\Sigma(x) \, = \, \frac{1}{\sqrt{2}}
\left[ v + h(x) \right] \; U(\varphi(x))\,,
\ee
where $h(x)$ is a $m_h=125$ GeV scalar singlet under the group $G$.  
Substituting eq.~(\ref{eq:polar}) in the scalar Lagrangian, eq.~(\ref{eq:l_sm}), one can rewrite it as \cite{Coleman:1969sm, Callan:1969sn, Appelquist:1980vg,Longhitano:1980iz,Longhitano:1980tm}
\be
\mL(\Sigma)\, =\, \frac{v^2}{4}\,
\langle\, D_\mu U^\dagger D^\mu U \,\rangle \, +\,
\cO\left( h/v \right) ,
\label{eq:sm_goldstones}
\ee
where all the terms involving the Higgs field are set apart.  Finally, when these contributions are removed, the last equation becomes the universal Goldstone Lagrangian. This Lagrangian is actually not exclusive for the electroweak sector but its usage is widely extended in several physical systems which are associated with the chiral symmetry breaking of eq.~(\ref{eq:ChiralSymmetry}). Particularly, this Lagrangian is the one employed to describe the QCD dynamics of pions in the two-flavour case. Actually, the changes to be performed so that this QCD meson Lagrangian is recovered are just $\vec\varphi \to \vec \pi$ and $v \to f_\pi$, where $f_\pi$ is the pion decay constant (see section \ref{sec:chpt}). 

This comparison with two-flavour QCD is a reference point for this work, not only because chiral perturbation theory was profundly studied during the last decades and this information helps to understand the dynamics of the EWSB, but also because of the presence of meson resonance states in the spectrum. 

\section{Characterization of the EWET}
As stated in section \ref{sec:EFTs}, the particle content, symmetries and organization are the three fundamental elements to properly define an EFT and this is also the case for the EWET. Since this low-energy theory aims to be as general as possible, every particle of the SM and their symmetries must be included.

In this section we deal with the technical details of the EWET. An exhaustive analysis of this EFT has already been done in \cite{Buchalla:2013rka,Buchalla:2012qq}. However, in this work the EWET is built in such a way that massive states can be comfortably incorporated in a  high-energy resonance theory, described in the following chapter.

\subsection{Symmetries}

Once the pattern of symmetry breaking is well defined, we incorporate all the bosonic and fermionic fields to the framework. On the one hand, the symmetry group $G$ in eq.~(\ref{eq:ChiralSymmetry}) has to be extended by the gauge subgroup $SU(3)_C$ in order to incorporate gluons and colored objects. As in the SM case, the vacuum is invariant under this symmetry so this subgroup remains unaltered after the EWSB too. On the other hand, embedding the SM fermion multiplets in the chiral group $SU(2)_L\otimes SU(2)_R$ demands the presence of an extra $U(1)_X$ auxiliary symmetry, being $X=({\rm B}-{\rm L})/2$ with $B$ and $L$ the baryon and lepton number, respectively, which is explicitly broken after the EWSB. Of course, baryon and lepton conservation, as well as Lorentz invariance, are respected in this EFT. Hence the complete symmetry group of the EWET is
\bel{eq:ewetSymmetry}
\mG \equiv SU(3)_C \otimes SU(2)_L\otimes SU(2)_R \otimes U(1)_X\quad \longrightarrow \quad SU(3)_C \otimes SU(2)_{L+R}\, .
\ee
In this work, both quarks and leptons are incorporated. However, only one family of each kind of fermionic field is considered. The analysis of a three family version of the EWET is postponed to future works.  

The Lagrangian of the theory is also assumed to be invariant under the $CP$ discrete symmetry, whereas the distinction among $C$ (and $P$) even or odd operators will be present along the discussion. Indeed, the sources of $CP$-violating processes in the SM are found to be very small and they are generally associated to flavor-changing interactions. Certainly, this assumption may be relaxed whenever all fermion families enter in the EWET game.

\subsection{Bosonic fields}
For this purpose, we reformulate the different fields in a more appropriate formalism, with slight modifications in the way they transform under $\mG$ and their properties. 

The Goldstone bosons originated due to the EWSB can be understood as the coordinates of the coset space $SU(2)_L\otimes SU(2)_R/SU(2)_{L+R}$, according to the Goldstone theorem (they are also singlets under the colour group and the $U(1)_X$ group, since they are colourless and carry $L=B=0$). In general, the Goldstone bosons transform like
\be
u_L^{\phantom{\dagger}}(\varphi)\,\longrightarrow\, g_L^{\phantom{\dagger}}\, u_L^{\phantom{\dagger}}(\varphi)\, g_h^\dagger(\varphi,g)\, ,
\qquad
u_R^{\phantom{\dagger}}(\varphi)\, \longrightarrow\, g_R^{\phantom{\dagger}}\, u_R^{\phantom{\dagger}}(\varphi)\, g_h^\dagger(\varphi,g)\, ,
\ee
through the combined action of $g\equiv (g_L^{\phantom{\dagger}},g_R^{\phantom{\dagger}})\in SU(2)_L\otimes SU(2)_R$ and $g_h^{\phantom{\dagger}}(\varphi,g)\equiv g_h^{\phantom{\dagger}}$.  Notice how they do not transform as conventional $SU(2)$ elements and, therefore, the action of $g$ must be compensated by $g_h(\varphi, g)$ since they are elements of the coset group \cite{Coleman:1969sm,Callan:1969sn}. Additionally, $g_h$ transforms identically in both chiral sectors,\footnote{This is justified due to the fact that parity exchanges left an right components.} leaving the custodial group $SU(2)_{L+R}$ invariant.  

Nevertheless, the selection of the coset coordinates has some freedom and it is customary to take the canonical\footnote{
	Other works in the field of $\chi$PT adopt the opposite convention $u_R^{\phantom{\dagger}}(\varphi)=u_L^\dagger(\varphi)=u(\varphi)$ ~\cite{Ecker:1988te}.}
 ones, $u_L^{\phantom{\dagger}}(\varphi)=u_R^\dagger(\varphi)=u(\varphi)$ \cite{Pich:2012dv, Pich:2013fea}. 
The advantages of this choice are both evident for our purpose: simpler transformation properties and an exponential representation,
\bel{eq:u_transf}
u(\varphi)=\exp{(i \vec \sigma \vec \varphi/(2v))}\, \longrightarrow\, g_L^{\phantom{\dagger}}\,  u(\varphi)\,  g_h^\dagger(\varphi,g) \; =\; g_h^{\phantom{\dagger}}(\varphi,g) \, u(\varphi) \, g_R^\dagger\, .
\ee
Besides, this result can be easily related with the Goldstone matrix $U(\varphi)$ in eq.~(\ref{eq:Goldstones}):
\bel{eq:U_u_rel}
U(\varphi)\,\equiv\, u_L^{\phantom{\dagger}}(\varphi)\, u_R^\dagger(\varphi)\, =\, u(\varphi)^2
\quad\longrightarrow\quad g_L^{\phantom{\dagger}}\, U(\varphi)\, g_R^\dagger \, .
\ee
Note that despite $U(\varphi)$ and $u(\varphi)$ have similar definitions formally (see eq.~(\ref{eq:Goldstones}) and eq.~(\ref{eq:U_u_rel})), they transform differently under $\mG$. 

Now we introduce all the bosonic gauge fields in the EWET associated with the symmetry group: the gauge bosons $\hat W_\mu$ and $\hat B_\mu$, the $\hat X_\mu$ and the gluon field, $\hat G_\mu$. 
In the first place, we recall the weak gauge bosons $\hat W_\mu$ and $\hat B_\mu$ as the $SU(2)_L$ and $SU(2)_R$ matrix associated fields, respectively, which transformation rules are already indicated in eq.~(\ref{eq:FakeTransform}).
They allow us to define properly the Goldstone gauge derivative in the context of the EWET:
\bel{eq:DU}
D_\mu U \, =\, \partial_\mu U -  i\, \hat{W}_\mu  U + i\, U \hat{B}_\mu
\,\longrightarrow\, g_L^{\phantom{\dagger}}\, D_\mu U \, g_R^\dagger
\, ,
\ee
Secondly, we define the $\hat X_\mu$ field,\footnote{Recall that $X=({\rm B} - {\rm L})/2$.} associated with the $U(1)_X$ symmetry, transforming as
\bel{eq:Xtransform}
\hat{X}^\mu\,\longrightarrow\, \hat{X}^\mu
+ i\, g_X^{\phantom{\dagger}}\, \partial^\mu g_X^\dagger\, \qquad \mbox{with}\qquad g_X\in U(1)_X\,.
\ee
Finally, as in the SM, we introduce the gluon field in the EWET, $\hat G_\mu=g_S G_\mu^a T^a$, being $g_S$ the strong coupling and $T^a=\frac{1}{2}\lambda^a$, $a=1,\ldots,8$ the $SU(3)$ generators. It transforms under this symmetry group as follows
\bel{eq:Gtransform}
\hat G_\mu \,\longrightarrow\, g_C^{\phantom{\dagger}}\, \hat G_\mu g_C^\dagger - \frac{i}{g_S}\,g_C^{\phantom{\dagger}}\, \partial_\mu g_C^\dagger \qquad \mbox{with} \qquad g_C^{\phantom{\dagger}}\in SU(3)_C\,.
\ee

Next to these definitions, it is convenient to set down the field strength tensors for all these bosonic fields together with their transformation rules under $\mG$. For the gauge bosons, we get
\bear
\hat{W}_{\mu\nu}  &=& \partial_\mu \hat{W}_\nu - \partial_\nu \hat{W}_\mu
- i\, [\hat{W}_\mu,\hat{W}_\nu]
\, \longrightarrow\, g_L^{\phantom{\dagger}}\, \hat{W}_{\mu\nu} \, g_L^\dagger
\, ,
\nn\\
\hat{B}_{\mu\nu}  &=& \partial_\mu \hat{B}_\nu - \partial_\nu \hat{B}_\mu
- i\, [\hat{B}_\mu,\hat{B}_\nu]
\,\longrightarrow\,
g_R^{\phantom{\dagger}}\, \hat{B}_{\mu\nu} \, g_R^\dagger
\, ;
\eear
while the $U(1)_X$ field strength tensor is found to be a singlet under the global group:
\be
\hat{X}_{\mu\nu}\, =\,\partial_\mu\hat{X}_\nu -\partial_\nu \hat{X}_\mu\, \longrightarrow \, \hat X_{\mu\nu}\,.
\ee
The gluon fields strength tensor is also defined as in the SM case. For completeness, 
\bear \label{eq:Gstrength}
G_{\mu\nu}^a & = & \partial_\mu G_\nu^a \, - \, \partial_\nu G_\mu^a \, + g_S\,f^{abc}\,G_\mu^b\,G_\nu^c \quad, \qquad a,b,c=1,\ldots,8\ ,
\nn\\
\hat G_{\mu\nu} &= & g_S\,G_{\mu\nu}^a\,T^a\,\longrightarrow\, g_C^{\phantom{\dagger}} \hat G_{\mu\nu} g_C^\dagger\,,
\eear
where $f^{abc}$ is the group structure constant defined as $[T^a,T^b]=i\,f^{abc}\,T^c$. More Lie-algebra relations and properties relevant for this work are listed down in the appendix \ref{app:liealgebra}.

At this point, the number of different combinations of fields, even at the bosonic level only, is countless and so it is the number of operators. For sure, a proper counting is still lacking (see section \ref{sec:powercounting}) but, prior to this, one realizes that the number of structures that are invariant under the full symmetry group $\mG$ is not that large. Indeed, we can reduce the so-called building blocks that we are going to use to a short list. Hence, every possible operator to be defined in the EWET is made out of a combination of these tensorial pieces. 

In order to introduce resonances in the high energy theory (chapter \ref{ch:resonancetheory}) we want the building blocks, $\mX$, and also their covariant derivatives, $\nabla_\mu \mX$ to transform as $SU(2)_{L+R}$ singlets or triplets and $SU(3)_C$ singlets or octets. Actually, these tensors follow the same transformation rules that the massive high energy resonances we will couple in the following chapter. 

Without loss of generality, let's consider a particular $SU(2)_{L+R}$ triplet and $SU(3)_C$ octet tensor, $\mX$. It transforms as 
\bear \label{eq:CovDev}
\mX &\longrightarrow& g_C^{\phantom{\dagger}}\,g_h^{\phantom{\dagger}} \, \mX\, g_h^\dagger\,g_C^\dagger\,,  \nonumber\\
\nabla_\mu \mX \, =\, \partial_\mu \mX \, +\, [\Gamma_\mu , \mX ]
 &\longrightarrow&  g_C^{\phantom{\dagger}}\,g_h^{\phantom{\dagger}} \, \nabla_\mu \mX\, g_h^\dagger\,g_C^\dagger
\, .
\eear
The tensor $\Gamma_\mu$ represents the physical connection and it is defined as
\bear \label{eq:connection}
\Gamma_\mu &=&\Frac{1}{2}\, u_L^\dagger(\varphi) \left(\partial_\mu \,-\, i\,\hat{W}_\mu \,-\, i \hat G_\mu\right) u_L^{\phantom{\dagger}}(\varphi) \nonumber \\
 &+&  \Frac{1}{2}\, u_R^\dagger(\varphi)  \left(\partial_\mu \,-\, i\,\hat{B}_\mu \,-\, i \hat G_\mu\right) u_R^{\phantom{\dagger}}(\varphi) 
 \nn\\
 &\equiv& \Gamma_\mu^L \,+\, \Gamma_\mu^R \,, 
\eear
which is generated by both left and right components of the coset representative of the Goldstone fields and by the Gluon field\footnote{Notice that the $u$ and $u^\dagger$ terms vanish for $SU(3)_C$.} too\cite{Pich:2012jv}. Besides, we have defined $\Gamma_\mu^L$ and $\Gamma_\mu^R$ as the first and second line of eq.~(\ref{eq:connection}), respectively.
For constructing a covariant derivative of a building block that is singlet under $SU(3)_C$ but still a triplet under $SU(2)_{L+R}$ the connection gets modified by removing the color fields from it. The same thing applies for the opposite situation, $SU(2)$ singlet and $SU(3)$ octet, where we get rid of the electroweak bosons. In the singlet-singlet case, the covariant derivative becomes a partial derivative.

We define the following bosonic building blocks, which are color-singlet and triplet under $SU(2)_{L+R}$, 
\bear\label{eq.cov-bosonic-tensors}
u_\mu  &=& i
\left( \Gamma_\mu^R - \Gamma_\mu^L\right) \,=\,
i\, u (D_\mu U)^\dagger u \, =\, -i\, u^\dagger D_\mu U\, u^\dagger
\, =\, u_\mu^\dagger\, \longrightarrow \, g_h^{\phantom{\dagger}}\, u_\mu\, g_h^\dagger\, ,
\nn\\[5pt]
f_\pm^{\mu\nu} &\! =&\!
u^\dagger \hat{W}^{\mu\nu}  u \,\pm\, u\, \hat{B}^{\mu\nu} u^\dagger \, \longrightarrow g_h^{\phantom{\dagger}}\, f_\pm^{\mu\nu} \, g_h^\dagger\,  .
\eear

Furthermore, we can introduce the spurion tensor field $\mT$, accounting for the source of custodial symmetry breaking which is not directly implemented in the definition of the EWET gauge sources (\!\ie there is no explicit $\sigma_3$ matrix breaking $SU(2)_R$). Thus, every time this term is present implies an explicit breaking of this symmetry. This procedure is well known since it has been used in \chpt\ to break explicitly the chiral symmetry through electromagnetic interactions \cite{Ecker:1988te,chpt+photons,EckerIMNP:2000}. It can be formally defined as
\bel{eq:T_covariant}
\mT\, =\, u\, \mT_R\, u^\dagger \quad\longrightarrow\quad g_h^{\phantom{\dagger}} \mT g_h^\dagger\, .
\ee
where $\mT_R\longrightarrow g_R^{\phantom{\dagger}}\, \mT_R\, g_R^\dagger\,$ is the right-handed spurion field.

In short, the list of bosonic building blocks that are useful for this work are the following:
\be \label{eq:buildingblocks}
\partial_\mu\frac{h}{v},\quad \hat X_{\mu\nu},\quad u_\mu,\quad f_\pm^{\mu\nu}, \quad \mT, \quad \hat G_{\mu\nu} \,.
\ee 
Further details regarding their transformation properties under discrete symmetries are placed in the appendix \ref{app:CPtransformations}.

Our last step within the bosonic fields in the EWET is the configuration of the scalar potential. Notice that the Higgs field is a singlet under $\mG$ and it can appear at the operator level as many times as desired. Indeed, we will see in the power counting section  that this field does not alter the dimensional counting of the EWET, and neither do the Goldstone bosons (see section \ref{sec:powercounting}). For this reason, one can conceive a more general scalar sector
\be \label{eq:LscalarEWET}
\mL_{\mbox{scalar}}\, =\, \frac{1}{2}\,
\partial_\mu h\,\partial^\mu h
\, -\,\frac{1}{2}\, m_h^2\, h^2 \, -\, V(h/v)
\, +\,
\frac{v^2}{4}\,\mF_u(h/v)\;\langle u_\mu u^\mu\rangle\, ,
\ee
with
\be\label{eq:Fhu_V}
V(h/v)\, = \, v^4\;\sum_{n=3} c^{(V)}_n \left(\frac{h}{v}\right)^n\, ,
\qquad\qquad
\mF_u(h/v)\, = \, 1\, +\, \sum_{n=1} c^{(u)}_n \left(\frac{h}{v}\right)^n\, ,
\ee
some arbitrary polynomials expanded in powers of $h/v$. One can also wonder whether it is possible to build an even more general configuration by adding a $\mF_h(h/v)$ extra function in the Higgs field kinetic term. Nonetheless, we show in the appendix \ref{app:SimpscalarLagrangian} that it is always possible to redefine this field in such a way that $\mF_h=1$, without losing generality \cite{SILH, Buchalla:2013rka}.

\subsection{Fermionic fields} \label{sec:fermionsewet}
The inclusion of fermionic fields in the core of the EWET is particularly interesting since there is no mirror to compare in \chpt\ where baryons carry a very different structure, unlike mesons in the bosonic case. This analysis is, therefore, original in the context of the electroweak-based EFTs and, concretely, in the EWET, being widely spelled out in \cite{Buchalla:2013rka}. However, here the fermions are introduced so that they can be easily coupled to high-energy states \cite{Fingerprints, Colorful} transforming as $SU(2)_{L+R}$ objects, as in the bosonic case.

Fermions in the EWET may be quarks (color triplet) or leptons (color singlet). The formalism that follows is meant to be for colored fields, even so one can simply study the color-singlet case by removing all $SU(3)_C$ objects from the next equations. Given a left and right-handed fermion\footnote{In general, flavor and color fermion indices, $\psi^q_a$, are omitted. For leptons, the color ones do not apply.}  multiplets $\psi_{L,R} = P_{L,R}\,\psi$, they transforms under $\mG$ in a different way according to their chirality,
\bear
\psi_L = \left( \begin{array}{c} t_L \\ b_L \end{array}\right)&\longrightarrow& g_C\,g_X\, g_L\,  \psi_L  \,, \nonumber\\
\psi_R  =\left( \begin{array}{c} t_R \\ b_R \end{array}\right)&\longrightarrow& g_C\,g_X\, g_R\,  \psi_R \, ,
\eear
with $g_C\in SU(3)_C$, $g_X\in U(1)_X$ and $g_{L,R} \in SU(2)_{L,R}$; while their covariant derivatives are
\bear \label{eq.dpsi}
D_\mu^L\psi_L\, &=&\,\left(\partial_\mu - i \,\hat G_\mu - i\,\hat{W}_\mu - i\,
\hat{X}_\mu \,
\Frac{( {\rm B}-{\rm L})}{2}
\right) \psi_L\, , \nonumber
\\
D_\mu^R\psi_R\, &=&\,\left(\partial_\mu - i \,\hat G_\mu- i\,\hat{B}_\mu\, - i\,
\hat{X}_\mu
\Frac{({\rm B- L})}{2}
\right) \psi_R\, .
\eear
Recall that $({\rm B} - {\rm L})/2$ is an operator and it depends on whatever object it acts on.
 
 However, it is more convenient to employ fermionic objects that transform covariantly with $g_h$ and not with $g_{L,R}$. For simplicity, we will refer to these structures as covariant fermions or just fermions, from now on. They transform like
\bear \label{eq:xi_transf}
\xi_{L} \,\equiv\, u_{L}^\dagger \, \psi_{L}\, =\, u^\dagger  \psi_L & \longrightarrow & g_C\,g_X \, g_h\; \xi_L\, , \nonumber \\
\xi_{R} \,\equiv\,  u_{R}^\dagger \, \psi_{R}  \, =\, u^{\phantom{\dagger}} \psi_R & \longrightarrow &
g_C\,g_X \, g_h\; \xi_R\, ;
\eear
with $\xi_{L,R}= u^\dagger_{L,R}\, P_{L,R}\, \psi\,$ and their covariant derivatives being $d_\mu \xi  =  d_\mu^R\xi_R + d_\mu^L\xi_L$, where
\bear \label{eq:dxi}
d_\mu^{L} \xi_{L}   & =&
u_L^\dagger \left(  \partial_\mu - i\,\hat G_{\mu} - i\, \hat{W}_\mu - i
\,\hat{X}_\mu
\Frac{( {\rm B}  -{\rm L})}{2}
\right) \psi_{L}
\, ,
\nn\\
d_\mu^{R} \xi_{R}   & =&
u_R^\dagger \left(  \partial_\mu - i\,\hat G_{\mu} - i\,\hat{B}_\mu - i
\, \hat{X}_\mu
\Frac{({\rm B}-{\rm L})}{2}
\right) \psi_{R}
\, .
\eear

Operator building, though, requires closed fermion spinorial indices. For this reason fermionic fields always appear in an even number at the Lagrangian level, forming covariant bilinear structures, $J_\Gamma$, which have well-defined Lorentz transformation properties. Given a pair of covariant fermions, $\eta^\alpha_n$ and $\zeta^\beta_m$, with $\alpha,\beta=1,\ldots,4$ their explicit spinor indices and $m,n$ their $SU(2)$ matricial indices, they form the general bilinear
\be
J^\Gamma_{mn}\, = \, \bar{\eta}^\alpha_n \Gamma^{\alpha\beta} \zeta^\beta_m  \, = \, - \zeta^\beta_m \bar{\eta}^\alpha_n \Gamma^{\alpha\beta}
\, =\, - \mathrm{Tr}_D \{ \zeta_m \bar{\eta}_n \Gamma \} \,  ,
\ee
being ${\rm Tr_D}$ the trace in the Dirac space and $\Gamma$ the representative Dirac matrix in the Clifford algebra, what characterizes their bilinear type. These matrices\footnote{Notice that the convention for these matrices may differ from other works, like \cite{Itzykson:1980rh}, for instance.} form a basis\footnote{The tensor element is defined as $\sigma^{\mu\nu}=\frac{i}{2}[\gamma^\mu,\gamma^\nu]$.}   $\{I, i \gamma_5, \gamma^\mu, \gamma_\mu\gamma_5, \sigma^{\mu\nu}\}$ and we will refer to the bilinears as scalar, pseudoscalar, vector, axial-vector or tensor, respectively, in the same way as it is usually done for the $\Gamma$ matrices. In the following equations we show the different relevant\footnote{It is possible to build more exotic bilinear combinations in general but operators including them can always be rearranged in terms of eq.~(\ref{eq:bilinearnotation}) bilinears, at the order analyzed in this work.} bilinear structures that are needed to build the EWET and the resonance theory:
\bear \label{eq:bilinearnotation}
(J_S)_{mn} &\equiv & \, -\, Tr_D\{ \xi_m \bar{\xi}_n \}
\,\,\, =\,\,\, \bar{\xi}_n \xi_m    \, ,
\nn\\
(J_P)_{mn}  &\equiv & \, -\, i\, Tr_D\{ \xi_m \bar{\xi}_n \gamma_5 \}
\,\,\,=\,\,\, i\,   \bar{\xi}_n \gamma_5 \xi_m  \, ,
\nn\\
(J_V^\mu)_{mn} &\equiv & \, -\, Tr_D\{ \xi_m \bar{\xi}_n \gamma^\mu \}
\,\,\, =\,\,\,  \bar{\xi}_n\gamma^\mu \xi_m \, ,
\nn\\
(J_A^\mu)_{mn} &\equiv & \, -\, Tr_D\{ \xi_m \bar{\xi}_n\gamma^\mu\gamma_5\}
\,\,\, =\,\,\, \bar{\xi}_n\gamma^\mu\gamma_5 \xi_m \, ,
\nn\\
(J_T^{\mu\nu})_{mn} &\equiv & \, -\, Tr_D\{ \xi_m \bar{\xi}_n\sigma^{\mu\nu} \}
\,\,\, =\,\,\,  \bar{\xi}_n\sigma^{\mu\nu} \xi_m \, .
\eear
In the Appendix \ref{app:CPtransformations} we show how these covariant fermionic bilinears transform under the action of discrete symmetries.

Furthermore, we can distinguish in this work three different categories of fermion bilinears, according to their particle composition and color structure,
\bear \label{eq:bilinearcolortype}
{\rm Lepton} &\qquad & J_\Gamma^{l} \equiv \bar \xi \, \Gamma \, \xi \, , \nn\\
{\rm Quark\ singlet} &\qquad & J_\Gamma^q \equiv \bar \xi_i\, \Gamma \, \delta_{ij} \, \xi_j\, \quad ,
\nn\\
{\rm Quark\ octet}\,\; &\qquad &  J_\Gamma^8 \equiv (J_{\Gamma}^{8,a})\, T^a  \equiv (  \bar\xi_i\, \Gamma \,T^{a}_{ij}\, \xi_j)\, T^{a} \, ;
\eear
where $i,j=1,\ldots,3$ are the elements in the $SU(3)_C$ color group while $a=1,\ldots,8$ points the color index in the adjoint representation. Note that in the quark singlet bilinear operators, the $SU(3)_C$ indices are contracted through the identity matrix $\delta_{ij}$, in contrast to the color octet ones in the third line of the equation, which contract through the $T^a$ Gell-Mann generators. In the lepton case, indicated with the superindex\footnote{This superindex may be omitted if the context is clear.} $l$ in the first line of eq.~(\ref{eq:bilinearcolortype}), color indices do not apply.

As well as the rest of building blocks, bilinears can couple to other objects $\mO \longrightarrow g_h^{\phantom{\dagger}} \mO g_h^\dagger$ transforming under $SU(2)_{L+R}$ in order to form bigger structures, also invariant under $\mG$,
\be
\bra J_\Gamma \, \mO \ket  \, =\, -  \zeta_m^\beta \bar{\eta}_n^\alpha \Gamma^{\alpha\beta}  \; \mO_{nm}
\,  =\,    \bar{\eta}  \, \Gamma \,   \mO \,  \zeta\, .
\ee

The last element to be introduced is the Yukawa spurion, $\mY$. This interaction is required in order to give raise to fermion masses, which otherwise would be massless as $\mG$ imposes. Thus, the Yukawa interaction breaks this group. Following the spirit of $\mT$ in the bosonic case, one introduces a right-handed spurion field, $\mY_R$, and accommodates it so that it transforms properly with $g_h$:
\bel{eq:Yspurion}
\mY_R\,\longrightarrow\, g_R^{\phantom{\dagger}}\,\mY_R\, g_R^\dagger\, ,
\qquad\qquad
\mY\, =\, u\,\mY_R\, u^\dagger
\,\longrightarrow\, g_h^{\phantom{\dagger}}\,\mY\, g_h^\dagger
\, .
\ee

Therefore, the list of building blocks in the EWET that are relevant for us incorporates
\be \label{eq:buildingblocksfermion}
\mY,\quad J_\Gamma^l, \quad J_\Gamma^q, \quad J_\Gamma^8\, ; 
\ee
together with the bosonic building block structures in eq.~(\ref{eq:buildingblocks}).

\section{The SM case in the EWET} \label{sec:SMinEWET}
In this section we show the SM as a particular case of the EWET. This is a necessary test for every particle physics EFT within a low-energy region matching the electroweak energy spectrum. The huge success of the SM and its very accurate description of nature requires this sort of theories to be able to reproduce its results.

Thus, the SM demands to get the same fundamental interactions at the Lagrangian level and also to be leading order in the power expansion. Note that this statement does not imply that the SM Lagrangian must be exactly the leading order of our EFT, but it must be contained at least. In section \ref{sec:powercounting} we show indeed how our Lagrangian is organized. It is consistent with this premise, being easily checkeable how the SM is contained in the leading order of the EWET, also denoted $\mL_2$. 

The next equation collects a set of operators from the EWET. In the following, it is proved how the SM is recovered (eq. \ref{eq:sm}), 
\bear \label{eq:L_SMinL_EWET}
\mL_2 &\supset & -\frac{1}{2 g^2} \bra \hat W_{\mu\nu} \hat W^{\mu\nu} \ket_2 
\,-\,\frac{1}{2 g'^2}\bra \hat B_{\mu\nu} \hat B^{\mu\nu} \ket_2 
\,-\,\frac{1}{2 g_S^2} \bra \hat G_{\mu\nu} \hat G^{\mu\nu} \ket_3  \nn\\ 
 && +\quad \frac{1}{2}\, \partial_\mu h\, \partial^\mu h - V(h) \,+\, \frac{v^2}{4} \bra u_\mu u^\mu \ket_2 \left(1+ \frac{h}{v}\right)^2 \nn\\
 && +\quad \sum_{\xi}i \,\bar{\xi}\gamma^\mu D_\mu \xi \,-\, v\, \bar{\xi}\, \mY\, \xi \,.
\eear
Note that the last sum is performed among all the fermions present in the theory and $V(h)$ stands for the scalar field potential, defined analogously as in eq.~(\ref{eq:scalarlagrangianPhi}). The first line of this equation corresponds to the Yang-Mills Lagrangian, $\mL_{YM}$ associated with the SM gauge symmetry $SU(3)_C\otimes SU(2)_L \otimes U(1)_Y$  identifications for the electroweak gauge fields\cite{Pich:2012jv} (the gluon field definition is already implicit in the EWET, although it is included in the following for the sake of completeness):
\bear \label{eq:SMgauge}
\hat{W}_\mu  &=& -g\;\frac{\vec{\sigma}}{2}\, \vec{W}_\mu \, , \nn\\
\hat{B}_\mu  &=& -g'\;\frac{\sigma_3}{2}\, B_\mu\, , \nn\\
\hat{X}_\mu  &=& -g'\; B_\mu \,, \nn\\
\hat{G}_\mu  &=& g_S \; G_\mu^a\, T^a\,.
\eear
Together with the covariant fermion kinetic term, $\sum_{\xi}i\,\bar{\xi}\gamma^\mu D_\mu \xi$, one recovers the unbroken massless SM Lagrangian, $\mL_{\mathrm{SM}}^{(0)}$. Notice, however, that the fermionic sector in the EWET is rewritten in a much more compact notation than in eq.~(\ref{eq:sm}), since covariant fermions within bilinear structures give raise to different kinds of operators. Vector and axial-vector bilinears do not mix the left and the right fermion doublets; whereas scalar, pseudoscalars and tensor bilinears combine both fermion chiral sectors, where Goldstone bosons arise naturally. In short,
\bear
\bar{\xi}\,\Gamma\, \xi' &\! =&\! \left\{ \bat
\overline{\psi}_L\Gamma\psi_L' \,+\, \overline{\psi}_R\Gamma\psi_R'
& \qquad\mbox{\small  ($\Gamma=\gamma^\mu, \, \gamma^\mu \gamma_5$)}\, ,
\\[10pt]
\overline{\psi}_L \Gamma\, U(\varphi)\,\psi_R'
\,+\, \overline{\psi}_R \Gamma \, U(\varphi)^\dagger \psi_L'
& \qquad\mbox{\small ($\Gamma=1,\, i\gamma_5,\, \sigma^{\mu\nu}$)} \, .
\ea\right.\quad
\eear
%

The second line of eq.~(\ref{eq:L_SMinL_EWET}) contains both the Goldstone Lagrangian, giving raise to the SM interactions analyzed in section \ref{sec:SSB}, and the scalar Lagrangian, once identified the $V(h)$ as the scalar potential of the SM. In comparison to the general structure of the scalar Lagrangian in the EWET in eq.~(\ref{eq:LscalarEWET}) and eq.~(\ref{eq:Fhu_V}), the SM interactions are recovered once
\begin{align} \label{eq:SMscalarrecover}
c_3^{(V)} &= \frac{1}{2} \frac{m_h^2}{v^2}\,,& \qquad c_4^{(V)} &= \frac{1}{8} \frac{m_h^2}{v^2}\,,& \qquad c_{n>4}^{(V)} &= 0 \,,& \nn\\ 
c_1^{(u)} &= 2\, ,& \qquad c_2^{(u)} &= 1\, ,&  \qquad c_{n>2}^{(u)} &= 0\, .& 
\end{align}

Finally, the Yukawa term in the third line, 
\bel{eq:Yukawas}
-v\,\bar \xi\, \mY \, \xi\, =\, -v\; \bar\xi_L\, \mY\, \xi_R \, +\, \mathrm{h.c.}
\, =\, -v\;\bar\psi_L\, U(\varphi)\,\mY_R\,\psi_R\, +\, \mathrm{h.c.}
\ee
becomes the SM Yukawa interaction when the right-handed spurion turns into \cite{ABCH:85,BEM:99}
\bel{eq:SM_Yspurion}
\mY\, =\, \hat{Y}_t \,\mP_+ + \hat Y_b \,\mP_-\, ,
\qquad\qquad
\mP_\pm \,\equiv\,\frac{1}{2}\,\left( I_2\pm\sigma_3\right)\, ,
\ee
for one fermion family with a $(t, b)$ fermion multiplet\footnote{For three fermion families, the scalar Yukawas are promoted to $3\times 3$ matrices \cite{Espriu:2000fq}.} This configuration incorporates explicitly the symmetry breaking present in the SM.

\section{The EWET power counting} \label{sec:powercounting}
The possible number of operators to be constructed in the EWET is infinite and, therefore, unaffordable. As a consequence, it is mandatory to establish a hierarchy among all the operators in order to separate those ones sensitive to the phenomenology from those who are not. For instance, the Standard Model effective field theory\footnote{Also called linear-EFT.} (SM-EFT) \cite{Grzadkowski:2010es}, one of the most renowned EFTs, is a weakly coupled bottom-up EFT which operators are ordered according to their canonical dimension. This power counting yields an expansion wherein the SM remains at leading order as a renormalizable theory, while the rest of operators are set correspondingly at higher orders. Besides, it assumes that any new physics contribution must be coupled linearly to the SM.

Nevertheless, this power counting cannot be used for the EWET since we are not assuming  any particular configuration for the Higgs field to couple to the Higgsless SM and, therefore, it is not embedded in a complex doublet scalar field. As a consequence, since the EWET aims not to make any assumption in regard to the way new physics couples, it yields a more general framework.

The EWET is organized according to the so-called chiral counting. This power counting reflects the infrared behavior of any operator at low momenta. Therefore, every Feynman diagram, $\Gamma$, and hence every operator, is assigned a chiral dimension, $\hat d$, which allows the theory to be power-expanded according to this key variable. Although the number of operators is still countless, now the number at a given chiral order is finite,\footnote{Be aware that this statement refers strictly for operators and not for functions or Wilson coefficients set below, which can be in general polynomials of the Higgs field, like in eq.~(\ref{eq:Fhu_V}).} leading to the following expansion of the EWET Lagrangian,
\bel{eq:EchL}
\mL_{\mathrm{EWET}}\, =\, \sum_{\hat d\ge 2}\, \mL_{\hat d}\, ,
\qquad \mbox{with} \quad 
\mL_{\hat d} = \mO (p^{\hat d})\, ,
\ee
organized increasingly in terms of their low-energy momenta (\ie, when $p\to 0$).

The computation of the chiral dimension of a given Feynman diagram, $\hat d_\Gamma$, is non-trivial and all the details are found in the appendix \ref{app:powercounting}. We obtain
that $\Gamma$ scales like $p^{\hat d_\Gamma}$ \cite{Fingerprints, Hirn:2006,Buchalla:2012qq,Buchalla:2013rka,Buchalla:2013eza}, being
\bel{eq:chiraldim}
\hat{d}_\Gamma\, =\, 2 + 2L + \sum_{\hat{d}} (\hat{d} -2)\, N_{\hat{d}} \, ,
\ee
with $L$ the number of loops of the diagram and $N_{\hat d}$ the number of vertices (operators) with a given chiral dimension, $\hat d$, defined as
\bel{eq:chiraldim2}
\hat d \, = \, d \, + \, \frac{j}{2}\,,
\ee
where $d$ makes reference to every explicit light scale or coupling present in the operator and $j$ is the number of fermionic fields.
In short, it is possible to assign a chiral dimension to any operator provided the following power counting rules:
\bear\label{eq:power_counting}
v\, ,\, \Frac{\varphi^i}{v}\,  , \, u(\varphi/v)\, ,\, U(\varphi/v)\, , \, \Frac{h}{v}\,  , \, \Frac{W^i_\mu}{v}\,  ,\, \Frac{B_\mu}{v}\, , \, \Frac{G_\mu^a}{v}
& \sim & \cO\left(p^0\right)\, ,
\nn\\
\xi\, ,\, \bar\xi\, ,\,\psi\, ,\,\bar\psi & \sim & \cO\left( p^{1/2} \right)\, ,
\nn\\
\partial_\mu\, , \, D_\mu\,,\,d_\mu\,,\, \nabla_\mu\,,\, u_\mu\,,\, \hat{W}_\mu\, , \, \hat{B}_\mu\, ,\, \hat{X}_\mu\, , \, \hat{G}_\mu\,, & & \nn\\
m_h\, , \, m_W\, , \, m_Z\, , \, m_\psi\, ,\, g\, ,\, g'\, ,\, g_S\,,\, \mT\,,\, \mY  & \sim & \cO\left( p\right)\, ,
\nn\\
\hat{W}_{\mu\nu}\, ,\, \hat{B}_{\mu\nu}\, ,\, \hat{X}_{\mu\nu}\, ,\, f_{\pm\, \mu\nu} \, ,\, c_n^{(V)}\, , \, (\bar{\eta}\, \Gamma\, \zeta)\,,\, J_\Gamma^f & \sim & \cO\left( p^2\right)\, , 
\nn\\
\partial_{\mu_1}\partial_{\mu_2} ... \partial_{\mu_n}\,\mF(h/v)  & \sim & \cO\left( p^n\right)\, .
\eear
This last equation is central since it allows the operator organization in the EWET and all the elements therein have been already defined.

In the following, there is a naive and non-technical explanation of the chiral power counting, which helps us get a deeper understanding of the nature of this EFT. For sure, as mentioned before, the mathematical derivation is placed in the Appendix \ref{app:powercounting}. 
Firstly, Goldstone bosons are compiled non-linearly in $U(\varphi/h)$ (see eq.~(\ref{eq:Goldstones})), which is only consistent if they 
carry chiral dimension 0 \cite{Buchalla:2013eza}. Actually, they also have got canonical dimension 0 because these Goldstone bosons are always accompanied by the vev in the denominator, $\varphi/v$. Consequently, $[u(\varphi/v),\, U(\varphi/v),\, v]_{\hat d} = 0$. Hence, vertices with one Goldstone leg may have the same chiral dimension as another one with a much higher number of them, just like the pion case in \chpt . 
The same assignment is done for the Higgs field, $[h/v]_{\hat d}= 0$, since no assumption is done about how this scalar field is coupled to the SM, keeping the model-independent essence ahead. Hence, every operator in the EWET should have an indefinite number of Higgs bosons coupled to it \cite{Grinstein:2007iv}. This fact is manifest whenever an operator is defined: the Wilson coefficient of a given operator must be understood as a polynomial of $h/v$, like
\bel{eq:wilsonhv}
c\,\mO\, \longrightarrow\, c(h/v)\,\mO \qquad \mbox{with}\quad c(h/v)= c_0 + c_1 (h/v) + c_2 (h/v)^2\ldots
\ee
Naturally, one can analyze some scenarios where the Higgs is weakly-coupled and the series expansion is suppressed by some weak coupling carrying some chiral dimension.

Since this power counting is determined by the low-energy momenta expansion, it is straightforward that derivatives carry chiral dimension 1, as well as the bosonic light masses, due to $p^2_{W,Z,h}=m^2_{W,Z,h}$. Actually, this consideration forces the gauge couplings to bring the same chiral dimension, since $m_W=g\,v/2$ and $m_Z= \sqrt{g^2+g'^2}\,v/2\,$. Therefore, $[\partial_\mu, \, m_{W,Z,h},\,g,\,g',\,g_S]_{\hat d} = 1$. In addition, all the gauge fields of the Yang-Mills Lagrangian must carry the same chiral dimension because they appear in the covariant derivative, $[\hat W_\mu,\,\hat B_\mu,\,\hat X_\mu,\,\hat G_\mu,\,D_\mu,\,d_\mu,\,\nabla_\mu]_{\hat d} = 1$, and hence their strength tensor fields, chiral dimension 2.

The power counting of fermionic fields works differently than the canonical dimension: they weight as $[\psi,\, \xi]_{\hat d}=1/2$ when considered at low momenta. Fermion masses and thus the Yukawas bring chiral dimension 1, $[m_\xi, \mY]_{\hat d}=1$, introduced through the spurion field. 

Notice that, up to this point, we have analyzed all the terms included in the SM Lagrangian, as can be checked in eq.~(\ref{eq:L_SMinL_EWET}), with the exception of the ones related to the scalar sector, corresponding to eqs.~(\ref{eq:LscalarEWET}) and eq.~(\ref{eq:Fhu_V}). The quadratic derivative and the mass terms in the Higgs Lagrangian require the Higgs potential to bring chiral dimension 2 as well, in order to be consistent. This implies that the coefficients of the series expansion of $V(h/v)$ keep this dimension too, being $[V(h/v),\, c_n^{(V)}]_{\hat d}=2$. The same reasoning also yields $[\mF_u(h/v)]_{\hat d}=0$.  

Equation~(\ref{eq:chiraldim}) reads that the presence of a chiral loop in a Feynman diagram makes the chiral counting gain two units. Yet, amplitudes including loops are suppressed by the geometrical factor times two inverse powers of the vev, $1/(4\pi v)^2$, in order to compensate the gained powers of momenta. This sets the electroweak chiral scale at $\Lambda_{\rm EWET} = 4\pi v \sim 3\,{\rm TeV}$. 

Additionally, operators may be also suppressed by some different new physics scale,\footnote{The interplay between these two scales is model-dependent, even so one is able to perform a double expansion taking into account these two scales \cite{Buchalla:2016bse}.} $\Lambda_{\rm NP}$, hidden in some local interaction. As an example, let's consider a two-fermion scattering. This interaction may arise from some local four-fermion operator generated by the exchange of some heavier states at short-distance. Yet, this process would have a factor $g^2_{\rm NP} / \Lambda^2_{\rm NP}$ and thus a two unit chiral suppression. In order to account for this situation, it is necessary to incorporate an extra chiral dimension whenever there is a fermion bilinear which origin is set beyond the SM scope, provided we assume that SM fermions are weakly-coupled to the strong sector \cite{Buchalla:2013eza}. Notice that this is the same underlying reason that generates the Yukawas to bring chiral dimension 1 and thus it does not apply to the fermion kinetic term. In short, fermion bilinears related to some new-physics coupling contribute to the power counting as $[\bar \eta\,\Gamma\,\zeta,\, J_\Gamma^f]_{\hat d} = 2$. 

Finally, the custodial symmetry-breaking spurion incorporates one power to the chiral counting too, $[\mT]_{\hat d}=1$, and then the operators incorporating this term are shifted to higher orders in the EWET Lagrangian \cite{Morales:94}. These structures are generated through loops that introduce $B_\mu$ internal fields. Since these fields are carriers of the $g'$ coupling, they get this extra dimension to the counting, provided that new physics conserves custodial symmetry at first approximation too.  Nevertheless, this consideration was not present in the original studies of the Higgsless EWET \cite{Longhitano:1980iz,Longhitano:1980tm}. As a consequence, the operator $\bra u_\mu \mT \ket_2\bra u^\mu \mT\ket_2$ carried chiral dimension 2, corresponding to the leading order Lagrangian at the same level than the SM operators, while the phenomenology establishes hard bounds over this operator. Fortunately, the counting of eq.~(\ref{eq:power_counting}) shifts this operator to higher orders.

\section{The EWET Lagrangian}
The power counting set in eq.~(\ref{eq:power_counting}) lets us organize the theory according to an ordered hierarchy of operators. Hence, now it is possible to distinguish which operators are dominant or subdominant depending on their chiral dimension assignment. Consequently, the first operators to be considered are the ones that belong to the leading order (LO) Lagrangian, $\mL_2$, which corresponds to chiral dimension 2.

As explained in section \ref{sec:SMinEWET}, all the SM Lagrangian is included in the EWET LO Lagrangian in eq.~(\ref{eq:L_SMinL_EWET}), as it is required from a well-defined EFT within the electroweak energy spectrum. Furthermore, these are not the only terms to be included in the $\hat d=2$ Lagrangian, since the chiral counting allows the SM scalar sector Lagrangian to be enlarged and generalized as in eq.~(\ref{eq:LscalarEWET}). To sum up, the SM Lagrangian together with the extended scalar sector form the leading order EWET Lagrangian. 

However, we are not going to focus so much in the first order Lagrangian and its operators. Beyond the extended scalar sector considerations, it essentially contains SM information. New physics searches require to expand this theory to higher orders so that the validity range of the theory is postponed to higher energies and it gets closer to the threshold, $\Lambda_{\rm EW} = 4\pi v$. Additionally, the resonance model to be studied along this work (see chapter \ref{ch:resonancetheory}) is only able to be analyzed if the EWET is considered until next-to-leading order (NLO), corresponding with the chiral dimension 4 Lagrangian.

NLO amplitudes receive contributions from two different sources. They are generated both by loop diagrams of chiral dimension 2 operators \cite{Guo:2015isa,Alonso:2015fsp,Alonso:2016oah,Espriu1,Espriu2,Espriu3,Dobado,Dobado2,Herrero,Filipuzzi:2012bv,Gavela} and by tree-level diagrams with one vertex corresponding to a chiral dimension 4 operator and the rest of them being dimension 2, as eq.~(\ref{eq:chiraldim}) states. The same idea applies for even higher orders: a given order Lagrangian is filled with tree-level interactions involving high chiral order operators, many loops diagrams of low chiral operators and all the possibilities in between. This $\hat d=4$ Lagrangian was studied some decades ago for the bosonic sector in an analogous Higgsless EFT \cite{Longhitano:1980iz,Longhitano:1980tm}. In this work, we extend this analysis and perform a full study of the EWET including all the SM particles for one single fermion generation, including one lepton and one quark family.

The renormalization in the EWET is not performed in the same way as the SM. Renormalization in the EWET occurs order by order. Contrary to the leading order amplitudes which involves just tree-level interactions, there are divergences generated by the loops at NLO \cite{Alonso:2013hga}. These divergent terms are absorbed in the Wilson-coefficients of the tree-level operators involving $\hat d=4$ local structures. At the next order, the process continues and new loop divergent contributions are hidden in new arising counterterms, and so on and so forth.

In the following tables, we detail the full set of operators\footnote{Some other works include a different number of operators (generally larger) due to different power counting considerations \cite{Alonso:2012px,Buchalla:2013rka,Buchalla:2012qq}.} with chiral dimension $\hat d = 4$. The collection of operators is divided into three different groups according to the number of fermion bilinears: 
\begin{itemize}
	\item 15 (12 + 3) operators with only bosonic fields collected in one subset (table \ref{tab:bosonic-Op4}),
	\item 21 (15 + 6) operators with one fermionic bilinear and bosonic fields collected in two subsets (table \ref{tab:2fermion}),
	\item 32 (25 + 7) operators with two fermionic bilinears collected in three subsets (table \ref{tab:4fermion}).	
\end{itemize}
%

\begin{table}[h!]
	\begin{center}
		\renewcommand{\arraystretch}{1.8}
		\begin{tabular}{|c|c|c|c|}
			\hline
			\multicolumn{4}{|c|}{Bosonic $P$-even operators, $\mO_i$}\\
			\hline
			$1$  &
			$\Frac{1}{4}\bra {f}_+^{\mu\nu} {f}_{+\, \mu\nu}
			- {f}_-^{\mu\nu} {f}_{-\, \mu\nu}\ket_2$ &
			$7$  &
			$\Frac{(\partial_\mu h)(\partial_\nu h)}{v^2} \,\bra u^\mu u^\nu \ket_2$
			\\ [1ex]
			\hline
			$2$  &
			$ \Frac{1}{2} \bra {f}_+^{\mu\nu} {f}_{+\, \mu\nu}
			+ {f}_-^{\mu\nu} {f}_{-\, \mu\nu}\ket_2$ &
			$8$ &
			$\Frac{(\partial_\mu h)(\partial^\mu h)(\partial_\nu h)(\partial^\nu h)}{v^4}$
			\\ [1ex]
			\hline
			$3$  &
			$\Frac{i}{2}  \bra {f}_+^{\mu\nu} [u_\mu, u_\nu] \ket_2$ & 
			$9$ &
			$\Frac{(\partial_\mu h)}{v}\,\bra f_-^{\mu\nu}u_\nu \ket_2$
			\\ [1ex]
			\hline
			$4$  &
			$\bra u_\mu u_\nu\ket_2 \, \bra u^\mu u^\nu\ket_2 $  &
			$10$ & 
			$\bra \mT u_\mu\ket_2\bra \mT u^\mu\ket_2$  
			\\ [1ex]
			\hline
			$5$  &
			$  \bra u_\mu u^\mu\ket_2\bra u_\nu u^\nu\ket_2$ &
			$11$ & 
			$ \hat{X}_{\mu\nu} \hat{X}^{\mu\nu}$ 
			\\ [1ex]
			\hline
			$6$ &
			$\Frac{(\partial_\mu h)(\partial^\mu h)}{v^2}\,\bra u_\nu u^\nu \ket_2$ &
			$12$ & 
			$\bra \hat G_{\mu\nu}\,\hat G^{\mu\nu} \ket_3 $
			\\ [1ex]
			\hline
			\multicolumn{3}{c}{} 
			\\ [-2ex]
			\hline
			\multicolumn{4}{|c|}{Bosonic $P$-odd operators, $\widetilde \mO_i$}\\
			\hline		
			$1$ &  $\Frac{i}{2} \bra {f}_-^{\mu\nu} [u_\mu, u_\nu] \ket_2$ &
			$2$ & $\bra {f}_+^{\mu\nu} {f}_{-\, \mu\nu} \ket_2 $
			\\ [1ex]
			\hline
			$3$ &
			$\Frac{(\partial_\mu h)}{v}\,\bra f_+^{\mu\nu}u_\nu \ket_2$ 
			\\ [1ex]
			\cline{1-2}			
			\multicolumn{3}{c}{}
		\end{tabular}
	\end{center}
	\vspace*{-1.cm}
	\caption{\small
		$CP$-invariant bosonic operators of the $\cO(p^4)$ EWET Lagrangian, $\mO_i\,( \widetilde \mO_i)$, where $i$ is the operator assigned number in the set.
		$P$-even ($P$-odd) operators are shown in the upper (lower) block.}
	\label{tab:bosonic-Op4}
\end{table}
%
\begin{table} [!b] 
	\begin{center}
		\renewcommand{\arraystretch}{1.8}
		\begin{tabular}{|c|c|c|c|c|c|c|}
			\hline
			\multicolumn{7}{|c|}{Two-fermion operators --- leptons or quarks} \\
			\hline
			\multicolumn{7}{c}{} \\	[-3ex] 
			\cline{1-4} \cline{6-7}
			\multicolumn{4}{|c|}{$P$-even, $\mO^{\psi^2_f}_i$} &  &
			\multicolumn{2}{|c|}{$P$-odd, $\widetilde \mO^{\psi^2_f}_i$}
			\\   \cline{1-4} \cline{6-7}
			1  & 
			$\bra J_S \ket_2 \bra u_\mu u^\mu \ket_2$  &
			5 & $\displaystyle\frac{\partial_\mu h}{v} \, \bra u^\mu J_P \ket_2 $ & &
			1 &
			$\bra J_T^{\mu \nu} f_{- \,\mu\nu} \ket_2 $			
			\\  [1ex] \cline{1-4} \cline{6-7}
			2 & 
			$ i \,  \bra J_T^{\mu\nu} \left[ u_\mu, u_\nu \right] \ket_2$ & 
			6 & 
			$\bra J_A^\mu \ket_2 \bra u_\mu \mathcal{T} \ket_2 $ &  &
			2 &
			$\displaystyle\frac{\partial_\mu h}{v} \, \bra u_\nu J^{\mu\nu}_T \ket_2 $
			\\  [1ex] \cline{1-4} \cline{6-7} 
			3 &
			$\bra J_T^{\mu \nu} f_{+ \,\mu\nu} \ket_2 $ &
			7 &
			$\Frac{(\partial_\mu h) (\partial^\mu h)}{v^2} \bra J_S\ket_2 $ & &
			3 &
			$\bra J_V^\mu \ket_2 \bra u_\mu \mathcal{T} \ket_2 $
			\\  [1ex] \cline{1-4} \cline{6-7}
			4 &
			$\hat{X}_{\mu\nu} \bra J_T^{\mu \nu} \ket_2 $ &
			8 &
			$\bra \hat G_{\mu\nu} J_{T}^{8\,\mu\nu} \ket_{2,3} \quad(\dagger)$
			\\  [1ex] \cline{1-4} 
			\multicolumn{7}{c}{} \\ [-4ex]
		\end{tabular}
		\caption{\small
			$CP$-invariant two-fermion operators of the $\mO (p^4)$ EWET Lagrangian, $\mO_i^{\psi^2_f}\, (\widetilde \mO_i^{\psi^2_f})$, where $i$ is the operator assigned number in the set and $f= l,\,q$ refers to the type of the fermion bilinear. There are two subsets of two-fermion operators: 1) two-fermion operators with a leptonic bilinear and 2) two-fermion operators with a quark bilinear. P-even (P-odd) operators are shown in the left (right) block. $(\dagger)$ Operator does not exist for a leptonic bilinear.}
		\label{tab:2fermion}
	\end{center}
\end{table}
%
%
\begin{table}[b] 
	\renewcommand{\arraystretch}{1.8}
	\begin{center}
		\begin{tabular}{|c|c|c|c|c|c|c|}
			\multicolumn{7}{c}{}\\ [5ex]
			\hline
			\multicolumn{7}{|c|}{Four-fermion operators --- lepton-lepton bilinears} \\
			\hline
			\multicolumn{7}{c}{} \\	[-3ex] 
			\cline{1-4} \cline{6-7}
			\multicolumn{4}{|c|}{$P$-even, $\mO^{\psi^2_l\psi^2_l}_i$} & &
			\multicolumn{2}{|c|}{$P$-odd, $\widetilde \mO^{\psi^2_l\psi^2_l}_i$}
			\\   \cline{1-4} \cline{6-7}
			%
			1 &
			$\bra J_{S} \ket_2 \bra  J_{S} \ket_2 $ &
			4 &
			$\bra J_A^\mu\ket_2 \bra J_{A,\mu}^{\phantom{\mu}}\ket_2 $ & &
			1 &
			$\bra J_V^\mu\ket_2 \bra J_{A,\mu}^{\phantom{\mu}}\ket_2 $ 			
			\\  [1ex] \cline{1-4} \cline{6-7}
			2 & 
			$\bra J_{P} \ket_2 \bra  J_{P} \ket_2 $ & 
			5 &
			$\bra J^{\mu\nu}_{T} \ket_2 \bra J_{T\,\mu\nu}^{\phantom{\mu}} \ket_2 $ 	
			\\  [1ex] \cline{1-4} 
			3 &
			$\bra J_V^\mu\ket_2 \bra J_{V,\mu}^{\phantom{\mu}}\ket_2 $ 
			\\  [1ex]  \cline{1-2}
			\multicolumn{7}{c}{} \\	
\end{tabular}
\end{center}
\end{table}
			%
%
\begin{table}[tbh] 
	\renewcommand{\arraystretch}{1.8}
		\begin{center}
			\begin{tabular}{|c|c|c|c|c|c|c|}			
			\hline
			\multicolumn{7}{|c|}{Four-fermion operators --- lepton-quark bilinears} \\
			\hline
			\multicolumn{7}{c}{} \\	[-3ex] 
			\cline{1-4} \cline{6-7}
			\multicolumn{4}{|c|}{$P$-even, $\mO^{\psi^2_l\psi^2_q}_i$} & &
			\multicolumn{2}{|c|}{$P$-odd, $\widetilde \mO^{\psi^2_l\psi^2_q}_i$}
			\\   \cline{1-4} \cline{6-7}
			%
			1 &
			$\bra J_{S} J_{S} \ket_2 $ &
			6 &
			$\bra J_A^\mu J_{A,\mu}^{\phantom{\mu}}\ket_2 $ & &
			1 &
			$\bra J_V^{\mu\,(l)} J_{A,\mu}^{(q)}\ket_2 $ 			
			\\  [1ex] \cline{1-4} \cline{6-7}
			2 & 
			$\bra J_{P} J_{P} \ket_2 $ & 
			7 &
			$\bra J_V^\mu\ket_2 \bra J_{V,\mu}^{\phantom{\mu}}\ket_2 $  & &
			2 &
			$\bra J_A^{\mu\,(l)} J_{V,\mu}^{(q)}\ket_2 $
			\\  [1ex] \cline{1-4} \cline{6-7}
			3 &
			$\bra J_{S} \ket_2 \bra  J_{S} \ket_2 $ & 
			8 &
			$\bra J_A^\mu\ket_2 \bra J_{A,\mu}^{\phantom{\mu}}\ket_2 $ & &
			3 &
			$\bra J_V^{\mu\,(l)}\ket_2 \bra J_{A,\mu}^{(q)}\ket_2 $
			\\  [1ex] \cline{1-4} \cline{6-7}
			4 &
			$\bra J_{P} \ket_2 \bra  J_{P} \ket_2 $ &
			9 &
			$\bra J^{\mu\nu}_{T} J_{T\,\mu\nu}^{\phantom{\mu}} \ket_2 $ & &
			4 &
			$\bra J_A^{\mu\,(l)}\ket_2 \bra J_{V,\mu}^{(q)}\ket_2 $
			\\  [1ex]  \cline{1-4} \cline{6-7}
			5 &
			$\bra J_V^\mu J_{V,\mu}^{\phantom{\mu}}\ket_2 $ &
			10 &
			$\bra J^{\mu\nu}_{T} \ket_2 \bra J_{T\,\mu\nu}^{\phantom{\mu}} \ket_2 $
			\\ [1ex] \cline{1-4}
			\multicolumn{4}{c}{} \\	
			%
			%
%
			\hline
			\multicolumn{7}{|c|}{Four-fermion operators --- quark-quark bilinears} \\
			\hline
			\multicolumn{7}{c}{} \\	[-3ex] 
			\cline{1-4} \cline{6-7}
			\multicolumn{4}{|c|}{$P$-even, $\mO^{\psi^2_q\psi^2_q}_i$} & &
			\multicolumn{2}{|c|}{$P$-odd, $\widetilde \mO^{\psi^2_q\psi^2_q}_i$}
			\\   \cline{1-4} \cline{6-7}
			%
			1 &
			$\bra J_{S} J_{S} \ket_2 $ &
			6 &
			$\bra J_A^\mu J_{A,\mu}^{\phantom{\mu}}\ket_2 $ & &
			1 &
			$\bra J_V^\mu J_{A,\mu}^{\phantom{\mu}}\ket_2 $ 			
			\\  [1ex] \cline{1-4} \cline{6-7}
			2 & 
			$\bra J_{P} J_{P} \ket_2 $ & 
			7 &
			$\bra J_V^\mu\ket_2 \bra J_{V,\mu}^{\phantom{\mu}}\ket_2 $  & &
			2 &
			$\bra J_V^\mu\ket_2 \bra J_{A,\mu}^{\phantom{\mu}}\ket_2 $
			\\  [1ex] \cline{1-4} \cline{6-7}
			3 &
			$\bra J_{S} \ket_2 \bra  J_{S} \ket_2 $ & 
			8 &
			$\bra J_A^\mu\ket_2 \bra J_{A,\mu}^{\phantom{\mu}}\ket_2 $ 
			\\  [1ex]  \cline{1-4}
			4 &
			$\bra J_{P} \ket_2 \bra  J_{P} \ket_2 $ &
			9 &
			$\bra J^{\mu\nu}_{T} J_{T\,\mu\nu}^{\phantom{\mu}} \ket_2 $
			\\  [1ex]  \cline{1-4}
			5 &
			$\bra J_V^\mu J_{V,\mu}^{\phantom{\mu}}\ket_2 $ &
			10 &
			$\bra J^{\mu\nu}_{T} \ket_2 \bra J_{T\,\mu\nu}^{\phantom{\mu}} \ket_2 $
			\\ [1ex] \cline{1-4}
			\multicolumn{4}{c}{} \\	[-3ex]			
\end{tabular}
\caption{\small
	$CP$-invariant four-fermion operators of the $\mO (p^4)$ EWET Lagrangian, $\mO_i^{\psi^4_{f\,f'}}\, (\widetilde \mO_i^{\psi^4_{f\,f'}})$, where $i$ is the operator assigned number in the set and $f,\,f'=l,\,q$ refers to the type of the fermion bilinears. There are three subsets of four-fermion operators: 1) four-fermion operators with two leptonic bilinears, shown in the upper block; 2) four-fermion operators with one leptonic bilinear and one quark bilinear, shown in the middle block. In this subset, the leptonic bilinear, $(L)$, is always written before the quark bilinear, $(Q)$, although indices may be omitted. 3) Four-fermion operators with two quark bilinears, shown in the lower block.	P-even (P-odd) operators are shown in the left (right) side of each block.}
\label{tab:4fermion}
\end{center}
\end{table}
%

\clearpage
\newpage

All the listed operators are invariant under the $CP$ discrete symmetry, as already mentioned. However, we make an explicit distinction on denoting operators that are $C$ and $P$ even, $\mO$, from operators that are $C$ and $P$ odd, $\widetilde \mO$, with the incorporation of $\,\sim\,$ over both the operator and their Wilson coefficients in the second case. Notice too that the trace $\bra \cdot \ket_2$ refers to the custodial symmetry group, $SU(2)_{L+R}$, and $\bra \cdot \ket_3$ to the color group, $SU(3)_C$.



Hence, the full Lagrangian of the chiral dimension $\hat d = 4$ local operators with one lepton and one quark family is
\bear \label{eq:L4EWET}
\mL_4 & = & \mL_4^{\rm bosonic} \ + \ \mL_4^{\psi^2_l} \ + \ \mL_4^{\psi^2_q}\ + \ \mL_4^{\psi^2_l \psi^2_l} \ + \ \mL_4^{\psi^2_l \psi^2_q}\ + \ \mL_4^{\psi^2_q \psi^2_q} \nn\\
& = &
 \left( \sum_{i=1}^{12} \mF_i\, \mO_i \,+\, \sum_{i=1}^{3} \widetilde \mF_i\, \widetilde \mO_i \right) \nn\\
& + & \left( \sum_{i=1}^{7} \mF_i^{\psi^2_l}\, \mO_i^{\psi^2_l} \,+\,
 \sum_{i=1}^{3} \widetilde \mF_i^{\psi^2_l}\, \widetilde \mO_i^{\psi^2_l} \right) \,+\, \left( \sum_{i=1}^{8} \mF_i^{\psi^2_q}\, \mO_i^{\psi^2_q} \,+\,
 \sum_{i=1}^{3} \widetilde \mF_i^{\psi^2_q}\, \widetilde \mO_i^{\psi^2_q}\right) \nn\\ 
& + & \left(\sum_{i=1}^{5} \mF_i^{\psi^2_l\,\psi^2_l}\, \mO_i^{\psi^2_l\,\psi^2_l} \,+\,
 \sum_{i=1}^{1} \widetilde \mF_i^{\psi^2_l\,\psi^2_l}\, \widetilde \mO_i^{\psi^2_l\,\psi^2_l} \right) \,+\, 
 \left( \sum_{i=1}^{10} \mF_i^{\psi^2_l\,\psi^2_q}\, \mO_i^{\psi^2_l\,\psi^2_q} \right. \nn\\
& + & \left. \sum_{i=1}^{4} \widetilde \mF_i^{\psi^2_l\,\psi^2_q}\,\widetilde \mO_i^{\psi^2_l\,\psi^2_q} \right)  \,+\,
 \left( \sum_{i=1}^{10} \mF_i^{\psi^2_q \psi^2_q}\, \mO_i^{\psi^2_q \psi^2_q} \,+\, \sum_{i=1}^{2} \widetilde\mF_i^{\psi^2_q \psi^2_q}\, \widetilde \mO_i^{\psi^2_q \psi^2_q} \right) \,. \nn\\
\eear
where all the Wilson coefficients must be understood as functions of the Higgs field,  $\mF_i^j=\mF_i^j(h/v)$. We will also refer to these parameters as low-energy coupling constants or simply low-energy constants (LECs).

Building an $\mO(p^4)$ operator is a straightforward procedure: one only has to combine some of the building blocks of eq.~({\ref{eq:buildingblocks}) and eq.~(\ref{eq:buildingblocksfermion}) in order to form $\hat d=4$ structures according to the power counting of eq.~(\ref{eq:power_counting}), provided the outcome is CP invariant and all the Lorentz indices are contracted. Nevertheless, the configuration of a basis of the order $\mO_{\hat d}(p^4)$ Lagrangian, $\mL_4$, is highly non-trivial since the full collection of all the possible valid $\hat d=4$ structures is redundant. One aims to obtain the minimum set of non-related operators and keep the exhaustiveness at the same time. 
	
For example, it is possible to construct the bosonic operator $\bra u_\mu u_\nu u^\mu u^\nu \ket_2$. However, using a basic $SU(2)$ relation one finds that
\begin{align} \label{eq:examplesimplification}
\bra u_\mu u_\nu u^\mu u^\nu \ket_2 \,&=\, \bra u_\mu u_\nu \ket_2 \bra u^\mu u^\nu \ket_2 \,-\, \frac{1}{2}\, \bra u_\mu u^\mu \ket_2\bra u_\nu u^\nu \ket_2  \nn\\
& =\, \mO_4 - \frac{1}{2}\, \mO_5\,.
\end{align}
Hence, this bosonic operator is redundant. Notice that one can decide which of the three involved operators is the one to be removed. As a consequence, the particular set of operators conforming the basis is a convention, while the total number of them is fixed. 

In the appendix \ref{app:simp}, the mathematical derivation of eq.~(\ref{eq:examplesimplification}) is found, as well as many relations and tools for operator simplification,\ie Cayley-Hamilton equations, $SU(n)$ algebraic relations, equations of motion, field redefinitions and Fierz identities.
Special mention is made to the operators 
\be \label{eq:order3}
\bra \mT J_S^{(f)} \ket\,, \quad \bra u_\mu J_V^{\mu\,(f)}\ket \,, \quad \bra u_\mu J_A^{\mu\,(f)}\ket \,,
\ee
for both leptons and quarks, $f=l,q$. These operators, not listed in the operator basis, are counted as chiral dimension $\hat d=3$, according to eq.~(\ref{eq:power_counting}). However, they can be reabsorbed in the $\mL_2$ and $\mL_4$ Lagrangian, eq.~(\ref{eq:L_SMinL_EWET}) and eq.~(\ref{eq:L4EWET}), respectively; provided the proper redefinitions of the Yukawa coupling $\mY$ and the electroweak gauge bosons $\hat W_\mu$ and $\hat B_\mu$.

The whole analysis performed in this chapter relies on many other previous related works, either related and not related to the electroweak theory, as already mentioned along this chapter. In order to clarify the different conventions and notation, we provide a dictionary so that the existing literature can be easily translated into the EWET, and vice versa. The collection of tables and operator relations is found in the appendix \ref{app:dictionary}.


\chapter{Resonance Theory } \label{ch:resonancetheory}

The resonance theory, also electroweak resonance theory, is a phenomenological Lagrangian of general heavy high-energy states coupled linearly to the light fields already present in the SM. We name these massive states as resonance states or, simply, resonances. Thus, the resonance theory contains the same particle content of the EWET, introduced in chapter \ref{ch:ewet}, together with the resonances. Furthermore, the same symmetry structure, eq.~(\ref{eq:ewetSymmetry}), is present in this EFT too, as it is expected, because the EWET is recovered once the massive states are decoupled. No additional symmetry group is imposed for the resonances since there are no hints for any of these states nor any new physical scale so far. Actually, LHC has not detected the presence of new physical particles below the 1 TeV threshold. This fact does not discourage the purpose of this work but enhances it because EFTs are proven to be one of the best frameworks to search new physics once there is a gap between the existing physics and the new one. 

In regard to this work, we refer to the resonance theory as the high-energy theory, compared to the EWET being the low-energy theory. The EWET is only valid for energies set below the masses of the resonances, which are expected to be of the same order as $\Lambda_{{\rm EWET}}=4\pi v$ or higher. However, the resonance theory is perfectly well-behaved and solid at these energies and its validity may be extended in the energy spectrum as far as the next heavy states non considered in the Lagrangian appear.

The resonance theory has its analogous partner in \chpt\ with the resonance chiral theory, R$\chi$T.
In this particular scenario, the first resonance states to be considered are the $\rho$ mesons. The two-flavor chiral Lagrangian, which is well-defined below the resonance masses, is indeed extended in order to account for these particles. Therefore, the techniques developed for this theory and the research performed several years ago in this field \cite{Ecker:1988te, Ecker:1989yg, Cirigliano:2006hb} are very useful for the construction of the electroweak resonance theory.

\section{The resonances}
The resonance states that we are going to incorporate embrace most of the general massive objects that can be coupled linearly to the EWET. They are classified according to the following criteria:
\begin{itemize}
	\item Bosonic (spin 0 or 1) or fermionic (spin 1/2),
	\item $C$ and $P$ discrete symmetries,\footnote{The heavy states are assumed to be eigenvalues of C and P. This is consistent with an underlying strongly-coupled theory that preserves these discrete symmetries.}
	\item $SU(2)_{L+R}$ custodial symmetry representation,
	\item $SU(3)_C$ color structure.
\end{itemize}

\subsection{Bosonic resonances}
In the bosonic case, if only the quantum numbers $J^{PC}$ are considered, the types of resonances are
\begin{align}\label{eq:bosonicRtypes}
&\mbox{scalar},\, S\quad (J^{PC}=0^{++}) \,, & \qquad  
&\mbox{pseudoscalar},\, P \quad (J^{PC}=0^{-+}) \,, & \nn\\
&\mbox{vector},\, V\quad (J^{PC}=1^{--})\,,& \qquad 
&\mbox{axial-vector},\, A \quad (J^{PC}=1^{++})\,.&
\end{align}
Details about the way these objects transform under $P$, $C$, and hermitian conjugation are found in the appendix \ref{app:CPtransformations}.
In addition, in the spin-1 case (vector and axial-vectors) the resonances are also sorted depending on their Lagrangian representation, Proca or Antisymmetric (or another), and it must be specified. However, this is not a property of the resonance but a Lagrangian representation. These considerations will be discussed in chapter \ref{ch:spin1}.

In regard to the way the bosonic resonances transform under the full symmetry group, eq.~(\ref{eq:ewetSymmetry}), we also find four different types of bosonic heavy states: $SU(2)_{L+R}$ singlets and triplets, and $SU(3)_C$ singlets or octets, transforming as 
\begin{align}
R_1^1\, & \longrightarrow\, R_{1}^1 \, , &
\qquad  \partial_\mu R^1_1 & \quad &
\nn\\
R^1_3 \, & \longrightarrow\, g_h^{\phantom{\dagger}}\, R^1_3\, g_h^\dagger\, ,&
\qquad \nabla_\mu R^1_3 & = \partial_{\mu} R^1_3 + [\Gamma_\mu, R^1_3]\,, &
\nn\\
R^8_{1}\,  & \longrightarrow\, g_C\,R^8_{1}\,g_C^{\dagger} \, , &
\qquad \nabla_\mu R^8_1 &  = \partial_{\mu} R^8_1 + i\, [\hat G_\mu, R^8_1]\,, &
\nn\\
R^8_3 \,  & \longrightarrow\, g_C\,g_h^{\phantom{\dagger}}\, R^8_3\, g_h^\dagger\,g_C^{\dagger}\, ,  & 
\qquad \nabla_\mu R^8_3 & = \partial_{\mu} R^8_3 +  i\, [\hat G_\mu, R^8_3] + [\Gamma_\mu, R^8_3]\,, &
\label{eq.R-transform}
\end{align}
where the representation of the resonances under these symmetries is explicit in the notation $R^{SU(3)}_{SU(2)}$ with $R$ standing for any of the resonance types of eq.~(\ref{eq:bosonicRtypes}). For instance, $A^8_1$ means an axial-vector resonance that is a color octet and a singlet under the custodial group. The covariant derivative of these objects is defined as usual, provided the connection of eq.~(\ref{eq:connection}). The normalization for the resonance triplets and octets is the following:
\begin{align}
R^{m}_3 &= \frac{1}{\sqrt{2}}\,\sigma^i\,R^m_{3,i}\,, & \,\, & \mbox{with}  & \quad  \bra R^{m}_3\,R^{m}_3 \ket_2 &= R^{m}_{3,i}\,R^{m}_{3,i}\,, & \,\, i=1,2,3\,, \quad m=1,8\,;
\nn\\
R_n^8 &=  T^a\, R_n^{8,a}\,, & \,\, & \mbox{with} & \quad  \bra R^8_{n}\,R^8_{n} \ket_3 &= \frac{1}{2} \,R^{8,a}_{n}\,R^{8,a}_{n}\,, & \,\,  a=1\ldots 8\,,\quad n=1,3\, .
\end{align}

\subsection{Fermionic resonances}
The characterization of the fermionic heavy states in the resonance theory is very similar to the light fermion ones in the EWET (see section \ref{sec:fermionsewet}). Fermionic resonances appear in doublet fields and transform as their light partners. Although it is possible to build different configurations for these massive particles, this is the only one that involves a single resonance that couples linearly to the EWET Lagrangian we are considering. Other exotic fermionic objects might arise when two or more resonances are considered simultaneously but they contribute to higher orders than the ones analyzed in this work.

Contrary to the bosonic heavy states, fermionic and antifermionic resonances are related through charge conjugation, $C$, and thus they are not eigenstates of this discrete symmetry. 
\begin{equation} \label{eq:fermionicRtypes}
\mbox{fermionic},\, \Psi\quad (J^P = 1/2^+)\,, \qquad\quad \mbox{anti-fermionic},\,\overline{\Psi}\quad (J^P = 1/2^-)\,. 
\end{equation}
Notice that all the properties related to $\overline \Psi$ are set by the action of $C$ over a given fermionic resonance, so it is pointless to make further distinctions among these two objects.

It is possible to build fermionic resonances with singlet and doublet representations under the custodial group $SU(2)_{L+R}$ and singlet and triplet under the color group, as well as the previous case,

\begin{align} \label{eq:Psitransform}
\Psi^1_1 \, &\longrightarrow\, g_X\,\Psi^1_1 \,,& \qquad \overline \Psi^1_1\, &\longrightarrow\, \overline \Psi^1_1 \,g_X^\dagger \,,& \nn\\
\Psi^1_2 \, &\longrightarrow\, g_h\,g_X\,\Psi^1_2 \,,& \qquad \overline \Psi^1_2\, &\longrightarrow\, \overline \Psi^1_2 \,g_X^\dagger\,g_h^\dagger \,,& \nn\\
\Psi^3_1 \, &\longrightarrow\, g_C\,g_X\,\Psi^3_1 \,,& \qquad \overline \Psi^3_1\, &\longrightarrow\, \overline \Psi^3_1 \,g_X^\dagger\,g_C^\dagger \,,& \nn\\
\Psi^3_2 \, &\longrightarrow\, g_C\,g_h\,g_X\,\Psi^3_2 \,,& \qquad \overline \Psi^3_2\, &\longrightarrow\, \overline \Psi^3_2 \,g_X^\dagger\,g_h^\dagger\,g_C^\dagger \,;& 
\end{align}
and the covariant derivative is defined as in eq.~(\ref{eq:dxi}). Nevertheless, we will show in the next section that the fermionic resonance notation can be simplified since most of the previous combinations do not contribute to the leading order high-energy Lagrangian.

Finally, it is convenient to introduce the fermionic resonances within a fermionic bilinear:\footnote{It is also possible to build bilinears involving two high-energy fermions. However, these objects involve two massive resonances and their low-energy implications are further suppressed.}
\begin{align} \label{eq:bilinearnotationR}
(\mmJ_S)_{mn} &= \bar\xi_n   \Psi_m + \overline{\Psi}_n  \xi_m\,, & 
(\mmJ_{S-})_{mn} &= i \left( \bar\xi_n   \Psi_m -  \overline{\Psi}_n  \xi_m\right)\,, & \nn\\
(\mmJ_P)_{mn} &= i\bar\xi_n  \gamma_5 \Psi_m +i \overline{\Psi}_n  \gamma_5  \xi_m\,, & 
(\mmJ_{P-})_{mn} &= i\left( i\bar\xi_n  \gamma_5 \Psi_m - i \overline{\Psi}_n  \gamma_5  \xi_m\right) \,, & \nn\\
(\mmJ_V^\mu)_{mn} &= \bar\xi_n \gamma^\mu \Psi_m + \overline{\Psi}_n \gamma^\mu \xi_m\,, & 
(\mmJ_{V-}^\mu)_{mn} &= i \left( \bar\xi_n \gamma^\mu \Psi_m -  \overline{\Psi}_n \gamma^\mu \xi_m\right) \,, & \nn\\
(\mmJ_A^\mu)_{mn} &= \bar\xi_n \gamma^\mu\gamma_5 \Psi_m + \overline{\Psi}_n \gamma^\mu\gamma_5 \xi_m\,, & 
(\mmJ_{A-}^\mu)_{mn} &= i\left( \bar\xi_n \gamma^\mu\gamma_5 \Psi_m -  \overline{\Psi}_n \gamma^\mu\gamma_5 \xi_m\right)\,, & \nn\\
(\mmJ_T^{\mu\nu})_{mn} &= \bar\xi_n  \sigma^{\mu\nu}  \Psi_m + \overline{\Psi}_n \sigma^{\mu\nu}  \xi_m\,, & 
(\mmJ_{T-}^{\mu\nu})_{mn} &= i
\epsilon^{\mu\nu\alpha\beta} \left( \bar\xi_n  \sigma_{\alpha\beta}  \Psi_m 
- i \overline{\Psi}_n \sigma_{\alpha\beta}  \xi_m\right) \,; 
&
\end{align}
which are analogous to the light fermionic bilinears defined in the EWET, in eq.~(\ref{eq:bilinearnotation}). Notice that in this case there are two different sets of them, with different behaviour under $C$. Further details are placed in the appendix \ref{app:CPtransformations}.

\section{High-energy Lagrangian}
Once introduced the resonance notation and properties, we are able to incorporate resonance operators in the action. Formally, one can split the contributions to the Lagrangian making an explicit distinction of operators involving resonances from those with light fields only, as follows:
\be \label{eq:formalRLagrangian}
\mL \quad = \quad \mL_{\rm Heavy\, Fields}[R, \Psi,\phi,\psi]\quad+\quad \mL_{\text{non-R}}[\phi,\psi]\, .
\ee
Notice that the non-resonance Lagrangian has the same operator content as the $\mL_{\rm EWET}$ but different Wilson coefficients, since they describe different EFTs. Recall that the EWET is no longer valid at the resonance mass energies.

The heavy-fields Lagrangian, $\mL_{\rm Heavy\, Fields}$, contains either interactions involving bosonic resonances, fermionic resonances and operators with both types, as well. However, in this work the single-resonance approximation is assumed.
This hypothesis states that the low-energy behavior of a given resonance theory is well-described by the only inclusion of the lightest resonances in the spectrum. Therefore, all the high-energy contributions are suppressed as powers of the momenta over the mass of the heavy states, $\mO(p/M_{\rm Res})$, when considered at low-momenta. Naturally, this consideration is also successful in R$\chi$T with the $\rho$ mesons for energies set below and at the same order as their masses. 
As a consequence, we are able to separate the Lagrangian contributions like 
\be \label{eq:formalRLagrangian2}
\mL_{\rm Heavy\, Fields}[R, \Psi,\phi,\psi] \quad = \quad \mL_{R}[R, \phi, \psi] \quad+\quad \mL_{\Psi}[\Psi, \phi, \psi]\,,
\ee
being $\mL_R$ and $\mL_{\Psi}$ the interaction Lagrangians involving one single bosonic and fermionic resonance, respectively.

Moreover, we also consider in the following subsections the spin-0 bosonic resonance Lagrangian and the spin-1 bosonic resonance Lagrangian with both the Proca-vector and the Antisymmetric formalism, in order to facilitate and clarify the reading. 

Keeping the spirit of the EWET NLO construction, we provide the minimum set of independent high-energy operators for any of the following Lagrangians. For this purpose, several techniques are applied (see appendix \ref{app:simp}).

\subsection{High-energy bosonic spin-0 Lagrangian}
The spin-0 resonance Lagrangian covers the tree-level interactions of one single bosonic scalar or pseudoscalar massive resonance with the rest of light fields of the EWET. This Lagrangian is formed by the kinetic and mass terms of the resonances and by all the resonance operators collected in $\bra R\,\chi_R \ket$, where $\chi_R$ is defined as a light-field tensor which groups all the possible tree-level interactions with a given resonance $R$. The Lagrangian reads
\begin{align}
\mL_{R^1_1} \, & = \, \Frac{1}{2}  \left(  \partial^\mu R_1^1\,  \partial_\mu R^1_1 \, -\, M_{R^1_1}^2\, (R_1^1)^2 \right)  \; +\;
R^1_1\, \chi_{R^1_1}^{\phantom{\mu}} \,,  
\nn\\
\mL_{R^1_3}\, & =\, \Frac{1}{2}\bra \nabla^\mu R^1_3\,  \nabla_\mu R^1_3 \, -\, M_{R^1_3}^2\, (R^1_3)^2\ket_2 \; +\; \bra R^1_3\, \chi_{R^1_3}\ket_2 \, , 
\nn\\
\mL_{R^8_1}\, & =\,  \bra  \nabla^\mu R^8_1\,  \nabla_\mu R^8_1 \, -\, M_{R_1^8}^{2}\, (R_1^8)^2 \ket_3  \; +\;
\bra R_1^8\, \chi_{R_1^8} \ket_3\,, 
\nn\\
\mL_{R^8_3}\, & =\, \bra \nabla^\mu R^8_3\,  \nabla_\mu R^8_3 \, -\, M_{R^8_3}^2\, (R^8_3)^2 \ket_{2,3} \; +\; \bra R^8_3\, \chi_{R^8_3} \ket_{2,3} \,, 
\label{eq:L0}
\end{align}
with $R^m_n = S^m_n, P^m_n$ and being the tensors for the the scalar and pseudoscalar defined as
\begin{align}
\chi_{S_1^1} \, & = \,
\lambda_{hS_1}  v \, h^2   +\,
\Frac{c_{d}}{\sqrt{2}}\, \bra u_\mu u^\mu \ket_2 \,+\, \Frac{c^{S_1^1}_f}{\sqrt{2}}\, \bra J^f_S \ket_2\,, & \quad\!\!
\chi_{P_1^1}\, & =\,
\Frac{c^{P_1^1}_f}{\sqrt{2}}\,  \bra  J^f_P\ket_2 \, , &
\nn\\
\chi_{S^1_3} \,& =\, c^{S^1_3}_f\,  J^f_S \, ,&
\quad \!\!
\chi_{P^1_3}\, & =\,
c^{P^1_3}_f\, J^f_P \, +\,
d_P\, \Frac{(\partial_\mu h)}{v}\, u^\mu \, , &
\nn\\
\chi_{S_1^{8}}\, & =\,  \Frac{c^{S_1^8}}{\sqrt{2}}\,\bra J^8_{S} \ket_2 \,,    &
\quad \!\!
\chi_{P_1^{8}}\, & =\,  \Frac{c^{P_1^8}}{\sqrt{2}}\,\bra J^8_{P} \ket_2 \,,    &
\nn\\
\chi_{S^8_3} \,& =\,  c^{S^8_3}\,J^8_{S}   \, , &
\quad \!\!
\chi_{P^8_3}\, & =\,   c^{P^8_3}\,J^8_{P}  \,;& 
\label{eq:L0tensor}
\end{align}
where $f=l,q$ indicates the fermion type of bilinears. As a general consideration, a resonance operator including the bilinear $J_\Gamma^f$ must be understood as two operators: one with the bilinear made out of leptons and the other with quarks. For the $J_\Gamma^8$ the fermion type index is omitted since it only applies for quarks. In addition, as well as the EWET LECs, the resonance couplings must be understood as polynomials of $h/v$:
\be
c_i^{R} \, =\,  \sum_{n=0}\, c_i^{R\, (n)}\; \left(\Frac{h}{v}\right)^n\, .
\ee

Notice that we have not introduced any power counting for the resonance theory. In order to know if one operator belongs to the LO high-energy Lagrangian it is necessary to integrate out the massive fields and then check whether the resulting operators belong to the LO or NLO EWET Lagrangians. If not, this means that the resonance operator is a high order operator and thus negligible at low-energies. All the operators included in eq.~(\ref{eq:L0tensor}) satisfy this consideration. Anyhow, the integrating out process will be properly analyzed when we match the high-energy theory with the low-energy one, in chapter \ref{ch:LECs}. A naive way to check if one resonance operator is LO in the high-energy theory consists in checking the chiral dimension of the tensor $\chi$. The interacting terms of this Lagrangian are found to be order $\mO(p^2)$, provided we assign to the coupling\footnote{This assignment will be justified in chapter \ref{ch:LECs}, too.} $\lambda_{hS_1}$ a chiral dimension $\hat d =2$. 

\subsection{High-energy bosonic spin-1 Proca Lagrangian} 
Opposite to spin-0 resonances, spin-1 heavy states carry some freedom with the selection of their Lagrangian representation. One may think that building the spin-1 high-energy Lagrangian with more than one different formalisms is redundant. However, it is not pointless because the usage of either one or the other formalism has been controversial in the existing literature and it is not clear which of them (if any) to use for this type of massive states. Moreover, an intelligent choice of the representation may lead to much simpler interactions and a cleaner phenomenology \cite{Ecker:1988te,Ecker:1989yg}.

The Proca-vector formalism, or just Proca formalism, describes the spin-1 heavy fields as 4-vector Proca fields, $\hat R_\mu$, being the most standard convention. We will use the hat $\wedge$ to make this representation choice explicit. The Proca spin-1 high-energy Lagrangian is formulated as follows:
\begin{align} \label{eq:L1proca}
\mL_{\hat R_1^1}  & =   -\Frac{1}{4}  \left(  \hat R^1_{1\,\mu\nu}\,  \hat R_1^{1\,\mu\nu}  \,-\, 2 M_{R_1^1}^2\,\hat R^1_{1\,\mu}\, \hat R_1^{1\,\mu} \right) \,+ \, \hat R^1_{1\,\mu}\, \chi_{\hat R_1^1}^{\mu}  \,+\,  \hat R^1_{1\,\mu\nu}\,\chi_{\hat R_1^1}^{\mu\nu} \,, 
\nn\\[1ex]
\mL_{\hat R^1_3}\, & = -\Frac{1}{4}  \bra  \hat R^1_{3\,\mu\nu}\,  \hat R^{1\,\mu\nu}_3  \,-\, 2 M_{R^1_3}^2\,\hat R^1_{3\,\mu}\, \hat R^{1\,\mu}_3 \ket_2 \, +\, \bra \hat R_{3\,\mu}^1 \, \chi_{\hat R^1_3}^{\mu}   \,+\,  \hat R^1_{3\,\mu\nu}\,\chi_{\hat R^1_3}^{\mu\nu}\ket_2 \,, 
\nn\\[1ex]
\mL_{\hat R^8_1}\, & =  -\Frac{1}{2}  \bra  \hat R_{1\,\mu\nu}^8\,  \hat R_1^{8\,\mu\nu}  \,-\, 2 M_{R_1^8}^2\,\hat R_{1\,\mu}^8\, \hat R_1^{8\,\mu} \ket_3 \,  + \,
\bra \hat R_{1\,\mu}^8\, \chi_{\hat R_1^8}^{\mu}  \,+\,  \hat R_{1\,\mu\nu}^8\,\chi_{\hat R_1^8}^{\mu\nu} \ket_3\,, 
\nn\\[1ex]
\mL_{\hat R^8_3}\, & = -\Frac{1}{2}  \bra  \hat R_{3\,\mu\nu}^8\,  \hat R^{8\,\mu\nu}_3  \,-\, 2 M_{R^8_3}^2\,\hat R_{3\,\mu}^8\, \hat R^{8\,\mu}_3 \ket_{2,3} \,+\,
\bra \hat R_{3\,\mu}^8\, \chi_{\hat R^8_3}^{\mu}  \,+\,  \hat R_{3\,\mu\nu}^8\,\chi_{\hat R^8_3}^{\mu\nu} \ket_{2,3} \,, 
\end{align}
with $\hat R^m_n = \hat V^m_n, \hat A^m_n$ and being $\hat R^m_{n\,\mu\nu} = \nabla_\mu \hat R^m_{n\,\nu} - \nabla_\nu \hat R^m_{n\,\mu}$. Note that we have split the interaction Lagrangian in light fields coupled to $\hat \chi^\mu_{\hat R}$ from those coupled to $\hat \chi^{\mu\nu}_{\hat R}$. However, it is always possible to rewrite all of them in terms of the first option by performing a partial integration. The resulting tensors are, respectively,   
\begin{align} \label{eq:proca-chimu}
\hat \chi_{\hat V^1_1}^{\mu} \, & = \,  \widetilde{c}_{\mathcal{T}}  \bra u^\mu \mathcal{T} \ket_2 \,+\,
\Frac{c_{{\hat{V}}^1_1}}{\sqrt{2}}
\, \bra J^{f\,\mu}_V\ket_2        \, ,&
\quad
\hat\chi_{\hat A_1^1}^{\mu}\, & =\,  c_{\mathcal{T}}  \bra u^\mu \mathcal{T} \ket_2 \,+\,
\Frac{  c^{{\hat{A}}^1_1}  }{\sqrt{2}}
\, \bra J^{f\,\mu}_A\ket_2  
\, , &
\nn\\[-5pt]
& \quad\; +\, \Frac{   \widetilde c^{{\hat{V}}^1_1}  }{\sqrt{2}}
\, \bra J^{f\,\mu}_A\ket_2   \,, &
\quad
& \quad +\,
\Frac{  \widetilde c^{{\hat{A}}^1_1}  }{\sqrt{2}}
\, \bra J^{f\,\mu}_V\ket_2 \,, & 
\nn\\[5pt]
\hat \chi_{\hat V^1_3}^{\mu} \,& =\,  c^{\hat{V}^1_3}\, J^{f\,\mu}_V  \,+\, \widetilde c^{\hat{V}^1_3}\, J^\mu_A       \, ,&
\quad
\hat \chi_{\hat A^1_3}^{\mu}\, & =\, c^{\hat{A}^1_3}\, J^{f\,\mu}_A  \,+\, \widetilde c^{\hat{A}^1_3}\, J^{f\,\mu}_V       \, , &
\nn\\[5pt]
\hat \chi_{\hat V_1^{8}}^\mu\, & =\,  \Frac{c^{{\hat{V}}_1^8}}{\sqrt{2}}
\, \bra J^{8\,\mu}_{V}\ket_2  \,+\,
\Frac{   \widetilde c^{{\hat{V}}_1^8}  }{\sqrt{2}}
\, \bra J^{8\,\mu}_{A}\ket_2       \,,&
\quad
\hat \chi_{\hat A_1^{8}}^\mu\, & =\, \Frac{  c^{{\hat{A}}_1^8}  }{\sqrt{2}}
\, \bra J^{8\,\mu}_{A}\ket_2  \,+\,
\Frac{  \widetilde c^{{\hat{A}}_1^8}  }{\sqrt{2}}
\, \bra J^{8\,\mu}_{V} \ket_2         \,,&
\nn\\[5pt]
\hat \chi_{\hat V^{8}_3}^\mu \,& =\, c^{\hat{V}^8_3}\, J^{8\,\mu}_{V}  \,+\, \widetilde c^{\hat{V}^8_3}\, J^{8\,\mu}_{A}       \,, &
\quad
\hat \chi_{\hat A^{8}_3}^\mu\, & =\,  c^{\hat{A}^8_3}\, J^{8\,\mu}_{A}  \,+\, \widetilde c^{\hat{A}^8_3}\, J^{8\,\mu}_{V}         \,,& 
\end{align}
\\
and
\begin{align} \label{eq:proca-chimunu}
\hat \chi_{\hat V^1_1}^{\mu\nu} & = 
f_{X}\, \hat X^{\mu\nu} 
\, +\, \Frac{c^{\hat V^1_1}_{0}}{\sqrt{2}}\,  \bra J_T^{f\,\mu\nu} \ket_2
\, , &
\quad
\hat \chi_{\hat A^1_1}^{\mu\nu}  & = 
\widetilde{f}_{X}\, \hat X^{\mu\nu} 
\, +\, \Frac{c^{\hat A^1_1}_{0}}{\sqrt{2}}\,  \bra J_T^{f\,\mu\nu} \ket_2
\, , &
\nn\\[5pt]
\hat \chi_{\hat V^1_3}^{\mu\nu}  & =
\Frac{f_{\hat V}}{2\sqrt{2}}\,  f_+^{\mu\nu}\, +\,
\Frac{i\, g_{\hat V}}{2\sqrt{2}}\, [u^\mu, u^\nu] + &
\quad 
\hat \chi_{\hat A^1_3}^{\mu\nu}  & =
\Frac{f_{\hat A}}{2\sqrt{2}}\,  f_-^{\mu\nu} \, + \, \Frac{\widetilde{f}_{\hat A}}{2\sqrt{2}}\, f_+^{\mu\nu}\,    &
\nn\\
& +\, \Frac{\widetilde{f}_{\hat V} }{2\sqrt{2}}\, f_-^{\mu\nu} \, +\, c_{0}^{\hat V^1_3}\, J_T^{f\,\mu\nu} \, + & 
\quad 
& +\, \Frac{i\, \widetilde{g}_{\hat A}}{2\sqrt{2}}\, [u^{\mu}, u^{\nu}]
\, +\,  \widetilde{c}_{0}^{\hat A^1_3}\,  J_{T}^{f\,\mu\nu}\,  &
\nn\\
& + \, \Frac{ \widetilde{\lambda}_1^{h \hat V} }{\sqrt{2}}\left[
(\partial^\mu h)\, u^\nu-(\partial^\nu h)\, u^\mu \right]
\, ,  & 
\quad
& +\, \Frac{ \lambda_1^{h\hat A} }{\sqrt{2}} \left[
(\partial^\mu h)\, u^\nu-(\partial^\nu h)\, u^\mu \right]\,, &
\nn\\[5pt]
\hat\chi^{\mu\nu}_{\hat V^8_1} & = \Frac{c_0^{\hat V_1^8}}{\sqrt{2}}\, \bra J^{8\,\mu\nu}_{T} \ket_2  \,+\, c_G\,\hat G^{\mu\nu}\,, &
\quad
\hat \chi^{\mu\nu}_{\hat A^8_1} & = \Frac{\widetilde c_0^{\hat A_1^8}}{\sqrt{2}}\, \bra J^{8\,\mu\nu}_{T} \ket_2   \,+\, \widetilde c_G\,\hat G^{\mu\nu}\,, &
\nn\\[5pt]
\hat \chi^{\mu\nu}_{\hat V^8_3} & =   c_0^{\hat V^8_3}\, J^{8\, \mu\nu}_{T}\,,   &
\quad
\hat \chi^{\mu\nu}_{\hat A^8_3} & =  \widetilde c_0^{\hat A^8_3}\, J^{8\,\mu\nu}_{T} \,,   & 
\end{align}
where $f=l,q$ is included in operators with both a lepton and a quark copy.

This resonance Lagrangian can be compared with some QCD models written in the Proca Lagrangian representation \cite{Ecker:1989yg,Op3-res-Lagrangian}, provided one removes color and P-odd operators from the action.

It should be also pointed out that we have not included the operators $\bra \hat V^\mu u_\mu \ket_2$
and $\bra \hat A^\mu u_\mu \ket_2$, which naively carry chiral dimension $\hat d =1$ light field structures interacting with the resonances (in contrast with the $\mO(p^2)$ operators of the previous equations). In the appendix \ref{app:simpresonances}, it is shown how this operators can be removed by performing appropriate field redefinitions.

\subsection{High-energy bosonic spin-1 Antisymmetric Lagrangian}
The antisymmetric formalism is also a well-established framework for studying the spin-1 fields, which utilization was very extended in the \chpt\ literature \cite{Ecker:1988te,Ecker:1989yg}. It describes the high-energy massive states as rank-2 antisymmetric tensors. Notice that the main difference between Proca and Antisymmetric remains in kinematics, since the resonances carry one and two Lorenz indices, respectively. Furthermore, antisymmetric objects are notated with no hat, $R_{\mu\nu}$, as opposite to Proca, $\hat R_\mu$. The technical details regarding this formalism are presented in the appendix \ref{app:spin1comparison}. 

The antisymmetric Lagrangian is found to be

\begin{align}
\mL_{R_1^1} \, &\! =  -\Frac{1}{2}  \left(  \partial^\lambda R^1_{1\,\lambda\mu}\,   \partial_\sigma R_1^{1\,\sigma\mu}  - \frac{1}{2} M_{R_1^1}^2\, R_{1\,\mu\nu}\,  R_1^{1\,\mu\nu} \right)\, + \,
R^1_{1\,\mu\nu}\, \partial^\mu \chi_{ R_1^1}^{\nu}  \,+\,   R^1_{1\,\mu\nu}\,\chi_{ R_1^1}^{\mu\nu}\,,
\nn\\[1ex]
\mL_{ R^1_3}\, &\! = -\Frac{1}{2} \, \bra   \nabla^\lambda R^1_{3\,\lambda\mu}\,  \nabla_\sigma R^{1\,\sigma\mu}_3  - \frac{1}{2} M_{R^1_3}^2\, R^1_{3\,\mu\nu}\,  R^{1\,\mu\nu}_3 \ket_2  \, + 
\nn\\
&\! \qquad\qquad + \,
\bra R^1_{3\,\mu\nu}\, \nabla^\mu \chi_{ R^1_3}^{\nu}   \,+\,   R^1_{3\,\mu\nu}\,\chi_{ R^1_3}^{\mu\nu}\ket_2 \,,  
\nn\\[1ex]
\mL_{ R^8_1}\, &\! =  -\,  \bra \nabla^\lambda R_{1\,\lambda\mu}^8\,   \nabla_\sigma R_1^{8\,\sigma\mu}  - \frac{1}{2} M_{R_1^8}^2\, R_{1\,\mu\nu}^8\,  R_1^{8\,\mu\nu} \ket_3  \, + 
\nn\\
&\! \qquad\qquad + \,
\bra  R_{1\,\mu\nu}^C\, \nabla^\mu \chi_{ R_1^8}^{\nu}  \,+\,   R_{1\,\mu\nu}^8\,\chi_{ R_1^8}^{\mu\nu} \ket_3 \,, 
\nn\\[1ex]
\mL_{ R^8_3}\, &\! = - \, \bra  \nabla^\lambda R_{3\,\lambda\mu}^8\, \nabla_\sigma  R^{8\,\sigma\mu}_3 - \frac{1}{2} M_{R^8_3}^2\, R_{3\,\mu\nu}^8\,  R^{8\,\mu\nu}_3 \ket_{2,3}   \, + 
\nn\\
&\! \qquad\qquad + \,
\bra   R_{\mu\nu}^C\, \nabla^\mu \chi_{ R^8_3}^{\nu}  \,+\,   R_{3\,\mu\nu}^C\,\chi_{ R^8_3}^{\mu\nu} \ket_{2,3} \, , 
\label{eq:L1anti}
\end{align}
with $ R^m_n =  V^m_n, A^m_n$. As well as in the Proca case, we make an explicit separation of the interaction tensors with one index, $\chi_R^\mu$, from the ones with two indices, $\chi_R^{\mu\nu}$. For sure, the first case is a subgroup of the second because they are always accompanied by a derivative. However, it is interesting to keep them apart for future considerations. The interaction tensors are, respectively,
\begin{align} \label{eq:antisym-chimu}
\chi_{V^1_1}^{\mu} \, & = \,  \widetilde{C}_{\mathcal{T}}  \bra u^\mu \mathcal{T} \ket_2 \,+\,
\Frac{C^{{{V}}^1_1}_f}{\sqrt{2}}
\, \bra J^{f\,\mu}_V\ket_2        \, ,&
\quad
\chi_{A_1^1}^{\mu}\, & =\,  C_{\mathcal{T}}  \bra u^\mu \mathcal{T} \ket_2 \,+\,
\Frac{  C^{{{A}}^1_1}_f  }{\sqrt{2}}
\, \bra J^{f\,\mu}_A\ket_2  
\, , &
\nn\\[-5pt]
& \quad\; +\, \Frac{   \widetilde C^{{{V}}^1_1}_f  }{\sqrt{2}}
\, \bra J^{f\,\mu}_A\ket_2   \,, &
\quad
& \quad +\,
\Frac{  \widetilde C^{{{A}}^1_1}_f  }{\sqrt{2}}
\, \bra J^{f\,\mu}_V\ket_2 \,, & 
\nn\\[5pt]
\chi_{V^1_3}^{\mu} \,& =\,  C^{{V}^1_3}_f\, J^{f\,\mu}_V  \,+\, \widetilde C^{{V}^1_3}_f\, J^{f\,\mu}_A       \, ,&
\quad
\chi_{A^1_3}^{\mu}\, & =\, C^{{A}^1_3}_f\, J^{f\,\mu}_A  \,+\, \widetilde C^{{A}^1_3}_f\, J^{f\,\mu}_V       \, , &
\nn\\[5pt]
\chi_{V_1^{8}}^\mu\, & =\,  \Frac{C^{{{V}}_1^8}}{\sqrt{2}}
\, \bra J^{8\,\mu}_{V}\ket_2  \,+\,
\Frac{   \widetilde C^{{{V}}_1^8}  }{\sqrt{2}}
\, \bra J^{8\,\mu}_{A}\ket_2       \,,&
\quad
\chi_{A_1^{8}}^\mu\, & =\, \Frac{  C^{{{A}}_1^8}  }{\sqrt{2}}
\, \bra J^{8\,\mu}_{A}\ket_2  \,+\,
\Frac{  \widetilde C^{{{A}}_1^8}  }{\sqrt{2}}
\, \bra J^{8\,\mu}_{V} \ket_2         \,,&
\nn\\[5pt]
\chi_{V^{8}_3}^\mu \,& =\, C^{{V}^8_3}\, J^{8\,\mu}_{V}  \,+\, \widetilde C^{{V}^8_3}\, J^{8\,\mu}_{A}       \,, &
\quad
\chi_{A^{8}_3}^\mu\, & =\,  C^{{A}^8_3}\, J^{8\,\mu}_{A}  \,+\, \widetilde C^{{A}^8_3}\, J^{8\,\mu}_{V}         \,,&
\end{align}
and\footnote{In \cite{Fingerprints} the LEC $\Lambda_1^{h A}$ is written in small letters for the antisymmetric formalism too, like $\lambda_1^{h A}$. }
\begin{align} \label{eq:antisym-chimunu}
\chi_{V^1_1}^{\mu\nu} & = 
F_{X}\, \hat X^{\mu\nu} 
\, +\, \Frac{C^{V^1_1}_{0\,f}}{\sqrt{2}}\,  \bra J_T^{f\,\mu\nu} \ket_2
\, , &
\quad
\chi_{ A^1_1}^{\mu\nu}  & = 
\widetilde{F}_{X}\, \hat X^{\mu\nu} 
\, +\, \Frac{C^{ A^1_1}_{0\,f}}{\sqrt{2}}\,  \bra J_T^{f\,\mu\nu} \ket_2
\, , &
\nn\\[5pt]
\chi_{ V^1_3}^{\mu\nu}  & =
\Frac{F_{ V}}{2\sqrt{2}}\,  f_+^{\mu\nu}\, +\,
\Frac{i\, G_{ V}}{2\sqrt{2}}\, [u^\mu, u^\nu] + &
\quad 
\chi_{ A^1_3}^{\mu\nu}  & =
\Frac{F_{ A}}{2\sqrt{2}}\,  f_-^{\mu\nu} \, + \, \Frac{\widetilde{F}_{ A}}{2\sqrt{2}}\, f_+^{\mu\nu}\,    &
\nn\\
& +\, \Frac{\widetilde{F}_{ V} }{2\sqrt{2}}\, f_-^{\mu\nu} \, +\, C_{0\,f}^{ V^1_3}\, J_T^{f\,\mu\nu} \, + & 
\quad 
& +\, \Frac{i\, \widetilde{G}_{ A}}{2\sqrt{2}}\, [u^{\mu}, u^{\nu}]
\, +\,  \widetilde{C}_{0\,f}^{ A^1_3}\,  J_{T}^{f\,\mu\nu}\,  &
\nn\\
& + \, \Frac{ \widetilde{\Lambda}_1^{h  V} }{\sqrt{2}}\left[
(\partial^\mu h)\, u^\nu-(\partial^\nu h)\, u^\mu \right]
\, ,  & 
\quad
& +\, \Frac{ \Lambda_1^{h A} }{\sqrt{2}} \left[
(\partial^\mu h)\, u^\nu-(\partial^\nu h)\, u^\mu \right]\,, &
\nn\\[5pt]
\chi^{\mu\nu}_{ V^8_1} & = \Frac{C_0^{ V_1^8}}{\sqrt{2}}\, \bra J^{8\,\mu\nu}_{T} \ket_2  \,+\, C_G\, G^{\mu\nu}\,, &
\quad
\chi^{\mu\nu}_{ A^8_1} & = \Frac{\widetilde C_0^{ A_1^8}}{\sqrt{2}}\, \bra J^{8\,\mu\nu}_{T} \ket_2   \,+\, \widetilde C_G\, G^{\mu\nu}\,, &
\nn\\[5pt]
\chi^{\mu\nu}_{ V^8_3} & =   C_0^{ V^8_3}\, J^{8\, \mu\nu}_{T}\,,   &
\quad
\chi^{\mu\nu}_{ A^8_3} & =  \widetilde C_0^{ A^8_3}\, J^{8\,\mu\nu}_{T} \,,   & 
\end{align}
where a sum over $f=l,q$ is implied.

Comparing the equations (\ref{eq:proca-chimu}) and (\ref{eq:antisym-chimu}), as well as equations (\ref{eq:proca-chimunu}) and (\ref{eq:antisym-chimunu}), one finds that the tensors $\chi^\mu$ and $\chi^{\mu\nu}$ are formally equivalent in both Proca and Antisymmetric formalisms. They contain the same operators but with different resonance couplings (small letters versus capital letters, respectively) since they represent two different EFTs. Nevertheless, it is crucial to point out that the particular tensor structure accompanied by a derivative is different for each formalism. On the one hand, the Proca tensor $\hat \chi^\mu_{\hat R}$ comes alone but the Proca resonance strength $ \hat R_{\mu\nu}$, which contains one derivative, couples to $\hat \chi_{\hat R}^{\mu\nu}$. On the other hand, resonances in the Antisymmetric formalism couple naturally to the two indices tensor $\chi_R^{\mu\nu}$ but demand one additional derivative in order to couple to $\chi_R^\mu$. Considering that all the light-field tensor structures have chiral dimension $\hat d=2$, those contributions that require one extra derivative in the Lagrangian get also one extra power to the counting. Therefore, attending to the naive counting for the light fields in the resonance Lagrangian,
\begin{itemize}
	\item $\hat \chi_{\hat R}^\mu$, in eq.~(\ref{eq:proca-chimu}), for the Proca Lagrangian; and
	$\chi_R^{\mu\nu}$, in eq.~(\ref{eq:antisym-chimunu}), for the Antisymmetric Lagrangian, count effectively as $\mO(p^2)$;
	\item  $\hat \chi_{\hat R}^{\mu\nu}$, in eq.~(\ref{eq:proca-chimunu}), for the Proca Lagrangian; and
	$\chi_R^{\mu}$, in eq.~(\ref{eq:antisym-chimu}), for the Antisymmetric Lagrangian, count effectively as $\mO(p^3)$.
\end{itemize} 

\subsection{High-energy fermionic Lagrangian}
The determination of the fermionic resonance Lagrangian is not as direct as for their bosonic heavy partners. 
The chiral nature of the resonance doublet structure lets the light and heavy fermions\footnote{Recall that other exotic representations for fermionic resonances are allowed but SM-like heavy fermions are the only ones that contribute to the physical amplitudes at NLO. Moreover, we will only consider interactions with one heavy resonance, as in the bosonic sector.} to mix naturally, while this situation does not happen in the bosonic sector.
Nonetheless, it is possible to diagonalize\footnote{
	In order to build the kinetic Lagrangian of fermionic resonances we embed the light, $\xi$, and the heavy, $\Psi$, fermionic fields in a doublet vector like
	\be \label{eq:baredoublet}
	V_B = \displaystyle \left( \begin{matrix} \xi_B \\[1ex] \Psi_B \end{matrix} \right)\,, \qquad 
	\overline V_B = \left( \begin{matrix} \overline \xi_B \\[1ex] \overline \Psi_B \end{matrix} \right)\,,
	\ee
	where the subindex $B$ stands for non-diagonalized bare fields. The most general form of the kinetic Lagrangian turns out to be
	\be \label{eq:fermionRkineticbare}
	\mL^{\rm kin,mass}_{\xi,\Psi} = \frac{i}{2} \left( \overline V_B \gamma^\mu\, A\, d_\mu V_B \,-\, \overline{(d_\mu V_B)} \gamma^\mu \, A\, V_B \right) \, - \, \overline V_B \, B \, V_B \,,
	\ee
	being $A$, $B$ non-diagonal orthogonal matrices that mix light and heavy fermionic resonance fields in a general way for both kinetic and mass terms, respectively. }
the fermion kinetic Lagrangian, following a standard procedure \cite{Escribano:2010wt}, and thus both light and heavy fermions are separable:
\be \label{eq:fermionRseparable}
\mL^{\rm kin,mass}_{\xi,\Psi} \quad=\quad \mL^{\rm kin,mass}_{\xi} \quad + \quad \mL^{\rm kin,mass}_{\Psi}\, .
\ee
Hence, the fermionic fields can be redefined so that one recovers the EWET kinetic Lagrangian for light fermionic fields in eq.~(\ref{eq:L_SMinL_EWET}), and an analogous structure for the resonance fields, being
\be \label{eq:fermionRkineticbare}
\mL^{\rm kin,mass}_{\Psi} = \frac{i}{2} \left( \overline \Psi \gamma^\mu\, d_\mu \Psi \,-\, \overline{(d_\mu \Psi)} \gamma^\mu \,  \Psi \right) \, - \, \overline \Psi \, M \, \Psi \,,
\ee
where $d_\mu$ and $M$ are the heavy fermion derivative and mass, respectively.

The interaction Lagrangian for fermionic resonances, however, is constructed like the previous ones for the bosonic heavy states:
\bear
\mL_\Psi &=& \mL_\Psi^{\rm kin,\, mass} 
+  \left( \overline\Psi \chi_\Psi + \overline{\chi}_\Psi \Psi \right)\, ,
\label{eq:Psi-L}
\eear
with 
\bear \label{eq:fermionictensor}
\chi_\Psi
&=&  u_\mu \gamma^\mu \left(\lambda^f_1\,\gamma_5+\widetilde \lambda^f_1\right) \xi^f
\,-\, i \Frac{(\partial_\mu h)}{v}  \gamma^\mu \left(\lambda^f_2+\widetilde \lambda^f_2\gamma_5\right)\xi^f
\nn\\
& &
+ \, \lambda^f_0 \mT\, \xi^f  \,+\, \widetilde{\lambda}_0^f \mT \gamma_5\, \xi^f \, ,   
\nn\\[1ex]
\overline{\chi}_\Psi
&=&
\bar\xi^f \gamma^\mu \left(\lambda^f_1\,\gamma_5+\widetilde \lambda^f_1\right)  u_\mu
\,+\, i\, \bar\xi^f \gamma^\mu \left(\lambda^f_2+\widetilde \lambda^f_2\,\gamma_5\right) \Frac{(\partial_\mu h)}{v}
\nn\\
&&
+ \, \lambda^f_0\, \bar{\xi}^f \mT  \,-\,  \widetilde{\lambda}_0^f\, \bar{\xi}^f \gamma_5 \mT  \, .   
\eear   
representing $\xi^f$ an implicit sum over the light fermion types (lepton and quark).
Therefore, the interacting tensors are defined to be chiral dimension $\hat d = 2$. 

Regarding the fermionic heavy sector, we have not mentioned the color and custodial symmetry representation of the fermionic resonances or the light fermion types involved up to this point. On the one side, notice that the fermionic resonance structure does not allow the presence of $SU(2)_{L+R}$ singlets for one single massive state only. Since they need to be contracted with a light fermion field, all the fermionic resonances are custodial doublets at the order studied. On the other side, the presence of just one single bilinear and ${\rm B} - {\rm L}$ conservation demand that SM leptons only mix with color singlet fermionic resonances and SM quarks just with colored ones. In short, there are two copies of fermionic resonance operators within the scope of this work at the order analyzed:
one for $SU(2)_{L+R}$ doublet, $SU(3)_C$ singlet resonances with SM leptons and another for $SU(2)_{L+R}$ doublet, $SU(3)_C$ triplet resonances with SM quarks. For this reason, $SU(n)$ indices are omitted in the fermionic heavy states.


	\chapter[Determination of the LECs]{Determination of the low-energy coupling constants} \label{ch:LECs}
	Direct searches of resonance heavy states have not revealed any new heavy state so far and thus the gap between the electroweak and the new physics scale becomes eventually larger. Considering the SM not to be the ultimate theory of particle physics, this fact only postpones high-energy states to higher energies in the spectrum. Hence, the only way to glimpse these heavy states at low-energies compared to their masses is through the low-energy coupling constants (LECs), i.e, the Wilson coefficients of the EWET NLO Lagrangian, studied in chapter \ref{ch:ewet}. In other words, one can perceive the presence of some high-energy state at low momenta if some SM prediction presents any significant deviation from the expected values. The precise pattern of deviations in some subset of LECs may give the information to know which particular heavy states are generating it.
	
	The procedure for obtaining the infrared behavior of the electroweak resonance theory is very well known \cite{Ecker:1988te,Ecker:1989yg,Pich:2002xy,Cirigliano:2006hb,Cirigliano:2004ue,Cirigliano:2005xn,RuizFemenia:2003hm,Rosell:2004mn,Rosell:2006dt,Pich:2008jm,Pich:2010sm}. It was employed in the context of the resonance chiral theory, ${\rm R}\chi{\rm T}$, some decades ago \cite{Weinberg:1978kz,Gasser:1983yg,Gasser:1984gg,Pich:1995bw,Ecker:1994gg,Bijnens:1999sh,Bijnens:1999hw,Bijnens:2014lea}. 
	The steps for integrating out the resonance fields from the action are the following:
	\begin{enumerate}
		\item Obtain the resonance EoM solutions and expand in terms of the light fields.
		\item Substitute these results in the resonance interaction Lagrangian.
		\item Project the outcome in the EWET NLO operator basis.
	\end{enumerate}
	
	\section{Integrating out the high-energy states}
	In the following, we perform the LEC obtention procedure for all the resonance interaction Lagrangians we have analyzed so far: the spin-0 bosonic in eq.~(\ref{eq:L0}), spin-1 bosonic Proca in eq.~(\ref{eq:L1proca}), spin-1 bosonic Antisymmetric in eq.~(\ref{eq:L1anti}) and fermionic in eq.~(\ref{eq:fermionRkineticbare}). This process must be done for all the $SU(2)_{L+R}$ and $SU(3)_C$ multiplet configurations. Therefore, instead of writing down all the technical details for every possible high-energy state, it is more interesting to do it for the custodial and color n-plets only, $R^8_3$ and $\Psi^3_2$, and properly adjust these results in order to determine the rest of possibilities.
	
	Using the interaction Lagrangians already mentioned, it is straightforward to obtain the EoMs of the high-energy resonance states,
	\begin{align} \label{eq:EoMs} 
	&\qquad\mbox{Spin-0},\quad (R = S,P)& \nn\\
	\nabla^2\,R \,+\, M_R^2\, R & \,=\, \chi_R \,-\, \frac{1}{2}\,\bra \chi_R \ket_2 \,-\,\frac{1}{3}\,\bra \chi_R \ket_3 \, +\, \frac{1}{6}\, \bra \chi_R \ket_{2,3}\,, & \nn\\[3ex]
	&\qquad\mbox{Spin-1 Proca},\quad (\hat R = \hat V, \hat A )& \nn\\
	\nabla_\mu \hat R^{\mu\nu} \,+\, M_R^2 \hat R^\nu & \,=\, - \left( \hat\chi^\nu_{\hat R} \,-\, 2\,\nabla_\mu \hat \chi_{\hat R}^{\mu\nu} \,-\, \frac{1}{2}\,\bra \hat\chi^\nu_{\hat R} \,-\, 2\,\nabla_\mu \hat \chi_{\hat R}^{\mu\nu} \ket_2 \right. & \nn\\
	& \qquad\quad \left.	-\,\frac{1}{3}\,\bra \hat \chi^\nu_{\hat R}\,-\, 2\,\nabla_\mu \hat \chi_{\hat R}^{\mu\nu} \ket_3 \, +\, \frac{1}{6}\, \bra \hat \chi^\nu_{\hat R}\,-\, 2\,\nabla_\mu \hat \chi_{\hat R}^{\mu\nu} \ket_{2,3} \right)\,, & \nn\\[3ex]
	&\qquad\mbox{Spin-1 Antisymmetric},\quad ( R =  V,  A )& \nn\\
	\nabla^{[\mu}\nabla_\rho  R^{\rho\,,\nu]} \,+\, M_R^2  R^\nu & \,=\, - \left( \chi^{\mu\nu}_{ R} \,+\,\nabla^\mu  \chi_{ R}^{\nu} \,-\, \frac{1}{2}\,\bra \chi^{\mu\nu}_{ R} \,+\, \nabla^\mu  \chi_{ R}^{\nu} \ket_2 \right. & \nn\\
	&  \qquad\quad\left.	-\,\frac{1}{3}\,\bra  \chi^{\mu\nu}_{ R}\,+\, \nabla^\mu  \chi_{ R}^{\nu} \ket_3 \, +\, \frac{1}{6}\, \bra  \chi^{\mu\nu}_{ R}\,+\,\nabla^\mu  \chi_{ R}^{\nu} \ket_{2,3} \right)\,, & \nn\\[3ex]
	&\qquad\mbox{Fermionic},\quad (\Psi)& \nn\\
	i\,\gamma^\mu d_\mu \Psi \,+\, M_\Psi\,\Psi & \,=\, \chi_\Psi\,, \qquad -i\,\gamma^\mu \overline{d_\mu \Psi} \,+\, M_\Psi\,\overline{\Psi}  \,=\, \chi_\Psi\, , & 
	\end{align}
	where the $R^8_3$ and $\Psi^3_2$ indices have been omitted for simplicity. In order to recover the EoMs for the rest of resonance states the next criteria applies: remove all $SU(2)_{L+R}$ traces for heavy states being a singlet under this group, remove all $SU(3)_C$ traces for colorless resonances and thus remove all traces for singlet-singlet massive states. 
	
	Right after the resonance EoMs are set, they must be power-expanded as powers of momentum over the resonance mass at first order. Naively, this is equivalent to get rid of every covariant derivative and isolate the heavy states in terms of the light fields \cite{Ecker:1988te}. This yields, respectively,
	\begin{align} \label{eq:EoMspowerexpand}
	R & \, = \, \frac{1}{M_R^2} \left( \chi_R \,-\, \frac{1}{2} \bra \chi_R \ket_2 \,-\, \frac{1}{3} \bra \chi_R \ket_3 \right. & \qquad &\mbox{Spin-0} &
	\nn\\
	& \qquad\qquad  \left. +\, \frac{1}{6} \bra \chi_R \ket_{2,3}\right) \,+\,  \mO_{\hat d}(p^{>2}) \,, &  \qquad & (R = S,P)\,, &
	\nn\\[3ex]
	\hat R^\nu & \, = \,-\, \frac{1}{M_R^2} \left( \hat \chi_{\hat R}^\nu \,-\, \frac{1}{2} \bra \hat \chi_{\hat R}^\nu \ket_2 \,-\, \frac{1}{3} \bra \hat \chi_{\hat R}^\nu \ket_3 \right. & \qquad &\mbox{Spin-1, Proca} &
	\nn\\
	& \qquad\qquad  \left. +\, \frac{1}{6} \bra \hat \chi_{\hat R}^\nu \ket_{2,3}\right) \,+\,  \mO_{\hat d}(p^{>2}) \,, & \qquad & (\hat R = \hat V, \hat A )\,,  &
	\nn\\[3ex]
	R^{\mu\nu} & \, = \,-\, \frac{1}{M_R^2} \left( \chi_{ R}^{\mu\nu} \,-\, \frac{1}{2} \bra  \chi_R^{\mu\nu} \ket_2 \,-\, \frac{1}{3} \bra  \chi_R^{\mu\nu} \ket_3 \right. & \qquad & \mbox{Spin-1 Antisym} &
	\nn\\
	& \qquad\qquad \left. \frac{1}{6} \bra  \chi_R^{\mu\nu} \ket_{2,3}\right) \,+\,  \mO_{\hat d}(p^{>2}) \,,  & \qquad & (\hat R = \hat V, \hat A )\,,  &
	\nn\\[3ex]
	\Psi & \, = \, \frac{1}{M_\Psi}\,\chi_\Psi\, + \, \mO_{\hat d} (p^{>2})\,, \qquad \overline \Psi \, = \, \frac{1}{M_\Psi}\,\overline \chi_\Psi\, + \, \mO_{\hat d} (p^{>2})\,, & \qquad & \mbox{Fermionic}\,\,(\Psi)\,.  &
	\end{align} 
	%
	%
	%
	%
	%
	When light and heavy fields are set apart, it is indeed the very precise point where we are allowed to assign an actual chiral power counting to the light fields. As eq.~(\ref{eq:EoMspowerexpand}) shows, all the high-energy fields are expressed in terms of chiral dimension $\hat d= 2$ operators\footnote{The expansion in canonical dimensions instead provides the same outcome for all the bosonic operators. However, the fermionic resonances get truncated at $\mO(p^2/M_\Psi^2)$. This is a consequence of the non-trivial assignment of an extra unit to the power counting for the light fermionic fields when coupled to the composite sector.} in addition to some other non-relevant higher chiral dimension terms that we will neglect.
	
	Furthermore, notice that either the two indices interacting tensors of the Proca spin-1 resonances and the single-index interacting tensors of the Antisymmetric ones are boosted to higher orders in the light field power counting. This is consistent with the naive counting performed in the last chapter where these terms counted as chiral dimension $\hat d=3$. Hence, only the $\hat \chi_{\hat R}^\mu$ and $\chi_R^{\mu\nu}$ tensor structures (for the Proca and Antisymmetric, respectively) remain.
	
	We replace these $\mO_{\hat d}(p^2)$ resonance EoM solutions into the interaction Lagrangian in order to get the resonance imprints in the EWET Lagrangian. Once each term of eq.~(\ref{eq:EoMspowerexpand}) is substituted in eqs.~(\ref{eq:L0}, \ref{eq:L1proca}, \ref{eq:L1anti}, \ref{eq:Psi-L}), respectively, we obtain the contributions of the massive states to the EWET NLO Lagrangian: 
	\begin{align} \label{eq:Lcontributions} 
	\Delta \mL_{R}^{\cO_{\hat d}(p^4)} &\! =  \Frac{1}{4 M_{R}^2} 
	\left(  \bra  \chi_{R}\, \chi_{R}\ket_{2,3}\, -\, \Frac{1}{2} \bra \bra \chi_{R} \ket_2 \, \bra \chi_{R} \ket_2 \ket_3 \right.& \quad &\mbox{Spin-0} &
	\nn\\
	& \; \left. -\, \Frac{1}{3} \bra \bra \chi_{R} \ket_3 \, \bra \chi_{R} \ket_3 \ket_2 \,+\, \Frac{1}{6} \bra \chi_{R}\ket_{2,3}\bra \chi_{R}\ket_{2,3} \right)\,,&  \quad & (R = S,\,P)\,, &
	\nn\\[3ex]
	\Delta \mL_{\hat R}^{\cO_{\hat d}(p^4)} &\! =  -\Frac{1}{4 M_{R}^2} 
	\left(  \bra  \chi^\mu_{\hat R}\, \chi_{\hat R\,\mu}\ket_{2,3}\, -\, \Frac{1}{2} \bra \bra \chi^\mu_{\hat R} \ket_2 \,\bra \chi_{\hat R\,\mu} \ket_2 \ket_3 \right. & \quad & \mbox{Spin-1 Proca} &
	\nn\\
	& \; \left. -\,\Frac{1}{2} \bra \bra \chi^\mu_{\hat R} \ket_3 \,\bra \chi_{\hat R\,\mu} \ket_3 \ket_2 \,+\, \Frac{1}{6} \bra  \chi_{\hat R\,\mu}\ket_{2,3}\bra  \chi_{\hat R}^\mu\ket_{2,3} \right)\,, & \quad & (\hat R = \hat V, \hat A )\,,  &
	\nn\\[3ex]
	\Delta \mL_{ R}^{\cO_{\hat d}(p^4)} &\! = -\Frac{1}{2 M_{R}^2} 
	\left(  \bra  \chi^{\mu\nu}_{ R}\, \chi_{ {R}\,\mu\nu}^{\phantom{\mu\nu}} \ket_{2,3}\, -\, \Frac{1}{2} \bra \bra \chi^{\mu\nu}_{ R} \ket_2 \, \bra \chi_{{ R}\,\mu\nu}^{\phantom{\mu\nu}} \ket_2 \ket_3 \right. &  \quad & \mbox{Spin-1 Antisym} &
	\nn\\
	& \; \left. -\,\Frac{1}{2} \bra \bra \chi^{\mu\nu}_{ R} \ket_3 \, \bra \chi_{{ R}\,\mu\nu}^{\phantom{\mu\nu}} \ket_3 \ket_2 \, + \, \Frac{1}{6}\bra  \chi_{ {R}\,\mu\nu}^{\phantom{\mu\nu}} \ket_{2,3}\bra  \chi_{ {R}}^{\mu\nu} \ket_{2,3} \right) \,,& \quad & ( R =  V,  A )\,, &
	\nn\\[3ex]
	\Delta\mL_{\Psi}^{\cO_{\hat d}(p^4)}&= 
	\Frac{1}{M_\Psi}\, \overline{\chi}_\Psi\,\chi_\Psi\,, & \quad & \mbox{Fermionic}\,\,(\Psi)\,.  &
	\end{align}
	Recall that the same trace criteria apply for this equation too: $SU(n)$ traces only apply for resonance Lagrangians charged under this symmetry. Furthermore, all these Lagrangian contributions to the LECs can be simplified. Notice that the second line of each  Lagrangian contribution is exactly zero for all the Lagrangian tensor structures from eqs.~(\ref{eq:L0tensor}, \ref{eq:proca-chimu}, \ref{eq:antisym-chimunu}, \ref{eq:fermionictensor}) since there are no color octet operators carrying closed color indices and thus they vanish when traced. Nonetheless, these Lagrangian contributions have been written for the sake of completeness. 
	
	\section{Resonance contributions to the EWET}
	In this section, we calculate all the resonance contributions obtained in eq.~(\ref{eq:Lcontributions}) and project them in the EWET NLO Lagrangian in order to obtain all the high-energy imprints for each of the LECs. These results (tables \ref{tab:LECbosonic}, \ref{tab:LEC2fermion}, \ref{tab:LEC4fermionlepton}, \ref{tab:LEC4fermionmixed}, \ref{tab:LEC4fermionquark}) constitute a milestone for this thesis since we are able to show the contributions to the EWET Lagrangian of the most general set of high-energy states (within one lepton and one quark family) invariant under the gauge symmetries of the SM. Note that we have included both spin-1 contributions coming from the Proca and the Antisymmetric Lagrangian. This will be justified in chapter \ref{ch:spin1}.
	
	\section{The scalar singlet resonance case}
	The scalar singlet massive state, $S^1_1$, is quite singular with respect to the rest of resonances. It is precisely its Higgs-like similarity and its uncharged nature under any gauge group what makes it so interesting. Indeed, along the last decades, there have been lots of works in the literature where the scalar sector is expanded with some heavy Higgs partner.
	
	In the last section, we collected all the resonance contributions to the EWET NLO Lagrangian in a general way. For sure, this statement includes the scalar singlet ones as well. However, this resonance is the only one that leave some imprints at the EWET LO Lagrangian, particularly to the Higgs sector. This happens through the interacting term, already included in eq.~(\ref{eq:L0tensor}), 
	\be \label{eq:S1potential}
	S_1\,\chi^h_{S^1_1} \,\equiv\, \lambda_{hS_1}\,v\,h^2\,S^1_1\,,
	\ee
	which couples directly to the Higgs field. 
	
	When the heavy state is integrated out from the action, it manifests at the $\mO_{\hat d}(p^2)$ Lagrangian like
	\be \label{eq:S1-p2}
	\Delta\mL_{S^1_1}^{\mO(p^2)}\, =\, \frac{1}{2 M_{S^1_1}^2} \left\{
	(\lambda_{hS_1})^2\, v^2\, h^4\, +\,
	\sqrt{2}\, \lambda_{hS_1} v\, h^2 \left[c_d\, \bra u_\mu u^\mu \ket_2
	+ c_f^{S^1_1}\,\bra J^f_S \ket_2 \right]\right\}\, ,
	\ee
	where, as usual, a sum over $f=l,q$ is implicit. Notice that this Lagrangian contribution contains light fields operators with different chiral dimension assignment. Hence, a  consistent power counting demands that the $\lambda_{hS_1}$ coupling brings chiral dimension\footnote{This counting was already assigned in the last chapter at the resonance level, so that all the elements in eq.~(\ref{eq:L0tensor}) carry the same naive counting.} $\hat d=2$.

	\begin{table}[b!]
		\begin{center}
			\renewcommand{\arraystretch}{2.2}
			\begin{tabular}{|c|c|c|c|}
				\hline
				\multicolumn{4}{|c|}{Bosonic $P$-even LECs, $\Delta\mF_i$}\\
				\hline
				$1$  &
				$- \Frac{F_V^2-\widetilde{F}_V^2}{4M_{V^1_3}^2} + \Frac{F_A^2-\widetilde{F}_A^2}{4M_{A^1_3}^2} $  &
				$7$  &
				$\Frac{ d_P^2}{2 M_{P^1_3}^2}+\Frac{\Lambda_1^{hA\,\, 2}v^2}{M_{A^1_3}^2} +  \Frac{\widetilde{\Lambda}_1^{hV\,\, 2}v^2}{M_{V^1_3}^2} $
				\\ [1ex]
				\hline
				$2$  &
				$- \Frac{F_V^2+{\widetilde{F}_V}^2}{8M_{V^1_3}^2} - \Frac{F_A^2+{\widetilde{F}_A}^2}{8M_{A^1_3}^2}$  &
				$8$ &
				0 
				\\ [1ex]
				\hline
				$3$  &
				$-  \Frac{F_VG_V}{2M_{V^1_3}^2} - \Frac{\widetilde{F}_A\widetilde{G}_A}{2M_{A^1_3}^2}$ & 
				$9$ &
				$  - \Frac{F_A \Lambda_1^{hA} v}{M_{A^1_3}^2} - \Frac{\widetilde{F}_V \widetilde{\Lambda}_1^{hV} v}{M_{V^1_3}^2}$
				\\ [1ex]
				\hline
				$4$  &
				$\Frac{G_V^2}{4M_{V^1_3}^2} + \Frac{{\widetilde{G}_A}^2}{4M_{A^1_3}^2} $  &
				$10$ & 
				$-\displaystyle\frac{\widetilde{c}_{\mathcal{T}}^2}{2M_{V^1_1}^2}-\displaystyle\frac{c_{\mathcal{T}}^2}{2M_{A^1_1}^2}$  
				\\ [1ex]
				\hline
				$5$  &
				$\Frac{c_{d}^2}{4M_{S^1_1}^2}-\Frac{G_V^2}{4M_{V^1_3}^2} - \Frac{{\widetilde{G}_A}^2}{4M_{A^1_3}^2}$ &
				$11$ & 
				$- \Frac{F_{X}^2}{M_{V^1_1}^2} - \Frac{\widetilde{F}_{X}^2}{M_{A^1_1}^2} $ 
				\\ [1ex]
				\hline
				$6$ &
				$ - \Frac{\widetilde{\Lambda}_1^{hV\,\, 2}v^2}{M_{V^1_3}^2} - \Frac{\Lambda_1^{hA\,\, 2}v^2}{M_{A^1_3}^2}$ &
				$12$ & 
				$  - \Frac{(C_G)^2}{2 M_{V^8_1}^2} - \Frac{(\widetilde{C}_G)^2}{2 M_{A^8_1}^2}  $
				\\ [1ex]
				\hline
				\multicolumn{3}{c}{} 
				\\ [-2ex]
				\hline
				\multicolumn{4}{|c|}{Bosonic $P$-odd LECs, $\Delta \widetilde \mF_i$}\\
				\hline		
				$1$ &
				$- \Frac{\widetilde{F}_VG_V}{2M_{V^1_3}^2} - \Frac{F_A\widetilde{G}_A}{2M_{A^1_3}^2}$   &
				$3$ & 
				$- \Frac{F_V \widetilde{\Lambda}_1^{hV} v}{M_{V^1_3}^2} - \Frac{\widetilde{F}_A \Lambda_1^{hA} v}{M_{A^1_3}^2}$
				\\ [1ex]
				\hline
				$2$ &
				$- \Frac{F_V \widetilde{F}_V}{4M_{V^1_3}^2} - \Frac{F_A \widetilde{F}_A}{4M_{A^1_3}^2}$ 
				\\ [1ex]
				\cline{1-2}			
			\end{tabular}
		\end{center}
		\caption{\small
			All resonance contributions (spin-0, spin-1 Proca, spin-1 Antisymmetric and fermionic) to the bosonic LECs of the EWET NLO Lagrangian (table \ref{tab:bosonic-Op4}), where $i$ is the operator assignment index for this set.  LECs corresponding to $P$-even ($P$-odd) operators are shown in the upper (lower) block.}
		\label{tab:LECbosonic}
	\end{table}
	%
	%
	\begin{table}[t!]
		\begin{center}
			\renewcommand{\arraystretch}{2.2}
			\begin{tabular}{|c|c|c|c|}
				\hline
				\multicolumn{4}{|c|}{Two-fermion $P$-even LECs, $\Delta \widetilde \mF^f_i \quad $ ($f\,=\, l,\,q$)}\\
				\hline
				1   &  
				$\displaystyle\frac{c_d\, c^{S^1_1}_f}{2M_{S^1_1}^2}\,+\, \frac{ (\widetilde \lambda_1^f)^2-(\lambda_1^f)^2}{2 M_\Psi}$ &
				5   &  
				$\quad\! \displaystyle\frac{d_P\, c^{P^1_3}}{M_{P^1_3}^2}\,+\, \Frac{    2(\lambda^f_1\lambda^f_2 -\widetilde \lambda^f_1\widetilde \lambda^f_2)   }{M_\Psi} \quad\!$ \\[1ex]
				\hline 
				\multirow{2}{*}{ 2 }   &
				$-\displaystyle\frac{G_V\, C_{0,f}^{V^1_3}}{\sqrt{2}M_{V^1_3}^2} \,-\, \displaystyle\frac{\widetilde{G}_A \widetilde{C}_{0,f}^{A^1_3}}{\sqrt{2}M_{A^1_3}^2}$ &      
				\multirow{2}{*}{ 6 }   &  
				$-\displaystyle\frac{\widetilde{c}_{\mathcal{T}}\, \widetilde{c}_f^{\hat{V}^1_1}    }{\sqrt{2} M_{V^1_1}^2} \,-\, \displaystyle\frac{c_{\mathcal{T}}\, c_f^{\hat{A}^1_1}    }{\sqrt{2} M_{A^1_1}^2}$     
				\\[1ex] 
				&
				$ -\,  \displaystyle \frac{(\widetilde \lambda_1^f)^2-(\lambda_1^f)^2}{2 M_\Psi} $ &
				&
				$ +\,  \displaystyle \frac{(\lambda^f_0\widetilde \lambda^f_1 +\widetilde \lambda^f_0  \lambda^f_1)}{M_\Psi} $
				\\[1ex] 
				\hline
				3   &  
				$-\displaystyle\frac{F_V C_{0,f}^{V^1_3}}{\sqrt{2}M_{V^1_3}^2} -  \displaystyle\frac{\widetilde{F}_A \widetilde{C}_{0,f}^{A^1_3}}{\sqrt{2}M_{A^1_3}^2} $  &      
				7   & 
				$\displaystyle \frac{(\lambda_2^f)^2 - (\widetilde \lambda_2^f)^2}{M_\Psi} $    \\[1ex] 
				\hline
				4   &  
				$-\displaystyle\frac{\sqrt{2}F_{X} \, C_{0,f}^{V^1_1}}{M_{V^1_1}^2} -\displaystyle\frac{\sqrt{2}\widetilde{F}_{X} \widetilde{C}_{0,f}^{A^1_1}}{M_{A^1_1}^2} $ & 
				8   &  
				$ -\Frac{C_G\, C_0^{V^8_1}}{\sqrt{2} M^2_{V^8_1}} - \Frac{\widetilde C_G\, \widetilde C_0^{A^8_1}}{\sqrt{2} M^2_{A^8_1}}  $ 
				\\[1ex] \hline
				\multicolumn{3}{c}{} 
				\\ [-2ex]
				\hline
				\multicolumn{4}{|c|}{Two-fermion $P$-odd LECs, $\Delta \widetilde \mF^f_i \quad $ ($f\,=\, l,\,q$)}\\
				\hline		
				1 &
				$-\Frac{\widetilde{F}_V\, C_{0,f}^{V^1_3}}{\sqrt{2}M_{V^1_3}^2}  -  \Frac{F_A\, \widetilde{C}_{0,f}^{A^1_3}}{\sqrt{2}M_{A^1_3}^2} $   &
				\multirow{2}{*}{ 3 } & 
				$-\displaystyle \frac{\widetilde{c}_{\mathcal{T}}\, c_f^{\hat{V}^1_1}    }{\sqrt{2} M_{V^1_1}^2} - \Frac{c_{\mathcal{T}}\, \widetilde{c}_f^{\hat{A}^1_1}    }{\sqrt{2} M_{A^1_1}^2}$
				\\ [1ex]
				\cline{1-2}
				\multirow{2}{*}{ 2 } &
				$-\Frac{2\sqrt{2}\,v\,\widetilde{\Lambda}_1^{hV}C_{0,f}^{V^1_3}}{M_{V^1_3}^2} - \Frac{2\sqrt{2}\,v\,\Lambda_1^{hA}\widetilde{C}_{0,f}^{A^1_3}}{M_{A^1_3}^2}$ &
				&
				$+ \, \displaystyle \frac{(\lambda^f_0\, \lambda^f_1 +\widetilde \lambda_0^f  \widetilde\, \lambda_1^f)}{M_\Psi} $  
				\\
				\cline{3-4} 
				& 
				$+\displaystyle \frac{   2(\widetilde \lambda_1^f\,\lambda_2^f -\lambda_1^f\,\widetilde \lambda_2^f)   }{M_\Psi} $
				\\ [1ex]
				\cline{1-2}			
				\multicolumn{3}{c}{}
			\end{tabular}
		\end{center}
		\vspace*{-1.cm}
		\caption{\small
			All resonance contributions (spin-0, spin-1 Proca, spin-1 Antisymmetric and fermionic) to the two-fermion LECs of the EWET NLO Lagrangian (table \ref{tab:2fermion}), where $i$ is the operator assignment index for this set. Resonance couplings with the index $f$ must be understood as one copy from a lepton coupling and another from a quark coupling,\ie a sum over $f=l,q$ is implied. LECs corresponding to $P$-even ($P$-odd) operators are shown in the upper (lower) block.}
		\label{tab:LEC2fermion}
	\end{table}
	%
	
	\begin{table}[t!]
		\begin{center}
			\renewcommand{\arraystretch}{2.2}
			\begin{tabular}{|c|c|}
				\hline
				\multicolumn{2}{|c|}{Four-fermion $P$-even LECs, $\Delta \mF^{l,l}_i\, $ | \ lepton--lepton}\\
				\hline
				\ 1 \   &  
				$\displaystyle  \Frac{(c_{l}^{S^1_1})^2}{4M_{S^1_1}^2} 
				- \Frac{3(c_{l}^{S^1_3})^2}{8 M_{S^1_3}^2} 
				+ \Frac{(c_l^{P^1_3})^2}{8 M_{P^1_3}^2}
				+ \Frac{(c_{l}^{\hat{V}^1_3})^2}{2M_{V^1_3}^2} 
				- \Frac{(\widetilde{c}_{l}^{\hat{V}^1_3})^2}{2M_{V^1_3}^2} 
				- \Frac{({c}_{l}^{\hat{A}^1_3})^2}{2M_{A^1_3}^2}
				+ \Frac{(\widetilde{c}_{l}^{\hat{A}^1_3})^2}{2M_{A^1_3}^2} $
				\\
				&
				$ \displaystyle + \frac{3\, (C_{0,l}^{V^1_3})^2}{M_{V^1_3}^2}
				+ \frac{3\, (\widetilde{C}_{0,l}^{A^1_3})^2}{M_{A^1_3}^2} 
				$ \\[1ex]
				\hline
				2	&
				$\displaystyle 
				\Frac{(c_l^{S^1_3})^2 }{8 M_{S^1_3}^2}
				+ \Frac{(c_{l}^{P^1_1})^2}{4M_{P^1_1}^2} 
				- \Frac{3(c_{l}^{P^1_3})^2}{8M_{P^1_3}^2}  
				+ \Frac{(c_{l}^{\hat{V}^1_3})^2}{2M_{V^1_3}^2} 
				- \Frac{(\widetilde{c}_{l}^{\hat{V}^1_3})^2}{2M_{V^1_3}^2}			 
				- \Frac{({c}_{l}^{\hat{A}^1_3})^2}{2M_{A^1_3}^2}
				+ \Frac{(\widetilde{c}_{l}^{\hat{A}^1_3})^2}{2M_{A^1_3}^2}
				$	\\
				&
				$\displaystyle - \frac{3\,(C_{0,l}^{V^1_3})^2}{M_{V^1_3}^2} 
				- \frac{3\,(\widetilde{C}_{0,l}^{A^1_3})^2}{M_{A^1_3}^2} $\\
				\hline
				3	&
				$\displaystyle 
				- \Frac{(c_l^{S^1_3})^2 }{8 M_{S^1_3}^2}
				- \Frac{(c_l^{P^1_3})^2}{8 M_{P^1_3}^2}
				- \Frac{({c}_{l}^{\hat{V}^1_1})^2}{4M_{V^1_1}^2} 
				- \Frac{(\widetilde{c}_{l}^{\hat{A}^1_1})^2}{4M_{A^1_1}^2}
				- \Frac{(\widetilde{c}_{l}^{\hat{V}^1_3})^2}{4 M_{V^1_3}^2}
				- \Frac{({c}_{l}^{\hat{A}^1_3})^2}{4 M_{A^1_3}^2}
				$
				\\ \hline
				4	&
				$\displaystyle 
				\Frac{(c_l^{S^1_3})^2 }{8M_{S^1_3}^2}
				+ \Frac{(c_l^{P^1_3})^2}{8 M_{P^1_3}^2}
				- \Frac{(\widetilde{c}_{l}^{\hat{V}^1_1})^2}{4M_{V^1_1}^2}  
				- \Frac{({c}_{l}^{\hat{A}^1_1})^2}{4M_{A^1_1}^2}
				- \Frac{(c_{l}^{\hat{V}^1_3})^2}{4 M_{V^1_3}^2}			
				- \Frac{(\widetilde{c}_{l}^{\hat{A}^1_3})^2}{4 M_{A^1_3}^2}
				$
				\\ \hline
				5	&
				$\displaystyle 
				- \Frac{(c_l^{S^1_3})^2 }{16 M_{S^1_3}^2}
				+ \Frac{(c_l^{P^1_3})^2}{16 M_{P^1_3}^2}
				- \Frac{(C_{0,l}^{V^1_1})^2}{2M_{V^1_1}^2}
				- \Frac{(\widetilde{C}_{0,l}^{A^1_1})^2}{2M_{A^1_1}^2}
				$
				\\ \hline
				\multicolumn{2}{c}{} 
				\\[-2ex] 
				\hline
				\multicolumn{2}{|c|}{Four-fermion $P$-odd LECs, $\Delta \widetilde \mF^{l,l}_i\, $ | \ lepton--lepton}\\
				\hline
				1	&
				$ \displaystyle 
				- \Frac{c_l^{\hat{V}^1_1}\,\widetilde{c}_l^{\hat{V}^1_1}}{2M_{V^1_1}^2} 
				+ \Frac{3\, c_l^{\hat{V}^1_3}\, \widetilde{c}_l^{\hat{V}^1_3}}{2M_{V^1_3}^2}
				- \Frac{  c_l^{\hat{A}^1_1}\, \widetilde{c}_l^{\hat{A}^1_1}}{2M_{A^1_1}^2} 
				+ \Frac{3\, c_l^{\hat{A}^1_3}\, \widetilde{c}_l^{\hat{A}^1_3}}{2M_{A^1_3}^2} 
				%
				$
				\\ \hline
			\end{tabular}
		\end{center}
		\caption{\small
			All resonance contributions (spin-0, spin-1 Proca, spin-1 Antisymmetric and fermionic) to the four-fermion LECs of the EWET NLO Lagrangian including two lepton bilinears (table \ref{tab:4fermion}), where $i$ is the operator assignment index for this set.  LECs corresponding to $P$-even ($P$-odd) operators are shown in the upper (lower) block.}
		\label{tab:LEC4fermionlepton}
	\end{table}

	\begin{table}[t!]
		\begin{center}
			\renewcommand{\arraystretch}{2.2}
			\begin{tabular}{|c|c|c|c|}
				\hline
				\multicolumn{4}{|c|}{Four-fermion $P$-even LECs, $\Delta \mF^{l,q}_i\, $ | \ lepton--quark}\\
				\hline
				\ 1 \   &  
				$\displaystyle \Frac{c_l^{S^1_3}\,c_q^{S^1_3} }{2M_{S^1_3}^2} $ &
				\ 6 \   &
				$\displaystyle - \Frac{\widetilde{c}_{l}^{\hat{V}^1_3}\, \widetilde{c}_{q}^{\hat{V}^1_3}}{2M_{V^1_3}^2} -\Frac{{c}_{l}^{\hat{A}^1_3}\,{c}_{q}^{\hat{A}^1_3}}{2M_{A^1_3}^2} $
				\\[1ex]
				\hline 			
				2	&
				$\displaystyle \Frac{c_l^{P^1_3}\,c_q^{P^1_3}}{2M_{P^1_3}^2} $ &
				7	&
				$\displaystyle -\,\Frac{{c}_{l}^{\hat{V}^1_1}\,{c}_{q}^{\hat{V}^1_1}}{4M_{V^1_1}^2} -\Frac{\widetilde{c}_{l}^{\hat{A}^1_1}\, \widetilde{c}_{q}^{\hat{A}^1_1}}{4M_{A^1_1}^2}
				+\Frac{{c}_{l}^{\hat{V}^1_3}\, {c}_{q}^{\hat{V}^1_3}}{4M_{V^1_3}^2} +\Frac{\widetilde{c}_{l}^{\hat{A}^1_3}\, \widetilde{c}_{q}^{\hat{A}^1_3}}{4M_{A^1_3}^2} $
				\\[1ex]
				\hline 	
				3	&
				$\displaystyle  \Frac{c_{l}^{S^1_1}\,c_{q}^{S^1_1}}{4M_{S^1_1}^2} \,-\, \Frac{c_{l}^{S^1_3}\,c_{q}^{S^1_3}}{4M_{S^1_3}^2} $ &
				8   &
				$\displaystyle -\,\Frac{\widetilde{c}_{l}^{\hat{V}^1_1}\, \widetilde{c}_{q}^{\hat{V}^1_1}}{4M_{V^1_1}^2}  -\Frac{{c}_{l}^{\hat{A}^1_1}\, {c}_{q}^{\hat{A}^1_1}}{4M_{A^1_1}^2}
				+ \Frac{\widetilde{c}_{l}^{\hat{V}^1_3}\, \widetilde{c}_{q}^{\hat{V}^1_3}}{4M_{V^1_3}^2} +\Frac{{c}_{l}^{\hat{A}^1_3}\, {c}_{l}^{\hat{A}^1_3}}{4M_{A^1_3}^2} $
				\\[1ex]
				\hline 	
				4	&
				$\displaystyle \Frac{c_{l}^{P^1_1}\,c_{q}^{P^1_1}}{4M_{P^1_1}^2} \,-\,\Frac{c_{l}^{P^1_3}\,c_{q}^{P^1_3}}{4M_{P^1_3}^2}  $ &
				9	&
				$-\displaystyle\frac{C_{0,l}^{V^1_3}\, C_{0,q}^{V^1_3}}{M_{V^1_3}^2} \,-\, \displaystyle\frac{\widetilde{C}_{0,l}^{A^1_3}\, \widetilde{C}_{0,q}^{A^1_3}}{M_{A^1_3}^2} $
				\\[1ex]
				\hline 	
				\multirow{2}{*}{5} &
				\multirow{2}{*}{
					$\displaystyle -\,\Frac{c_{l}^{\hat{V}^1_3}\,c_{q}^{\hat{V}^1_3}}{2M_{V^1_3}^2} \,-\,\Frac{\widetilde{c}_{l}^{\hat{A}^1_3}\,\widetilde{c}_{q}^{\hat{A}^1_3}}{2M_{A^1_3}^2}  $ }  &
				\multirow{2}{*}{10}   &
				$\displaystyle -\, \Frac{C_{0,l}^{V^1_1}\, C_{0,q}^{V^1_1}}{2M_{V^1_1}^2}
				\,+\, \Frac{C_{0,l}^{V^1_3}\, C_{0,q}^{V^1_3}}{2M_{V^1_3}^2}  $
				\\
				&
				&
				& 
				$- \Frac{\widetilde{C}_{0,l}^{A^1_1}\, \widetilde{C}_{0,q}^{A^1_1}}{2M_{A^1_1}^2}
				\,+\, \Frac{\widetilde{C}_{0,l}^{A^1_3}\, \widetilde{C}_{0,q}^{A^1_3}}{2M_{A^1_3}^2} $
				\\[1ex]
				\hline 	
				\multicolumn{4}{c}{} \\[-2ex]
				\hline
				\multicolumn{4}{|c|}{Four-fermion $P$-odd LECs, $\Delta \widetilde \mF^{l,q}_i\, $ | \ lepton--quark}\\
				\hline
				\ 1 \   &  
				$ -\Frac{c_{l}^{\hat{V}^1_3}\, \widetilde{c}_{q}^{\hat{V}^1_3}} {M_{V^1_3}^2} -\Frac{\widetilde{c}_{l}^{\hat{A}^1_3}\,  c_{q}^{\hat{A}^1_3}}{M_{A^1_3}^2}$ &
				\ 3 \   &
				$ \displaystyle -\,\Frac{c_l^{\hat{V}^1_1}\,\widetilde{c}_q^{\hat{V}^1_1}}{2M_{V^1_1}^2} 
				+ \Frac{c_l^{\hat{V}^1_3}\, \widetilde{c}_q^{\hat{V}^1_3}}{2M_{V^1_3}^2}
				-\Frac{ \widetilde c_l^{\hat{A}^1_1}\, {c}_q^{\hat{A}^1_1}}{2M_{A^1_1}^2} +\Frac{\widetilde c_l^{\hat{A}^1_3}\, {c}_q^{\hat{A}^1_3}}{2M_{A^1_3}^2} $
				\\[1ex]
				\hline 			
				2	&
				$ -\Frac{\widetilde c_{l}^{\hat{V}^1_3}\, {c}_{q}^{\hat{V}^1_3}} {M_{V^1_3}^2} -\Frac{{c}_{l}^{\hat{A}^1_3}\,  \widetilde c_{q}^{\hat{A}^1_3}}{M_{A^1_3}^2}$ &
				4	&
				$ \displaystyle -\,\Frac{\widetilde c_l^{\hat{V}^1_1}\, {c}_q^{\hat{V}^1_1}}{2M_{V^1_1}^2} 
				+ \Frac{\widetilde c_l^{\hat{V}^1_3}\,{c}_q^{\hat{V}^1_3}}{2M_{V^1_3}^2}
				-\Frac{ c_l^{\hat{A}^1_1}\, \widetilde{c}_q^{\hat{A}^1_1}}{2M_{A^1_1}^2} +\Frac{c_l^{\hat{A}^1_3}\, \widetilde{c}_q^{\hat{A}^1_3}}{2M_{A^1_3}^2} $
				\\[1ex]
				\hline 			
			\end{tabular}
		\end{center}
		\caption{\small
			All resonance contributions (spin-0, spin-1 Proca, spin-1 Antisymmetric and fermionic) to the four-fermion LECs of the EWET NLO Lagrangian including one lepton bilinear and one quark bilinear (table \ref{tab:4fermion}), where $i$ is the operator assignment index for this set.  LECs corresponding to $P$-even ($P$-odd) operators are shown in the upper (lower) block.}
		\label{tab:LEC4fermionmixed}
	\end{table}

	\begin{table}[t!]
		\begin{center}
			\renewcommand{\arraystretch}{2.2}
			\begin{tabular}{|c|c|}
				\hline
				\multicolumn{2}{|c|}{Four-fermion $P$-even LECs, $\Delta \mF^{q,q}_i \, $  | \ quark--quark }\\
				\hline
				\multirow{3}{*}{1}   &  
				$ \Frac{c_q^{S^1_3}\,c_q^{S^1_3} }{2M_{S^1_3}^2} \,+\,\displaystyle \frac{1}{128}\! \left(  - \Frac{(c^{S_1^8})^2}{ M^2_{S_1^8}} - \Frac{5(c^{S^8_3})^2}{3 M^2_{S^8_3}}  + \Frac{(c^{P_1^8})^2}{ M^2_{P_1^8}} -  \Frac{(c^{P^8_3})^2}{ M^2_{P^8_3}}    \right) $  
				\\	
				&
				$ +\, \displaystyle \Frac{1}{32}\! \left(  \Frac{(c^{\hat V^8_1})^2}{ M^2_{V^8_1}}   - \Frac{(\widetilde c^{\hat V^8_1})^2}{ M^2_{V^8_1}} - \Frac{(c^{\hat V^8_3})^2}{ M^2_{V^8_3}} + \Frac{(\widetilde c^{\hat{V}^8_3})^2}{ M^2_{V^8_3}} - \Frac{(c^{\hat A^8_1})^2}{ M^2_{A^8_1}}   + \Frac{(\widetilde c^{\hat A^8_1})^2}{ M^2_{A^8_1}}  +\Frac{(c^{\hat A^8_3})^2}{ M^2_{A^8_3}} \right. $	
				\\ 
				&
				$ \displaystyle \left.  -\, \Frac{(\widetilde c^{\hat A^8_3})^2}{ M^2_{A^8_3}}  \right)
				\,+\, \displaystyle \Frac{3}{16} \left(  \Frac{ (C_0^{V^8_1})^2}{ M^2_{V^8_1}} - \Frac{ (C_0^{V^8_3})^2}{ M^2_{V^8_3}}  +  \Frac{ (\widetilde C_0^{A^8_1})^2}{ M^2_{A^8_1}}    - \Frac{ (\widetilde C_0^{A^8_3})^2}{ M^2_{A^8_3}} \right)  $	
				\\[2ex]
				\hline 			
				\multirow{3}{*}{2}	&
				$ \Frac{c_q^{P^1_3}\,c_q^{P^1_3}}{2M_{P^1_3}^2} \,+\, \displaystyle\frac{1}{128}\! \left( \Frac{(c^{S_1^8})^2}{ M^2_{S_1^8}} - \Frac{(c^{S^8_3})^2}{ M^2_{S^8_3}} - \Frac{(c^{P_1^8})^2}{ M^2_{P_1^8}}  -  \Frac{5(c^{P^8_3})^2}{3 M^2_{P^8_3}} \right) $
				\\	
				&
				$ +\, \displaystyle \Frac{1}{32}\! \left(\Frac{(c^{\hat V^8_1})^2}{ M^2_{V^8_1}}   - \Frac{(\widetilde c^{\hat V^8_1})^2}{ M^2_{V^8_1}} -\Frac{(c^{\hat V^8_3})^2}{ M^2_{V^8_3}} + \Frac{(\widetilde c^{\hat V^8_3})^2}{ M^2_{V^8_3}}  - \Frac{(c^{\hat A^8_1})^2}{ M^2_{A^8_1}}  + \Frac{(\widetilde c^{\hat A^8_1})^2}{ M^2_{A^8_1}} +\Frac{(c^{\hat A^8_3})^2}{ M^2_{A^8_3}} \right. $
				\\ 
				&
				$ \displaystyle \left. -\, \Frac{(\widetilde c^{\hat A^8_3})^2}{ M^2_{A^8_3}} \right) \,+\, \displaystyle \Frac{3}{16} \left(  -\Frac{ (C_0^{V^8_1})^2}{ M^2_{V^8_1}} + \Frac{ (C_0^{V^8_3})^2}{ M^2_{V^8_3}} - \Frac{ (\widetilde C_0^{A^8_1})^2}{ M^2_{A^8_1}}   + \Frac{ (\widetilde C_0^{A^8_3})^2}{ M^2_{A^8_3}} \right)    $
				\\[1ex]
				\hline 	
				\multirow{2}{*}{3}	&
				$  \Frac{c_q^{S^1_1}\,c_q^{S^1_1}}{4M_{S^1_1}^2} - \Frac{c_q^{S^1_3}\,c_q^{S^1_3}}{4M_{S^1_3}^2} \,+\, \displaystyle\frac{1}{64}\! \left( - \Frac{2(c^{S_1^8})^2}{3 M^2_{S_1^8}} - \Frac{(c^{S^8_3})^2}{ 3 M^2_{S^8_3}}+ \Frac{(c^{P^8_3})^2}{M^2_{P^8_3}} \right)  $ 
				\\ 
				&
				$ \Frac{(c^{\hat V^8_3})^2}{16 M^2_{V^8_3}}  - \Frac{( \widetilde c^{\hat V^8_3})^2}{16 M^2_{V^8_3}}  -\Frac{(c^{\hat A^8_3})^2}{16 M^2_{A^8_3}} + \Frac{(\widetilde c^{\hat A^8_3})^2}{16 M^2_{A^8_3}} +  \Frac{3 (C_0^{V^8_3})^2}{8 M^2_{V^8_3}} + \Frac{3 (\widetilde C_0^{A^8_3})^2}{8 M^2_{A^8_3}}$
				\\[1ex]
				\hline 	
				\multirow{2}{*}{4}	&
				$\Frac{c_q^{P^1_1}\,c_q^{P^1_1}}{4M_{P^1_1}^2} -\Frac{c_q^{P^1_3}\,c_q^{P^1_3}}{4M_{P^1_3}^2} \,+\, \displaystyle \frac{1}{64}\! \left(  \Frac{(c^{S^8_3})^2}{ M^2_{S^8_3}} - \Frac{2(c^{P_1^8})^2}{3 M^2_{P_1^8}} - \Frac{(c^{P^8_3})^2}{3 M^2_{P^8_3}} \right)  $ 
				\\ 
				&
				$\Frac{(c^{\hat V^8_3})^2}{16 M^2_{V^8_3}}  - \Frac{( \widetilde c^{\hat V^8_3})^2}{16 M^2_{V^8_3}}  -\Frac{(c^{\hat A^8_3})^2}{16 M^2_{A^8_3}} + \Frac{(\widetilde c^{\hat A^8_3})^2}{16 M^2_{A^8_3}}  - \Frac{3 (C_0^{V^8_3})^2}{8 M^2_{V^8_3}} - \Frac{3 (\widetilde C_0^{A^8_3})^2}{8 M^2_{A^8_3}}$	
				\\[1ex]
				\hline 	
				\multirow{3}{*}{\ 5\ }	&
				$ -\Frac{c_q^{\hat{V}^1_3}\,c_q^{\hat{V}^1_3}}{2M_{V^1_3}^2} -\Frac{\widetilde{c}_q^{\hat{A}^1_3}\,\widetilde{c}_q^{\hat{A}^1_3}}{2M_{A^1_3}^2}  \,+\, \displaystyle \frac{1}{128}\! \left( \Frac{(c^{S^8_3})^2}{ M^2_{S^8_3}}- \Frac{(c^{S_1^8})^2}{ M^2_{S_1^8}}+ \Frac{(c^{P^8_3})^2}{ M^2_{P^8_3}}- \Frac{(c^{P_1^8})^2}{ M^2_{P_1^8}} \right)$ 
				\\ 
				&
				$+\,\displaystyle \frac{1}{64}\! \left( - \Frac{(c^{\hat V^8_1})^2}{ M^2_{V^8_1}} - \Frac{(\widetilde c^{\hat V^8_1})^2}{ M^2_{V^8_1}} + \Frac{7(c^{\hat V^8_3})^2}{3 M^2_{V^8_3}} + \Frac{(\widetilde c^{\hat V^8_3})^2}{ M^2_{V^8_3}}  \right.  $
				\\
				&
				$\displaystyle \left. \, -\, \Frac{(c^{\hat A^8_1})^2}{ M^2_{A^8_1}} - \Frac{(\widetilde c^{\hat A^8_1})^2}{ M^2_{A^8_1}} + \Frac{(c^{\hat A^8_3})^2}{ M^2_{A^8_3}}  + \Frac{7(\widetilde c^{\hat A^8_3})^2}{3 M^2_{A^8_3}}  \right)	$	
				\\[1ex]
				\hline 	
			\end{tabular}
		\end{center}
	\end{table}
	\begin{table}[t!]
		\begin{center}
			\renewcommand{\arraystretch}{2.2}
			\begin{tabular}{|c|c|}
				\hline	
				\multirow{3}{*}{6}	&
				$\,\quad- \Frac{(\widetilde{c}_q^{\hat{V}^1_3})^2}{2M_{V^1_3}^2} -\Frac{({c}_q^{\hat{A}^1_3})^2}{2M_{A^1_3}^2}\,+\, \displaystyle \frac{1}{128}\! \left(  -\Frac{(c^{S^8_3})^2}{ M^2_{S^8_3}} + \Frac{(c^{S_1^8})^2}{M^2_{S_1^8}}  - \Frac{(c^{P^8_3})^2}{ M^2_{P^8_3}} + \Frac{(c^{P_1^8})^2}{ M^2_{P_1^8}}  \right) \quad\,$ 
				\\ 
				&
				$+\displaystyle \frac{1}{64}\! \left(  - \Frac{(c^{\hat V^8_1})^2}{ M^2_{V^8_1}} - \Frac{(\widetilde c^{\hat V^8_1})^2}{ M^2_{V^8_1}} + \Frac{(c^{\hat V^8_3})^2}{ M^2_{V^8_3}} + \Frac{7(\widetilde c^{\hat V^8_3})^2}{3 M^2_{V^8_3}} \right.  $
				\\
				&
				$\displaystyle \left. -\, \Frac{(c^{\hat A^8_1})^2}{ M^2_{A^8_1}} - \Frac{(\widetilde c^{\hat A^8_1})^2}{ M^2_{A^8_1}} +\Frac{7(c^{\hat A^8_3})^2}{3 M^2_{A^8_3}}  + \Frac{(\widetilde c^{\hat A^8_3})^2}{ M^2_{A^8_3}}  \right)	$
				\\[1ex]
				\hline 	
				\multirow{2}{*}{7}	&
				$-\Frac{({c}_q^{\hat{V}^1_1})^2}{4M_{V^1_1}^2} -\Frac{(\widetilde{c}_q^{\hat{A}^1_1})^2}{4M_{A^1_1}^2}
				+\Frac{({c}_q^{\hat{V}^1_3})^2}{4M_{V^1_3}^2} +\Frac{(\widetilde{c}_q^{\hat{A}^1_3})^2}{4M_{A^1_3}^2}
				\,-\,\Frac{(c^{S^8_3})^2}{64 M^2_{S^8_3}}- \Frac{(c^{P^8_3})^2}{64 M^2_{P^8_3}} $
				\\ 
				&
				$ + \Frac{( c^{\hat V^8_1})^2}{96 M^2_{V^8_1}} -\Frac{(c^{\hat V^8_3})^2}{24 M^2_{V^8_3}}  - \Frac{(\widetilde c^{\hat V^8_3})^2}{32 M^2_{V^8_3}} + \Frac{( \widetilde c^{\hat A^8_1})^2}{96 M^2_{A^8_1}}   - \Frac{(c^{\hat A^8_3})^2}{32 M^2_{A^8_3}}  - \Frac{(\widetilde c^{\hat A^8_3})^2}{24 M^2_{A^8_3}}   $
				\\[1ex]
				\hline 	
				\multirow{2}{*}{8}	&
				$-\Frac{(\widetilde{c}_q^{\hat{V}^1_1})^2}{4M_{V^1_1}^2}  -\Frac{({c}_q^{\hat{A}^1_1})^2}{4M_{A^1_1}^2}
				+ \Frac{(\widetilde{c}_q^{\hat{V}^1_3})^2}{4M_{V^1_3}^2} +\Frac{({c}_q^{\hat{A}^1_3})^2}{4M_{A^1_3}^2}
				+\Frac{(c^{S^8_3})^2}{64 M^2_{S^8_3}}+ \Frac{(c_1^{P^8_3})^2}{64 M^2_{P^8_3}} $
				\\ 
				&
				$ + \Frac{(\widetilde c^{\hat V^8_1})^2}{96 M^2_{V^8_1}}- \Frac{( c^{\hat V^8_3})^2}{32 M^2_{V^8_3}}   -\Frac{(\widetilde c^{\hat V^8_3})^2}{24 M^2_{V^8_3}}  +  \Frac{(c^{\hat A^8_1})^2}{96 M^2_{A^8_1}}   - \Frac{(c^{\hat A^8_3})^2}{24 M^2_{A^8_3}} - \Frac{( \widetilde c^{\hat A^8_3})^2}{32 M^2_{A^8_3}}   $
				\\[1ex]
				\hline 	
				\multirow{2}{*}{9}	&
				$ \displaystyle \frac{1}{256}\! \left( - \Frac{(c^{S_1^8})^2}{ M^2_{S_1^8}} + \Frac{(c^{S^8_3})^2}{ M^2_{S^8_3}} +  \Frac{(c^{P_1^8})^2}{ M^2_{P_1^8}} - \Frac{(c^{P^8_3})^2}{ M^2_{P^8_3}}  \right) $
				\\ 
				&
				$-\displaystyle\frac{(C_{0,q}^{V^1_3})^2}{M_{V^1_3}^2} - \displaystyle\frac{(\widetilde{C}_{0,q}^{A^1_3})^2}{M_{A^1_3}^2} - \Frac{ (C_0^{V^8_1})^2}{32 M^2_{V^8_1}}  + \Frac{ 7( C_0^{V^8_3})^2}{96 M^2_{V^8_3}}   -  \Frac{ (\widetilde C_0^{A^8_1})^2}{32 M^2_{A^8_1}}  + \Frac{7 (\widetilde C_0^{A^8_3})^2}{96 M^2_{A^8_3}}  $	
				\\[1ex]
				\hline 	
				\multirow{2}{*}{10}	&
				$-\displaystyle \Frac{(c^{S^8_3})^2}{128 M^2_{S^8_3}}
				+ \Frac{(c^{P^8_3})^2}{128 M^2_{P^8_3}} 
				- \Frac{(C_{0,q}^{V^1_1})^2}{2M_{V^1_1}^2}
				+ \Frac{(C_{0,q}^{V^1_3})^2}{2M_{V^1_3}^2}   
				- \Frac{(\widetilde{C}_{0,q}^{A^1_1})^2}{2M_{A^1_1}^2}
				+ \Frac{(\widetilde{C}_{0,q}^{A^1_3})^2}{2M_{A^1_3}^2} $
				\\ 
				&
				$ - \Frac{ (C_0^{V^8_3})^2}{12 M^2_{V^8_3}} + \Frac{ (C_0^{V^8_1})^2}{48 M^2_{V^8_1}} - \Frac{ (\widetilde C_0^{A^8_3})^2}{12 M^2_{A^8_3}} +  \Frac{ (\widetilde C_0^{A^8_1})^2}{48 M^2_{A^8_1}}$	
				\\[1ex]
				\hline 	
				
				%
\end{tabular}
\end{center}
\end{table}				
	
\clearpage
				
\begin{table}[t!]
	\begin{center}
		\renewcommand{\arraystretch}{2.2}
		\begin{tabular}{|c|c|}
				\hline
				\multicolumn{2}{|c|}{Four-fermion $P$-odd LECs, $\Delta \widetilde \mF^{q,q}_i \, $ | \ quark--quark }\\
				\hline
				1	&
				$ \displaystyle 
				-\,\Frac{c_q^{\hat{V}^1_3}\, \widetilde{c}_q^{\hat{V}^1_3}} {M_{V^1_3}^2} 
				-\Frac{ c_q^{\hat{A}^1_3}\, \widetilde{c}_q^{\hat{A}^1_3}}{M_{A^1_3}^2}
				-\Frac{ c^{\hat V^8_1}\, \widetilde c^{\hat V^8_1}}{16 M^2_{V^8_1}} 
				+ \Frac{5 c^{\hat V^8_3}\, \widetilde c^{\hat V^8_3}}{48 M^2_{V^8_3}} 		
				- \Frac{ c^{\hat A^8_1}\, \widetilde c^{\hat A^8_1}}{16 M^2_{A^8_1}}
				+ \Frac{5\, c^{\hat A^8_3}\, \widetilde c^{\hat A^8_3}}{48 M^2_{A^8_3}} 
				$
				\\[1ex] \hline
				\multirow{2}{*}{2}	&
				$ \displaystyle  
				-\,\Frac{c_q^{\hat{V}^1_1}\, \widetilde{c}_q^{\hat{V}^1_1}}{2M_{V^1_1}^2}
				+\Frac{c_q^{\hat{V}^1_3}\, \widetilde{c}_q^{\hat{V}^1_3}}{2M_{V^1_3}^2}
				-\Frac{c_q^{\hat{A}^1_1}\, \widetilde{c}_q^{\hat{A}^1_1}}{2M_{A^1_1}^2} 
				+\Frac{c_q^{\hat{A}^1_3}\, \widetilde{c}_q^{\hat{A}^1_3}}{2M_{A^1_3}^2} $
				\\
				&
				$ \displaystyle
				+\, \Frac{ c^{\hat V^8_1}\, \widetilde c^{\hat V^8_1}}{48 M^2_{V^8_1}} 
				-\Frac{7\, c^{\hat V^8_3}\, \widetilde c^{\hat V^8_3}}{48 M^2_{V^8_3}}
				+ \Frac{ c^{\hat A^8_1}\, \widetilde c^{\hat A^8_1}}{48 M^2_{A^8_1}}
				- \Frac{7\, c^{\hat A^8_3}\, \widetilde c^{\hat A^8_3}}{48 M^2_{A^8_3}} $
				\\[1ex] \hline
			\end{tabular}
		\end{center}
		\caption{\small
			All resonance contributions (spin-0, spin-1 Proca, spin-1 Antisymmetric and fermionic) to the four-fermion LECs of the EWET NLO Lagrangian including two quark bilinears (table \ref{tab:4fermion}), where $i$ is the operator assignment index for this set.  LECs corresponding to $P$-even ($P$-odd) operators are shown in the upper (lower) block.}
		\label{tab:LEC4fermionquark}
	\end{table}

	The Lagrangian terms of eq.~(\ref{eq:S1-p2}) can be further acknowledged as precise contributions to the EWET scalar Lagrangian and the Yukawas, which were defined in eq.~(\ref{eq:LscalarEWET}) and eq.~(\ref{eq:Yukawas}) as a general extension of the SM ones.
	On the one hand, the first term in this equation can be rearranged in the Higgs potential, $V(h/v)$, while 
	the bosonic term involving the $c_d$ resonance coupling is included in $\mF^{(u)}(h/v)$ as 
	\be \label{eq:p2corr-h}
	\Delta c^{(V)}_{n\ge 4}\, =\, -\frac{v^2}{2 M_{S^1_1}^2}\;\sum_{k=0}^{n-4}
	\lambda_{hS_1}^{(k)} \,\lambda_{hS_1}^{(n-k-4)}\, ,
	\qquad\qquad
	\Delta c^{(u)}_{n\ge 2}\, =\, \frac{2\sqrt{2} v}{M_{S^1_1}^2}\;
	\sum_{k=0}^{n-2} \lambda_{hS_1}^{(k)}\, c_d^{(n-k-2)}\, .
	\ee
	On the other hand, the fermionic contribution of this expression joins the Yukawa coupling of the LO fermionic Lagrangian, either in the lepton or in the quark case, like
	\be \label{eq:DYukawa-p2}
	\Delta\mY^f\, =\, -\frac{1}{\sqrt{2} M_{S^1_1}^2}\; h^2\, \lambda_{hS_1}(h/v)\; c_f^{S_1^1}(h/v)\,,
	\ee
	although its effects are only present at $\mO(h^2)$.

Finally, it is interesting to discuss whether some resonance coupling might be enhanced in some particular scenario. So far, we have considered that all the resonances are linearly coupled to the EWET fields and that all the resonance couplings are of $\mO(1)$. However, there are models including uncharged color-singlet scalar states like $S_1^1$ that mix with a light Higgs and they are able to produce a significant enhancement of the scalar resonance coupling, up to order $\mO(M_{S_1^1}^2/f^2)\gg \mO(1)$, being $f$ the typical breaking scale\footnote{There are some models that make an explicit distinction between this scale and the vev.} associated to the heavy scalar. This circumstance is studied in \cite{Buchalla:2016bse}. From the phenomenology point of view, these scalar objects can generate contributions even at the EWET LO Lagrangian, contrary to the resonance interactions studied in the previous sections which leave their imprints at NLO or higher orders. As a consequence, the analysis of these models within the electroweak resonance theory requires to perform a more careful integration of the high-energy fields, where the NLO contributions of the resonance EoMs must be considered too.
		
	One can also wonder whether this resonance coupling enhancement can happen for the color-singlet custodial-triplet resonance, $S_3^1$, too. For this purpose, we analyze two types of models. First, models that assume the scalars to be elementary: a Higgs-like scalar boson and other scalar heavy states (like $S_1^1$ and $S_3^1$).\footnote{It is shown in \cite{Chanowitz:1985ug} that this kind of resonance models are not discarded phenomenologically.} It is required, however, that these scalars are custodial symmetry invariant so that they are stable under loop corrections and the $\rho$-parameter is protected. As a consequence, there is no mixing between the scalar triplet and the scalar singlets \cite{Gunion:1989ci}, unless it happens at the loop level where custodial breaking sources can arise.  Second, we can study models where the previous scalars are composite, like in \cite{Georgi:1985nv}. In this type of models, referred as ultracolor (similar to Technicolor), these particles are pseudo-Goldstone bosons generated by a dynamical breaking of the chiral symmetry at a high-energy scale. The Higgs boson and the resonance states are bond states of other more fundamental particles which actually produce the SSB. Certainly, the scalar triplet, $S^1_3$, can induce some mixing in this case \cite{Georgi:1985nv}. Nonetheless, the masses of the scalar particles are not generated from the Higgsing, but from a dynamical breaking. Hence, scalar triplet resonances, either elementary or composite, do not generate effects of order $\mO(M_{S_3^1}^2/f^2)$.


\chapter{Proca vs. Antisymmetric formalism} \label{ch:spin1}

As long as high-energy massive states remain undiscovered, it is not possible to know what are the ones that nature prefers nowadays or, for instance for the spin-1 resonances, how to properly describe them whatsoever. While for the first situation we have analyzed a complete set of resonance states which cover all the possible ways they can couple linearly to the SM;
for the second problem, we simply cannot decide a priori which spin-1 field formalism will be better adjusted to the properties of a given set of heavy states. Actually, spin-1 objects carry some freedom at the Lagrangian level in the way they are implemented in the resonance theory. Therefore, there are several available high-energy formalisms that are able to describe the same physical reality and thus it should be enough to analyze one single formalims in order to introduce these kind of heavy states, like the 4-vector Proca formalism or the rank-2 Antisymmetric tensor one introduced in the last section.

However, the results shown in the tables \ref{tab:LECbosonic}, \ref{tab:LEC2fermion}, \ref{tab:LEC4fermionlepton}, \ref{tab:LEC4fermionmixed}, \ref{tab:LEC4fermionquark} in chapter \ref{ch:LECs} deliver quite uneven predictions for the EWET. Naively, one may think that these differences could arise from the freedom of selecting the operator basis, from some incompleteness or redundancy or any other ambiguity. Nonetheless, if this were the problem, discrepancies between both formalism would occur for some operators only, since a given physical process just receives contributions for a particular subset of LEC. Furthermore, the Proca and Antisymmetric patterns of LECs are discrepant enough for not being some mistakes causing them. For instance, the resonance exchange of vector and axial-vector colorless triplet states, $V^1_3$ and $A^1_3$, respectively, contributes to the bosonic EWET Lagrangian as $\mO_{1,2,3,4,5,6,7,9}$ when resonances are expressed in the Antisymmetric representation, whereas they only affect $\mO_{11}$ in the Proca case. 

The source of this problem certainly comes from their dynamical structure which happens to be opposite for these spin-1 representations, as explained in chapter \ref{ch:resonancetheory}. Proca vector resonances contribute to the EWET NLO Lagrangian when interact with light fields collected in one single Lorentz index tensor\footnote{Recall that Proca elements bring a hat, $\wedge$, and Antisymmetric ones come alone.} $\hat \chi_{\hat R}^\mu$, while Antisymmetric resonances only leave their imprints with two Lorentz index operators included in $\chi_R^{\mu\nu}$. Nevertheless, we still do not know which is the right representation (if any) and this is a crucial issue, because all the LEC building formulated before would fall down unless we are able to deeply understand the spin-1 representation problem.

\section{Equivalence of the Proca and the Antisymmetric formalisms} \label{sec:equivalence}
\markboth{CHAPTER 6.\hspace{1mm} PROCA VS. ANTISYMMETRIC FORMALISM}{6.1.\hspace{1mm} FORMALISM EQUIVALENCE}

The particular mathematical prescription used to describe a spin-1 object should lead to the same results, and so it has to be for the Proca and the Antisymmetric prescriptions. Hence, it is evident that we are passing something over.

The answer to this puzzle can be found when performing a change of variables in the path integral \cite{Bijnens:1995,Kampf:2006}, shown explicitly in the appendix \ref{app:spin1equivalence}. Here we address a full calculation where we manage to rewrite the high-energy Antisymmetric Lagrangian including one single resonance coupled linearly in terms of the Proca Lagrangian, like
\be \label{eq:equivalence}
\mL_R^{(P)} \, + \, \mL_{{\rm non}-R}^{(P)} \, = \, \mL_R^{(A)} \, + \, \mL_{{\rm non}-R}^{(A)} \,,
\ee
where $P$ and $A$ stand for Proca and Antisymmetric, respectively. The key point of the formalism equivalence stays in the non-resonant contributions, which are not equal. Notice that the resonance Lagrangians  (omitting the non-resonant terms), and thus the interacting high-energy operators, are not equivalent alone. They require the remaining pieces not involving the high-energy fields in order to fulfill the equivalence. Therefore, Proca contributions to the LECs that were not generated by the Antisymmetric ones must be originated by some light field process in this representation and vice versa.

The precise relation between these two resonance implementations is found to be
\begin{align} \label{eq:equivalencetensor}
& \displaystyle \chi_{R^m_n}^{\mu\nu} \, + \, \Frac{1}{2} \left(\nabla^\mu \chi_{R^m_n}^\nu \,-\, \nabla^\nu \chi_{R^m_n}^\mu\right) &  & &\nn\\
& \qquad = \, 
\Frac{1}{2 M_{R^m_n}} \left(\nabla^\mu \hat{\chi}_{\hat{R}^m_n}^\nu
\,-\, \nabla^\nu \hat{\chi}_{\hat{R}^m_n}^\mu\right)
\, +\, M_{R^m_n}\, \hat{\chi}_{\hat{R}^m_n}^{\mu\nu} & \qquad
& (R=V,\, A)\, , &
\end{align}
with $m,n$ their $SU(3)_C$ and $SU(2)_{L+R}$ representations,\footnote{For the singlet-singlet case the covariant derivative becomes the partial derivative.} respectively. In addition, the non-resonant Lagrangians are related through
\begin{align} \label{eq.equivP-nonR}
\displaystyle \mL^{\rm (A)}_{\text{non-R}} & \,=\,
\sum_{R=V,A} \left\{ \sum_{\hat R^1_1,\hat R^1_3,\hat R^8_1,\hat R^8_3} \left[ \left( \bra \hat{\chi}_{\hat{R}^m_n\, \mu\nu}\,
\hat{\chi}_{\hat{R}^m_n}^{\mu\nu}\ket_{2,3}
\, -\, \Frac{1}{2}\, \bra \bra  \hat{\chi}_{\hat{R}^m_n}^{\mu\nu}\ket_2
\bra  \hat{\chi}_{\hat{R}^m_n \,\mu\nu}^{\phantom{\mu}}\ket_2 \ket_3 \right. \right. \right.
\nn\\
\displaystyle 
& \qquad \,-\, \left. \Frac{1}{3}\, \bra \bra  \hat{\chi}_{\hat{R}^m_n}^{\mu\nu}\ket_3 
\bra  \hat{\chi}_{\hat{R}^m_n \,\mu\nu}^{\phantom{\mu}}\ket_3 \ket_2 \,+\, \Frac{1}{6}\,\bra \hat{\chi}_{\hat{R}^m_n\, \mu\nu}\ket_{2,3}^2 \right)
\nn\\
\displaystyle 
& \,-\, 
\Frac{1}{2M_{R^m_n}^2}\, \left(  \bra  \hat{\chi}_{\hat{R}^m_n}^{\mu}\,
\hat{\chi}_{\hat{R}^m_n\,\mu}^{\phantom{\mu}} \ket_{2,3}
-\Frac{1}{2}\, \bra \bra \hat{\chi}_{\hat{R}^m_n}^{\mu}\ket_2
\bra \hat{\chi}_{\hat{R}^m_n\, \mu}^{\phantom{\mu}}\ket_2 \ket_3 \right. 
\nn\\
\displaystyle 
& \qquad \,-\, \left.  \Frac{1}{3}\, \bra \bra  \hat{\chi}_{\hat{R}^m_n}^{\mu}\ket_3 
\bra  \hat{\chi}_{\hat{R}^m_n \,\mu}^{\phantom{\mu}}\ket_3 \ket_2 \,+\, \Frac{1}{6}\,\bra \hat{\chi}_{\hat{R}^m_n\, \mu}\ket_{2,3}^2 \right) \bigg]
\Bigg\}
\,\, +\,\, \mL^{\rm (P)}_{\text{non-R}}
\, .
\end{align}
where we perform a sum over vector and axial-vector spin-1 resonances in all the custodial symmetry and color representations. The notation of this equation has been simplified\footnote{It can be further simplified attending to the fact that all the Proca tensors $\hat \chi_{\hat R}^\mu$ of eq.~(\ref{eq:proca-chimu}) are color traceless. Hence, lines 2 and 4 in eq.~(\ref{eq.equivP-nonR}) vanish (within the key parenthesis). Nevertheless, we keep the whole algebraic form of the equation for the sake of completeness.} and not all the terms in eq.~(\ref{eq.equivP-nonR}) hold for all the resonances types in the sum: 
\begin{itemize}
	\item $SU(n)$ traces containing two $\hat \chi_{\hat R}$ tensors have to be understood as traces when the given resonance forms a multiplet under $SU(n)$.
	\item $SU(2)$ single-traced objects,\ie $\bra \hat \chi^\mu_{\hat R^m_n} \ket_2$ or $\bra \hat \chi^{\mu\nu}_{\hat R^m_n} \ket_2$, must be removed when $n=1$.
	\item $SU(3)$ single-traced objects,\ie $\bra \hat \chi^\mu_{\hat R^m_n} \ket_3$ or $\bra \hat \chi^{\mu\nu}_{\hat R^m_n} \ket_3$, must be removed when $m=1$.	
\end{itemize}

The two equations above become essential for the spin-1 high-energy states understanding  and also for the pattern of LECs results obtained in chapter \ref{ch:LECs}, as we proceed to explain. Indeed, one can check that the tensors $\hat \chi_{\hat R}^\mu$, $\hat \chi_{\hat R}^{\mu\nu}$ and $\chi_{R}^\mu$, $\chi_{R}^{\mu\nu}$ of eqs.~(\ref{eq:proca-chimu}, \ref{eq:proca-chimunu}) and eqs.~(\ref{eq:antisym-chimu}, \ref{eq:antisym-chimunu}), respectively, satisfy this mathematical relation, as expected, provided the following identifications between the resonance couplings of both formalisms are made: 
\begin{align} \label{eq:LECsrelation}
F_X \, &=\, f_X\, M_{V^1_1}\,, & \quad  
\widetilde F_X \, &=\, \widetilde f_X\, M_{A^1_1}\,, & \quad
&	&
\nn\\
F_R \, &=\, f_{\hat R}\, M_{R^1_3}\,, &	\quad
\widetilde F_R \, &=\, \widetilde f_R\, M_{R^1_3}\,, & \quad
& (R\,=\,V,\, A) &
\nn\\
G_V \, &=\, g_{\hat V}\, M_{V^1_3}\,, & \quad
\widetilde G_A \, &=\, g_{\hat A}\, M_{A^1_3}\,, & \quad
&	&
\nn\\
\widetilde\Lambda^{h V}_1 \, &=\, \widetilde\lambda^{h \hat V}_1\, M_{V^1_3}\,, & \quad
\Lambda^{h A}_1 \, &=\, \lambda^{h \hat A}_1\, M_{A^1_3}\,, & \quad
&	&
\nn\\
C_{0,f}^{V^1_n} \, &=\, c_{0,f}^{\hat V^1_n}\, M_{V^1_n}\,, & \quad
\widetilde C_{0,f}^{A^1_n} \, &=\, \widetilde c_{0,f}^{\hat A^1_n}\, M_{A^1_n}\,, & \quad
& (n\,=\,1,\,3\,;\,f\,=\, l,\,q) &
\nn\\
C_G \, &=\, c_G\, M_{V^8_1}\,, & \quad 
\widetilde C_G \, &=\, \widetilde c_G\, M_{A^8_1}\,, & \quad 
&	&
\nn\\
C_{0}^{V^8_n} \, &=\, c_{0}^{\hat V^8_n}\, M_{V^8_n}\,, & \quad
\widetilde C_{0}^{A^8_n} \, &=\, \widetilde c_{0}^{\hat A^8_n}\, M_{A^8_n}\,, & \quad
& (n\,=\, 1,\, 3) &
\nn\\ \nn\\
\widetilde C_\mT \, &=\, \widetilde c_\mT\, /\, M_{V^1_1}\,, & \quad 
C_\mT \, &=\, c_\mT\, /\, M_{A^1_1}\,, & \quad 
&	&
\nn\\
C_f^{R^1_n} \, &=\, c_f^{\hat R^1_n} \,/ M_{R^1_n}\,, &\quad 
\widetilde C_f^{R^1_n} \, &=\, \widetilde c_f^{R^1_n} \,/ M_{R^1_n}\,, &\quad 
&(R\,=\, V,\, A\, ; \, n\,=\,1,\,3\,;\,f\,=\, l,\,q)&
\nn\\
C^{R^8_n} \, &=\, c^{\hat R^8_n} \,/ M_{R^8_n}\,, & \quad 
\widetilde C^{R^8_n} \, &=\, \widetilde c_f^{\hat R^8_n} \,/ M_{R^8_n}\,.& \,\quad 
&(R\,=\, V,\, A\, ; \, n\,=\,1,\,3)& 
\end{align}
where the upper (lower) block of equations in eq.~(\ref{eq:LECsrelation}) come from the resonance couplings from the double-index tensors, $\chi_R^{\mu\nu}$ and $\hat \chi_{\hat R}^{\mu\nu}$ (single-index tensors, $\chi_R^{\mu}$ and $\hat \chi_{\hat R}^{\mu}$).

Notice the very interesting behavior reflected by these coupling parameters: all the resonance couplings in the two formalisms respond to the same pattern since they are related through a mass factor which comes multiplying or dividing depending only on the chiral tensor they come from. When they come from the two Lorentz index ones (upper case), $\chi_R^{\mu\nu}$ and $\hat \chi_{\hat R}^{\mu\nu}$, couplings in the Antisymmetric representation (capital letters) carry one extra dimension of mass with respect to the Proca parameters (small letters). This accounts for the additional derivative present in $\hat R_{\mu\nu}$ with respect to $R_{\mu\nu}$ where it is absent. Hence, this crucial difference makes the chiral structures of the Antisymmetric formalism carry one power of momenta less than the Proca formalism, what makes the tree-level exchanges of resonances to count as $\mO_{\hat d}(p^4)$ and thus contribute to the EWET NLO LECs in the Antisymmetric case, while the Proca tensors contribute at $\mO_{\hat d}(p^6)$ instead, not leaving any imprint in the studied pattern of LECs. However, it happens exacty the reverse situation for those resonance couplings coming from the single Lorentz index tensors (lower case), $\chi_R^{\mu}$ and $\hat \chi_{\hat R}^{\mu}$. The extra derivative required for antisymmetric couplings is compensated with an inverse mass factor, $1/M_R$, paying the price of an extra unit in the chiral power counting in comparison with Proca couplings. The consequence is that only resonances in this last representation contribute to the $\mO_{\hat d}(p^4)$ EWET NLO Lagrangian. 

When a given resonance coupling does not generate any contribution to the LECs in one formalism but it does in the other, the discrepancy gets restored by the non-resonant Lagrangian, in order to acomplish the equivalence, eq.~(\ref{eq:equivalence}), between both formalisms. Every spin-1 formalism must predict the same pattern of LECs for the EWET NLO Lagrangian. The different representations only differ in the way the explicit resonance Lagrangian and the local non-resonant one are split and this is only a choice of a functional field representation of the effective Lagrangian in the path-integral formulation of the generating functional. Therefore, the particular splitting between resonance and local Lagrangians is unphysical \cite{Ecker:1989yg}.

Nevertheless, non-resonant Lagrangians still present some ambiguity. Despite the fact that eq.~(\ref{eq.equivP-nonR}) brings back the resonance contributions that were boosted to higher orders, it only sets the difference between the two local Lagrangians $\mL^{\rm (A)}_{\text{non-R}} -\mL^{\rm (P)}_{\text{non-R}}$. In order to unequivocally establish which are the right contributions further inputs are still required. This will be fixed with the so-called short-distance constraints in the next section. In short, additional requirements over the non-resonant local pieces are imposed so that the resonance theory is consistent and behaves properly in the short-distance regime,\ie at large energies. These considerations assume notwithstanding that the background fundamental UV theory is well-behaved in the UV, too.

We mainly conclude for the equivalence analysis that any given spin-1 formalism gives all the right predictions for the LECs, as naively expected. However, it is not always easy to pin down all the contributions just by studying one single resonance formalism. For this reason, we will analyze the spin-1 field high-energy interactions using a mixed procedure that combines both the Proca and the Antisymmetric formalisms, following these prescriptions: 
\begin{enumerate} 
	\item We should use the Proca resonance Lagrangian when dealing with interactions that can be collected in the single Lorentz index chiral tensor $\hat \chi_{\hat R}^\mu$.
	\item We should employ the Antisymmetric formalism in the case that a given interaction can be included in the double Lorentz index chiral tensor $\chi_R^{\mu\nu}$.
	\item The full set of resonance predictions for the EWET LECs is exhaustive when considered together both contributions obtained from both the Proca and the Antisymmetric resonance Lagrangians.
	\item Short-distance constraints must be imposed in order to set the remaining local non-resonant Lagrangian which do not add any further prediction to the LECs than the Proca and Antisymmetric combined ones.
\end{enumerate}
Of course, the proof of the last statement still is lacking, so this is what we aim for the next section.

As a consequence, the results for the LECs obtained in tables \ref{tab:LECbosonic}, \ref{tab:LEC2fermion}, \ref{tab:LEC4fermionlepton}, \ref{tab:LEC4fermionmixed}, \ref{tab:LEC4fermionquark} are not redundant at all for the spin-1 fields, but they are the full complete set of resonance contributions to the EWET NLO Lagrangian. The Proca and the Antisymmetric formalisms are found to be complementary and whenever one resonance Lagrangian in one representation fails to give the right predictions for a given interaction, the other Lagrangian representation gets it.

\section{Short-distance constraints} \label{sec:SD}
The resonance theory is an EFT including not only high-energy states coupled linearly to the light fields, but additionally the same degrees of freedom that the EWET has got, as eq.~(\ref{eq:formalRLagrangian}) shows. It contains the most general set of light field operators, formally identical to the EWET ones, but not necessarily with the same couplings, since they belong to different EFTs. Hence, for every EWET coupling, $\mF_i$, there is as well an analogous short-distance coupling, $\mF^{{\rm SD}}_i$, for the local resonance Lagrangian, with the same operator structure as tables \ref{tab:bosonic-Op4}, \ref{tab:2fermion} and \ref{tab:4fermion}. Particularly for the spin-1 fields, we denote $\mF_i^{{\rm SDA}}$ and $\mF_i^{{\rm SDP}}$ to the light-field couplings of the Proca and 
Antisymmetric short-distance effective theories, SDET-A and SDET-P, respectively. Therefore, the high-energy local non-resonant Lagrangian reads
\be \label{eq.SDA-P}
\mL_{\text{non-R}}^{(A)} \; =\; \sum_i \, \mF_i^{\,\mathrm{SDA}}\, \cO_i[\phi,\psi] \, ,
\qquad\quad
\mL_{\text{non-R}}^{(P)} \; =\; \sum_i \, \mF_i^{\,\mathrm{SDP}}\, \cO_i[\phi,\psi] \, ,
\ee
where the sum acts over all the light-field operators.

The determination of these couplings is performed by an analysis of a wide and exhaustive set of Green functions. The assumed good behavior of these structures at large energies will supply for the required constraints so that we are able to fix the local Lagrangians in eq.~(\ref{eq.SDA-P}). In this section we only analyze those operators sensitive to spin-1 resonance contributions, whereas operators only related to spin-0 and fermionic resonances are not studied here, because their resonance field implementation is unique.

\subsection{Double Lorentz index tensor structures, $\boldsymbol{\hat \chi_{\hat R}^{\mu\nu}}$ and $\boldsymbol{\chi_{R}^{\mu\nu}}$}
In the first place, we are going to determine the short-distance non-resonant couplings (or simply SD couplings) for those operators receiving their resonance contributions from chiral tensor structures including two Lorentz indices,\ie $\bra \hat \chi_{\hat R}^{\mu\nu} \hat \chi_{\hat R\,\mu\nu}\ket_{(2,3)}$ for the Proca formalims and $\bra \chi_{R}^{\mu\nu} \chi_{R\,\mu\nu}\ket_{(2,3)}$ for the Antisymmetric one. This classification of spin-1 operators include in this section all purely bosonic operators (except $\mO_{10}$) and most of two-fermionic operators, found in tables \ref{tab:bosonic-Op4} and \ref{tab:2fermion}, respectively.

The full set of Green-function calculations required to work out the local non-resonant contributions to the LECs can be found either in this chapter or in appendix $\ref{app:SD}$.
As an example of how the SD Lagrangian works for this set of operators, we address the explicit determination of the $\mF^{{\rm SD}}_3$ and $\widetilde \mF^{{\rm SD}}_1$ local, non-resonant contributions to the LECs through the calculation of the two-Goldstones matrix elements, which can be characterized in terms of the vector and axial-vector form factors, $\mathbb{F}^\mJ_{\varphi\varphi} (s)$, as follows:
\be \label{eq:2Goldstone}
\bra \varphi^+ (p_1)\, \varphi^- (p_2) \,|\, \mJ^\mu_3 \,|\, 0 \ket\; =\;
(p_1 - p_2)^\mu\; \mathbb{F}^\mJ_{\varphi\varphi} (s)\,, \qquad\quad (\mJ = \mV,\,\mA )\, ,
\ee
where $s=(p_1+p_2)^2$ is the Mandelstam variable and $\mJ^\mu_a$ the vector and axial-vector currents, being defined as the functional derivatives of the action with respect to the corresponding external sources,
\bear \label{eq:VAcurrent}
\mV^\mu_a &\equiv & \Frac{\partial S}{\partial  v_\mu^a}\, ,
\qquad\qquad
\hat v_\mu\; =\; \frac{1}{2}\,\left( \hat{B}^\mu + \hat{W}^\mu\right)
\; =\; \frac{1}{2}\,\vec\sigma\, \vec v_\mu\, ,
\nn\\
\mA^\mu_a &\equiv & \Frac{\partial S}{\partial  a_\mu^a}\, ,
\qquad\qquad
\hat a_\mu\; =\; \frac{1}{2}\,\left( \hat{B}^\mu - \hat{W}^\mu\right)
\; =\; \frac{1}{2}\,\vec\sigma\, \vec a_\mu\, .
\eear

The tree-level calculation of the form functions turns out to be
\begin{align} \label{eq:formfunctions}
\mathbb{F}^\mV_{\varphi\varphi}(s) & = \! \left\{  \!\! \bat
1\, +\,\Frac{F_V\,G_V}{v^2}\,\Frac{s}{M_V^2-s}\,
+\,\Frac{\widetilde{F}_A\,\widetilde{G}_A}{v^2}\,\Frac{s}{M_A^2-s}
\,  -\, 2\, \mF_3^{\,\mathrm{SDA}}\,\Frac{s}{v^2}\,,
& \quad\mbox{\small  (SDET-A)}\, ,\!\!
\nonumber \\[10pt]
\!\!\!\!\! 1\, +\,\Frac{f_{\hat{V}}\, g_{\hat{V}}}{  v^2  }\,\Frac{ s^2 }{M_V^2-s}\,
+ \,\Frac{\widetilde{f}_{\hat{A}}\,\widetilde{g}_{\hat{A}}   }{  v^2 }\,\Frac{  s^2   }{M_A^2-s}\,
-\, 2\, \mF_3^{\,\mathrm{SDP}}\,\Frac{s}{v^2}\,,
& \quad \mbox{\small (SDET-P)} \, , \!\!
\ea\right.
\\[15pt]
\mathbb{F}^\mA_{\varphi\varphi}(s) & = \! \left\{ \bat
\! \! \Frac{\widetilde{F}_V\, G_V}{v^2}\,\Frac{s}{M_V^2-s} \, +\,
\Frac{F_A\,\widetilde{G}_A}{v^2}\,\Frac{s}{M_A^2-s}
\, -\, 2\, \widetilde\mF_1^{\,\mathrm{SDA}}\,\Frac{s}{v^2}\,,
& \qquad\quad\mbox{\small  (SDET-A)}\, ,\!\!
\\[10pt]
\!\!\!\!\! \Frac{\widetilde{f}_{\hat{V}}\, g_{\hat{V}} }{v^2}\,\Frac{s^2}{M_V^2-s}
\, +\,
\Frac{f_{\hat{A}}  \,\widetilde g_{\hat{A}}}{ v^2 }\,\Frac{s^2}{M_A^2-s}
\, -\, 2\, \widetilde\mF_1^{\,\mathrm{SDP}}\,\Frac{s}{v^2}\,,\!\!
& \qquad\quad \mbox{\small (SDET-P)} \, ;
\ea\right.
\end{align}
receiving contributions as well from the resonance Lagrangian as from the SD Lagrangian. Proca and Antisymmetric resonance contributions are found to be apparently similar but they show a crucial discrepancy in the way they behave at short-distances. While the resonance contributions for the form factors in the Antisymmetric formalism do not grow with energy,\footnote{Terms including the fraction $s/(M_R^2-s)$ tend to -1 at large energies.} Proca resonance terms grow unacceptably. Furthermore, vector and axial-vector form factors are sensitive to the $\mO_3$ and $\widetilde \mO_1$ local operators too, being proportional to $s$ as well. Imposing the proper short-distance contraint, form factors are demanded not to grow with energy, yielding
\begin{align} \label{eq:F3SD}
\mF_3^{\,\mathrm{SDA}} & \,=\,    0\, , & 
\widetilde\mF_1^{\,\mathrm{SDA}}  &\, = \, 0\,, &
\nn\\
\mF_3^{\,\mathrm{SDP}} &\, =\,  -\,\Frac{f_{\hat V}\,g_{\hat V}}{2}\,
- \, \Frac{\widetilde f_{\hat A}\,\widetilde g_{\hat A}}{2}\, , &
\qquad 
\widetilde\mF_1^{\,\mathrm{SDP}} & \,=\,
-\,\Frac{\widetilde{f}_{\hat V}\, g_{\hat V}}{2}\,
- \, \Frac{f_{\hat A}\, \widetilde g_{\hat A}}{2}\, .&
\end{align}
On the one hand, antisymmetric SD terms are forced to be 0; otherwise form factors would blow up at short-distances. On the other hand, Proca SD contributions need to fix the resonance energy growing and thus their values are actually set. Moreover, the differences $\mF_3^{\,\mathrm{SDA}} - \mF_3^{\,\mathrm{SDP}}$ and $\widetilde\mF_1^{\,\mathrm{SDA}}-\widetilde\mF_1^{\,\mathrm{SDP}}$ are the precise ones predicted by eq.~(\ref{eq.equivP-nonR}). As already known, both formalisms produce the same outcome for the form factors despite the fact that the relevant contributions come from different sides, provided the identifications
\begin{align} \label{eq:FG}
& F_V\, G_V \; = \; f_{\hat V}\,g_{\hat V}\, M_V^2\, ,
\qquad
& \widetilde{F}_A\, \widetilde{G}_A \; = \; \widetilde{f}_{\hat A}\,\widetilde{g}_{\hat A}\, M_A^2\, ,
\nn\\
& \widetilde{F}_V\, G_V \; = \; \widetilde{f}_{\hat V}\,g_{\hat V}\, M_V^2\, ,
\qquad
& {F}_A\, \widetilde{G}_A \; = \; {f}_{\hat A}\,\widetilde{g}_{\hat A}\, M_A^2\, ;
\end{align}
which agree with the resonance couplings relations, found in eq.~(\ref{eq:LECsrelation}), as it was expected. The values of $\mF_3$ and $\widetilde \mF_1$ are unambiguously determined with the only consideration of the resonance contributions from the Proca and the Antisymmetric formalisms combined.

The remaining bosonic operators work similarly, except for the operator $\mO_{10}$ that will be explained in the next section. The analysis of the proper set of Green functions show that tree-level exchanges of resonances in the Antisymmetric formalism carry the right high-energy behavior and thus no additional SD contribution is allowed. However, the same processes for the Proca formalism show resonance contributions that grow with energy so local contributions are required to repair this bad short-distance behavior. Particularly, the Green functions employed for the determination of the local SD couplings are the two-Goldstone scattering for the $\mO_4$ and $\mO_5$ operators, the $hh\to\varphi\varphi,\,hh$ scattering for $\mO_{6,7,8}$. 
The $\mO_9$ and $\widetilde \mO_3$ SD contributions are fixed with the three-point functions of the vector and axial-vector currents with a Higgs-Goldstone pair 
while two-point correlators are required to set $\mO_{1,2,11}$ and $\widetilde\mO_2$ with the vector and axial-vector currents, and $\mO_{12}$ with the gluon one. Hence,
\begin{align} \label{eq:FSDbosonic}
&\mF_i^{\,\mathrm{SDA}} \, =\, 0\,, & \qquad\quad  &\widetilde{\mF}_i^{\,\mathrm{SDA}} \,=\, 0\,, & \qquad &	&
\nn\\
&\mF_i \,=\, \mF_i^{\,\mathrm{SDP}}\, , & \qquad\quad &\widetilde{\mF}_i \,=\, \widetilde{\mF}_i^{\,\mathrm{SDP}}\,, &
\qquad &(i\not= 10)\, . &
\end{align}

Moreover, most of two-fermion operators (combinations of one fermion leptonic or quark bilinear with bosonic operators) also behave in the same way. 
According to the different Green-functions to be studied in order to set the SD contributions, we split the analysis of these operators in three groups: form factors involving the fermionic tensor bilinear ($\mO^{\psi^2_f}_{3,4,8}$ and $\widetilde \mO^{\psi^2_f}_1$), $\psi h$ or $\psi \varphi$ scattering amplitudes ($\mO^{\psi^2_f}_{1,2,5,7}$ and $\widetilde \mO^{\psi^2_f}_2$) and custodial symmetry breaking operators ($\mO^{\psi^2_f}_{6}$ and $\widetilde \mO^{\psi^2_f}_3$). While the first and the second group of LECs receive resonance contributions from structures collected in double Lorentz index tensors, $\chi^{\mu\nu}$, as the purely bosonic operators studied just before; the third group only gets spin-1 high-energy imprints from single Lorentz index structures, $\chi^\mu$, and will be studied in the next section.

The analysis of two-fermion form factors, particularly the so-called magnetic form factor, $\mmF^\mJ_{2} (q^2)$ with $\mJ = \mV_3,\,\mA_3,\,\mV_{(0)}$, is performed through the calculation of the two-fermion matrix elements $\bra \psi (p_1) \,|\, \mJ^\mu \,| \, \psi (p_2)\,  \ket$ (see appendix \ref{app:SD}). Imposing the magnetic form factor to vanish at short-distances, it yields
\begin{align} \label{eq:FSDbosonic}
&\mF_3^{\,\mathrm{SDA}} \, =\, 0\,, & \qquad\quad  
& \mF_4^{\,\mathrm{SDA}} \, =\, 0\,, & \qquad\quad 
& \widetilde{\mF}_1^{\,\mathrm{SDA}} \,=\, 0\,, &
\nn\\
&\mF_3 \,=\, \mF_3^{\,\mathrm{SDP}}\, , & \qquad\quad 
&\mF_4 \,=\, \mF_4^{\,\mathrm{SDP}}\, , & \qquad\quad
&\widetilde{\mF}_1 \,=\, \widetilde{\mF}_1^{\,\mathrm{SDP}}\,; &
\end{align}
and the same results apply for $\mF_8^{{\rm SDA}}$ and $\mF_8^{{\rm SDP}}$ for the two-fermion form factor involving a gluon current.

For the second group of two-fermionic operators, one should analyze the following scattering processes: 
\begin{align*}
\psi_f (p_1)\; \varphi (p_2) & \,\, \longrightarrow \,\, \psi_f (p_3)\; \varphi (p_4) \,, \\
\psi_f (p_1)\; \varphi (p_2) & \,\, \longrightarrow \,\, \psi_f (p_3)\; h (p_4) \,, \\
\psi_f (p_1)\; h (p_2) & \,\, \longrightarrow \,\, \psi_f (p_3)\; h (p_4) \,, \qquad\qquad (f\,=\,l,\,q)\,,
\end{align*}
which receive spin-0 resonance contributions from the operators
$\,\mO^{\psi^2_f}_{1,5,7}\,$, mediated by the scalar and pseudoscalar fermionic bilinears, $J^f_S$ and $J^f_P$; while exchanges of high-energy massive spin-1 states contribute to $\, \mO^{\psi^2_f}_{2}\,$ and $\,\widetilde \mO^{\psi^2_f}_{2}\,$, involving the tensor fermionic bilinear instead, $J^{f\,\mu\nu}_T$. 
In the first case, for the spin-0 objects not explicitly studied here, this kind of scattering amplitudes grow with energy as $\mM_{\psi \varphi \to \psi \varphi ,\,  \psi  h}\sim E$ but do not violate the Froissart bound on the cross section since they satisfy $\sigma(s)< C\, \ln^2(s/s_0)$ \cite{Fingerprints}. For sure, it is possible to set stronger bounds over these calculations by computing partial-wave projections or forward scatterings, although it is not required for our purposes. 
However, this is not always the case for spin-1 resonance exchanges. For instance, the custodial triplet color singlet vector resonance exchange, $V^1_3$, through both a vertex including a two-fermion tensor bilinear and a two-Goldstone vertex, shows a very different high-energy behavior depending on the spin-1 resonance implementation. Hence, Antisymmetric and Proca formalisms bring in scattering amplitudes proportional to
\be
\mM_{\psi \varphi \to \psi \varphi  } \bigg|_{\rm V\, through\, J_T} \; \sim \; \left\{ \bat
\Frac{C_{0,f}^V\, G_V}{v^2}\;  E\,,  & \qquad\quad\mbox{\small  (SDET-A)}\, ,
\\[10pt]
\Frac{c_{0,f}^{{\hat{V}}}\, g_{\hat{V}} }{v^2}\;  E^3\,,
& \qquad\quad\mbox{\small (SDET-P)} \, .
\ea\right.\quad
\ee
While the local term in the SDET-A fulfill the Froisart bound and is well-behaved at the cross section level, its large energy behavior in the SDET-P is unacceptable.
This is also the situation for two-Higgs or Goldstone-Higgs scattering amplitudes involving the tensor bilinear $J^{f\,\mu\nu}_T$. In order to fix it, local non-resonant contributions to the LECs must adopt the values
\begin{align} \label{eq:FSDbosonic}
&\mF_2^{\,\mathrm{SDA}} \, =\, 0\,, & \qquad\quad   
& \widetilde{\mF}_2^{\,\mathrm{SDA}} \,=\, 0\,, &
\nn\\
&\mF_2 \,=\, \mF_2^{\,\mathrm{SDP}}\, , & \qquad\quad 
&\widetilde{\mF}_2 \,=\, \widetilde{\mF}_2^{\,\mathrm{SDP}}\,, &
\end{align}
following the same pattern of the rest of local contributions to the LECs analyzed so far.

It is actually more clear to understand the origin of all these differences in the SDET-A and SDET-P local Lagrangians involving two Lorentz index chiral structures attending to the formal structure of their resonance propagators. Considering the effective action $S^{(X)}\, (X=A,P)$ for the resonance exchange written compactly \cite{Ecker:1989yg} as
\be \label{eq:S_X}
S^{(X)}\; =\;-\frac{1}{2}\,\int d^4x\, d^4y\;\,
\bra \chi^{\mu\nu}_{(X)}(x)\,\Delta^{(X)}_{\mu\nu,\rho\sigma}(x-y)\,\chi^{\rho\sigma}_{(X)}(y)\ket\, , \quad\qquad (X\,=\, A, P)\,,
\ee
where $\chi^{\mu\nu}_{(P)}$ and $\chi^{\mu\nu}_{(A)}$ stand for the two Lorentz index chiral tensors $\hat \chi_{\hat R}^{\mu\nu}$ and $\chi_{R}^{\mu\nu}$ in the Proca and the Antisymmetric formalism, respectively. Their corresponding resonance propagators are found to be 
\begin{align} \label{eq:Propagatorsdouble}
\Delta^{(P)}_{\mu\nu,\rho\sigma}(x) &\, = \, \int\frac{d^4 k}{(2\pi)^4}\;\frac{\mathrm{e}^{-ikx}}{M_R^2-k^2}\; \left[g_{\mu\rho}\, k_\nu k_\sigma -g_{\mu\sigma}\, k_\nu k_\rho - (\mu\leftrightarrow\nu)\right]\, ,
\nn\\[1ex]
\Delta^{(A)}_{\mu\nu,\rho\sigma}(x) &\, = \, \frac{1}{M_R^2}\;\left\{
\Delta^{(P)}_{\mu\nu,\rho\sigma}(x)\, +\, \delta^{(4)}(x)\;\left(g_{\mu\rho}\, g_{\nu\sigma} -
g_{\mu\sigma}\, g_{\nu\rho} \right)\right\}\, .
\end{align}
More details regarding the Antisymmetric formalism construction are extended in appendix \ref{app:spin1comparison}. Notice that the two propagators are similar except for a local contribution in the Antisymmetric propagator. This precise additional piece is the one responsible of exhausting the $\mO_{\hat d}(p^4)$ EWET Lagrangian when considering interactions reflected in two Lorentz index tensor structures. Otherwise, it is absent in the Proca case and therefore resonance contributions to the LECs come from some local non-resonant Lagrangian not naturally incorporated in the effective action. Furthermore, one is able to check that the resonance couplings can be easily related with the identification of the pole residues at $k^2=M_R^2$, yielding identical couplings except for some mass factors, as already shown in eq.~(\ref{eq:LECsrelation}).

\subsection{Single Lorentz index tensor structures, $\boldsymbol{\hat \chi_{\hat R}^{\mu}}$ and $\boldsymbol{\chi_{R}^{\mu}}$}

The local SD interactions to be analyzed here are those that can be grouped in single Lorentz index structures, as opposite to the last section. They will contribute to the LECs of the EWET NLO Lagrangian like $\bra \hat \chi_{\hat R}^\mu\, \hat \chi_{\hat R\,\mu} \ket_{(2,3)}$ and $\bra \chi_{ R}^\mu\, \chi_{ R\,\mu} \ket_{(2,3)}$ in the Proca and Antisymmetric resonance implementations, respectively. EWET operators sensitive to spin-1 tree-level resonance exchanges fitting these prescriptions can be categorized in three different groups: four-fermion operators (table \ref{tab:4fermion}), custodial breaking two-fermion operators,  $\mO^{\psi^2_f}_{6}$ and $\widetilde \mO^{\psi^2_f}_3$, (table \ref{tab:2fermion}), and the leftover purely bosonic operator $\mO_{10}$ (table \ref{tab:bosonic-Op4}). 

As it occurs with the resonance imprints in the LECs, the short-distance local Lagrangian works in a complete different way when dealing with single Lorentz tensor structures rather than the two index ones. This purely kinematical discrepancy turns out to be crucial as well for the non-resonant Lagrangian and modifies critically the SD behavior with respect to the previous case. Therefore, it is convenient to look into the Proca and the Antisymmetric resonance propagators when the given vertices carry only one Lorentz index
\begin{align} \label{eq:Propagatorssingle}
\Delta^{(P)}_{\mu\nu}(x) &\, = \, \int\frac{d^4 k}{(2\pi)^4}\;\frac{\mathrm{e}^{-ikx}}{M_R^2-k^2}\; \left( g_{\mu\nu} - \Frac{k_\mu k_\nu}{M_R^2} \right)\, ,
\nn\\[2ex]
\Delta^{(A)}_{\mu\nu}(x) &\, = \, -\, \nabla^\alpha \nabla^\beta \Delta^{(A)}_{\mu\alpha,\nu\beta}(x) 
\nn\\
&\,=\, \int\frac{d^4 k}{(2\pi)^4}\;\Frac{2\, k^2\, \mathrm{e}^{-ikx}} {M^2_R\,(M_R^2-k^2)}\;  \left( k^2\,g_{\mu\nu} - k_\mu k_\nu \right)
\, +\, \Frac{\delta^{(4)}(x)}{M^2_R} \left(k^2\,g_{\mu\nu} - k_\mu k_\nu \right) \,.
\end{align}

On the one side, we consider for the spin-1 Proca Lagrancian a generic Green function involving all the possible resonance and SD local interactions able to be collected in $\bra \hat \chi_{\hat R}^\mu \hat \chi_{\hat R\, \mu} \ket_{(2,3)}$ and sensitive to the operators mentioned before. Making use of the effective propagator of the first line of eq.~(\ref{eq:Propagatorssingle}), it formally reads at tree-level as
\begin{align} \label{eq:4F-GreenP}
G^{(P)}(x)& \, = \,  \int
\frac{d^4 k}{(2\pi)^4}\,\mathrm{e}^{-ikx} 
\left\{ 
\Frac{1}{2}\sum_{R=V,A}  \left[ \sum_{R^1_1,R^1_3,R^8_1,R^8_3} 
\Frac{g_{\mu\nu}-k_\mu k_\nu/M_{R^m_n}^2}{k^2-M_{R^m_n}^2}
\,\times \right. \right. 
\nn\\
& \, \times\, \bigg( \bra \hat{\chi}_{\hat R^m_n}^{\mu}(x)\,\hat{\chi}_{\hat R^m_n}^\nu(x)\ket_{2,3}
\,-\,\frac{1}{2} \,\bra \bra \hat{\chi}_{\hat R^m_n}^{\mu}(x)\ket_2 \bra \hat{\chi}_{\hat R^m_n}^\nu(x) \ket_2 \ket_3 
\nn\\
& \qquad  \,-\, \frac{1}{3} \bra \bra \hat{\chi}_{\hat R^m_n}^{\mu}(x)\ket_3 \bra \hat{\chi}_{\hat R^m_n}^\nu(x) \ket_3 \ket_2 \,+\, \frac{1}{6}\,\bra \hat{\chi}_{\hat R^m_n}^{\mu}(x) \ket_{2,3}  \, \bra \hat{\chi}_{\hat R^m_n}^\nu(x)\ket_{2,3}\bigg) \Bigg] 
\nn\\[1ex]
&  \,+\, \left( \sum_{i=5}^{8} \mF_i^{\psi^{2}_l\psi^{2}_q,\, {\rm SDP}}\,\mO^{\psi^{2}_l\psi^{2}_q}_i(x)
\,+\, \sum_{i=1}^{4} \widetilde\mF_i^{\psi^{2}_l\psi^{2}_q,\, {\rm SDP}}\,\widetilde\mO^{\psi^{2}_l\psi^{2}_q}_i(x)
\,+\, \ldots \right)
\nn\\ 
& \left. \, + \, \sum_{f=l,q} \left(\mF_6^{\psi^{2}_f,\, {\rm SDP}}\,\mO^{\psi^{2}_f}_6(x)
\,+\, \widetilde\mF_3^{\psi^{2}_f,\, {\rm SDP}}\,\widetilde\mO^{\psi^{2}_f}_3(x) \right) \,+\, \mF_{10}^{\rm SDP}\,\mO_{10}(x)
\right\} \,.
\end{align}
For simplicity, only lepton-quark bilinear four-fermions, $\mO^{\psi^2_l\psi^2_q}_i$ and $\widetilde \mO^{\psi^2_l\psi^2_q}_i$, have been listed in the Green function computation. Lepton-lepton and quark-quark bilinear four-fermion operators are implicitly present too (in the dots). These local subsets contribute to the Green function with all the list of SD non-resonant operators due to the fact that they are fierzed and thus all the contributions are highly mixed. Nonetheless, all the results derived for this subset are totally equivalent for the other ones.
Additionally, the same $SU(n)$ trace criteria as in eq.~(\ref{eq.equivP-nonR}) has been employed in order to make the notation more compact.

Attending to canonical dimensions only, the momentum contraction of a fermion bilinear yields $k_\mu\,J_{V/A}^{\mu\,f} \sim m_f$ and thus all four-fermion operators and $\bra \hat \chi_{\hat R}^\mu \hat \chi_{\hat R\,\mu} \ket_{(2,3)}$ resonance contributions scale as $p^2$. Custodial breaking two-fermion operators and $\mO_{10}$ do it too, since the spurion field $\mT$ do not increase the canonical dimension (it does for the chiral counting though). Non-local resonance exchange interactions compensate these momenta enhancement with the propagator momentum suppression $(k^2-M_R^2)^{-1}$ and therefore they reflect a good high-energy behavior. However, local SD contributions grow quadratically with energy and do not get balanced. In short, different contributions to the Green function in eq.~(\ref{eq:4F-GreenP}) scale at the amplitude level like
\be \label{eq:GFbehavior}
\mM \Big|_{\rm resonance\ contrib.}  \sim \; 1 \,, \qquad\quad 
\mM \Big|_{\rm local\ SD\ contrib.}  \sim \; E^2 \,.
\ee
Hence, a consistent well-behaved short-distance regime for the resonance theory requires all the SD local contributions to vanish,
\begin{align} \label{eq:FSDProcasingle}
&\mF^{\psi^2_l \psi^2_l\,,{\rm SDP}}_{1,2,3,4,5}  \,=\, 
\widetilde \mF^{\psi^2_l \psi^2_l\,,{\rm SDP}}_{1} \,=\, 0 \,, & \qquad
&\mF^{\psi^2_f\,,{\rm SDP}}_6  \,=\, 
\widetilde \mF^{\psi^2_f\,,{\rm SDP}}_3 \,=\, 0 \,, & \qquad
\nn\\
&\mF^{\psi^2_l \psi^2_q\,,{\rm SDP}}_{5,6,7,8}  \,=\,
\widetilde \mF^{\psi^2_l \psi^2_q\,,{\rm SDP}}_{1,2,3,4}\,=\, 0 \,, & \qquad 
&\mF^{{\rm SDP}}_{10}  \,=\, 0 \,, &
\nn\\
&\mF^{\psi^2_q \psi^2_q\,,{\rm SDP}}_{1,\ldots,10}  \,=\, 
\widetilde \mF^{\psi^2_q \psi^2_q\,,{\rm SDP}}_{1,2}  \,=\, 0 \,, & \qquad
&	& 
\end{align}

On the other side, it is easy to appreciate in eq.~(\ref{eq:Propagatorssingle}) that the Antisymmetric propagator bring more powers of momenta than the Proca one. Kinematics requires the presence of two extra derivatives in order to contract the inner two Lorentz index structure of Antisymmetric spin-1 resonances, as opposite to Proca vectors which comfortably adjust to vertices like $\hat \chi_{\hat R}^\mu$. 
Indeed, the one index Antisymmetric propagator in eq.~(\ref{eq:Propagatorssingle}) is equivalent to the two indices Antisymmetric propagator in eq.~(\ref{eq:Propagatorsdouble}), except for those indices contracted with the covariant derivative.
Notice that both the chiral and the canonical dimensions increase in two units with the incorporation of the derivatives. As a consequence, this formalism is not appropriate to work with $\bra \chi_R^{\mu} \chi_{R\,\mu} \ket_{(2,3)}$ interactions since every Green function to be analyzed would reflect an undesirable high-energy behavior. Therefore, compensating non-trivial local terms need to arise so that the short-distance regime of the theory is fixed, like
\begin{align} \label{eq:FSDProcasingle}
& \mF^{\psi^2_l \psi^2_l\,,{\rm SDA}}_{1,2,3,4,5}  \,=\, \mF^{\psi^2_l \psi^2_l}_{1,2,3,4,5} \,, &  \quad 
& \widetilde \mF^{\psi^2_l \psi^2_l\,,{\rm SDA}}_{1} \,=\, \widetilde \mF^{\psi^2_l \psi^2_l}_{1} \,, & \qquad
& \mF^{\psi^2_f\,,{\rm SDA}}_6  \,=\, \mF^{\psi^2_f}_6\,,&
\nn\\
& \mF^{\psi^2_l \psi^2_q\,,{\rm SDA}}_{5,6,7,8}  \,=\, \mF^{\psi^2_l \psi^2_q}_{5,6,7,8}\,, & \quad
& \widetilde \mF^{\psi^2_l \psi^2_q\,,{\rm SDA}}_{1,2,3,4}\,=\, \widetilde \mF^{\psi^2_l \psi^2_q}_{1,2,3,4}\,, & \qquad
& \widetilde \mF^{\psi^2_f\,,{\rm SDA}}_3 \,=\, \widetilde \mF^{\psi^2_f}_3 \,, &
\nn\\
& \mF^{\psi^2_q \psi^2_q\,,{\rm SDA}}_{1,\ldots,10} \,=\, \mF^{\psi^2_q \psi^2_q}_{1,\ldots,10}\,, & \quad 
& \widetilde \mF^{\psi^2_q \psi^2_q\,,{\rm SDA}}_{1,2}  \,=\, \widetilde \mF^{\psi^2_q \psi^2_q}_{1,2}\,, & \qquad
&\mF^{{\rm SDA}}_{10}  \,=\, \mF_{10} \,. &
\end{align}

As it had been anticipated at the beginning of the section, either Proca or Antisymmetric formalism give the right predictions for the LECs, depending only on the interaction to be written as $\bra \hat \chi_{\hat R}^\mu \hat \chi_{\hat R\,\mu} \ket_{(2,3)}$ or $\bra \chi_{R}^{\mu\nu} \chi_{R\,\mu\nu} \ket_{(2,3)}$, respectively. Moreover, no further local contributions to the LECs appear when selecting the proper resonance field representation since short-distance constraints force them to vanish. On the contrary, when choosing the inconvenient (but also valid) resonant implementation, non-trivial SD local contributions restore the disturbing high-energy bad behavior.

\section{Gauge-like spin-1 formalism} 
Proca-vector and rank-2 Antisymmetric tensors are not the only field implementations for spin-1 massive states. Indeed, there could be other possible redefinitions of the quantum fields in the generating functional, but they may not show any remarkable property nor guarantee a simpler phenomenology. 

We are going to analyze a specific spin-1 $SU(2)_{L+R}$ triplet color-singlet vector model where these high-energy objects are incorporated as massive Yang-Mills fields in a gauge-invariant way. They are also called hidden local symmetry (HLS) gauge vectors \cite{BKUY:85,BKY:88,HY:92,HY:03,Casalbuoni:1985kq,Feruglio:1988,Casalbuoni:93,ME:88}, represented by the field $\bar V_\mu$, which transforms under $\mG$ as
%
%
\be \label{eq:hlstransform}
\bar{V}_\mu \,\longrightarrow\, g_h \,\bar{V}_\mu\, g_h^\dagger
\, +\, \Frac{i}{g_\rho} \, g_h\, \partial_\mu g_h^\dagger\, ,
\ee
being $g_h\in SU(2)_{L+R}$ and $g_\rho$ the HLS gauge coupling. The gauge vector resonance Lagrangian \cite{Ecker:1989yg} is found to be
\be \label{eq:L_HLS}
\mL_V^{\rm (H)} \; =\;
-\Frac{1}{4}\, \bra \bar{V}_{\mu\nu} \bar{V}^{\mu\nu}\ket_2
\, +\, \Frac{1}{2}\, M_{V^1_3}^2\; \bra \left(\bar{V}_\mu - \Frac{i}{g_\rho}\, \Gamma_\mu \right) \left(\bar{V}^\mu - \Frac{i}{g_\rho}\, \Gamma^\mu \right)\ket_2\, ,
\ee
where the gauge field strength tensor is defined as $\bar{V}_{\mu\nu}=\partial_\mu \bar{V}_\nu \,-\,  \partial_\nu \bar{V}_\mu \,-\, ig_\rho\, [\bar{V}_\mu,\bar{V}_\nu ]$ and the connection corresponds to the first and second line in eq.~(\ref{eq:connection}). Notice that this is a Higgsless and colorless
purely bosonic gauge model. One can always extend it with all the particles and symmetries provided they interact with resonances in a non-gauge invariant way, as in the resonance theory evaluated in chapter \ref{ch:resonancetheory}. It is also possible to distinguish the left and the right parts of the covariant connection, as in eq.~(\ref{eq:connection}) and eq.~(\ref{eq.cov-bosonic-tensors}), so that the hidden gauge field transformations are chiral-dependent.\footnote{In \cite{Contino}, a $SU(2)_L$ triplet gauge field was introduced, following the prescriptions of the hidden gauge field.}

The Lagrangian in eq.~(\ref{eq:L_HLS}) is well-behaved in the UV regime. One can identify the first term as the dimension-4 Yangs-Mills Lagrangian and thus it is renormalizable. The second term introduces the triplet gauge vector mass preserving gauge-invariance together with Goldstone and gauge field interactions through the covariant connection which is absent in the already analyzed spin-1 formalisms. These additional interacting terms are generated by the inner local symmetry of the gauge model in such a way that good short-distance properties are just softly modified.  

Indeed, it is straightforward to recover the Proca formalism just by performing the following field redefinition
\be \label{eq.HLS-Proca}
\bar{V}_\mu \; =\; \hat{V}^1_{3\,\mu} \,+\, \Frac{i}{g_\rho}\,\Gamma_\mu \,.
\ee
Therefore, it yields for the strength tensor 
\be
\bar{V}_{\mu\nu} \; =\; \hat{V}^1_{3\,\mu\nu}
\, +\, \Frac{i}{g_\rho}\, \Gamma_{\mu\nu} \, - i\, g_\rho\, [\hat{V}^1_{3\,\mu} ,\,\hat{V}^1_{3\,\nu}]\, ,
\ee
where we define
\be
\hat{V}^1_{3\,\mu\nu} \,=\, \nabla_\mu \hat{V}^1_{3\,\nu} -\nabla_\nu \hat{V}^1_{3\,\mu}\,, \qquad\quad \Gamma_{\mu\nu} \,=\, \frac{1}{4}\, [u_\mu,u_\nu]\, -\, \frac{i}{2}\, f_{+\, \mu\nu}\,.
\ee

Replacing in eq.~(\ref{eq:L_HLS}) all the HLS gauge field expressions in terms of triplet vector resonances, the previous Lagrangian is rewritten as \cite{Ecker:1989yg}
\begin{align}  \label{eq.redefined-HLS-Lagr}
\mL_V^{\rm (H)} & \,=\, \bigg(
-\,\Frac{1}{4}\, \bra \hat{V}^1_{3\,\mu\nu} \hat{V}_3^{1\,\mu\nu}\ket_2
\,  +\, \Frac{1}{2}\, M_{V^1_3}^2\, \bra \hat{V}^1_{3\,\mu} \hat{V}^{1\,\mu}_3 \ket_2
\, -\, \Frac{i}{2 g_\rho}\, \bra \hat{V}_3^{1\,\mu\nu}\, \Gamma_{\mu\nu}\ket_2 
\nn\\
&  
\, +\, \Frac{1}{4 g_\rho^2}\, \bra \Gamma_{\mu\nu}\, \Gamma^{\mu\nu} \ket_2 \bigg) \,+\, \bigg(
- \Frac{1}{2}\,\bra \Gamma_{\mu\nu}\, [\hat{V}^{1\,\mu}_3,\hat{V}^{1\,\nu}_3] \ket_2
\, +\,\Frac{i g_\rho}{2}\, \bra  \hat{V}^1_{3\,\mu\nu}\, [\hat{V}_3^{1\,\mu}, \hat{V}_3^{1\,\nu} ]\ket_2 
\nn\\
&  \, +\,\Frac{g_\rho^2}{4}\,\bra  [\hat{V}^1_{3\,\mu},\hat{V}^1_{3\,\nu}]\, [\hat{V}^{1\,\mu}_3,\hat{V}^{1\,\nu}_3]\ket_2 \bigg)\, . 
\end{align}
The computation displays several types of operators which we split in two different brackets. While the second one shows operators with two or more vector resonances and hence should be dropped (we are just interested in single-resonance interactions); the first bracket, however, leads to the Proca kinetic and mass term Lagrangian together with well-determined interactions fixed by the HLS gauge coupling. These additional terms are both local and non-local operators associated to the SD non-resonance Lagrangian and the resonance Lagrangian. The key fact here is that they are not only reflected in the LECs in terms of some set of resonance couplings (expected to be $\mO(1)$), but all the predicted resonance couplings and local non-resonant couplings are unequivocally set in terms of one single parameter: the HLS gauge coupling $g_\rho$. The precise value for these couplings is
\begin{align} \label{eq.HLS-Proca-couplings}
f_{\hat V} &\,=\, 2\, g_{\hat V}\, =  \, -\, \Frac{1}{\sqrt{2}\, g_\rho}\, ,
\nn\\
\mF_1^{\rm SDP} &\,=\,
2 \,\mF_2^{\rm SDP} \, =\,\mF_3^{\rm SDP} \, =\,
\,-\, 4\,\mF_4^{\rm SDP} \, =\, 4 \,\mF_5^{\rm SDP} \, =\, - \Frac{1}{8g_\rho^2}\, ,
\end{align}
while the rest of couplings do not receive any contribution from the custodial triplet color-singlet heavy vector, $\hat V^1_{3\,\mu}$.

The resonance interactions in eq.~(\ref{eq.redefined-HLS-Lagr}) are a particular case of the more general Proca custodial-triplet color-singlet Lagrangian in eqs.~(\ref{eq:L1proca}, \ref{eq:proca-chimu} and \ref{eq:proca-chimunu}), provided fermions, the Higgs field and P-odd terms are detached from the interaction Lagrangian. Besides, the hidden gauge symmetry supplies the additional constraint $f_{\hat V} = 2\, g_{\hat V}$, which is not generated by $\mG$.

In regard to the local SDET-P values in eq.~(\ref{eq.HLS-Proca-couplings}), they satisfy as well all the short-distance constraints obtained in the preceding section, in eq.~(\ref{eq:F3SD}) and in the appendix \ref{app:SD}. The hidden local symmetry guarantees the model to be well-behaved for the local non-resonance Lagrangian too. Actually, the structure $\mL_{\text{non-R}}^{\rm (P)} =(4 g_\rho^2)^{-1}\, \bra \Gamma_{\mu\nu}\, \Gamma^{\mu\nu} \ket$ arises naturally from its gauge invariance nature, in such a way that the bad short-distance behavior of the Proca resonance Lagrangian is precisely repaired by the presence of this local operator. Furthermore, since the tree-level exchange of resonances leave their imprints for $g_{\hat V}$ and $f_{\hat V}$ at the (higher order) chiral dimension $\hat d = 6$ Lagrangian, all the contributions to the  $\mO_{\hat d}(p^4)$ EWET NLO come indeed from these local interactions. These results also agree with the purely bosonic contributions to the LECs, in table \ref{tab:bosonic-Op4}.

The HLS model can be extended with the inclusion of any other (global) symmetry invariant as in eqs.~(\ref{eq:L1proca}, \ref{eq:proca-chimu}, \ref{eq:proca-chimunu}) making use of the field redefinition $\hat{V}^1_{3\,\mu} \, =\, \bar{V}_\mu \,-\, i g_\rho^{-1}\,\Gamma_\mu$. For sure, these linearly coupled interactions are not introduced in a gauge-invariant way so the inclusion of local terms from the SDET-P are required to fix the high-energy behaviour of the resonant theory. One would need to include the most general set of Green functions to set the short-distance regime, just as well as in the last section.


\chapter{Phenomenology in the bosonic sector} \label{ch:pheno}

We have proved the resonance theory to be well-defined and consistent both at the  electroweak scale and in the short-distance regime. When integrated out, the resonance states introduced in the previous sections are shown to contribute to the $\mO_{\hat d}(p^4)$ EWET Lagrangian (or to an extended scalar potential at $\mO_{\hat d}(p^2)$), according to the chiral power counting. Moreover, the massive objects that we have implemented lead to a well-behaved high-energy regime, provided the inclusion of some unavoidable and necessary short-distance constraints.

Nevertheless, the analysis of the parameter space for these high-energy states is still missing. Particularly, we will focus in this chapter on the parameter space allowed by indirect searches,\ie the permitted energy region for the resonances so that their imprints in the LECs are set within the experimental uncertainties. Being aware that all the low-energy coupling constants associated to the operators in tables \ref{tab:bosonic-Op4}, \ref{tab:2fermion} and \ref{tab:4fermion} are related to physical observables,
we can directly link the massive states to these measurable quantities. An experimental deviation for a given LEC (or observable) with respect to the SM predictions is an indicator of the possible existence of a resonance at some near scale. Likewise, the absence of deviations from the expected values within certain experimental error determines either a wider or a more narrow parameter space region for the presence of high-energy resonances.

There are several well-known models that predict these heavy states to exist within the one and several TeVs range. On the one hand, many of the strongly coupled models in the literature, like technicolor \cite{Weinberg:1975gm, Susskind:1978ms}
and Walking Technicolor \cite{Holdom:1984sk, Appelquist:1986an, Bando:1986bg}, call the bound states to be around the TeV scale. As well, composite fermions and technibaryons are estimated to be present above this threshold \cite{Ryttov:2008xe}. On the other hand, near-conformal theories expect spin-1 resonances with masses from 1 TeV on \cite{Pich:2012dv, Pich:2012jv, Orgogozo:2011kq, Foadi:2007ue, Foadi:2007se}.

\section{Shortening the high-energy theory}
One of the main problems of the resonance theory is the amount of LECs involved in the phenomenological analyses. According to tables \ref{tab:bosonic-Op4}, \ref{tab:2fermion} and \ref{tab:4fermion}, there are 68 different low-energy coupling constants in the $\mO_{\hat d}(p^4)$ EWET NLO Lagrangian, as well as those additional parameters coming from the generic scalar Lagrangian we introduced in eqs.~(\ref{eq:LscalarEWET}, \ref{eq:Fhu_V}) entering at LO. Furthermore, the number grows when one considers three lepton and three quark families, which we postpone to future works. Notice that in this case the number of fermionic operators not only is increased by a factor 3, but it blows up due to the flavor mixing operators that are currently not present in our computation. 
Therefore, the required task to cover all the phenomenology related to all these parameters is enormous and out of the reach of this thesis. Moreover, it would be certainly more interesting to implement consistently the three fermion generation structure in the EWET and in the resonance theory, prior to analyze the fermion phenomenology with one single-fermion family physical observables.

Instead, we will study a reduced version of the resonance theory for a phenomenological analysis. Despite the fact that studying the available parameter space for all the heavy states considered so far would be better, it is also interesting to obtain this information for a subset of them, allowing us to glimpse where can the resonance scale be set or which are the typical values for these massive states. Keeping this idea in mind, we restrict our phenomenological search to P-even colorless high-energy bosonic objects. Resonance operators satisfying these prescriptions will only contribute to the P-even EWET purely bosonic operators found in the upper side block of table \ref{tab:bosonic-Op4}. In fact, the EWET NLO operators sensitive to this reduced model will be $\mO_{1,\ldots,9}$. We will discard for this analysis the operators $\mO_{10,11}$ because they are custodial symmetry breaking and also $\mO_{12}$ because it only receives contributions from color-octet heavy states. 

The resonance operators that accomplish these restrictions are collected in the following Lagrangian:
\begin{align} \label{eq:Lpheno}
\mL & \,=\, \Frac{v^2}{4}\, \bra u_\mu u^\mu \ket_2 \left( 1\,+\, \Frac{2\, \kappa_W}{v}\, h \,+\, \Frac{4\, c_d}{\sqrt{2}\,v^2}\, S_1^1 \right) \,+\, \lambda_{hS_1} \, v \, h^2\,S^1_1 
\nn\\[1ex]
&\,+\, \Frac{F_{ V}}{2\sqrt{2}}\,  \bra V^1_{3\,\mu\nu}\, f_+^{\mu\nu} \ket_2 \,+\, \Frac{i\, G_{ V}}{2\sqrt{2}}\, \bra V^1_{3\,\mu\nu}\,[u^\mu, u^\nu] \ket_2 \,+\,
\Frac{F_{ A}}{2\sqrt{2}}\,  \bra A^1_{3\,\mu\nu}\, f_-^{\mu\nu} \ket_2
\nn\\[1ex]
& \,+\, \sqrt{2}\,\Lambda_1^{hA}\,(\partial_\mu h)\, \bra A^{1\,\mu\nu}_3 u_\nu \ket_2 \,+\, 
\Frac{d_P}{v}\, (\partial_\mu h)\, \bra P^1_3 u^\mu \ket_2 \,.
\end{align}
It can be easily checked from eqs.~(\ref{eq:L0}, \ref{eq:L0tensor}) and eqs.~(\ref{eq:L1anti}, \ref{eq:antisym-chimunu}) when all the other interactions are left out. We also include a light Higgs coupled to the Goldstones in the bracket of the first line in eq.~(\ref{eq:Lpheno}), parametrized by $\kappa_W$. This coupling accounts for how SM-like the interaction is, being unity in the SM. One can easily relate this value with the linear coefficient of the power series of $\mF^{(u)}(h/v)$, in eq.~(\ref{eq:Fhu_V}). The first line of the Lagrangian is completed with the Goldstone Lagrangian and scalar-singlet resonance interactions contributing at chiral dimension $\hat d=2$ to the LO EWET Lagrangian.\footnote{Recall that the coupling $\lambda_{hS_1}$ scales as $\mO_{\hat d}(p^2)$ in the chiral power counting.} The rest of interactions are generated through the triplet resonance fields, $V^1_{3\,\mu\nu}$, $A^1_{3\,\mu\nu}$ and $P^1_3$, where only the interactions in the third line in eq.~(\ref{eq:Lpheno}) are Higgs-related. Notice too that spin-1 fields are represented in the Antisymmetric formalism since the light interacting fields carry two Lorentz indices, attending to the criteria established in the last chapter.

\section{Further constraints for the bosonic model}

In section \ref{sec:SD}, short-distance constraints were introduced in order to make the resonance theory stable and well-behaved at high energies. These relations only demanded that the underlying fundamental UV theory was consistent in the same sense,\ie it should not break unitarity, cross sections must not grow dangerously with energy, etc. In other words, the only assumption was that the electroweak resonance theory needs to work consistently like quantum field theories do. However, one can also incorporate additional constraints which are not strictly required by all the UV theories but they cover a wide subset of them provided a given specification is satisfied. For sure, the tighter some assumptions are, the more predictive a model becomes. Up to this chapter, we have stayed as general and model-independent as possible in our considerations and the results were valid for any high-energy theory including massive states coupled linearly to the SM. In the following, we will show how the reduced model in eq.~(\ref{eq:Lpheno}) gets constrained when assuming some fairly reasonable short-distance conditions.

\subsection{Form factors}
In the characterization of the local non-resonance Lagrangian (see chapter \ref{ch:spin1}) form factors were demanded not to grow with energy in order to bring unitarity conserving cross sections. This consideration allowed us to fix the short-distance behavior of the resonance theory. However, further constraints can be imposed to the form factors in order to be well-behaved in the UV regime.

For this purpose, we consider the two Goldstone boson \cite{Ecker:1988te, Ecker:1989yg} and the scalar-Goldstone axial-vector form factors \cite{Pich:2012dv, Pich:2013fea}. The vector form factor is already described in eq.~(\ref{eq:2Goldstone}) while the axial-vector one is defined as
\be \label{eq:Goldstonescalar}
\bra h(p_1)\, \varphi(p_2) |\, \mJ^\mu_A\, | 0\ket\, =\, (p_1-p_2)^\mu\; \mmF_{h\,\varphi}^A (s) \,,
\ee
Their computation in the context of the simplified resonance theory in eq.~(\ref{eq:Lpheno}) brings, respectively,
\begin{align} \label{eq:FF}
\mmF_{\varphi\varphi}^V (s) \, = \, 1\,+\, \frac{F_V\, G_V}{v^2}\,\frac{s}{M_{V^1_3}^2-s} \,.
\qquad\;
\mmF_{h\,\varphi}^A (s)  \; =\;\kappa_W \left( 1\,+\,\frac{F_A\, \Lambda_1^{hA}}{\kappa_W v}\,\frac{s}{M_{A^1_3}^2-s} \right) \,.
\nn\\
\end{align}
This set of form factors should fall as $\mO(1/s)$ at large energies in order to guarantee a proper UV behavior \cite{Brodsky:1974vy, Lepage:1980fj}. Therefore, it yields
\be \label{eq:FFSD}
F_V\, G_V\; =\; v^2 \, ,\qquad\qquad F_A\, \Lambda_1^{hA}\; =\; \kappa_W\, v \,. 
\ee

\subsection{Weinberg sum rules}

We analyze now the off-diagonal correlator $\Pi_{30}(s)$ with external legs $W^3$ and $B$, which can be written as \cite{Peskin:90, Peskin:92}
\be \label{eq:corrPi30}
\Pi_{30}(s) \,=\, \Frac{g^2\,\tan^2\theta_W}{4}\;s\left( \Pi_{VV}(s) \,-\, \Pi_{AA}(s) \right) \,,
\ee
in terms of the vector and axial-vector currents two-point Green functions. Provided chiral symmetry is preserved\footnote{Notice that it is implied in eq.~(\ref{eq:corrPi30}) that parity and weak isospin are good symmetries of the background UV theory too.} in the underlying fundamental theory, the non-zero difference $\Pi_{VV} \,-\, \Pi_{AA}$ is found to be an order parameter of the EWSB, as this quantity is not invariant under the unbroken symmetry group, $\mG$.

The analysis of $\Pi_{30}$ is performed through an Operator Product Expansion (OPE) \cite{Pich:2013fea}. Attending to what are the assumptions made over the UV theory, this calculation leads to high-energy super-convergence relations which derive in additional short-distance constraints for the considered scenario. For instance, if the background theory is assumed to be asymptotically-free, like QCD, the difference $ \Pi_{VV}-\Pi_{AA}$  is required to fall at high-energies as $1/s^3$ at least \cite{Bernard:1975cd}.

A priori, we cannot assume this behavior for the resonance theory. However, the reduced Lagrangian in eq.~(\ref{eq:Lpheno}) is invariant under chiral symmetry and contributions from operators that break $SU(2)_L\otimes SU(2)_R$ can only be proportional to order parameters of the EWSB. 
This implies the super-convergent relations called Weinberg Sum Rules (WSRs) \cite{Weinberg:1967kj}, which also set some constraints over the correlators. They conform a set of two sum rules being the first WSR (or 1WSR) more general than the second WSR (or 2WSR) in such a way that every model satisfying this last one automatically accomplishes the first one. Their definitions are, respectively,
\begin{align} \label{eq:WSRs}
\displaystyle \Frac{1}{\pi} \int_0^\infty {\rm d}t\,\left[ {\rm Im}\,\Pi_{VV}(s) \,-\, {\rm Im}\,\Pi_{AA}(s) \right] &= v^2 \,, & &[\text{1WSR}]\,,
\nn\\[1ex]
\displaystyle \Frac{1}{\pi} \int_0^\infty {\rm d}t\,t\,\left[ {\rm Im}\,\Pi_{VV}(s) \,-\, {\rm Im}\,\Pi_{AA}(s) \right] &= 0 \,, & &[\text{2WSR}]\,.
\end{align}
The first WSR is found to be valid in all the possible models to be built, even for those gauge scenarios with non-trivial UV fixed points \cite{Peskin:90, Peskin:92}. Otherwise, the application of the 2WSR is not allowed in some Conformal Technicolor models \cite{Orgogozo:2011kq} and is not clear for Walking Technicolor ones \cite{Appelquist:1998xf}. However, most of the models in the literature fulfill both conditions, since they satisfy 
\be
\lim_{s\to\infty} s^2\,[\Pi_{VV}(s)-\Pi_{AA}(s)] \,=\, 0\,. 
\ee

The next step is to apply the sum-rules of eq.~({\ref{eq:WSRs}) to the resonance reduced model in eq.~(\ref{eq:Lpheno}). Imposing the high-energy theory to satisfy the first WSR at LO yields
	\be \label{eq:1WSRLO}
	F_V^2 \,-\, F_A^2 \, = \, v^2 \,,
	\ee
	while computing the 2WSR at LO implies the relation
	\be \label{eq:2WSRLO}
	F_V^2\,M_{V^1_3}^2 \, - \, F_A^2\,M_{A^1_3}^2 \, = \, 0 \,.
	\ee
Notice that the combined action of these two sum relations sets the axial-vector resonance mass above the vector one, $M_{V^1_3} < M_{A^1_3}$, and also they allow to rewrite the resonance couplings $F_V$ and $F_A$ in terms of the spin-1 resonance masses, as follows
	\be \label{eq:12WSRLO}
	F_V \,=\, v^2\, \Frac{M_{V^1_3}^2}{M_{A^1_3}^2\,-\, M_{V^1_3}^2} \,, \qquad 
	F_A \,=\, v^2\, \Frac{M_{A^1_3}^2}{M_{A^1_3}^2\,-\, M_{V^1_3}^2} \,.
	\ee

	In addition, it is possible to get further SD constraints from the WSRs when they are analyzed at NLO. Particularly, considering the 2WSR at NLO together with the constraints obtained from the vector and axial-vector form factors in eq.~(\ref{eq:FF}), one can express the Higgs coupling to the Goldstone bosons as \cite{Pich:2013fea}
	\be \label{eq:2WSRNLO}
	\kappa_W \,=\, \Frac{M_{V^1_3}^2}{M_{A^1_3}^2} \,.
	\ee

	\subsection{Prediction for the LECs}
	
	The reduced resonance bosonic theory collected in eq.~(\ref{eq:Lpheno}) contains a total of 12 free parameters: 8 independent resonance couplings and 4 unknown heavy masses. So far, the incorporation of the SD constraints from the form factors and the two WSRs reduce the available parameter space with 5 modest and fairly general conditions. Hence, there are only 7 independent parameters in this case which let us glimpse some very interesting properties about these massive states. We select the following set: the vector and axial-vector heavy masses, $M_{V^1_3}$ and $M_{A^1_3}$, and the scalar and pseudoscalar masses and couplings, commonly expressed through the ratios $c_{d}/M_{S^1_1}$ and $d_P/ M_{P^1_3}$, and $\lambda_{hS^1_1}$ (which will be disregarded since it does not further contribute to the analysis).
	Furthermore, this picture is even better when only the spin-1 couplings and masses are involved. Then, it is possible to parametrize all their high-energy couplings and masses in terms of the two spin-1 masses alone (and also the Goldstone-Higgs coupling $\kappa_W$ if the 2WSR does not apply).
	

	The predictions for the LECs are collected in table \ref{tab:pheno}. There we can find the imprints of the resonance couplings of the reduced bosonic theory (just a compilation of the resonance imprints from table \ref{tab:LECbosonic} in chapter \ref{ch:LECs}) that are sensitive to the scope of the P-even colorless bosonic Lagrangian in eq.~(\ref{eq:Lpheno}). In addition, the same low-energy couplings are analyzed when all the studied SD constraints are applied (including both WSRs). Therefore, the number of independent parameters shrinks remarkably, in particular those regarding the spin-1 fields, as already suggested before. 

	\begin{table}[h!]
		\vspace{0.5cm}
		\begin{center}
			\renewcommand{\arraystretch}{2.2}
			\begin{tabular}{|c|c|c|}
				\hline
				$i$ & Bosonic LECs,  $\Delta\mF_i(h/v)$ & with SD constraints, $\Delta\mF_i(0)$
				\\[1ex]
				\hline
				$1$  &
				$- \Frac{F_V^2}{4M_{V^1_3}^2} + \Frac{F_A^2}{4M_{A^1_3}^2} $  &
				$ - \,\displaystyle \Frac{v^2}{4} \, \left( \Frac{1}{M_{V^1_3}^2} \,+\, \Frac{1}{M_{A^1_3}^2} \right) $
				\\ [1ex]
				\hline
				$2$  &
				$- \Frac{F_V^2}{8M_{V^1_3}^2} - \Frac{F_A^2}{8M_{A^1_3}^2}$  &
				$ \Frac{v^2\, \left(M_{V^1_3}^4 + M_{A^1_3}^4\right)}{8\,M_{V^1_3}^2\,M_{A^1_3}^2\, \left(M_{A^1_3}^2 - M_{V^1_3}^2\right)} $
				\\ [1ex]
				\hline
				$3$  &
				$-  \Frac{F_VG_V}{2M_{V^1_3}^2} $ & 
				$-\,\Frac{v^2}{2\,M_{V^1_3}^2} $
				\\ [1ex]
				\hline
				$4$  &
				$\Frac{G_V^2}{4M_{V^1_3}^2}  $  &
				$ \Frac{v^2 \left(M_{A^1_3}^2 - M_{V^1_3}^2\right)}{4\,M_{V^1_3}^2\,M_{A^1_3}^2} $
				\\ [1ex]
				\hline
				$5$  &
				$\Frac{c_{d}^2}{4M_{S^1_1}^2}\,-\,\Frac{G_V^2}{4M_{V^1_3}^2} $ &
				$\Frac{c_{d}^2}{4M_{S^1_1}^2}\,-\,\Frac{v^2 \left(M_{A^1_3}^2 - M_{V^1_3}^2\right)}{4\,M_{V^1_3}^2\,M_{A^1_3}^2} $
				\\ [1ex]
				\hline
				$6$ &
				$ - \Frac{\Lambda_1^{hA\,\, 2}v^2}{M_{A^1_3}^2}$ &
				$ - \Frac{v^2\, M_{V^1_3}^2\,\left(M_{A^1_3}^2 - M_{V^1_3}^2\right)}{M_{A^1_3}^6} $
				\\ [1ex]
				\hline
				$7$  &
				$\Frac{ d_P^2}{2 M_{P^1_3}^2}\,+\, \Frac{\Lambda_1^{hA\,\, 2}v^2}{M_{A^1_3}^2} $	&
				$\Frac{ d_P^2}{2 M_{P^1_3}^2} \,+\, \Frac{v^2\, M_{V^1_3}^2\,\left(M_{A^1_3}^2 - M_{V^1_3}^2\right)}{M_{A^1_3}^6} $
				\\ [1ex]
				\hline
				$8$ &
				0 	&
				0
				\\ [1ex]
				\hline
				$9$ &
				$  - \Frac{F_A \Lambda_1^{hA} v}{M_{A^1_3}^2}$		&
				$ \Frac{v^2\,M_{V^1_3}^2}{M_{A^1_3}^4} $
				\\ [1ex]
				\hline
			\end{tabular}
		\end{center}
		\caption{\small
			Unconstrained resonance contributions (column 2) and resonance contributions with the implementation of the form factor constraints and WSRs (column 3) from the reduced bosonic theory in eq.~(\ref{eq:Lpheno}) to the $P$-even bosonic LECs of the EWET NLO Lagrangian (table \ref{tab:bosonic-Op4}), where $i$ is the operator assignment index for this set. LECs in column 2 must be understood as functions of $h/v$. In column 3, SD constraints only apply for the first term of the series expansion, $\mF_i \equiv \mF_i(0)$. \vspace{0.5cm} }
		\label{tab:pheno}
	\end{table}
	
	One should be aware that, once form factors and WSRs are imposed, all the LECs must be no longer understood as polynomial functions of $h/v$, as opposite to unconstrained low-energy couplings, since the SD relations only attach to the $\mO(h^0)$ terms in their series expansion,\ie when all the $h/v$ powers of the polynomial are set to zero. Hence, the bounded LECs must be understood as $\mF_i = \mF_i(0)$.

	\section{Oblique parameters}
	
	The oblique parameters, also called the Peskin-Takeuchi parameters, consist of a set of three physical quantities that account for the differences between new physics and the SM predictions in regard to electroweak radiative corrections \cite{Peskin:90, Peskin:92, Kennedy:1990ib, Altarelli:1990zd, Barbieri:2004qk},\ie the photon and gauge bosons self-energies.  These observables are sensitive to the resonance theory analyzed in this work and they get contributions even at tree-level, depending on the particular process to be analyzed.
		This is also in agreement with the resonance interactions subset we are analyzing in eq.~(\ref{eq:Lpheno}). Hence, the study of the oblique parameters happens to be the best framework to study the phenomenology of the considered heavy states.
	
	The oblique parameters are a set of three physical observables, usually\footnote{Also named $\varepsilon_1$, $\varepsilon_2$ and $\varepsilon_3$.} denoted as $S$, $T$ and $U$, although the last one is usually disregarded from the phenomenological analysis since it is negligible for almost all the models present in the literature. 	
	They are defined as difference quantities with respect to the SM. This means that no deviations from SM expected values imply these parameters to be zero. 
	
	In the first place, the $S$ parameter represents the SM discrepancy of the correlator  $\Pi_{30}$ in eq.~(\ref{eq:corrPi30}). It is defined as 
	\be \label{eq:Sparameter}
	S \,=\, \Frac{16 \pi}{g^2}\, \left( e_3\,-\,e_3^{{\rm SM}} \right) \qquad\quad \mbox{with} \qquad e_3 \,=\, \Frac{g}{g'}\, \widetilde \Pi_{30} (0) \,,
	\ee
	being $\widetilde \Pi_{30}$ 
	proportional to the correlator function once the Goldstone tree-level contribution is substracted \cite{Peskin:90, Peskin:92}:
	\be \label{eq:Pi30correlator}
	\Pi_{30}(q^2) \,=\,-\, \Frac{g^2\, v^2\, \tan \theta_W}{4} \, + \,  q^2\,\widetilde \Pi_{30}(q^2) \,.
	\ee
	In the second place, the $T$ parameter measures the custodial breaking and it is introduced as
	\be \label{eq:Tparameter}
	T \,=\, \Frac{4\pi}{g^2\,\sin^2 \theta_W} \, \left( e_1 \,-\, e_1^{{\rm SM}} \right) \qquad\quad \mbox{with} \qquad e_1 \,=\, \Frac{\Pi_{33}(0) \,-\, \Pi_{WW}(0)}{M_W^2}\,.
	\ee
	For the sake of completeness, we should mention the $U$ parameter, which is defined analogously to the $T$ parameter, but involving the quadratic terms of the correlator expansion momentum series, instead of the leading terms. As already mentioned, we will discard this quantity for our analysis because it just accounts for higher order new physics contributions with respect to the other two oblique parameters.

	Nevertheless, the calculation of the correlator functions is challenging and awkward sometimes. It is usually more convenient to determine these parameters making use of the following dispersion relations for the $S$ and $T$ parameters \cite{Peskin:90, Peskin:92, Pich:2013fea}, correspondingly 
	\begin{align} \label{eq:dispersionrels}
	S & \,=\, \displaystyle  \Frac{16\pi}{g^2 \tan^2 \theta_W}\, \int^\infty_0 \Frac{{\rm d}t}{t} \left[ \rho_S(t) \,-\, \rho_S^{{\rm SM}}(t) \right] \,,
	\nn\\[1ex]
	T & \,=\, \displaystyle  \Frac{4\pi}{g'^2 \cos^2 \theta_W}\, \int^\infty_0 \Frac{{\rm d}t}{t} \left[ \rho_T(t) \,-\, \rho_T^{{\rm SM}}(t) \right] \,,
	\end{align}
	being expressed as integrals of their spectral function representations \cite{Peskin:90, Peskin:92, Pich:2013fea}, defined respectively as
	\be \label{eq:spectralfunctions}
	\rho_S(t) \,=\, \Frac{1}{\pi} \, {\rm Im}\left( \widetilde \Pi_{30}(t)\right) \,, \qquad \rho_T(t) \,=\, \Frac{1}{\pi} \, {\rm Im}\left( \Sigma^{(0)}(t) \,-\, \Sigma^{(+)}(t) \right) \,,
	\ee
	being $\Sigma(t)$ the Goldstone self-energy. Naively, one can understand that the $1/t$ integration weight factors suppress the highest UV contributions to the spectral functions and only the lightest heavy states are significant for the calculation \cite{Pich:2008jm}. Therefore, there is no need of imposing arbitrary cut-offs. Anyhow, convergence is guaranteed and further details can be found in \cite{Peskin:90, Peskin:92, Pich:2013fea}. In addition, one of the main advantages of this method is to avoid the computation of non-absorptive loop diagrams. 

	%
	
	\section[Oblique parameters experimental bounds]{Bounds on the reduced bosonic resonance theory from the oblique parameters}
	
	The global fit to electroweak global precision data \cite{Baak:2012kk} has set the experimental value for the $S$ ant $T$ parameter with their respective uncertainties bounds, provided the Higgs mass to be $m_h=125\,{\rm GeV}$. They are currently
	\be \label{eq:obliquevalues}
	S \,=\, 0.03 \,\pm\, 0.10\,, \qquad\quad T\,=\, 0.05 \,\pm\, 0.12 \,.
	\ee

	We compare these experimental values of the oblique parameters with their predictions for the reduced resonance model proposed in eq.~(\ref{eq:Lpheno}). For this purpose, we apply the form factors and WSRs short-distance contraints from eq.~(\ref{eq:FF}) and eqs.~(\ref{eq:1WSRLO}, \ref{eq:2WSRLO}, \ref{eq:2WSRNLO}), respectively. In the following, there is an explicit distinction for the results where only the 1WSR is imposed from those where both WSRs are incorporated. The calculation is performed at LO and at NLO. 
	
	The LO computation of the $S$ and $T$ oblique parameters brings
	\be \label{eq:obliqueST}
	S_{{\rm LO}}\, =\, 4\pi\,\left( \frac{F_V^2}{M_{V^1_3}^2}\,-\,\frac{F_A^2}{M_{A^1_3}^2} \right)\,, \qquad\quad\; T_{LO}\, =\, 0\, ,
	\ee
	where no SD constraints have been implemented. The $T$ parameter is exactly zero since the Goldstone self-energies vanish at this order; while the $S$ parameter only receives contributions from the tree-level exchanges of vector and axial-vector massive states \cite{Peskin:90, Peskin:92}. Notice that only spin-1 vector and axial-vector massive states, $V^1_3$ and $A^1_3$, are involved in this calculation, whereas the spin-0 high-energy resonances $P^1_3$ and $S^1_1$ are not involved in these processes, given the interaction Lagrangian of eq.~(\ref{eq:Lpheno}). Incorporating the SD constraints to the computation, it yields
	\begin{align} \label{eq:obliqueSLO}
	\Frac{4\pi v^2}{M_{V^1_3}} &\,<\, S_{{\rm LO}} \,=\,  4\pi \left[ \frac{v^2}{M_{V^1_3}^2} + F_A^2\left( \frac{1}{M_{V^1_3}^2} - \frac{1}{M_{A^1_3}^2} \right)\right]\,,&    \qquad &(\mbox{1WSR},\,M_{V^1_3} < M_{A^1_3})& 
	\nn\\[1ex]
	\Frac{4\pi v^2}{M_{V^1_3}} &\,<\, S_{{\rm LO}} \,=\, 4\pi v^2 \left( \Frac{1}{M_{V^1_3}^2}\,+\, \Frac{1}{M_{A^1_3}^2} \right)\,<\, \Frac{8\pi v^2}{M_{V^1_3}} \,,& \qquad &(\mbox{2WSRs})&
	\end{align}
	giving rise to a lower bound when the 1WSR is applied and both lower and upper bounds when the 2WSR is considered too.
	
	The NLO calculation\footnote{It is performed at leading orger in the gauge couplings $g$ and $g'$.} requires to make use of the dispersion relations from eq.~(\ref{eq:dispersionrels}). Only the lightest two-particles cuts, $\varphi\varphi$ and $\varphi h$, have been computed for the spectral functions loop calculations, eq.~(\ref{eq:spectralfunctions}), since higher-energy states entering the loops are found to be very suppressed \cite{Pich:2012jv}. As well as before, only spin-1 resonances have been considered. The calculation of the $S$ oblique parameter at NLO leads to \cite{Pich:2012dv, Pich:2013fea, Pich:2012jv,  Rosell:2006dt, Pich:2008jm},
	\begin{align} \label{eq:obliqueSNLO}
	S_{{\rm NLO}} & \,\geq\, \Frac{4\pi v^2}{M_{V^1_3}^2} \, + \, \frac{1}{12\pi}\left[ \log \frac{M_V^2}{m_h^2} \,-\, \frac{11}{6}  \,-\, \kappa_W^2\left( \log\frac{M_A^2}{m_h^2} \,-\, \frac{17}{6} \,+\, \frac{M_A^2}{M_V^2}\right)\right]\,, 
	\nn\\
	& \hspace{7cm}  \qquad (\mbox{1WSR},\,M_{V^1_3} < M_{A^1_3})\,,
	\nn\\[1ex]
	S_{{\rm NLO}} & \,=\, 4\pi v^2 \left( \Frac{1}{M_{V^1_3}^2}\,+\, \Frac{1}{M_{A^1_3}^2} \right) \, + \, \Frac{1}{12\pi} \left[ \left( 1\,-\, \Frac{M_{V^1_3}^4}{M_{A^1_3}^4} \right) \! \left(\log \Frac{M_{V^1_3}^2}{m_h^2} \,-\, \Frac{11}{6} \right) \right.
	\nn\\
	& \left. \, +\, \left( \Frac{M_{V^1_3}^2}{M_{A^1_3}^2}\,-\, \Frac{M_{V^1_3}^4}{M_{A^1_3}^4} \right) \log \Frac{M_{A^1_3}^2}{M_{V^1_3}^2} \right]\,,  \qquad\qquad\qquad (\mbox{2WSRs})\,,
	\end{align}
	depending on which SD constraints are fulfilled.
	On the other hand, the $T$ parameter is non-trivial when computed at NLO \cite{Pich:2013fea}, being
		\begin{align} \label{eq:obliqueTNLO}
		T_{{\rm NLO}} &\,=\, \Frac{3}{16\pi \cos^2 \theta_W} \left[1 \,+\, \log\frac{m_h^2}{M_{V^1_3}^2} \,-\, \kappa_W^2\left(1\,+\,\log\frac{m_h^2}{M_{A^1_3}^2}\right)\right] \,,
		\end{align}
		when the 1WSR is applied. If the 2WSR is imposed too, one has the additional relation for the resonance masses quotient in terms of the Higgs-Goldstone coupling, already introduced in eq.~(\ref{eq:2WSRNLO}).
	As it is expected, this oblique parameter vanishes when $M_{V^1_3}=M_{A^1_3}$ and $\kappa_W=1$,\ie the SM setting for this last coupling. Actually, these conditions are equivalent and they mean the same when the two WSRs are imposed, according to eq.~(\ref{eq:2WSRNLO}). Finally, both $S$ and $T$ NLO calculations in eqs.~(\ref{eq:obliqueSNLO}, \ref{eq:obliqueTNLO}) have been truncated at $\mO(m_h^2/M^2_{R^1_3})$, since these contributions are very suppressed, as it is seen in what follows. 
	
	\subsection{Resonance masses and $\boldsymbol{\kappa_W}$ parameter space}
	
	
	The determination of the $S$ and $T$ oblique parameters for the resonance bosonic model allows to obtain the permitted energy region for the vector and axial-vector spin-1 masses and their resonance couplings, provided the implementation of the SD constraints. In the following, we will assume the normal-ordering for the spin-1 masses: $M_{V^1_3}\leq M_{A^1_3}$. Nevertheless, inverted-order is not discarded yet, but it is more unlikely.\footnote{More details can be found in \cite{Pich:2013fea}.} Attending to which SD constraints are imposed, two different scenarios are analyzed  \cite{Pich:2013fea, Santos:2015mqa, Pich:2012dv}: 
	\begin{enumerate}
		\item form factors and 1WSR only (figure \ref{fig:ST1WSR}).
		\item form factors, 1WSR and 2WSR (figure \ref{fig:ST2WSR}).
	\end{enumerate}
	%
	
	\begin{figure}[t!] 
		\centering
		\includegraphics[width=0.9\textwidth]{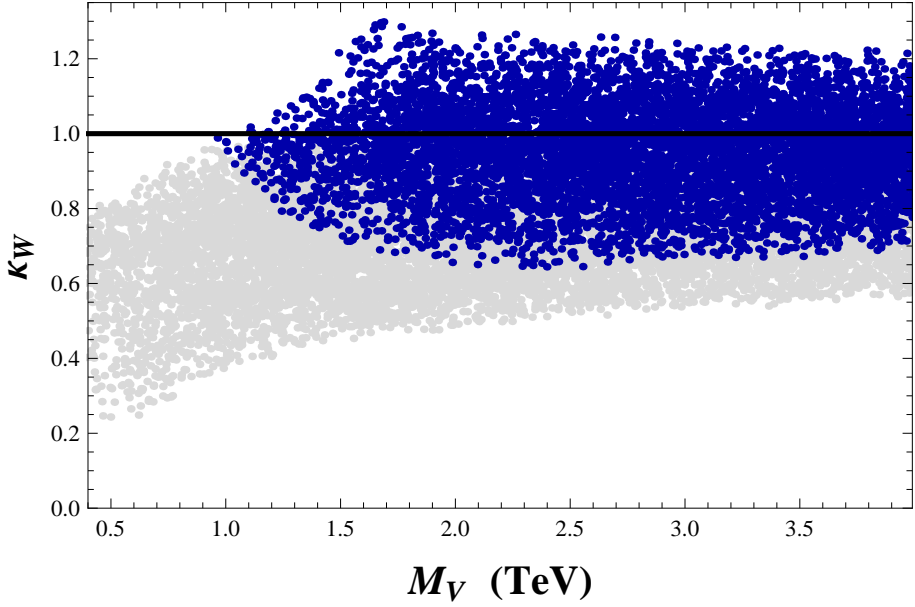}
		\small{
			\caption{
				Scatter plot for the allowed 68\% region for the coupling $\kappa_W$ and the triplet vector mass, $M_{V^1_3}$ associated to the $S$ and $T$ oblique parameters analysis at NLO, provided the implementation of the form factors and 1WSR short-distance constraints. Different points stand for the ratio of the vector mass and the axial-vector mass, $M_{V^1_3}/M_{A^1_3}$, remaining this last one implicit. The gray region depicts heavy mass splitting, $0.02 < M_{V^1_3}/M_{A^1_3} < 0.2$, while the blue region illustrates $0.2 < M_{V^1_3}/M_{A^1_3} < 1$.}
			\label{fig:ST1WSR}
		}
	\end{figure}
	%
	%
	\begin{figure}[t!] 
		\centering
		\includegraphics[width=0.75\textwidth]{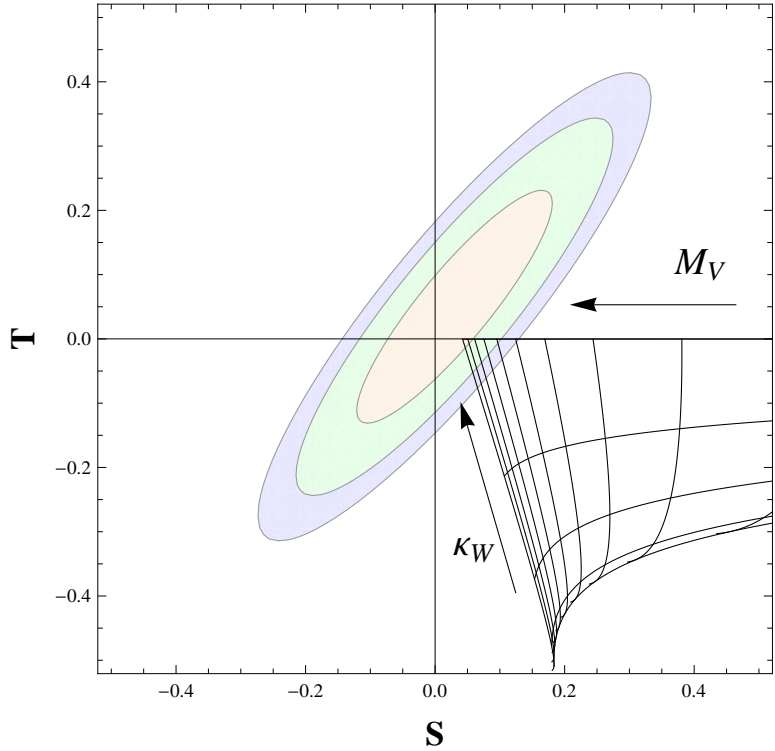}
		\small{
			\caption{$S$ and $T$ parameter space plot at NLO imposing form factors and both WSRs short-distance constraints for a Higgs mass $m_h=125$ GeV. The allowed region for the oblique parameters describes three ellipses with 68\% C.L. (orange), 95\% C.L (green) and 99\% C.L.. Horizontal lines represent  $\kappa_W = \{0,\, 0.25,\, 0.5,\, 0.75,\, 1\}$ (from down to up). Vertical lines describe different values of $M_{V^1_3}$ from 1.5 TeV to 6.0 TeV.}
			\label{fig:ST2WSR}
		}
	\end{figure}
	%
	The first scheme displays for the oblique parameter analysis a scatter plot of the one-sigma confidence level (C.L.) region for the coupling $\kappa_W$ versus the heavy vector mass, $M_{V^1_3}$, for different mass splittings between 0 and 1,\ie the mass ratio
	$M_{V^1_3}/M_{A^1_3}$ (figure \ref{fig:ST1WSR}). The graph shows that non SM-like values for $\kappa_W$ correspond generally to large mass splitting and there are no compatible solutions for this coupling from 1.3 and above. Nonetheless, if one assumes a moderate splitting for this mass ratio, $0.5 < M_{V^1_3}/M_{A^1_3} < 1$, we get \cite{Pich:2013fea, Santos:2015mqa, Pich:2012dv}
	\be \label{eq:pheno1WSR}
	0.86 \,<\, \kappa_W \,<\, 1.16\,, \qquad\quad M_{V^1_3} \,>\, 1.5\,{\rm TeV}\,,\qquad (68\%\, {\rm C.L.})\,.
	\ee

	The second scenario is displayed in Figure \ref{fig:ST2WSR} that shows a $S$--$T$ diagram where the allowed values for these quantities in the resonance model framework are drawn inside the three ellipses attending to the confidence level (C.L.) (1$\sigma$, $2\sigma$ or $3\sigma$ from the smaller to the greater, respectively). On the bottom right side of the figure, there is a grid which lines describe different values for the spin-1 vector resonance mass (vertical lines) and axial-vector mass through the representation of $\kappa_W$ (horizontal lines), according to eq.~(\ref{eq:2WSRNLO}).
	
	The oblique parameters analysis shows that the Higgs-Goldstone coupling $\kappa_W$ is very SM-like and thus small values for this parameter are discarded, as expected from LHC data \cite{Khachatryan:2016vau}.	
	Nevertheless, tighter bounds for this couplings are found to be imposed from the oblique parameter analysis \cite{Pich:2013fea}:
	\be \label{eq:phenokappaW2WSR}
	\kappa_W \,\in\, [0.97\,,\, 1] \quad 68\%\, \mbox{C.L.} \qquad 
	\left( \kappa_W \,\in\, [0.94\,,\, 1] \quad 95\%\, \mbox{C.L.} \right) \,.
	\ee
	Therefore, the spin-1 vector and axial-vector masses are expected to be very close in the energy spectrum. The computation of the available parameter space with both WSRs shows that
	\be \label{eq:phenoMV2WSR}
	M_{A^1_3} \gsim M_{V^1_3} \,>\, 5\,\mbox{TeV} \quad 68\%\, \mbox{C.L.} \quad
	( M_{A^1_3} \gsim M_{V^1_3} \,>\, 4\,\mbox{TeV} \quad 95\%\, \mbox{C.L.} )  \,.
	\ee

	\subsection{Bounds on the bosonic LECs}
	
	The determination of the $S$--$T$ parameter space and the bounds set over the resonance couplings and heavy masses let us analyze the allowed parameter region for the EWET bosonic LECs too, which are directly entangled with observables concerning new physics.
	
	Considering the predictions for the $P$-even bosonic LECs from table \ref{tab:pheno}, once imposed all the SD constraints, we match these results with the computation of the $S$ and $T$ oblique parameters for the reduced resonance theory, provided the experimental inputs of eq.~(\ref{eq:obliquevalues}). The available parameter space for the bosonic LECs is shown in figures \ref{fig:LECs1} and \ref{fig:LECs2}.
	So far, all the bounds for these LECs are compatible with SM predictions.
	
	Figure \ref{fig:LECs1} assembles five graphs displaying in the darker regions the allowed values for the spin-1 related LECs $\mF_{1,3,4,6,9}$ in terms of the heavy vector triplet mass, $M_{V^1_3}$, provided the SD constraints are imposed. The axial-vector mass is implicit in $\kappa_W$ (see eq.~(\ref{eq:2WSRNLO})) which is portrayed through blue, red and green lines\footnote{This consideration does not apply for $\mF_3$, which only depends on the vector resonance mass, $M_{V^1_3}$.} depending on the acquired value, 0.8, 0.9 and 0.95, respectively. Notice that there is no graph for $\mF_2$ since current experimental data are not precise enough to bound it. Neither is $\mF_8$ because it does not receive any contributions from the bosonic resonance exchange (nor from any resonance state analyzed in chapter \ref{ch:resonancetheory}).  The following bounds are set (95\% C.L.) \cite{Lowenergy}:
	%
	%
	%
	%
	\begin{align} \label{eq:LECsbounds}
	&-2\,\times\,10^{-3} \,<\, \mF_1 \,<\, 0 \,,& \qquad
	&-9\,\times\,10^{-5} \,<\, \mF_6 \,<\, 0 \,,&
	\nn\\
	&-2\,\times\,10^{-3} \,<\, \mF_3 \,<\, 0 \,,& \qquad
	&-4\,\times\,10^{-3} \,<\, \mF_9 \,<\, 0 \,.&
	\nn\\
	&0 \,<\, \mF_4 \,<\, 2.5\,\times\,10^{-5} \,.&	&	&
	\end{align}
	The results reveal rather strong limits for these LECs, much more constrained that the current experimental bounds, shown in the next section.

	The remaining two LECs, $\mF_5$ and $\mF_7$, are analyzed through $\mF_4 + \mF_5$ and $\mF_6 + \mF_7$ in figure \ref{fig:LECs2} because the sums permit to  isolate all the spin-0 couplings and masses from spin-1 ones. In these two graphs, we simply represent these combinations of LECs in terms of the ratios $M_{S^1_1}/c_d$, for the singlet scalar, and $M_{P^1_3}/d_P$, for the triplet pseudoscalar heavy state. \\
	
	\newpage		
	%
	\begin{figure*}[h]
	
	\vspace{10mm}
		
	\centering
	\begin{subfigure}[h]{1\textwidth}
		\centering
	\includegraphics[width=0.85\textwidth]{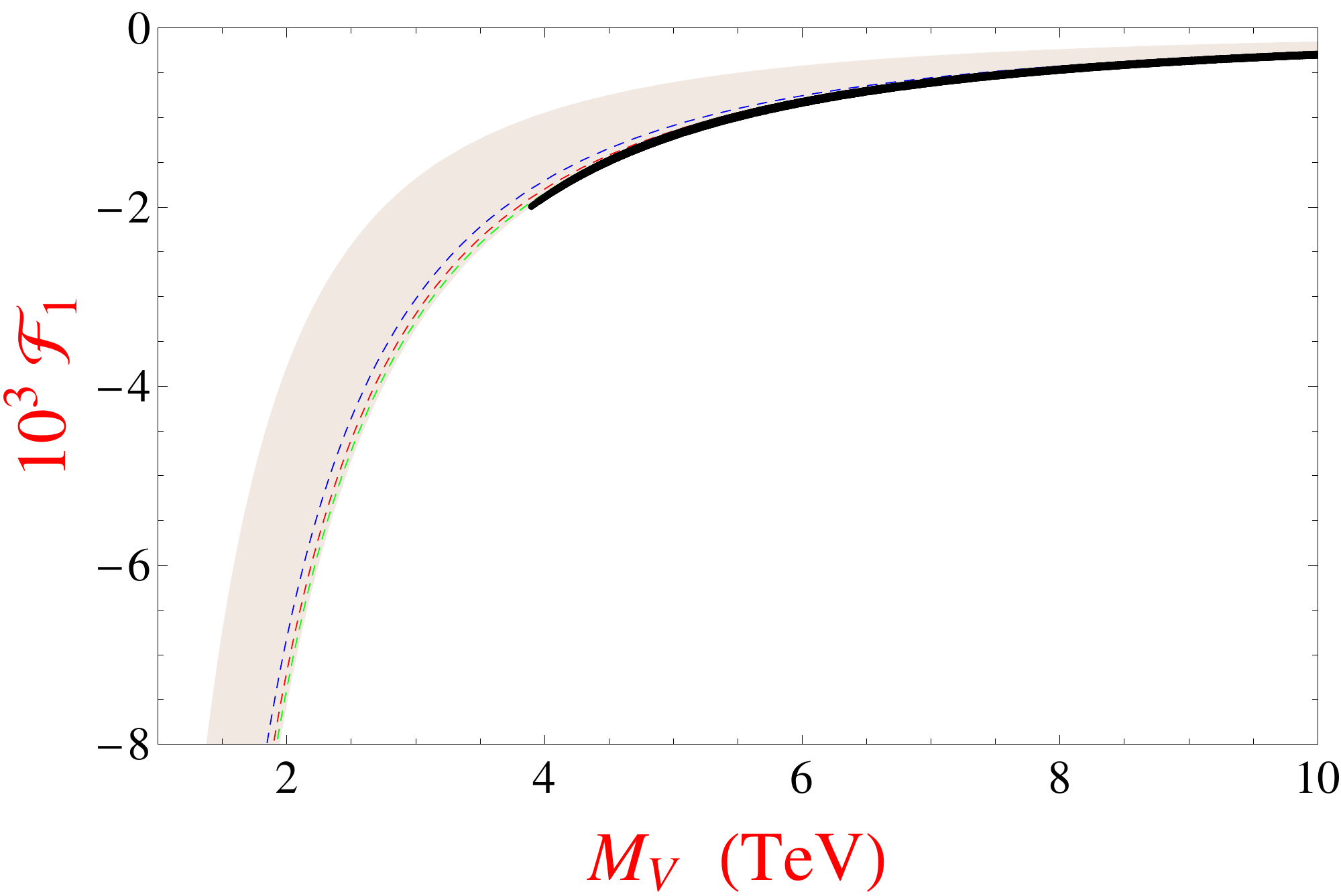}
	\caption*{(a) Predicted value for the purely bosonic LEC $\mF_1$.}
	\end{subfigure}
	
	\vspace{10mm}
	
	\begin{subfigure}[h]{1\textwidth}
		\centering
		\includegraphics[width=0.85\textwidth]{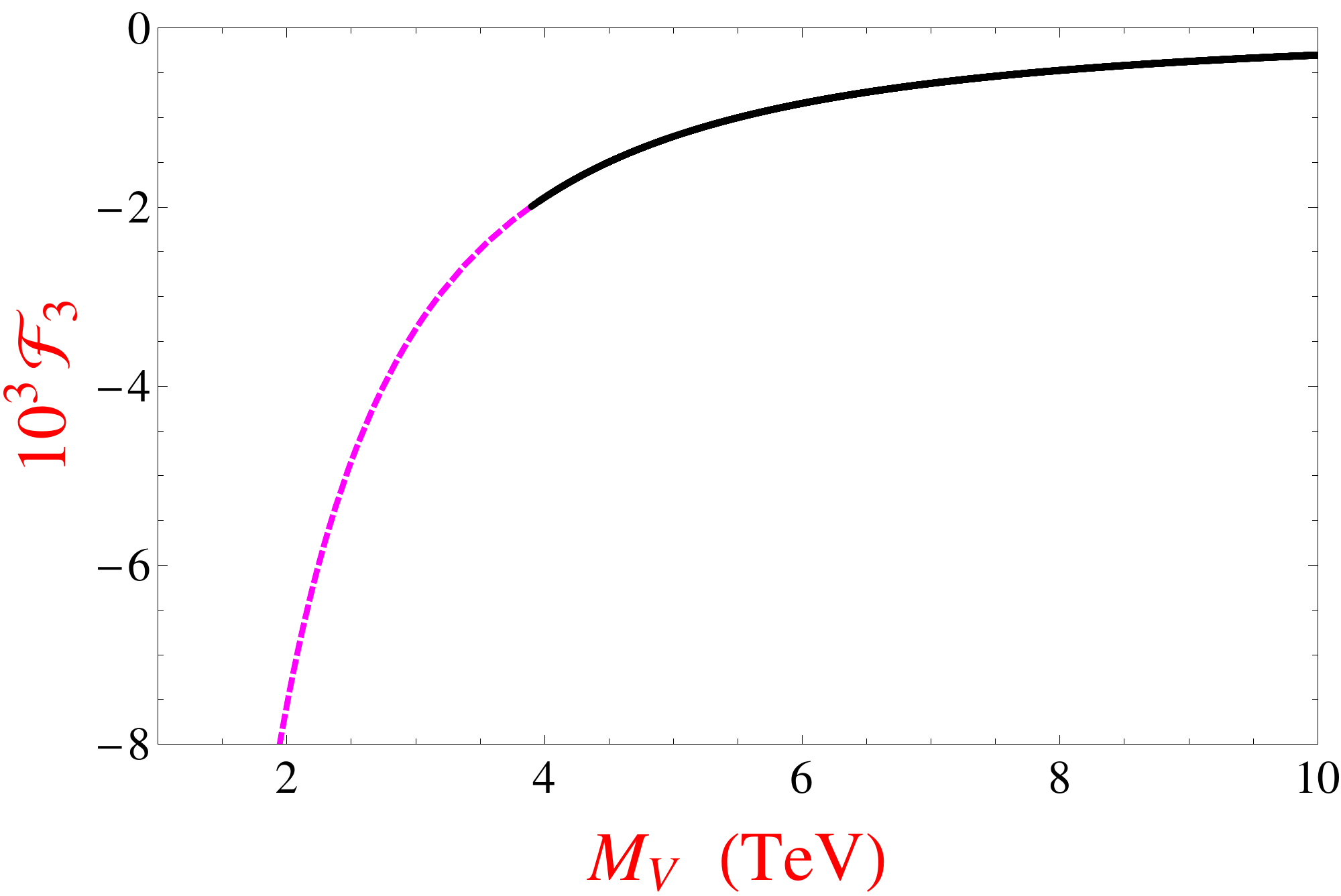}
		\caption*{(b) Predicted value for the purely bosonic LEC $\mF_3$.}
	\end{subfigure}
	\end{figure*}

	\begin{figure*}[h]
		
	\vspace{10mm}
	
	\centering	
	\begin{subfigure}[h]{1\textwidth}
		\centering
		\includegraphics[width=0.85\textwidth]{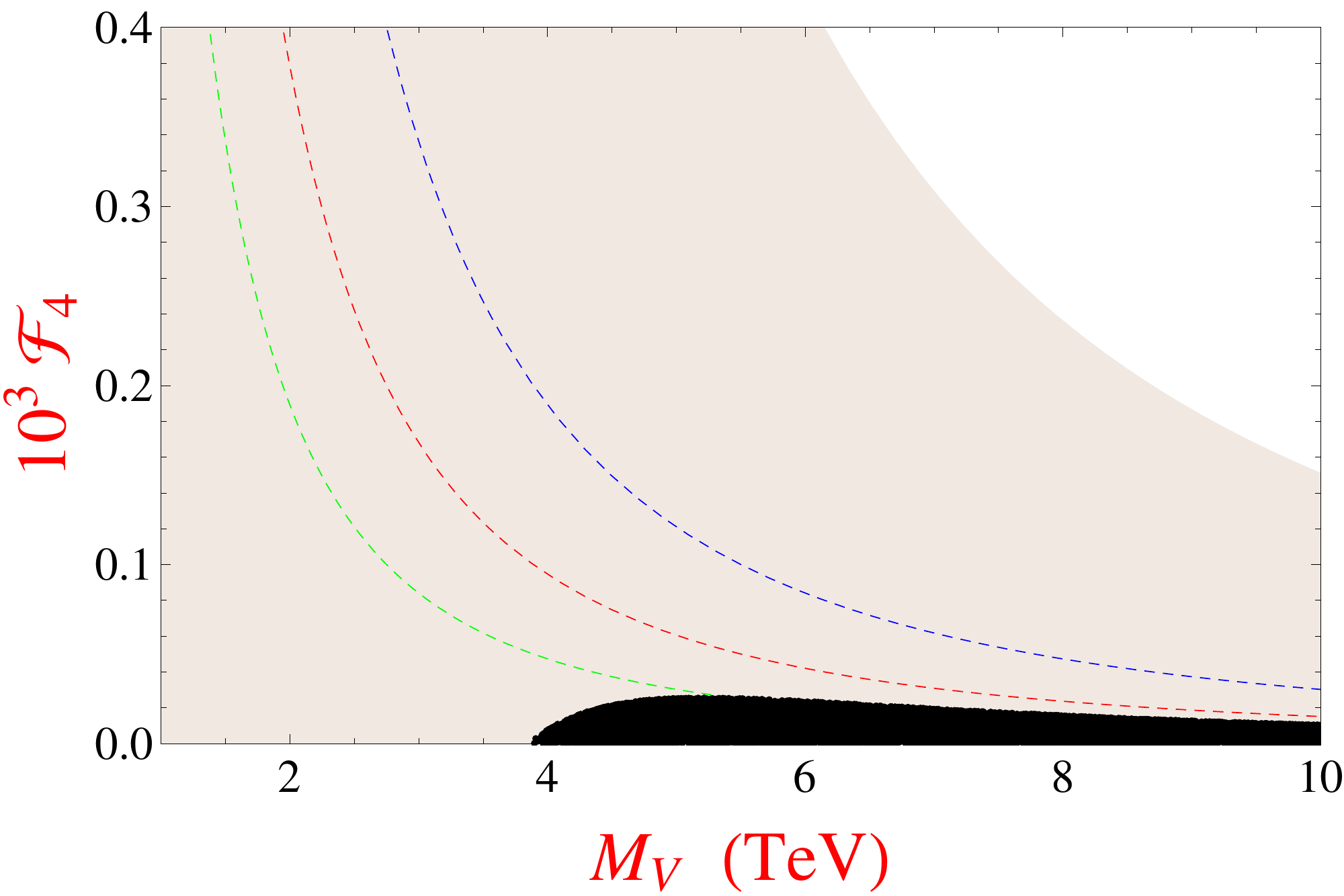}
		\caption*{(c) Predicted value for the purely bosonic LEC $\mF_4$.}
	\end{subfigure}
	
	\vspace{10mm}
	
	\begin{subfigure}[h]{1\textwidth}
		\centering
		\includegraphics[width=0.85\textwidth]{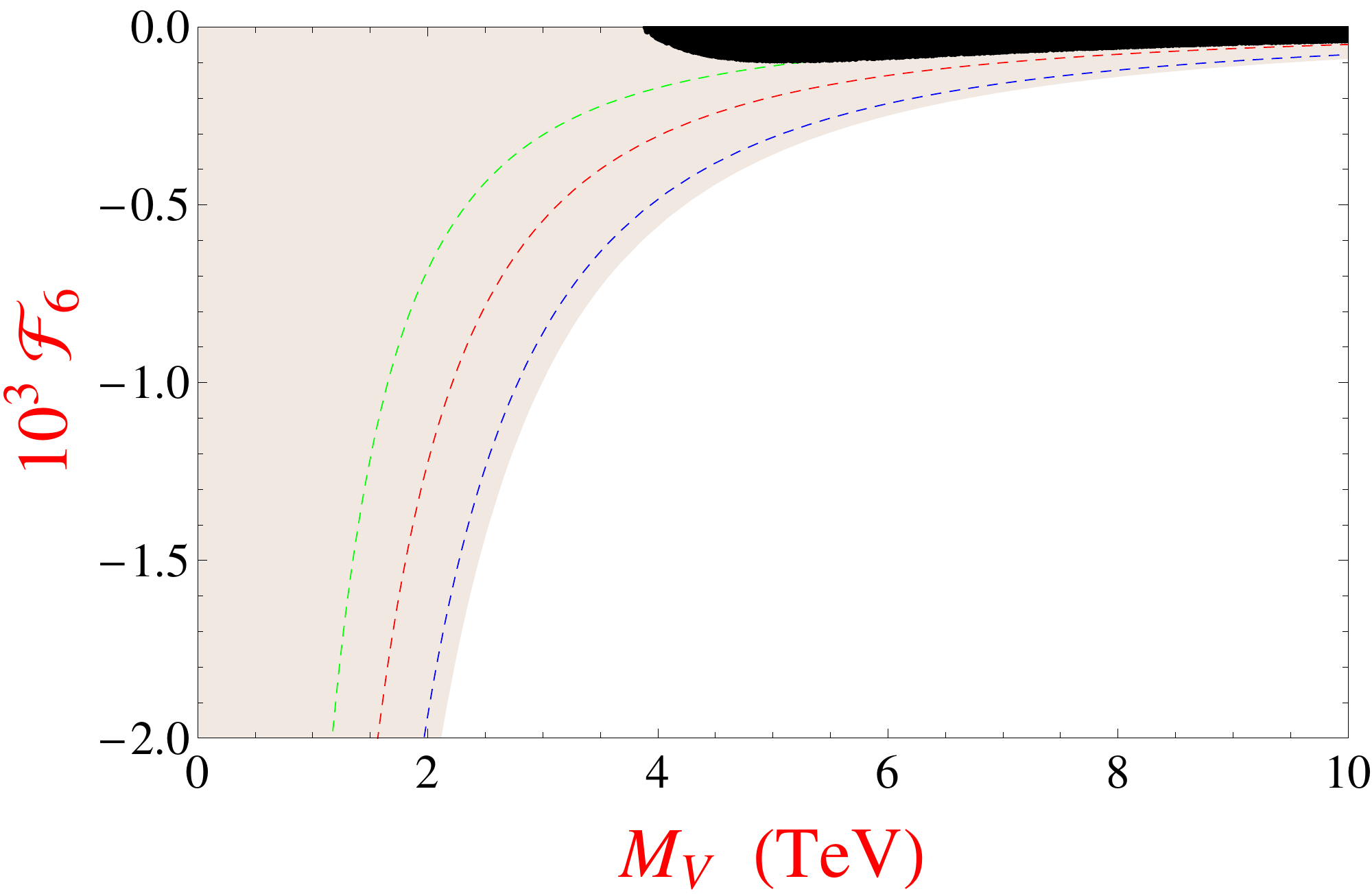}
		\caption*{(d) Predicted value for the purely bosonic LEC $\mF_6$.}
	\end{subfigure}
	\end{figure*}

	\begin{figure}[h]
	\centering
	
	\vspace{10mm}
	
	\begin{subfigure}[h]{1\textwidth}
		\centering
		\includegraphics[width=0.85\textwidth]{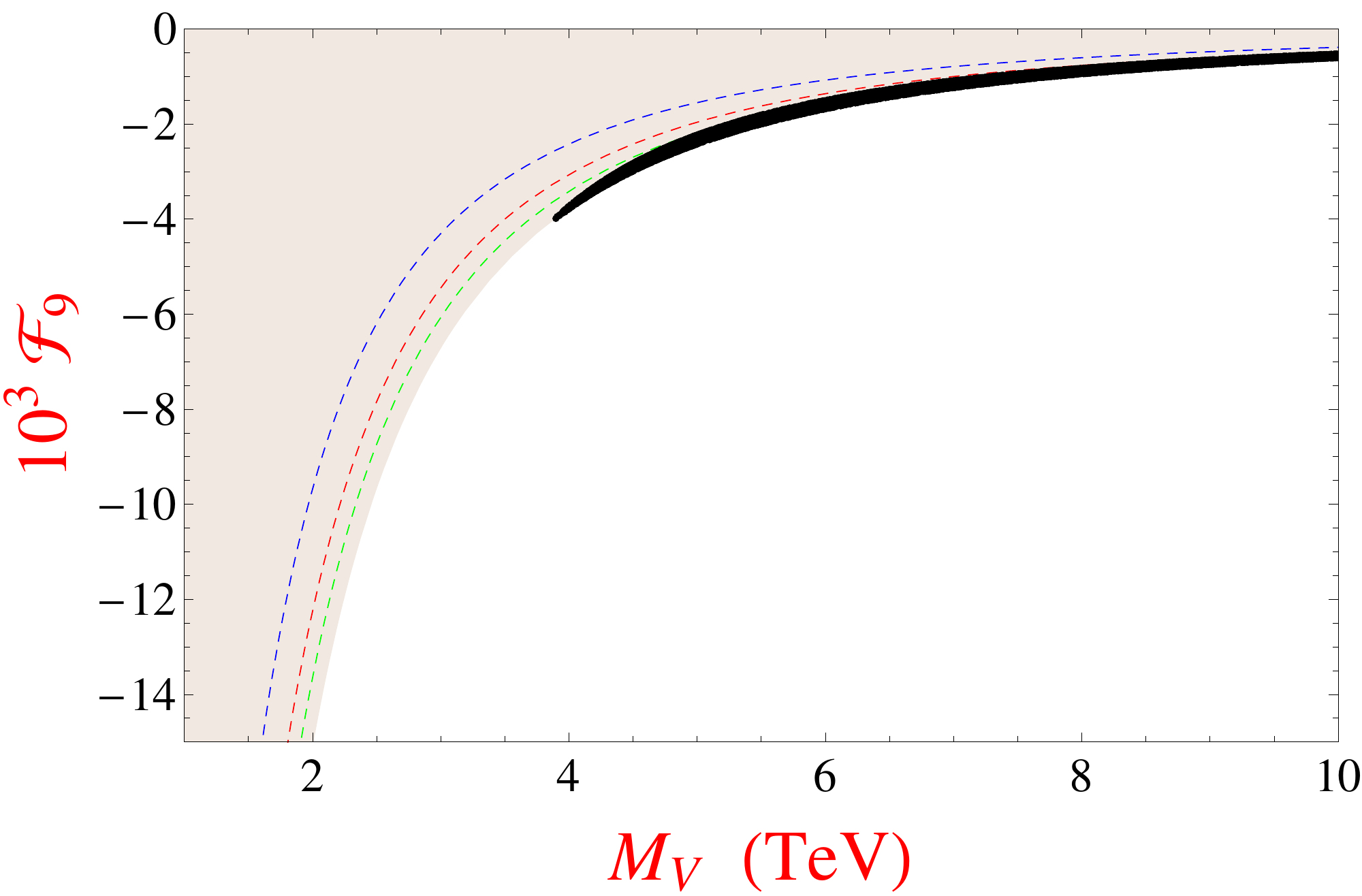}
		\caption*{(e) Predicted value for the purely bosonic LEC $\mF_6$.}
	\end{subfigure}
	\small{
	\caption{Vector and axial-vector predicted values for the LECs from the P-even bosonic Lagrangian expressed in terms of the custodial triplet heavy vector mass, $M_{V^1_3}$ provided form factors and the two WSRs are applied. The axial-vector resonance mass, $M_{A^1_3}$ is implicit in the blue, red and green dashed lines through $\kappa_W = M_{V^1_3}/M_{A^1_3}$ with the values 0.8, 0.9 and 0.95, respectively. Light-shaded grey region corresponds to the parameter space where $M_{V^1_3}<M_{A^1_3}$ (not present in $\mF_3$ since is not $M_{A^1_3}$-dependent). Black regions yield the $S$ and $T$ oblique parameter $95\%$ C.L. allowed region.}
	\label{fig:LECs1}
	}
	\end{figure}

	\begin{figure}[h]
		\centering
		\includegraphics[width=0.85\textwidth]{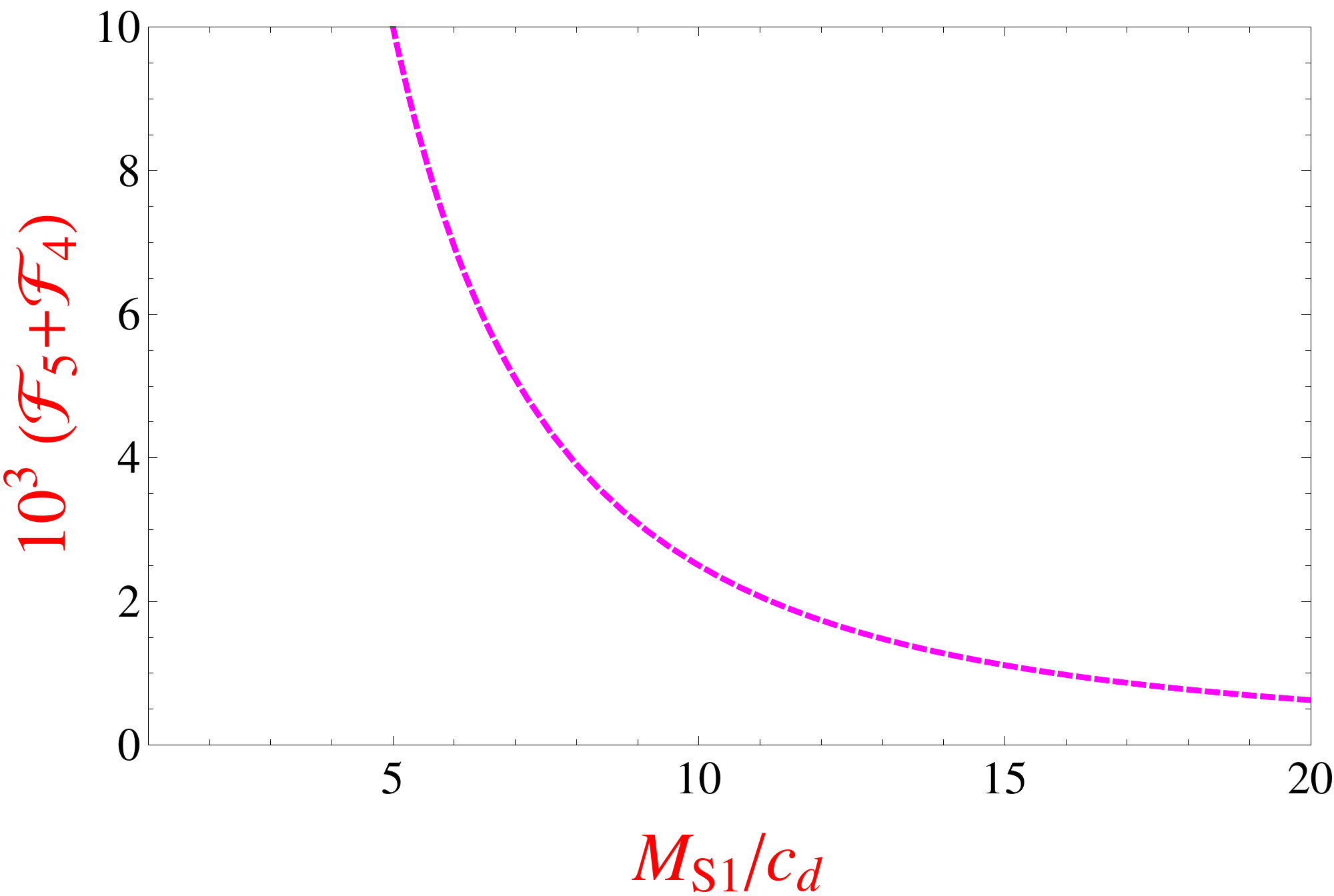}
	\end{figure}
	
	\vspace{10mm}
	
	\begin{figure}[h]
		\centering
		\includegraphics[width=0.85\textwidth]{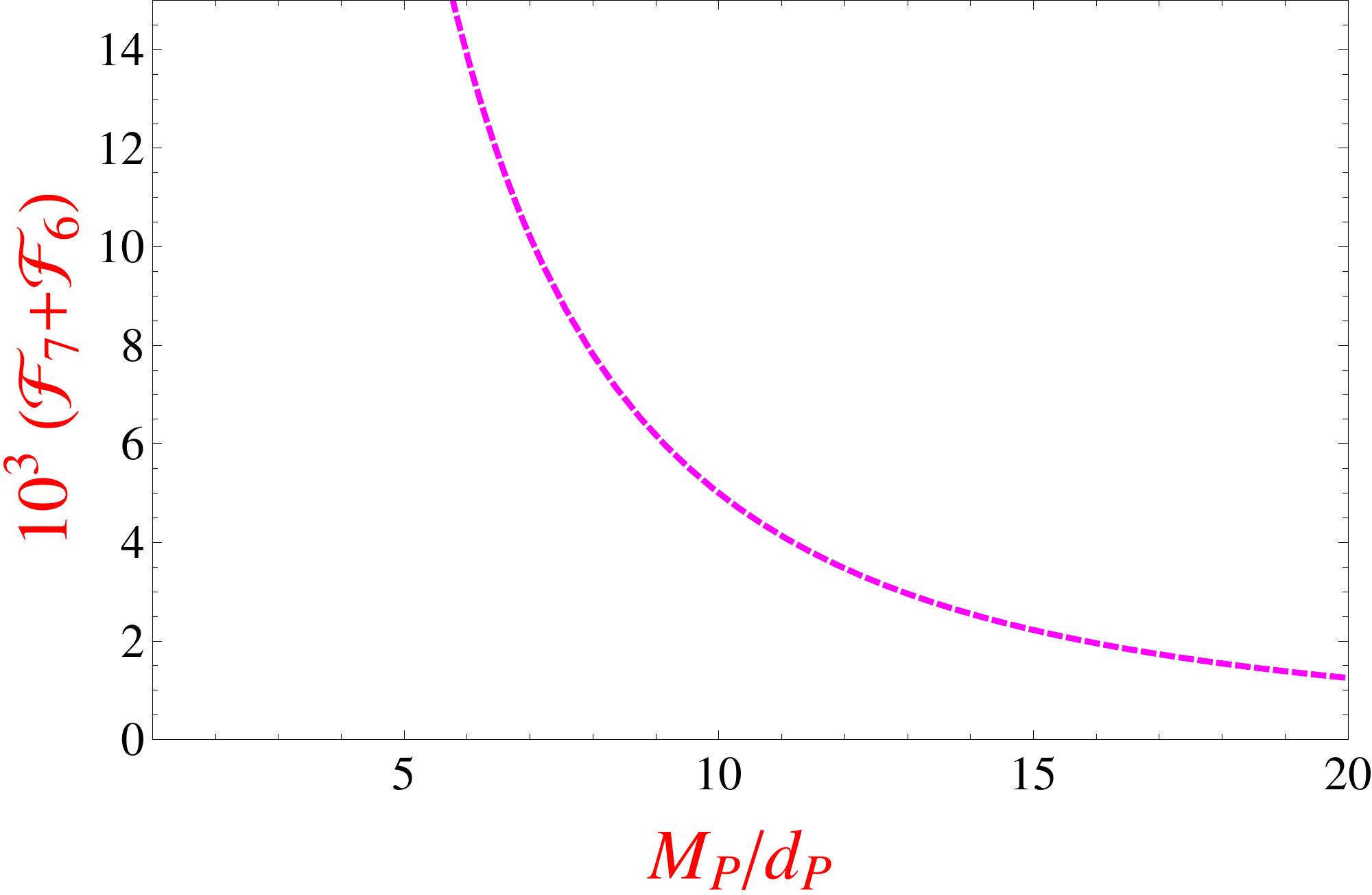}
		\small{
			\caption{Predicted values for the scalar and pseudoscalar contributions to the LECs from the P-even bosonic Lagrangian through the combinations $\mF_4 + \mF_5$ and $\mF_6 + \mF_7$, in terms of the ratios $M_{S^1_1}/c_d$ and $M_{P^1_3}/d_P$, respectively.}
			\label{fig:LECs2}
		}
	\end{figure}
	
	\clearpage

	\section{Status of the bosonic LECs experimental determination}
	\markboth{CHAPTER 7.\hspace{1mm} PHENOMENOLOGY IN THE BOSONIC SECTOR}{7.5.\hspace{1mm} STATUS OF THE BOSONIC LECS}
	
	As well as in the oblique parameters, the smaller the LECs experimental uncertainties become, the more stringent bounds arise for the resonance parameter space, yielding two possible scenarios. In the first one, data are still consistent with SM expectations. Yet, the allowed region for the resonances is getting smaller and their existence is presumably postponed to higher energies. The second one, and more interesting, data manifest significant deviations from SM predictions eventually, giving raise to a specific pattern for the LECs which reveals new physics states. This picture would become essential in order to infer the underlying UV theory.
	
	We focus on the current experimental bounds for the P-even bosonic LECs and the Higgs-Goldstone coupling, $\kappa_W$, and we analyze their prospects on a more precise determination. In general, most of the measurements of these couplings yield soft bounds and relatively large uncertainties, even more for Higgs related LECs since Higgs precision data are missing yet. This is the case of bosonic LECs $\mF_{6,7,8,9}$, which involve Higgs fields in their external legs. So far, there are no searches regarding these LECs and yet remain unknown, neither from single-Higgs nor multi-Higgs production processes. In the next decades, some information on these Higgs related couplings is expected to come from the LHC and still not built future colliders. 
	
	Nowadays, it seems more interesting to analyze the remaining bosonic couplings, $\mF_{1,2,3}$ and $\mF_{4,5}$, which are associated with the anomalous triple gauge couplings (aTGC) and the anomalous quartic gauge-boson couplings (aQGC), respectively.
	The most experimentally constrained coupling is found to be $\mF_{1}$, through LEP-I and LEP-II data. This coupling, as well as $\mF_3$, contributes directly to the oblique $S$ parameter \cite{Pich:2012dv, Pich:2013fea, Pich:1998xt} and the $\gamma\gamma \rightarrow WW$ scattering amplitude \cite{Herrero, Chatrchyan:2013akv}. Nevertheless, $\mF_1$ sets a much stronger bound than $\mF_3$ since it is reflected in the tree-level computation of the $S$ parameter,\footnote{This relation is equivalent to eq.~(\ref{eq:dispersionrels}) and eq.~(\ref{eq:spectralfunctions}).} $\Delta S = -16\pi \mF_1$, and the proporcionality can be observed indeed in figure \ref{fig:LECs1}.

	Therefore, the most restrictive limits for the $\mF_1$ low-energy coupling are obtained from the NLO analysis of the oblique $S$ parameter, once two WSRs are incorporated. In order to collect the status of all the bosonic LECs in this section, we rewrite the result for $\mF_1$ from eq.~(\ref{eq:LECsbounds}) and the exclusion limits (95 \% C.L.) that it imposes over the vector and axial-vector custodial triplet resonance masses from eq.~(\ref{eq:phenoMV2WSR}):
	\be \label{eq:phenoF1b}
	-2\,\times\,10^{-3} \,<\, \mF_1 \,<\, 0 \,, \qquad\qquad 	M_{A^1_3} \gsim M_{V^1_3} \,>\, 4\,\mbox{TeV} 
	\ee
 It is expected that $e^+e^-$ future colliders, like the ILC or CLIC, improve the experimental uncertainties by a factor 5. This remarkable enhancement in the precision would be enough to reduce the $S$ oblique parameter indetermination to $\delta S \sim \pm 0.02$ \cite{Falkowski:2015jaa}. It would imply an uncertainty for the coupling of order $\delta \mF_1 \sim 4 \times 10^{-4}$.

	The anomalous triple gauge-boson coupling $\mF_3$ is, however, softly constrained experimentally \cite{Lowenergy}. Despite the fact that it also contributes to the $S$ parameter, it does at NLO, unlike the previous case. Currently, the most stringent bounds come from the combined analysis of LEP and other collider data \cite{Schael:2004tq, Achard:2004ji, Abbiendi:2003mk, Schael:2013ita}.
	The results of the fit usually are given in terms of the SM-EFT Lagrangian parameters $\delta g_1^Z$, $\delta \kappa_\gamma$ and $\lambda_Z$ \cite{Falkowski:2015jaa}, defined in 
	\begin{align} \label{eq:tgc2}
	\mL_{\text{aTGC}} \,&=\, ie\,A^\nu(W_{\mu\nu}^+W^{-\,\mu} - W_{\mu\nu}^-W_{+}^\mu)
	\,+\, ie\Frac{c_\theta}{s_\theta}\,(1+\delta g_1^Z)\, Z^\nu\,(W^-_{\mu\nu} W^{+\,\mu} \nn\\
	& -W^+_{\mu\nu} W^{-\,\mu}) \,+\, ie\, (1 + \delta \kappa_\gamma)\, W^+_\mu W^-_\nu A^{\mu\nu} \,+\, ie\Frac{c_\theta}{s_\theta}\,(1 + \delta \kappa_Z)\,  W^+_\mu W^-_\nu Z^{\mu\nu} \nn\\
	&+ \, i\Frac{\lambda_Z\, e}{M_W^2} \left[ W_{\mu\nu} W^{-\,\nu}_\rho A^{\rho\mu}+ \Frac{c_\theta}{s_\theta}\, W_{\mu\nu} W^{-\,\nu}_\rho Z^{\rho\mu} \right]\,,
	\end{align}
	where $\delta \kappa_Z$ is related to the other couplings through 
	\be \label{eq:tgc3}
	\delta \kappa_Z \,=\, \delta g_1^Z - \Frac{s_\theta^2}{c_\theta^2}\, \delta \kappa_\gamma\,,
	\ee
	and the coupling of the first term in eq.~(\ref{eq:tgc2}) is fixed due to gauge invariance. Furthermore, the operator proportional to $\lambda_Z$ does not contribute to the NLO EWET Lagrangian, but to higher orders, and thus it is not considered for this analysis. 
	
	The current experimental bounds (95\% CL) for the aTGC parameters are \cite{Schael:2013ita}	
	\begin{align} \label{eq:tgc1}
	-0.054 \, < & \, \delta g_1^Z \, < \, 0.021 \,,  \nn\\
	-0.099 \, < & \, \delta \kappa_\gamma \, < \, 0.066 \,,  \nn\\
	-0.059 \, < & \,\; \lambda_Z \; < \, 0.017 \,, 
	\end{align}
	being all of them 0 in the SM. For the measurement of any of these observables, it has been that assumed the other ones are fixed to the SM. 
	A simple identification with the EWET LECs of table \ref{tab:bosonic-Op4} yields
	\begin{align} \label{eq:tgc3}
	\Delta g_1^Z\,&=\, \Frac{g'^2}{c_\theta^2 - s_\theta^2}\, \mF_1 \,-\, \Frac{2g^2}{c_\theta^2}\, (\mF_3 + \widetilde \mF_1) \,, \nn\\[1ex]
	\Delta \kappa_\gamma \,&=\,- g^2\, (\mF_1 + \mF_3 +  3\,\widetilde \mF_1) \,. 
	\end{align}
	Using the previous result for $\mF_1$ in eq.~(\ref{eq:phenoF1b}), one obtains for the EWET triple gauge couplings the following bounds (95\% CL):\footnote{Note that $\widetilde \mF_1$ corresponds to an $P$-odd operator which is not included in the eq.~(\ref{eq:Lpheno}) Lagrangian. Nonetheless, it is considered for giving a general view of the aTGC operators contributing at $\mO_{\hat d}(p^4)$.}
	\begin{align} \label{eq:LECspheno31}
	-0.11 \,& < \, \mF_3 \, < \, 0.12 \,, \nn\\
	-0.09 \,& < \, \widetilde \mF_1 \, < \, 0.12 \,. 
	\end{align}
	It is also common to present the same results expressed in the Longhitano basis (see appendix \ref{app:dictionary}). In this case, we obtain $-0.23 < a_2 < 0.21$ and $-0.053 < a_3 < 0.020$.

	However, these uncertainties are not small enough to discriminate the presence of resonance states in the TeV range. Indeed, these results only introduce new exclusion limits for the vector and axial-vector resonance masses far below the 1 TeV range. Prospects on a precision for aTGC couplings similar to the analysis in figure \ref{fig:LECs1} are not expected before the LC searches \cite{Baak:2013fwa, Baer:2013cma}, with  $|\mF_3| \sim |\Delta \kappa_\gamma |/g^2 \sim 5\times 10^{-4}$. Nevertheless, a significant precision enhancement is presupposed with the evaluation of LHC run-II data.
	
	The anomalous quartic gauge coupling search, associated to the LECs $\mF_{4,5}$, is currently at the same order of precision ($\mO(10^{-1})$) as the triple gauge couplings \cite{Aad:2014zda}.	
	Nowadays, the most stringent bounds on these couplings come from $WW$ and $WZ$ production via vector boson scattering from LHC (run-I) data \cite{Aaboud:2016uuk}, being, 
	%
	%
	\begin{align} \label{eq:LECspheno32}
	-0.024 \,& < \, \mF_4 \, < \, 0.030 \,,\nn\\
	-0.028 \,& < \, \mF_5 \, < \, 0.033 \,, \nn\\
	-0.037 \, < \; & \mF_4\,+\,\mF_5 \, < \, 0.045 \,.
	\end{align}
	As in the previous case, the precision for the LECs is not sufficient to establish relevant bounds for the vector and axial-vector resonance masses.
	LHC run-II  at $\sqrt{s}=13\,\text{TeV}$ expects to reach $\mO(10^{-3})$ precision on these LECs \cite{Fabbrichesi:2015hsa}.  In addition, the combination $\mF_4 + \mF_5$ (also shown in figure \ref{fig:LECs2}) is found to be proportional to the ratio $c_d/M_{S^1_1}$. We find that values above 0.4 are excluded at $95\,\%$ C.L.. This is still a very poor bound for these ratio, since $c_d$ is expected to be $\mO(1)$.
	
	Finally, the current direct bounds for the Higgs-Goldstone coupling, $\kappa_W$, are set by LHC data \cite{Khachatryan:2016vau}, being ($68\,\%$ C.L.)
	\be \label{eq:phenokappaW}
	0.79\,<\, \kappa_W \,<\, 1.01\,.
	\ee
	%
	Therefore, the analysis of $\kappa_W$ with the $S-T$ oblique parameters in the previous section is consistent with the current uncertainties, which are found to be of the same order. More Higgs precision data are still missing, although more narrow bounds are expected with the LHC run-II data analysis.


	\chapter{Summary and conclusions} \label{ch:conclusions}
	
	The Standard Model (SM) of particle physics is proved to be the best framework for the description of subatomic physics. It has meant the discovery of a whole new layer of nature, which was culminated with the Higgs announcement at the Large Hadron Collider (LHC) some years ago. So far, there is not experimental evidence of any new heavy state nor any significant deviation from its predictions (neutrino masses can be easily embedded in the model) and many fashionable UV models have been discarded or postponed to higher energies. However, it is also well known that the SM is not the definitive particle physics theory. There are still some fundamental matters and questions that are not explained or cannot be answered within this model.
	
	In this thesis, we seek new physics using indirect methods and from a model-independent approach, since direct searches are unfruitful at the moment. As a consequence, bottom-up effective field theories (EFTs) are the best strategy for this purpose and their usage is fundamental for the elaboration of this work. Their main advantage is undoubtedly the fact that we do not need to make any assumption about the underlying UV theory, so the final conclusions must be valid for every model satisfying the EFT premises. Hence, staying as general as possible is indeed a priority in this work and we will accomplish this objective as long as we do not make any unnecessary assumption. 
	
	We find another big motivation for this work in the analogy with chiral perturbation theory (\chpt) and the resonance chiral theory (R$\chi$T). Whereas quantum chromodynamics (QCD) is proved to be the theory of the strong interactions, its often cumbersome computation at low momenta makes the previous chiral theories the best tool and the more appropriate framework to deal with strong physics within this energy range. Furthermore, QCD is not even required for their formulation. Therefore, mesons can be considered as the actual variables that matters, instead of the more fundamental quarks and gluons. We wonder if a similar scenario can also happen in the context of electroweak physics. In this situation, the SM plays the role of \chpt\ as the low-energy theory, whereas the electroweak resonances are analogous to R$\chi$T. Certainly there is an underlying UV theory that explains the unexplored physics, as QCD does in the comparison, but this is unimportant for the purpose of this thesis and the conclusions and the analysis of the hypothetical high-energy states do not rely on it.	
		
	Being the EFT essence present along all the thesis, our starting point is, indeed, the construction of the electroweak effective theory (EWET). It is formulated as the most general EFT containing all the SM (light) degrees of freedom and also its symmetries, which accounts for both strong and electroweak interactions. Therefore, quarks and leptons are included in the formalism, although only one generation is considered. Gluons are invariant under the color group, $SU(3)_C$, while the electroweak left and right gauge bosons transform under the extended group $SU(2)_L\otimes SU(2)_R \otimes U(1)_X$, where $L$ and $R$ stand for their respective chiralities and $X\equiv (B-L)/2$ for the baryon minus lepton number. In addition, there is a Higgs-like $m_h=125$ GeV scalar boson, singlet under the previous groups. It should be mentioned that this field within the EWET is not embedded in a complex doublet scalar field as the SM Higgs is, since we aim not to make any dispensable assumption and  to be as model-independent as possible. 	
	Actually, the only assumption we make is the precise pattern of symmetry breaking, which is consistent with the current phenomenology too. The electroweak group is spontaneously broken, $SU(2)_L\otimes SU(2)_R \rightarrow SU(2)_{L+R}$, leading to three Goldstone bosons that account for the longitudinal polarization and give raise to the masses of the gauge bosons. 

	The EWET is organized according to the chiral power counting, which is based on the momenta contributions at low-energies. Thus, it is possible to assign to each operator and, particularly, to each building block (the pieces to form operators) a given chiral dimension. Finally, it establishes a hierarchy among all the interactions at the Lagrangian level and renormalization is guaranteed order by order. A particular feature of the power counting is that neither the Higgs nor the Goldstone bosons increase the chiral dimension of a diagram. This fact implies, first, that we do not assume that the Higgs is weakly coupled to the SM and, second, that the EWET is non-linear in the Goldstone bosons. For this thesis, we only consider operators that remain invariant under the $CP$ discrete symmetry, since $CP$ violating processes belong to higher orders and are found to be phenomenologically suppressed,  and we only require the EWET leading order (LO) and next-to-leading order (NLO) Lagrangians, which correspond, respectively, to chiral dimension $\hat d=2$ and $\hat d=4$. 
	
	The LO interactions, shown in eq.~(\ref{eq:LscalarEWET}) and eq.~(\ref{eq:L_SMinL_EWET}), consist in the SM Lagrangian with an extended scalar sector. Otherwise, the NLO Lagrangian, referred in eq.~(\ref{eq:L4EWET}), is formed by 12 $P$-even (3 $P$-odd) purely bosonic operators, 15 (6) two-fermion structures and 25 (7) four-fermion operators, collected respectively in tables \ref{tab:bosonic-Op4}, \ref{tab:2fermion} and \ref{tab:4fermion}. The precise number of structures is highly non-trivial: many redefinitions and simplification techniques have been required in order to minimize this basis. Particularly, we prove that all chiral dimension $\hat d=3$ operators can be reabsorbed or re-expressed in terms of other ones. Furthermore, we make an explicit conversion of the EWET basis into other bases also existing in the literature 	\cite{Buchalla:2012qq,Buchalla:2013rka,Krause:2016uhw, Longhitano:1980iz, Longhitano:1980tm}.
	%
	%

	All the operators are parametrized in terms of their corresponding Wilson coefficients, so-called low-energy coupling constants (LECs). 
	These parameters are precisely the best indirect indicators of the presence of heavy states at low-energies. If any LEC shows a significant deviation (more than five standard deviations) from the SM predicted value (zero in the case of the $\mO_{\hat d}(p^4)$ LECs), the existence of new physics will be claimed. There are a lot of works that have analyzed the pattern of LECs for particular models in the linear framework, where the Higgs boson is introduced through a complex doublet scalar field and the high-energy states are weakly coupled in most of the cases. Some examples of these models can be found at tree-level \cite{Brehmer:2015rna, deBlas:2014mba, delAguila:2010mx, delAguila:2008pw, delAguila:2000rc, Corbett:2015lfa, Bar-Shalom:2014taa, Pappadopulo:2014qza} or even at the one-loop level 
	\cite{HLM:16, FRW:15, DEQ:15, Drozd:2015rsp, Huo:15, Huo:15b, dAKS:16, Boggia:2016asg, Fuentes-Martin:2016uol, BS:94}, perturbatively expanded in some small coupling. In this work, we analyzed the contributions to the LECs in the non-linear framework, which does not make any assumption about the way the Higgs and the new physics couple to the SM \cite{Fingerprints, Colorful, Lowenergy, Buchalla:2016bse, pseudovector-Cata}.

	With an eye on \chpt\ and the resonance chiral theory (R$\chi$T) of mesons, we construct the most general and model-independent theory of electroweak resonances, so-called resonance chiral theory, as the tool for searching new massive objects in the few TeV range. We denote as resonance to any high-energy object associated to the electroweak interaction and not predicted by the SM. The resonance chiral theory includes spin-0 and spin-1 bosonic high-energy states with quantum numbers $J^{PC}=0^{++},\,0^{-+},\,1^{--},\,1^{++}$ and spin-(1/2) fermionic heavy resonances. All of them can be found forming singlets or n-plets under the $SU(2)_{L+R}$ custodial and $SU(3)_C$ color groups. Although the term resonance remind us an underlying strongly coupled theory, it can also be applied to weakly coupled models, as well. Indeed, the conclusions of this work are valid for every model containing resonance states. In general, the resonance interaction Lagrangian is formed by one massive state combined with light fields from the EWET, whereas the interaction of more than one resonance is not considered in this work. Only resonance interactions that leave their low-energy imprints in the LO or NLO EWET Lagrangians are analyzed in this work. As well as in the EWET case, a big effort has been done in order to get rid of all the dependencies among all the resonance interactions and simplify the high-energy Lagrangian as much as possible.
	
	The methodology for obtaining the low-energy contributions of the resonances to the EWET Lagrangian is very well known from the resonance chiral theory studies \cite{Ecker:1988te, Ecker:1989yg, Cirigliano:2006hb}. Formally, it consist in integrating out the massive resonance states from the high-energy action. We obtain the solutions of the resonance equations of motion (EoM) to a first approximation and we substitute them in the (high-energy) resonance Lagrangian. The outcome, expressed in terms of the resonance couplings, is projected into the EWET LO and NLO (low-energy) Lagrangian. The generic form of any resonance contribution to the LECs is 
	\be
	\Delta \mF_i \, \sim \Frac{1}{M_R^2} \times g_1 \times g_2 \,,
	\ee
	where $M_R$ is the mass of the given heavy resonance and $g_1$ and $g_2$ are two resonance couplings from the high-energy Lagrangian that parametrize some particular interaction. The complete set of these structures configures the resonance imprints in the LECs,\ie the signature of the heavy fields in the low-energy theory. The whole contributions to the purely bosonic, two-fermion and four-fermion LECs is found in tables \ref{tab:LECbosonic}, \ref{tab:LEC2fermion} and \ref{tab:LEC4fermionlepton}--\ref{tab:LEC4fermionquark}, respectively.
	
	%
	%
	
	We make special emphasis in the analysis of the spin-1 heavy states (vector and axial-vector). Contrary to the rest of massive resonances, these ones can be implemented in the model with different field representations. In this thesis, we focus on the Proca vector and the rank-2 Antisymmetric tensor formalisms, which are also the most used in the literature. The utilization of one or other representation in the description of spin-1 heavy states has been controversial during the last decades since they naively lead to very different low-energy predictions. 
	However, the selection of the representation is unphysical and it should not alter physics. Likewise, we observe the same problem for all the tree-level exchanges of spin-1 massive states: the low-energy predictions from the resonance Lagrangians depend heavily on the employed formalism. 
	
	We prove, nonetheless, that both spin-1 field representation are equivalent (as expected), by means of a change of variables in the general functional at the path integral level. We show that the confusion arises from the false premise that both resonance interaction Lagrangians should produce the same results. Instead, the equivalence occurs once considered the whole EFT description,\ie
	\be \label{eq:equivalenceconclusions}
	\mL_R^{(P)} \, + \, \mL_{{\rm non}-R}^{(P)} \, = \, \mL_R^{(A)} \, + \, \mL_{{\rm non}-R}^{(A)} \,
	\ee
	where $P$ and $A$ stand for Proca and Antisymmetric; and $\mL_R$ and $\mL_{\text{non}-R}$, for the resonance and non-resonant interactions of the high-energy theory.
	As well as the resonance Lagrangians are different in both representations, short-distance non-resonant operators carry different Wilson coefficients, since they do not belong to the same EFTs. As a consequence, we determine the explicit algebraic identities for how both Lagrangians in eq.~(\ref{eq:equivalenceconclusions}) and their couplings are related. 
	%
	%
	
	Nonetheless, the resonance theory at high-energy is not completely fixed yet. The previous results only state how local and non-local Proca operators transform into Antisymmetric structures, and vice versa. Therefore, it is a priori unknown which are the required light-field contributions involved in a given resonance process at short-distances. For this purpose, short-distance constraints are required. They consist in a set of high-energy requirements so that the resonance theory is well-behaved in the UV, like conservation of unitarity, form factors vanishing at large energies or correlator properties. We impose these necessary conditions over a definite set of Green functions in order to fix the Wilson coefficients of the local non-resonant Lagrangian.

	Indeed, we prove explicitly in this work that the spin-1 Proca and Antisymmetric formalisms lead to the same results at low-energy. Moreover, we show which of these representations is more convenient in order to study a given physical process involving spin-1 massive states. Actually, a remarkable feature of these representations is the fact that they are complementary. This means that whenever the resonance exchange of one of them provides non-vanishing contributions to a given order in the EWET, the other one is exactly zero, and vice versa. Yet, local non-resonant interactions compensate this behavior, although their computation is quite non-trivial. We also find that the Antisymmetric formalism is more suitable when a resonance interaction involves light fields carrying two Lorentz indices
	; whereas Proca resonances are more favorable when dealing with single-index ones. In fact, the underlying reason is found to be purely kinematic.

	Finally, we perform a phenomenological analysis of the resonance theory, restricted to the $P$-even purely bosonic sector (and not including custodial breaking operators). The reduced high-energy Lagrangian is collected in eq.~(\ref{eq:Lpheno}) and it contains 8 independent resonance couplings ($\kappa_W$, $c_d$, $\lambda_{hS_1}$, $d_P$, $F_V$, $G_V$, $F_A$ and $\Lambda_1^{hA}$) related to 4 distinct massive states ($S^1_1$, $P^1_3$, $V^1_{3\,\mu\nu}$ and $A^1_{3\,\mu\nu}$). Therefore, together with the resonance masses, there are a total of 12 free parameters, which yield a moderately wide parameter space yet. In order to reduce the number of independent variables, we incorporate to the reduced resonance theory some quite reasonable short-distance constraints. Indeed, we impose the vector and axial-vector form factors to fall properly at large energies and we also incorporate two super convergence sum rules, so-called Weinberg sum rules (WSRs). Unlike the relations introduced to fix the local non-resonant contributions to the LECs, the first and the second (more restrictive) WSRs require some soft UV conditions to be fulfilled, although they are valid for almost all the models analyzed in the literature, nonetheless. The predictions for the LECs associated to this restricted resonance Lagrangian are collected in table \ref{tab:pheno}. All the results are expressed in terms of 7 independent resonance couplings or masses, once applied the SD constraints. Furthermore, the picture turns even better when only the vector and axial-vector resonances are considered in the framework, since all the low-energy effects of the massive states can be parametrized in terms of their masses alone, $M_{V^1_3}$ and $M_{A^1_3}$, respectively.
	
	We contrast the predictions from the resonance theory with experimental data from the $S$ and $T$ oblique parameters and the current available information for the LECs. The oblique parameters are a set of observables that account for the variations of new physics with respect to the SM predictions in regard to electroweak radiative corrections. We study these physical quantities for the previous phenomenological Lagrangian up to NLO, being sensitive to the spin-1 massive states only. For this computation, we include the last short-distance constraints, making an explicit distinction on the application of the first and the second WSRs \cite{Pich:2013fea}. 
	In this last scenario, when both sum rules are satisfied, we find at 68\% C.L (95\% C.L.)
	\be \label{eq:phenomassesconclusions}
	M_{A^1_3} \gsim M_{V^1_3} \,>\, 5\, (4)\,\mbox{TeV}\,, \qquad\qquad \kappa_W = \Frac{ M_{V^1_3}}{M_{A^1_3}} \,\in\, [0.97\, (0.94)\,,\, 1]\,.
	\ee
	where $\kappa_W$, the Higgs-Goldstone coupling, is also fixed due to the application of the WSRs. Next to the resonance masses, we perform a similar $S$-$T$ parameter computation for the LECs implicated in the purely bosonic high-energy Lagrangian. The resonances impose quite strong limits over these couplings, of order $|\mF_i| \lsim 10^{-3}$ (95\% C.L.; with $i=1,3,4,6,9$),  as it is shown in eq.~(\ref{eq:LECsbounds}) and figure \ref{fig:LECs1}. 
	
	Nevertheless, the current bounds from collider data, concretely LHC run-I and LEP, reveal much less constrained LECs, specially Higgs related couplings, where precision data is still lacking. We show in eqs.~(\ref{eq:phenoF1b}}, \ref{eq:LECspheno31}, \ref{eq:LECspheno32}) the present experimental uncertainties for the studied LECs. All their bounds are found to be of order $\mF_i \sim 10^{-1}$, except for the case of $\mF_1\sim 10^{-3}$. However, it is not possible to discern the presence of any massive state from any of these low-energy couplings. Nowadays, new experimental data from LHC is being analyzed and important improvements are expected, particularly for the triplet and quartic anomalous gauge couplings. Incoming future $e^+e^-$ colliders will improve the precision on $\mF_1$ too, up to the reach of the resonances in the TeV range. Resonance interactions with the Higgs field are still far from the experimental access.
	
	There are still some upgrades to be addressed in the resonance theory for future works. It is necessary to promote both the EWET and the resonance Lagrangian to a three fermion generation theory. This highly non-trivial step will require an intense algebraic analysis in order to set the operator basis for both the low and the high-energy theories. This improvement will allow to properly analyze the phenomenology in the fermionic sector, which is still lacking. For this purpose, new flavor constrains must be considered. Hence, a global phenomenological study of the LECs is still left to be done.  Additionally, ongoing LHC run-II data will be incorporated to our analysis in order to update the current bounds. Certainly, it would be also interesting to translate our general language for electroweak resonances to already existing models that include high-energy states.


\chapter*{Acknowledgements}	
	
I would like to thank my thesis supervisor Antonio Pich for his constant support, his patience with me and surely for receiving a little piece of his physics knowledge. 
I am also very grateful to Claudius for all the discussions and his very helpful comments about this work. Together with them, I would like to thank my other collaborators Natxo and Juanjo too. Within this group, I have grown up professionally and I am very proud of all of our works, still ongoing. 

For sure, the personal success that this thesis means to me would have been impossible without the ceaseless presence of my parents, who have been always by my side. As well, thanks a lot to Blanca, my family and my friends, who have taken care of me no matter what.
Last but not least, I am very thankful with the opportunity that Fundaci\'o La Caixa has given to me during these years, being able to develop my professional career in whatever field I desired, unconditionally.

Work supported by the Spanish Government and ERDF funds from the European Commission (FPA2014-53631-C2-1-P), by the Spanish Centro de Excelencia Severo Ochoa Programme (SEV-2014-0398)
and La Caixa (Ph.D. grant for Spanish universities).


	\begin{appendices}
		
		\chapter{Relations and properties}
		
		\section{Lie algebra} \label{app:liealgebra}
		A Lie algebra is the algebra associated to a particular Lie group with a well-defined inner product, so-called the Lie bracket. These groups are specially important in the context of particle physics because they properly describe symmetries of the analytic structures in the theory. 
		
		Therefore, the SM or the EWET are invariant under symmetries that are mathematically realized through Lie groups, being represented in eq.~(\ref{eq:SMssb}) and eq.~(\ref{eq:ewetSymmetry}), respectively. In both cases, the full symmetry groups are compound by external products of the elementary Lie groups $U(1)$, $SU(2)$ and $SU(3)$. 
		
		The $U(1)$ is the most trivial Lie group and it is formed by all the complex numbers with absolute value 1 and thus its elements can be expressed as ${\rm e}^{i\, \theta}$ with $\theta \in [0, 2\pi[$. As it seems clear, the group has dimension one with only one generator, so it is an abelian group. Electromagnetism is the most famous physical interaction which transforms under this symmetry.
		
		On the contrary, the $SU(2)$ and $SU(3)$ Lie groups are non-abelian and bring three and eight independent generators, respectively. Actually, the dimension of a $SU(n)$ Lie group is $n^2-1$, and yet their elements can be displayed as $n\times n$ traceless matrices. Furthermore, the Lie bracket works like the matrix commutator for these group.
		
		In the $SU(2)$ case, the infinitesimal group generators can be written in terms of the so-called Pauli matrices, $T^i_{SU(2)} = \sigma_i/2$, with 
		\be \label{eq:sigmamatrices}
		\sigma_1\,=\, \left( \begin{matrix} 0 & 1 \\ 1 & 0 \end{matrix} \right)\,, \qquad
		\sigma_2\,=\, \left( \begin{matrix} 0 & -i \\ i & 0 \end{matrix} \right)\,, \qquad
		\sigma_3\,=\, \left( \begin{matrix} 1 & 0 \\ 0 & -1 \end{matrix} \right)\,.
		\ee
		satisfying the following relations
		\be \label{eq:commutatorSU2}
		[\sigma_i\,,\,\sigma_j] \,=\, 2i\,\epsilon_{ijk}\, \sigma_k \,, \qquad
		\sigma_i\,\sigma_j \,=\, \delta_{ij} I\,+\, i\, \epsilon_{ijk}\, \sigma_k\,,
		\ee
		where $\epsilon_{ijk}$ is the Levi-Civita symbol.
		
		Analogously, the $SU(3)$ generators, $T^a_{SU(3)} = \lambda_a/2$, are described through the Gell-Mann matrices
		\begin{align}
		&\lambda_1\,=\, \left( \begin{matrix}	 
		0 & 1 & 0 \\ 
		1 & 0 & 0 \\
		0 & 0 & 0 
		\end{matrix} \right)\,, & \qquad
		&\lambda_2\,=\, \left( \begin{matrix}	 
		0 & -i & 0 \\ 
		i & 0 & 0 \\
		0 & 0 & 0 
		\end{matrix} \right)\,,& \qquad
		&\lambda_2\,=\, \left( \begin{matrix}	 
		1 & 0 & 0 \\ 
		0 & -1 & 0 \\
		0 & 0 & 0 
		\end{matrix} \right)\,,& \nn\\[1ex]
		&\lambda_4\,=\, \left( \begin{matrix}	 
		0 & 0 & 1 \\ 
		0 & 0 & 0 \\
		1 & 0 & 0 
		\end{matrix} \right)\,, & \qquad
		&\lambda_5\,=\, \left( \begin{matrix}	 
		0 & 0 & -i \\ 
		0 & 0 & 0 \\
		i & 0 & 0 
		\end{matrix} \right)\,,& \qquad
		&\lambda_6\,=\, \left( \begin{matrix}	 
		0 & 0 & 0 \\ 
		0 & 0 & 1 \\
		0 & 1 & 0 
		\end{matrix} \right)\,,& \nn\\[1ex]
		&\lambda_7\,=\, \left( \begin{matrix}	 
		0 & 0 & 0 \\ 
		0 & 0 & -i \\
		0 & i & 0 
		\end{matrix} \right)\,,& \qquad
		&\lambda_8\,=\, \Frac{1}{\sqrt{3}} \left( \begin{matrix}	 
		1 & 0 & 0 \\ 
		0 & 1 & 0 \\
		0 & 0 & -2 
		\end{matrix} \right)\,, & 
		&	&
		\end{align}
		and the commutator can be expressed in terms of the fully antisymmetric structure constants $f_{abc}$, 
		\be \label{eq:commutatorSU3}
		[\lambda_a\,,\,\lambda_b] \,=\, 2i\,f_{abc}\, \lambda_c \,,
		\ee
		with $f_{123}=1$, $f_{147}=f_{165}=f_{246}=f_{257}=f_{345}=f_{346}=1/2$ and $f_{458}=f_{678}=\sqrt{3}/2$. In addition, we include some well-known $SU(3)$ relations for the group generators that appear to be very useful 
		\bear \label{eq:relationsSU3}
		Tr[T^a\,T^b] = T_R\,\delta^{ab}\,, &\qquad & \sum_{c,d} f^{acd}\,f^{bcd}= C_A\,\delta^{ab} \, , \nn\\
		\\
		\sum_a T^a_{ij}\,T^a_{jk} = C_F\,\delta_{ik} \,, &\qquad & T^a_{ij} T^a_{kl} = T_R\,(\delta_{jk}\,\delta_{il} - \frac{1}{N_C}\,\delta_{ij}\,\delta_{kl})\,, \label{eq:sun}
		\eear
		with $T_R=\frac{1}{2}$, $C_F=\frac{4}{3}$, $C_A=N_C=3$.

		\section{Transformation properties under discrete symmetries} \label{app:CPtransformations}
		\markboth{APPENDIX A.\hspace{1mm} RELATIONS AND PROPERTIES}{A.2.\hspace{1mm} TRANSFORMATION PROPERTIES}
		
		Operator building and classification requires to know how their constituent elements and building blocks transform under discrete symmetries ($C$, $P$, and $CP$) and hermitian conjugation. The following tables show the results of this transformations for all the ingredients conforming both for the EWET and the resonance theory. Table \ref{tab:CPGoldstoneDirac} reflects how quantum fields and algebraic basic terms (like Dirac algebra matrices) shift when applying these symmetries. This leads to the building blocks of the EWET listed in eq.~(\ref{eq:buildingblocks}) and eq.~(\ref{eq:buildingblocksfermion}) shown in table \ref{tab:CPbuildingblocks}. As already stated, the precise combination of these latter terms is essential for the construction of the EWET Lagrangian. Finally, table \ref{tab:CPresonances} reveals the transformation properties for the massive states that conform the resonance theory.

		\begin{table}[ht]
			\begin{center}
				\begin{tabular}{ |c||c|c|c|c| }
					\multicolumn{5}{c}{} \\
					\hline
					\rule{0pt}{3ex}
					Field & $P$    &     $C$ & $CP$ & h.c.
					\\[5pt] \hline
					\rule{0pt}{3ex}
					$U$ & $U^\dagger$  & $U^t$ & $U^*$ & $U^\dagger$
					\\[5pt] \hline
					\rule{0pt}{3ex}
					$u$ & $u^\dagger$  & $u^t$ & $u^*$ & $u^\dagger$
					\\[5pt] \hline
					\rule{0pt}{3ex}
					$\partial^\mu \cdot$ & $\partial_\mu \cdot$ & $ (\partial^{\mu} \cdot)^t$ & $ (\partial_\mu\cdot)^t$ & $\partial^{\mu} \cdot$
					\\[5pt] \hline 
					\multicolumn{5}{c}{} \\
					\hline 
					\rule{0pt}{3ex}
					Dirac & $P$    &  $C$  & $CP$  & h.c. 
					\\[3pt]
					\rule{0pt}{2ex}
					structure, $\Gamma$ &  ( $\gamma^0\Gamma \gamma^0$ )   &     ( $-\gamma^0\gamma^2\Gamma^t \gamma^2\gamma^0$ ) &
					( $-\gamma^2\Gamma^t \gamma^2$ )  & ( $\gamma^0\Gamma^\dagger \gamma^0$ )
					\\[5pt] \hline
					\rule{0pt}{3ex}
					1 & 1  & 1 & 1 & 1
					\\[5pt] \hline
					\rule{0pt}{3ex}
					$i\gamma_5$  & $- i\gamma_5$ &  $i\gamma_5$ & $-i\gamma_5$ & $i\gamma_5$
					\\[5pt] \hline
					\rule{0pt}{3ex}
					$\gamma^{\mu}$ & $\gamma_{\mu}$ &  $-\gamma^{\mu }$ & $-\gamma_\mu $ & $\gamma^{\mu}$
					\\[5pt] \hline
					\rule{0pt}{3ex}
					$\gamma^{\mu}\gamma_5$ & $- \gamma_{\mu }\gamma_5$ &  $\gamma^{\mu}\gamma_5$ & $-\gamma_{\mu}\gamma_5$
					& $\gamma^{\mu}\gamma_5$
					\\[5pt] \hline
					\rule{0pt}{3ex}
					$\sigma^{\mu\nu}$ & $\sigma_{\mu\nu }$ &  $-\sigma^{\mu\nu}$ & $-\sigma_{\mu\nu} $ & $\sigma^{\mu\nu}$
					\\[5pt] \hline
				\end{tabular}
				\caption{\small Transformation properties for quantum fields (upper block) and Dirac algebra (lower block) with definite $C$, $P$, $CP$ and h.c. symmetry transformation. The superindex $t$ denotes matrix transposition.
				}
				\label{tab:CPGoldstoneDirac}
			\end{center}
		\end{table}

		\begin{table}[ht]
			\begin{center}
				\begin{tabular}{ |c||c|c|c|c| }
					\hline
					\rule{0pt}{4ex}
					$\begin{matrix} \mbox{Bosonic} \\ \mbox{building blocks}\end{matrix}$ & $P$    &  $C$ & $CP$ & h.c.
					\\[5pt] \hline\hline
					\rule{0pt}{3ex}
					$u^\mu$ & $- u_\mu$ &  $u^{\mu\, t}$ & $-u_{\mu}^{t}$ & $u^\mu$
					\\[5pt] \hline
					\rule{0pt}{3ex}
					$f_\pm^{\mu\nu} $ & $\pm f_{\pm\, \mu\nu}$ &  $ \mp f_{\pm}^{\mu\nu\,  t} $ &   $- f_{\pm\, \mu\nu}^t $  & $f_\pm^{\mu\nu}$
					\\[5pt] \hline 
					\rule{0pt}{4ex}
					$\displaystyle \frac{\partial^\mu h}{v}$ & $\displaystyle \frac{\partial_\mu h}{v}$ & $\displaystyle \frac{(\partial^\mu h)^t}{v}$ & $\displaystyle \frac{(\partial_\mu h)^t}{v}$ & $\displaystyle \frac{\partial^\mu h}{v}$
					\\[5pt] \hline
					\rule{0pt}{3ex}
					$ \hat X^{\mu\nu}$ & $\hat X_{\mu\nu}$ & $\hat X^{\mu\nu\, t}$ & $\hat X^t_{\mu\nu}$ & $ \hat X^{\mu\nu} $
					\\[5pt] \hline
					\rule{0pt}{3ex}
					$ \hat G^{\mu\nu}$ & $\hat G_{\mu\nu}$ & $\hat G^{\mu\nu\, t}$ & $\hat G^t_{\mu\nu}$ & $ \hat G^{\mu\nu} $
					\\[5pt] \hline
					\rule{0pt}{3ex}
					$\mT$ & $\mT$ &	$\mT^t$ &	$\mT^t$ &	$\mT$ 
					\\[5pt] \hline
					\multicolumn{5}{c}{} \\
					\hline
					\rule{0pt}{3ex}
					$\begin{matrix} \mbox{Fermionic} \\ \mbox{building blocks}\end{matrix}$& $P$    &  $C$ & $CP$ & h.c.
					\\[5pt] \hline
					\hline
					\rule{0pt}{3ex}
					$\mY$ & $\mY$ & $\mY^t$ & $\mY^t$ & $\mY$ 
					\\[5pt] \hline
					\rule{0pt}{3ex}
					$J_S$ & $J_S$  & $J_S^t$ & $J_S^t$ & $J_S$
					\\[5pt] \hline
					\rule{0pt}{3ex}
					$J_P$ & $- J_P$ &  $J_P^t$ & $-J_P^t$ & $J_P$
					\\[5pt] \hline
					\rule{0pt}{3ex}
					$J_V^{\mu}$ & $J_{V\, \mu}$ &  $-J_V^{\mu \, t}$ & $-J_{V\mu}^{t}$ & $J_V^{\mu}$
					\\[5pt] \hline
					\rule{0pt}{3ex}
					$J_A^{\mu}$ & $- J_{A\, \mu }$ &  $J_A^{\mu \, t}$ & $-J_{A\, \mu}^{t}$ & $J_A^{\mu}$
					\\[5pt] \hline
					\rule{0pt}{3ex}
					$J_T^{\mu\nu}$ & $J_{T\, \mu\nu }$ &  $-J_T^{\mu\nu \, t}$ & $-J_{T\, \mu\nu}^{t}$ & $J_T^{\mu\nu}$
					\\[5pt] \hline
				\end{tabular}
				\caption{\small
					Transformation properties under discrete symmetries ($C$, $P$, $CP$) and hermitian conjugation for the building blocks of the EWET in eqs.~(\ref{eq:buildingblocks}, \ref{eq:buildingblocksfermion}). The same transformations apply for the fermion bilinears independently of their type: lepton ($J_\Gamma^l$), quark singlet ($J_\Gamma^q$) or quark octet bilinear ($J_\Gamma^8$). Fermion bilinear definitions are set in eqs.~(\ref{eq:bilinearnotation}, \ref{eq:bilinearcolortype}).
				}
				\label{tab:CPbuildingblocks}
			\end{center}
		\end{table}

		\begin{table}[!t]
			\begin{center}
				\begin{tabular}{ |c||c|c|c|c| }
					\hline
					\rule{0pt}{3ex}
					Resonance fields & $P$    & $C$ & $CP$ & h.c.
					\\[5pt] \hline\hline
					\rule{0pt}{3ex}
					$S$ & $S$  & $S^t$ & $S^t$ & $S$
					\\[5pt] \hline
					\rule{0pt}{3ex}
					$P$ & $- P$ &  $P^t$ & $-P^t$ & $P$
					\\[5pt] \hline
					\rule{0pt}{3ex}
					$V^{\mu\nu}$ & $V_{\mu\nu}$ &  $-V^{\mu\nu\, t}$ & $-V_{\mu\nu}^{t}$ & $V^{\mu\nu}$
					\\[5pt] \hline
					\rule{0pt}{3ex}
					$A^{\mu\nu}$ & $- A_{\mu\nu}$ &  $A^{\mu\nu\, t}$ & $-A_{\mu\nu}^{t}$ & $A^{\mu\nu}$
					\\[5pt] \hline
					\multicolumn{5}{c}{}
					\\[5pt] \hline
					\rule{0pt}{4ex}
					$\begin{matrix} \mbox{Fermionic} \\ \mbox{resonance bilinears}\end{matrix}$ & $P$    & $C$ & $CP$ & h.c.
					\\[5pt] \hline\hline
					\rule{0pt}{3ex}
					$\mmJ_S$ & $\mmJ_S$  & $\mmJ_S^t$ & $\mmJ_S^t$ & $\mmJ_S$
					\\[5pt] \hline
					\rule{0pt}{3ex}
					$\mmJ_P$ & $-\mmJ_P$ &  $\mmJ_P^t$ & $-\mmJ_P^t$ & $\mmJ_P$
					\\[5pt] \hline
					\rule{0pt}{3ex}
					$\mmJ_V^{\mu}$ & $\mmJ_{V\mu}$ &  $-\mmJ_V^{\mu\,t}$ & $-\mmJ_{V\mu}^{t}$ & $\mmJ_V^{\mu}$
					\\[5pt] \hline
					\rule{0pt}{3ex}
					$\mmJ_A^{\mu}$ & $-\mmJ_{A\mu}$ &  $\mmJ_A^{\mu\,t}$ & $-\mmJ_{A\mu}^{t}$ & $\mmJ_A^{\mu}$
					\\[5pt] \hline
					\rule{0pt}{3ex}
					$\mmJ_T^{\mu\nu}$ & $\mmJ_{T\,\mu\nu}$ &  $-\mmJ_T^{\mu\nu\,t}$ & $-\mmJ_{T\,\mu\nu}^t$ & $\mmJ_T^{\mu\nu}$
					\\[5pt] \hline
					\rule{0pt}{3ex}
					$\mmJ_{S-}$ & $\mmJ_{S-}$  & $-\mmJ_{S-}^t$ & $-\mmJ_{S-}^t$ & $\mmJ_{S-}$
					\\[5pt] \hline
					\rule{0pt}{3ex}
					$\mmJ_{P-}$ & $-\mmJ_{P-}$ &  $-\mmJ_{P-}^t$ & $\mmJ_{P-}^t$ & $\mmJ_{P-}$
					\\[5pt] \hline
					\rule{0pt}{3ex}
					$\mmJ_{V-}^{\mu}$ & $\mmJ_{V-\,\mu}$ &  $\mmJ_{V-}^{\mu\,t}$ & $\mmJ_{V-\,\mu}^{t}$ & $\mmJ_{V-}^{\mu}$
					\\[5pt] \hline
					\rule{0pt}{3ex}
					$\mmJ_{A-}^{\mu}$ & $-\mmJ_{A-\,\mu}$ &  $-\mmJ_{A-}^{\mu\,t}$ & $\mmJ_{A-\, \mu}^{t}$ & $\mmJ_{A-}^{\mu}$
					\\[5pt] \hline
					\rule{0pt}{3ex}
					$\mmJ_{T-}^{\mu\nu}$ & $\mmJ_{T-\,\mu\nu}$ &  $\mmJ_{T-}^{\mu\nu\,t}$ & $\mmJ_{T-\,\mu\nu}^t$ & $\mmJ_{T-}^{\mu\nu}$
					\\[5pt] \hline
				\end{tabular}
				\caption{\small
					Transformation properties for resonance heavy fields and resonance fermionic bilinears under discrete symmetries ($C$, $P$ and $CP$) and hermitian conjugarion. The definition of heavy fermionic bilinears is set in eq.~(\ref{eq:bilinearnotationR}).
				}
				\label{tab:CPresonances}
			\end{center}
		\end{table}

		
		\chapter{Lagrangian simplification}\label{app:simp}
		
		The number of operators which we can construct is actually higher than the one listed 
		in chapters \ref{ch:ewet} and \ref{ch:resonancetheory}, either for the EWET and the resonance Lagrangians. However the operator sets are not incomplete since they constitute a basis respectively. This is because they are related among themselves and they are sometimes redundant and, therefore, removed from the lists, many of them by making use of field redefinitions, partial integration, the classical EoMs and some algebraic identities \cite{Ecker:1988te,Bijnens:1999sh,odd-Op6-chpt}. Unfortunately, not all the simplifications are trivial and it is very complicated to discern some Lagrangian to be eventually not redundant. In the following, we illustrate some of these processes and also collect some of the not so obvious mechanisms, tricks and simplifications for the EWET and the resonance theory.

		\section{Basic relations and algebraic identities}\label{app:simprelations}
		In this section we collect some well-known algebraic relations and field properties that we used in order to eliminate all the possible operator redundancies in the EWET.
		
		Firstly, for the Goldstone bosons, the following equations  turn out to be very useful for the determination of the purely bosonic $\mO_{\hat d}(p^4)$ NLO operators \cite{Ecker:1988te,Bijnens:1999sh,odd-Op6-chpt}:
		\begin{gather}
		\nabla^\nu u^\mu - \nabla^\mu u^\nu \; =\; f_-^{\mu \nu} \, ,
		\nn\\
		\left[ \nabla_\mu ,\nabla_\nu  \right] X \; =\;  [\Gamma_{\mu\nu} , X]\, ,
		\qquad\qquad
		\Gamma_{\mu\nu} \; =\; \frac{1}{4} [u_\mu ,u_\nu ] - \frac{i}{2} f_{+\, \mu\nu}\, ,
		\nn\\
		\nabla_\rho \nabla_\mu u^\rho \; =\; \nabla_\mu (\nabla_\rho u^\rho) \,
		+\, [\Gamma_{\rho \mu} , u^\rho ]\, ,
		\nn\\[5pt]
		\nabla^2 u_\mu \; =\; \nabla_\mu (\nabla^\rho u_\rho) \,
		+\, \nabla^\rho f_{-\, \mu\rho} \, +  [\Gamma_{\rho \mu} ,u^\rho]\, .
		\end{gather}
		As a general rule in table \ref{tab:bosonic-Op4}, we have chosen for our basis those operators proportional to the gauge fields, expressed in terms of $f_\pm^{\mu\nu}$.
		
		Other kind of common simplifications arises from the interplay of traces and $SU(2)$ properties, which are present in almost every operator considered so far. Given four different $SU(2)$ objects, $x=a,b,c,d$ with $x=x^j\sigma^j$ and $\sigma^j$ the Pauli matrices (see eq.~(\ref{eq:sigmamatrices})) one obtains
		\bel{eq:SU2trace}
		2\,\bra a b c d \ket_2
		\, =\, \bra a b \ket_2\, \bra c d \ket_2 \, -\,
		\bra a c \ket_2\, \bra b d \ket_2 \, +\,\bra a d \ket_2\, \bra b c \ket_2 \, .
		\ee
		Moreover, we have made use of Cayley-Hamilton relations for $2\times 2$ matrices, like in
		\be \label{eq:CayleyHamilton}
		a^2\, -\, a\, \bra a \ket_2 \, +\, \frac{1}{2} \left( \bra a \ket_2\bra a \ket_2 - \bra a^2 \ket_2 \right)\,=\, 0\, ,
		\ee
		which also implies the following equations for two or three $2\times 2$ matrices within traces
		\bear \label{eq:CayleyHamilton2}
		\left\{ a , b \right\}\, &=&\, a\,\bra b \ket_2\, +\, \bra a \ket_2\, b\, +\, \bra a b \ket\, - \,
		\bra a \ket_2\,\bra b \ket_2\, . \nn\\
		\bra \{a, b\}\, c \ket_2  \,&=&\, \bra a \ket_2 \, \bra b c\ket_2\, +\,
		\bra b \ket_2 \, \bra a c\ket_2\, +\,\bra c \ket_2 \, \bra a b\ket_2 \nn\\
		&& -\, \bra a\ket_2 \, \bra b\ket_2 \, \bra c\ket_2\, .
		\eear
		In addition, if any of the $SU(2)$ elements happen to be traceless, like $u_\mu$ or any custodial triplet resonance, the number of redundant operators is even greater. For instance, if $b$ and $c$ are traceless objects in the second expression in eq.~(\ref{eq:CayleyHamilton2}), it simplifies to
		\bear
		\bra \{a, b\}\, c\ket_2  &=& \bra a \ket \, \bra b c\ket_2 \, .
		\eear

		As a consequence, $\mO_{\hat d}(p^2)$ resonance terms like $\bra S^1_3 u_\mu u^\mu \ket_2$ are identically 0 due to Cayley-Hamilton properties, and thus operators like this are absent\footnote{This operator is, however, present in the Resonance Chiral Theory \cite{Ecker:1988te}, since the underlying symmetry is $U(n)$ instead of $SU(n)$.} in the EWET. These relations can be also be applied to the fermionic operators, like in
		\begin{align}
		\bra S^1_3\, \{ J_A^{f\,\mu} , u_\mu\}\ket_2 &\,=\, \bra S^1_3\, u_\mu \ket_2 \,\, \bra J_A^{f\,\mu}\ket_2\, ,& \quad &(f=l,\,q)\,,
		\nn\\
		\bra S^1_3\, \{ J_V^{f\,\mu} , u_\mu\}\ket_2 &\,=\, \bra S^1_3\, u_\mu \ket_2 \,\, \bra J_V^{f\,\mu}\ket_2\, ,& \quad &(f=l,\,q)\,.
		\end{align}

		Another kind of simplification, so-called Schouten identity \cite{Schouten}, is performed when dealing with Levi-Civita tensors in the odd-intrinsic parity sector:
		\be
		A^\rho\, \epsilon^{\mu\nu\alpha\beta}
		\; =\;  A^\mu\,  \epsilon^{\rho\nu\alpha\beta}
		+ A^\nu\,  \epsilon^{\mu\rho\alpha\beta}
		+ A^\alpha\,  \epsilon^{\mu\nu\rho\beta}
		+ A^\beta\,  \epsilon^{\mu\nu\alpha\rho}  \, .
		\ee
		Finally, Fierz identities are the last type of general procedure to be studied and it is postponed to the next appendix section.

		\section{Fierz identities} \label{app:fierzidentities}
		
		Fierz identities are a set of equations that relate different four-fermion structures among themselves. They represent the main source of simplification for the four-fermion EWET Lagrangian in table \ref{tab:4fermion}. A generic four-fermion structure reads
		\be \label{eq:4fermion}
		(\bar s_k\, \Gamma^a_{kl}\,t_l)\,(\bar u_i\, \Gamma_{b\,ij}\,v_j)\,,
		\ee
		where $\Gamma^a$ and $\Gamma_b$ are any contravariant or covariant $4\times 4$ matrix from the Clifford algebra basis $\Gamma^a=\{1,\,i\gamma_5,\, \gamma^\mu,\, \gamma^\mu\gamma_5,\, \sigma^{\mu\nu}\}$ or $\Gamma_a=\{1,\,-i\gamma_5,\, \gamma_\mu,\, \gamma_5\gamma_\mu,\, \sigma^{\mu\nu}\}$ (with $\mu<\nu$ in $\sigma^{\mu\nu}$), respectively, and $i,\,j,\,k,\,l$ are Dirac indices. Fierz identities yield \cite{Itzykson:1980rh} 
		\be \label{eq:fierzidentities}
		(\bar s\, \Gamma_a\,t)\,(\bar u\, \Gamma^b\,v) \,=\, C^{bc}_{ad}\,(\bar s\, \Gamma_c\,v)\,(\bar u\, \Gamma^d\,t)\,,
		\ee
		being a sum over $c$ and $d$ implicit in the right-hand side, with
		%
		%
		\renewcommand{\arraystretch}{2.1}
		\begin{table}[t]
			\centering
			\begin{tabular}[t]{|l|c|c|c|c|c|}
				\cline{2-6}
				\multicolumn{1}{c|}{} & $SS'$ & $PP'$ & $VV'$ & $AA'$ & $TT'$  \\
				\hline
				$SS=(\bar s t)(\bar u v)$ & $-\displaystyle \frac{1}{4}$ & $\displaystyle\frac{1}{4} $& $-\displaystyle\frac{1}{4}$&$\displaystyle\frac{1}{4}$&$-\displaystyle\frac{1}{8}$ \\
				\hline
				$PP=(\bar s i\gamma_5 t)(\bar u i\gamma_5 v)$ &$\displaystyle\frac{1}{4}$ & $-\displaystyle\frac{1}{4} $& $-\displaystyle\frac{1}{4}$&$\displaystyle\frac{1}{4}$&$\displaystyle\frac{1}{8}$ \\
				\hline
				$VV=(\bar s \gamma^\mu t)(\bar u \gamma^\mu v)$ &$-1$ & $-1 $& $\displaystyle\frac{1}{2}$&$\displaystyle\frac{1}{2}$&$0$ \\
				\hline
				$AA=(\bar s \gamma^\mu\gamma_5 t)(\bar u \gamma^\mu\gamma_5 v)$ &$1$ & $1$& $\displaystyle\frac{1}{2}$&$\displaystyle\frac{1}{2}$&$0$ \\
				\hline
				$TT=(\bar s \sigma^{\mu\nu} t)(\bar u \sigma^{\mu\nu} v)$ & $-3$ & $3 $& $0$&$0$&$\displaystyle\frac{1}{2}$ \\
				\hline 
				\multicolumn{6}{c}{} \\[-2ex]
				\cline{2-5}
				\multicolumn{1}{c|}{}	 & $SP'$ & $PS'$& $VA'$ & $AV'$ \\
				\cline{1-5}
				$VA=(\bar s \gamma^\mu t)(\bar u \gamma^\mu\gamma_5 v)$ & $-1$ & $1$ &$-\displaystyle\frac{1}{2}$ & $-\displaystyle\frac{1}{2} $ \\
				\cline{1-5}
				$AV=(\bar s \gamma^\mu\gamma_5 t)(\bar u \gamma^\mu v)$ & $1$ & $-1$ &$-\displaystyle\frac{1}{2}$ & $-\displaystyle\frac{1}{2} $ \\
				\cline{1-5}
			\end{tabular}
			\caption{Relevant Fierz identities for the four-fermion EWET operators with $\Gamma=(1, i\gamma^{5},\gamma^{\mu}, \gamma^{\mu}\gamma^{5},\sigma^{\mu\nu}=\tfrac{i}{2}[\gamma^{\mu},\gamma^{\nu}])$. A given element of the first column is equivalent to the sum of all the terms in its row multiplied by their corresponding bilinears. Four-fermions carrying $'$ are defined as their analogous partners once exchanged $t$ and $v$.}
			\label{tab:fierzidentities}
		\end{table}
		\renewcommand{\arraystretch}{1}\\
		\be 
		C^{bc}_{ad} = -\Frac{1}{16} \bra \Gamma_a \Gamma_d \Gamma^b \Gamma^c \ket\,,
		\ee
		being the minus sign a result of fermion switching.\footnote{If $s$, $t$, $u$ and $v$ are not fermions but just spinors this sign does not apply.} In particular, the convention employed in this work is slightly different from \cite{Itzykson:1980rh} since the fermion bilinears are always written in a contravariant way.\footnote{There are extra minus signs when considering pseudoscalar and axial vector elements in $C^{ab}_{cd}$.} In table \ref{tab:fierzidentities}, we show the tensor coefficients for only the relations that are used in the present work, although there are many more. The four-fermion operators are expressed in terms of two fermion pairs, denoted as scalar ($S$), pseudoscalar ($P$), vector ($V$), axial-vector ($A$) or tensor ($T$). Notice that it is a generic label and it just specifies the Dirac structure and not the fermion content (as opposite to $J_\Gamma^{(f)}$ where the particular fermions forming the bilinear are known). In the following, we show the scalar-scalar four-fermion Fierz identity as an example of how to interpret table \ref{tab:fierzidentities}:
		\begin{align}
		(\bar s t)(\bar u v) \,&=\, -\frac{1}{4} (\bar s v)(\bar u t) \, + \, \frac{1}{4} (\bar s\,i\gamma_5\, v)(\bar u\,i\gamma_5\, t) \,-\, \frac{1}{4} (\bar s\,\gamma_\mu\, v)(\bar u \,\gamma^\mu \, t) \nn\\
		& -\,\frac{1}{4} (\bar s\,\gamma_\mu\gamma_5\, v)(\bar u \,\gamma^\mu\gamma_5 \, t) \,-\,\frac{1}{8} (\bar s\,\sigma_{\mu\nu}\, v)(\bar u \,\sigma^{\mu\nu} \, t) \,.
		\end{align}

		The configuration of the four-fermion NLO EWET operator basis depends, however, on the fermions being leptons, quarks or mixed types. On the one side, lepton and baryon number conservation forces these structures to incorporate an even number of these fields: four leptons, two leptons and two quarks or four quarks. Hence, there are several possible ways to contract them, and many more when $\xi^{(f)}$ structures are considered, which are $SU(2)_{L+R}$ objects and may be $SU(3)_C$ elements too, in the case of quarks. On the other side, Fierz identities (table \ref{tab:fierzidentities}) and $SU(n)$ identities (eq.~(\ref{eq:commutatorSU2}) and eq.~(\ref{eq:relationsSU3})) set some relations on the different ways of contracting indices. The simplification criteria are 
		\begin{itemize}
			\item Spinors are contracted within the fermion bilinear, yielding $J_\Gamma^l$, $J_\Gamma^q$ or $J_\Gamma^8$ fermionic structures only.
			\item Explicit Gell-Mann matrices (color generators) are avoided.
			\item Whenever possible, $SU(2)_{L+R}$ traces for a  single bilinear are preferred to double bilinear traces.
		\end{itemize}
		These considerations for the four-fermion operators lead to the results found in table \ref{tab:4fermion}.
		
		\section{Scalar Lagrangian} \label{app:SimpscalarLagrangian}
		
		The $\mO_{\hat d}(p^2)$ EWET scalar Lagrangian in eq.~(\ref{eq:LscalarEWET}) represents a general extension of the SM scalar Lagrangian and it counts as $\hat d=2$ because the addition of $h/v$ fields to any operator does not alter its chiral dimension, as already explained. Therefore, one incorporates the polynomial functions $V(h/v)$ and $\mF^{(u)}$ to account for a more general scalar potential and Goldstone-Higgs interactions. Naively, it may be also possible to add the arbitrary function $\mF(h/v)$ coupled to the kinetic term of the Higgs, as in
		\be \label{eq:L2Fh}
		\widetilde\mL_2\, =\, \frac{1}{2}\,\mF_h(h/v)\;
		\partial_\mu h\,\partial^\mu h\, ,
		\qquad\qquad
		\mF_h(h/v)\, = \, 1\, +\, \sum_{n=1} c^{(h)}_n
		\left(\frac{h}{v}\right)^n\, .
		\ee
		However, we can get rid of this contribution by redefining the Higgs field in a non-linear way \cite{Fingerprints}. Shifting the Higgs field like
		\bel{eq:h-redefinition}
		h(x) \,\longrightarrow \, h'(x) \, =\, v\; \sum_{n=1} a_n\left(\frac{h}{v}\right)^n\, ,
		\qquad \mbox{with} \quad a_1 = 1\, ,
		\ee
		so that the derivative kinetic term yields $\widetilde\mL_2 = \frac{1}{2}\,\partial_\mu h' \partial^\mu h'$, it is possible to obtain the value for the $a_n$ parameters provided the following iterative relation for a given set of $c_n^{(h)}$ values,
		\be \label{eq:a_n}
		c^{(h)}_n\, =\,\sum_{k=1}^{n+1}\, k\, (n+2-k)\, a_k\, a_{n+2-k}\, .
		\ee

		Additionally, the scalar Lagrangian gets modified too when the heavy scalar singlet $S^1_1$ is considered. In general, one should incorporate to the resonance Lagrangian the following mixing operators
		\be \label{eq:S-h_mixing}
		\Delta\mL_{S_1h}\, =\, a\; S_1\, h \, +\, b\;\partial_\mu S_1\,\partial_\mu h \, +\, S_1\,\partial_\mu h \,\partial^\mu h \;\sum_{n=0} c_n\,\left(\frac{h}{v}\right)^n\, ,
		\ee
		apart from the ones written in eq.~(\ref{eq:L0}). The mixing between $h$ and $S^1_1$ is therefore reflected through the parameters $a$, $b$ and $c_n$. Nevertheless, all these operators can be set aside. An adequate redefinition of these two scalar quantum fields and their masses gets rid of the first two terms\footnote{Indeed, the removal of the operator proportional to $b$ can be generalized to any resonance operator, since wherever the resonance field carries a derivative, $\bra \partial_\mu R\,\chi^\mu\ket$, it can be rewritten into $-\bra R\,\partial_\mu \chi^\mu\ket$, using partial integration.} in eq.~(\ref{eq:S-h_mixing}); while the third one is eliminated by performing the following partial integration,
		\be \label{eq:O_n}
		S_1\, h^n\,\partial_\mu h \,\partial^\mu h
		\; =\; \frac{1}{(n+1)(n+2)}\, h^{n+2}\,\Box S_1\, -\,
		\frac{1}{(n+1)}\, h^{n+1}\, S_1\,\Box h\, .
		\ee
		The resulting operators can be rewritten in terms of other operators already present in the Lagrangian, provided the usage of EoMs.
		
		\section[Order ${P^3}$ fermionic operators]{$\boldsymbol{\mO_{\hat d}(p^3)}$ fermionic operators} \label{app:simpOp3} \label{app:simpreabsorb}
		
		According to the power counting in eq.~(\ref{eq:power_counting}) the EWET LO Lagrangian brings chiral dimension $\hat d=2$, happening to contain all the operators of the SM. Next to this, there are the NLO operators shown in eq.~(\ref{eq:L4EWET}), subdominant with respect to the previous ones, which carry chiral dimension $\hat d =4$. Nevertheless, we can naively find between these two sets some rare fermionic operators contributing to the Lagrangian with $\hat d=3$ which remain not analyzed so far:
		\be \label{eq:L3F}
		\mL_3^{\mathrm{Fermionic}}\; =\; \beta^f_S\,\bra\mT J^f_S\ket_2
		\,+\, \beta^f_V\,\bra u_\mu J_V^{f\,  \mu  }  \ket_2 \,+\, \beta^f_A\,\bra u_\mu J_A^{f\,  \mu  }  \ket_2\, ,
		\ee
		with $f=l,q$ being a implicit sum over the lepton and quark singlet fermion bilinears\footnote{Quark octet bilinears, $J_\Gamma^8$, are not allowed in eq.~(\ref{eq:L3F}) due to symmetry requirements.} and raising the number of $\mO_{\hat d}(p^3)$ structures to 6.
		
		First, we focus on the operators proportional to $\beta^f_S$. These structures carrying the custodial spurion can be easily eliminated, however, by reabsorbing them into the definition of the Yukawa spurion fields, either for the lepton and the quark fermion fields,
		\be
		\mY^f\; =\; \mY'^f + \frac{\beta_S^f}{v}\,\mT\, .
		\ee
		When these redefinitions are included in the EWET Lagrangian in eq.~(\ref{eq:L_SMinL_EWET}) it reads
		\be
		\sum_f \left[-v\,\left(\bar\xi^f_L\,\mY^f\,\xi^f_R + \mathrm{h.c.}\right) + \beta^f_S\,\bra\mT J^f_S\ket_2 \right]
		\; =\; \sum_f \left[-v\,\left(\bar\xi^f_L\,\mY'^f\,\xi^f_R + \mathrm{h.c.}\right)\right]\, ,
		\ee
		and thus the LO picture is restored.
		
		Second, operators proportional to $\beta_V^f$ and $\beta_A^f$ corresponding to the second and third term of eq.~(\ref{eq:L3F}) are more involved. In order to illustrate how these chiral dimension $\hat d = 3$ operators are removed we first consider the lepton case.
		The procedure consist on redefining the gauge sources so that the new terms arising in the kinetic term can compensate these additional operators \cite{Fingerprints}. For the lepton case only, we redefine the $\hat W_\mu$ and $\hat B_\mu$ gauge chiral sources:
		\bear \label{eq:WBtransfP3}
		\hat W_\mu\,& =&\, \hat W'_\mu \,-\,
		(\beta^l_V - \beta^l_A)
		\, u u_\mu u^\dagger\, , \nn\\
		\hat B_\mu\, &=&\, \hat B'_\mu \,-\,
		(\beta^l_V + \beta^l_A)
		\, u^\dagger u_\mu u\, .
		\eear
		When performing this redefinition, new terms appear in the lepton kinetic Lagrangian due to the presence of these gauge fields in the lepton derivative, defined in eq.~(\ref{eq:dxi}), leading to 
		\be \label{eq:Op3redefinition}
		i\, \bar\xi^l\gamma^\mu d_\mu\xi^l
		+ \beta^l_V\,\bra u_\mu J_V^{l\,  \mu  }  \ket_2
		+ \beta^l_A\,\bra u_\mu J_A^{l\,  \mu  }  \ket_2
		\; =\; i\, \bar\xi^l\gamma^\mu d'_\mu\xi^l\,  .
		\ee
		Furthermore, the redefinition of the gauge sources alters other parts of the Lagrangian as well. The Yang-Mills Lagrangian gets shifted and it generates contributions to the purely bosonic $\mO_{\hat d}(p^4)$ EWET Lagrangian in table \ref{tab:bosonic-Op4}, being these terms suppressed by powers of $(\beta^l_V)^n$ or $(\beta^l_A)^n$ with $1\leq n\leq 4$. In addition, $\beta_A^l$ also modifies the definition of the $u_\mu$ building block like
		\be \label{eq:u_mod}
		u_\mu = u'_\mu/(1+2\beta_A)\,.
		\ee
		Indeed, only contributions that enter in $\hat W_\mu$ and $\hat B_\mu$ with different values affect the definition of $u_\mu$, according to eq.~(\ref{eq.cov-bosonic-tensors}). For consistency, the axial coupling is assigned a chiral dimension $[\beta_A^l]_{\hat d} = 1$ in order to account for the extra unit in the power counting associated to the $J_A^{l\,\mu}$ bilinear. Hence, the Goldstone Lagrangian gets also shifted, 
		\begin{align}
		\frac{v^2}{4}\,\mF_u(h/v)\,\bra u_\mu u^\mu\ket_2\, &=\,
		\frac{v^2}{4\, (1+2\beta_A)^2}\,\mF_u(h/v)\,\bra u'_\mu u'^{\,\mu}\ket_2\, \nn\\
		& =\,
		\frac{v'^{\, 2}}{4}\,\mF'_u(h/v')\,\bra u'_\mu u'^{\, \mu}\ket_2\, ,
		\end{align}
		requiring the vev, the Goldstone\footnote{Notice that this field is redefined so that $u(\varphi/v) = u(\varphi'/v')$.} and the $c_n^{(u)}$ coefficients from $\mF_{u}(h/v)$ to be rescaled as
		\be
		v'\, =\, \frac{v}{1+2\beta_A}\, ,\qquad
		\varphi'\, =\, \frac{\varphi}{1+2\beta_A}\, , \qquad
		c'^{\, (u)}_n\, =\, \frac{c_n^{(u)}}{(1+2\beta_A)^n}\, ,
		\ee
		in order to remain the Lagrangian formally the same.
		
		We also want to remove quark $\mO_{\hat d}(p^3)$ operators proportional to $\beta_V^f$ and $\beta_A^f$ in eq.~(\ref{eq:L3F}). For this purpose, we need to redefine additional gauge sources that appear in the fermion derivative. Apart from the $\hat W_\mu$ and $\hat B_\mu$ gauge fields, the derivative from the kinetic term brings $\hat X_\mu$ and $\hat G_\mu$  fields as well, attending to eq.~(\ref{eq:dxi}), being the last one not included in the lepton derivative. However, it is not possible to redefine these fields in order to remove the remaining $\mO_{\hat d}(p^3)$ operators since neither $\hat X_\mu$ nor $\hat G_\mu$ are axial sources and thus the axial operator $\bra u_\mu J_A^{q\,\mu}\ket_2$ cannot be generated through the rescaling. 
		%
		%
		
		Otherwise, in order to reabsorb $\beta_V^q\,\bra u_\mu J_V^{q\,\mu}\ket_2$ and $\beta_A^q\,\bra u_\mu J_A^{q\,\mu}\ket_2$, we analyze the $\mO_9$ and $\widetilde \mO_3$ bosonic operators
		\be \label{eq:O9O3}
		\mO_9 \,=\, \left(\Frac{\partial_\mu h}{v}\right)\bra f_-^{\mu\nu} u_\nu \ket_2 \,,\qquad\quad
		\widetilde \mO_3 \,=\, \left(\Frac{\partial_\mu h}{v}\right)\bra f_+^{\mu\nu} u_\nu \ket_2 \,.
		\ee 
		Performing an integration by parts, one obtains the following relation 
		\begin{align}\label{eq:dic14}
		\left(\Frac{\partial_{\mu} h}{v}\right) \bra f^{\mu\nu}_{\pm}u_{\nu}\ket_2 &\longrightarrow\,  -\,\bra (u^{\dagger}D_{\mu}\hat W^{\mu\nu}u \,\pm\, u D_{\mu}\hat B^{\mu\nu}u^{\dagger} ) u_{\nu}\ket_2 \nn\\
		& \hspace{0.8cm} - \frac{i}{2}\bra f^{\mu\nu}_{\mp}[u_{\mu},u_{\nu}]\ket_2 \,+\, \frac{1}{2}\bra f^{\mu\nu}_{\pm}f_{- \, \mu\nu}\ket_2 \nn\\
		&\hspace{-2cm} = -\,\frac{v^{2}}{4}(g^{2}\pm g'^{2})\bra u_{\mu}u^{\mu}\ket_2 \,-\, \frac{i}{2}\bra f^{\mu\nu}_{\mp}[u_{\mu},u_{\nu}]\ket_2\,+\, \frac{1}{2}\bra f^{\mu\nu}_{\pm}f_{- \, \mu\nu}\ket_2 \nn\\
		& \hspace{-2cm}+ \,\frac{g^{2}}{4} \bra u_{\mu}(J^{l\,\mu}_{V}-J^{l\,\mu}_{A})\ket_2 \,\pm\, \frac{g'^{2}}{4} \bra u_{\mu}(J^{l\,\mu}_{V}+J^{l\,\mu}_{A})\ket_2  \nn\\
		& \hspace{-2cm}+ \,\frac{g^{2}}{4} \bra u_{\mu}(J^{q\,\mu}_{V}-J^{q\,\mu}_{A})\ket_2 \,\pm\, \frac{g'^{2}}{4} \bra u_{\mu}(J^{q\,\mu}_{V}+J^{q\,\mu}_{A})\ket_2 \,,
		\end{align}
		where some Higgs-dependent functions multiplying the right-hand side of the arrow have been omitted for simplicity. For computing the integration the following EoMs have been employed
		\begin{align}\label{eq:dic5}
		(D_{\mu}\hat W^{\mu\nu})^{a} &= (\partial_{\mu}\hat W^{\mu\nu}-i[\hat W_{\mu},\hat W^{\mu\nu}])^{a} \nn\\
		&= -g^{2} \left(\sum_{\psi=l,\,q} \bar\psi_{L}\gamma^{\nu}T^{a}\psi_{L}\right) +\,\frac{g^{2}v^{2}}{4}(1+F_{U}(h))(u u^{\nu} u^{\dagger})^{a}\,,\nn\\[1ex]
		(D_{\mu}\hat B^{\mu\nu})^{a} &= (\partial_{\mu}\hat B^{\mu\nu}-i[\hat B_{\mu},\hat B^{\mu\nu}])^{a} \nn\\
		& = -g'^{2} \left(\sum_{\psi=l,\,q} \bar\psi_{R}\gamma^{\nu}T^{a}\psi_{R}\right) -\,\frac{g'^{2}v^{2}}{4}(1+F_{U}(h))(u^{\dagger} u^{\nu} u)^{a}\,.
		\end{align}

		The last three lines of the calculation of eq.~(\ref{eq:dic14}) show a mixture of structures that we have classified as chiral dimension 2, 3 and 4, although all of them count as $\mO_{\hat d}(p^4)$ according to the power counting in eq.~(\ref{eq:power_counting}) once considered their preceding $g$ and $g'$ couplings. Indeed, the third line in eq.~(\ref{eq:L3F}) correspond to purely bosonic operators that belong to the LO and NLO EWET Lagrangians already listed in chapter \ref{ch:ewet}. One must pay special attention to the remaining operators in lines four and five, since they are naively counted as chiral dimension $\hat d=5$ (squared gauge couplings plus $\mO_{\hat d}(p^3)$ operators). However, the fermionic bilinears have to be considered as chiral dimension $\hat d =1$ since the gauge coupling is already explicit.
		
		
		Therefore, we can re-express the two quark $\mO_{\hat d}(p^3)$ operators in terms of the $\mO_9$ and $\widetilde \mO_3$ NLO EWET operators and other purely bosonic operators from table \ref{tab:bosonic-Op4}. In conclusion, all the chiral dimension $\hat d=3$ operators from eq.~(\ref{eq:L3F}) can be reabsorbed or rewritten in terms of the other operators considered previously.
		
		\section{$\boldsymbol{\mO_{\hat d}(p)}$ resonance operators} \label{app:simpresonances}
		
		Analogously to the previous EWET operators bringing one less unit to the chiral counting than usual, we can also find a few resonance operators accompanied with chiral tensors, $\bra R\,\chi_R \ket$, with chiral dimension $[\chi_R]_{\hat  d} = 1$. We find the following operators:
		\be \label{eq:Opresonanceoperators}
		\bra S^1_3 \mT \ket_2 \,,\qquad \bra \hat V^{1\,\mu}_3 u_\mu \ket_2 \,,\qquad  \bra \hat A^{1\,\mu}_3 u_\mu \ket_2 \,,
		\ee
		where the last two of them (the spin-1 ones) are written in the Proca formalism.\footnote{These two operators can also be written in the Antisymmetric formalism although they are chiral dimension $\hat d=2$ in this representation.} Notice that, when integrated out, these resonance structures yield precisely the $\mO_{\hat d}(p^3)$ EWET operators studied in the previous section (or the Goldstone Lagrangian $\bra u^\mu u_\mu \ket_2$). Therefore, their imprints can be related to other EWET operators and there are no further problems. However, there is no need for this because the resonance operators from eq.~(\ref{eq:Opresonanceoperators}) can be reabsorbed as well in the heavy fields $S^1_3$, $\hat V^1_{3\,\mu}$ and $\hat A^1_{3\,\mu}$ respectively. The particular resonance rescaling reads
		\begin{align} \label{eq:Oprescaling}
		S^1_3 &\quad\longrightarrow \quad S'^1_3 \,=\, S^1_3 \,+\, \alpha_S\,\mT \,, \nn\\
		\hat V^{1\,\mu}_3 &\quad\longrightarrow\quad \hat V'^{1\,\mu}_3\, =\,\hat V^{1\,\mu}_3 \,+\, \alpha_{\hat V}\, u^\mu \,,\nn\\
		\hat A^{1\,\mu}_3 &\quad\longrightarrow\quad \hat A'^{1\,\mu}_3\, =\,\hat A^{1\,\mu}_3 \,+\, \alpha_{\hat A}\, u^\mu \,;
		\end{align}
		where $\hat{R}'_{\mu\nu} = \hat{R}_{\mu\nu}  - \alpha_{\hat R}\, f_{-\,\mu\nu}$ with $R=V,A$  are the resonance strength tensor redefinition for the spin-1 massive fields above. The value adopted for $\alpha_S$ in the heavy scalar case must be such that one is allowed to shift $\bra S^1_3 \mT\ket_2$ to $\bra \nabla_\mu S^1_3 \nabla^\mu \mT \ket_2$, which is a higher order resonance operator.

		\chapter{Power counting of the EWET} \label{app:powercounting}
		
		The characterization of the hierarchy of the operators within an EFT is essential and it defines the core of the theory. In section \ref{sec:powercounting}, there is a qualitative and consistent explanation of the chiral power counting for the EWET. Nonetheless, a precise mathematical derivation of the counting rules is also in order. 
		
		Given a generic operator containing EWET fields it can be decomposed as 
		\be \label{eq:operatordecomposition}
		\Delta\mL_{djk}  \,\sim\,  f_\ell \,\, \, p^d \,\left(\Frac{\psi }{v}\right)^j\,
		\left(\Frac{\phi}{v}\right)^k\, ,
		\ee
		being $\psi$ a fermionic field and $\phi$ whatever bosonic field able to couple: Goldstone bosons ($\varphi$), the Higgs field ($h$), or any gauge source ($\hat W_\mu$, $\hat B_\mu$, $\hat X_\mu$ or $\hat G_\mu$). Otherwise, light scales coming from derivatives ($\partial_\mu$), light masses ($m_\psi$, $m_h$, $m_W$ or $m_Z$) or couplings ($g$, $g'$ or $y_\xi$) count in eq.~(\ref{eq:operatordecomposition}) as $p$; while $f_l$ stands for the operator associated LEC.
		
		Once parametrized the fields for a single operator, we extend the computation for a generic (connected) Feynman diagram, $\Gamma$. It is formed by $L$ loops, $I$ internal lines (being $I_B$ bosonic and $I_F$ fermionic) and $E$ external lines (being $E_B$ bosonic and $E_F$ fermionic). It is precisely the particular combinations of vertices, $N_{djk}$, with $V = \sum_{d,j,k} N_{djk}$ the total number of them, what defines the topology of the diagram. These pieces are proven to fulfill the following diagram identities \cite{Fingerprints}:
		\bear \label{eq:topological-rel}
		\sum_{d,j,k}  \, j\, N_{djk}  &\! =&\! 2 I_F \,+\, E_F\, ,
		\nn\\
		\sum_{d,j,k}  \, k\, N_{djk}  &\! =&\! 2 I_B \,+\, E_B\, ,
		\nn\\
		L\,=\, I\,+\,1 \,-\,V &\! =&\! I_B\,+\,I_F \,+\,1  \,-\, \sum_{d,j,k} N_{djk}\, .
		\eear
		In order to compute the Feynman diagram one should substitute all the external lines by the given external fields. Besides, all the loop calculations are managed with the mass-independent dimensional regularization scheme, which imposes no cut-offs. The procedure consists in counting naively with an standard dimensional analysis \cite{Weinberg:1978kz,Georgi-Manohar} all the particular momentum contributions of each of the elements conforming the diagram for obtaining its global low-momenta scaling \cite{Weinberg:1978kz, Georgi:1994qn, Pich:1998xt, Buchalla:2013eza}, also called superficial degree of divergence. Therefore, a generic Feynman diagram can be decomposed as
		\bear \label{eq:scalingGamma}
		\Gamma & \sim & \Int \left( \Frac{d^4p}{(2\pi)^d}  \right)^L   \,\,\,
		\Frac{1}{(p^2)^{I_B}\, (p)^{I_F}}  \,\,\,
		\left(\prod_{d,j} (p^d)^{N_{dj}} \right)  \,\,\,
		\left( p^\frac{1}{2} \right)^{E_F}
		\nn\\[10pt]
		&\sim & p^{4L - 2I_B -I_F  + \sum_{d,j,k} d\, N_{djk} +\frac{1}{2} E_F}
		\quad = \quad
		p^{2+2L  + \sum_{d,j,k} (d-2)\,  N_{djk} + I_F + \frac{1}{2} E_F  }
		\nn\\[10pt]
		& = &
		p^{2+2L  +  \sum_{d,j,k} \left(d+\frac{1}{2} j -2\right)\, N_{djk}} \, .
		\eear
		%
		The last line of eq.~(\ref{eq:scalingGamma}) reveals that only the number of loops and the number of vertices weighted by the quantity $d + \frac{1}{2}j - 2$ contribute to the momenta power. Hence, it is possible to obtain the dimension of a Feynman diagram just by summing the momenta scaling of each vertex and the number of loops. For this reason, we are able to assign to any operator the so-called chiral dimension
		\be \label{eq:chiraldimension}
		\displaystyle \hat d \,=\, d \,+\, \frac{j}{2} \,.
		\ee
		Therefore, the power counting of a given operator only depends on the explicit light scales or couplings and the number of fermions involved, being independent of the interacting bosonic fields. Hence, we can also define the chiral dimension for a Feynman diagram as
		\be \label{eq:powercountingdiagram}
		\hat{d}_\Gamma\, =\, 2 + 2L + \sum_{\hat{d}} (\hat{d} -2)\, N_{\hat{d}} \, ,
		\ee
		analogous to the \chpt\ power counting \cite{Weinberg:1978kz, Donoghue:1992dd}, where $N_{\hat d}$ is the total number of operators with chiral dimension $\hat d$. 
		Notice that eq.~(\ref{eq:powercountingdiagram}) yields chiral dimension $\hat d=2$ operators to not increase the power counting, and hence they can be as many as needed.

		Finally, using the identities in eq.~(\ref{eq:topological-rel}), we address all the remaining scales and geometrical factors missing in eq.~(\ref{eq:scalingGamma}) 
		\bear
		\Gamma &\sim & \Frac{1}{(16\pi^2)^L}\,\,\,
		\prod_{d,j,k} \left( \,\,  \Frac{f_\ell}{v^{j+k} }\,\,  \right)^{N_{djk}}
		\nn\\
		&=&
		\Frac{1}{(16\pi^2)^L}\,\,\,
		\left[ \prod_{d,j,k} \left(\Frac{f_\ell}{v^2}\right)^{N_{djk}}\, \right]
		\,\,\, \left(\Frac{1}{v}\right)^{\sum_{d,j,k} (j+k-2) N_{djk}} \,
		\nn\\[10pt]
		& = &
		\Frac{1}{(16\pi^2 v^2)^L}\,\,\,
		\left[ \prod_{d,j,k} \left(\Frac{f_\ell}{v^2}\right)^{N_{djk}}\, \right]
		\,\,\, \Frac{1}{v^{E-2} } \, ,
		\eear
		so the final contribution to a given Feynman diagram amplitude turns out to be
		\be\label{eq:counting_amplitude}
		\Gamma \,\sim\, \Frac{p^2}{v^{E-2}} \,\,\,
		\left(\Frac{p^2}{16\pi^2 v^2}\right)^L  \,\,\,
		\prod_{\hat{d}} \left(
		\Frac{f_\ell\,\,  p^{\, \hat{d}-2}   }{v^2}\right)^{N_{\hat{d}}}
		\, .
		\ee

		\chapter{EWET relation to other basis} \label{app:dictionary}
		
		All simplifications, redundancies and relations among the operators performed so far lead to the necessity of choosing which operators have to be included in the EWET basis and which not. The particular selection depends on several factors like providing an easier phenomenology, simplicity or even keeping some other basis criteria. All of them are valid provided they lead to an independent set of operators. In the following, we compare our particular operator basis, which will be referred as EWET \cite{Fingerprints, Colorful}, with the basis of \cite{Buchalla:2012qq,Buchalla:2013rka,Krause:2016uhw}, denoted as MUC, and the Longhitano basis \cite{Longhitano:1980iz, Longhitano:1980tm, Morales:94}, which only applies for the Higgsless bosonic sector. 
		
		On the one hand, the EWET and the Longhitano basis are constructed in terms of custodial invariant operators which transforms under $SU(2)_{L+R}$. The main advantage of these building blocks setting is that they couple directly to the resonances \cite{Ecker:1988te, Fingerprints}. On the other hand, the MUC basis is built under $SU(2)_L \otimes SU(2)_R$ invariant operators. As a consequence, this is a more extended basis since operators belonging to it may be custodial suppressing as well. The comparison to our basis can only apply, therefore, to those MUC operators or combinations of them being invariant under this symmetry. The same consideration must be done for $CP$ discrete symmetry, which is not imposed in this basis. Hence, $CP$ odd or breaking operators are disregarded.  
		
		In the EWET and Longhitano basis, however, custodial symmetry breaking is only introduced through the custodial spurion field, $\mT$ and the fermion Yukawas, $\mY$, in the EWET case. The main difference between these sets is precisely the explicit chiral suppression assigned to these spurion fields in EWET, accounting for some weak coupling from BSM physics, as opposite to the Longhitano basis, where this building block is considered to be chiral dimension $\hat d = 0$. In any case, we only deal with those operators that can be related to the LO and NLO EWET Lagrangians and we set apart those terms matching higher orders.

		\section{Conversion tools from EWET to MUC}
		In this section, a transformation guide for the EWET and MUC basis is shown. The elements and building blocks from these two basis are constructed under the premises of invariance under the Lie groups $SU(2)_{L+R}$ and $SU(2)_L \otimes SU(2)_R$, respectively, as already commented. The conventions adopted yield moderately different definitions, attached as follows.
		
		In order to analyze the MUC structures, it is necessary to introduce the $SU(2)$ projectors:
		\begin{align} \label{eq:dic1}
		P_{+}  = \frac{1}{2} + T^{3} = \bigg(\, \begin{matrix}1 & 0 \\[.5ex] 0 & 0 \end{matrix}\, \bigg)\,, &\qquad\qquad P_{-}  = \frac{1}{2} - T^{3} = \bigg( \, \begin{matrix}0 & 0 \\[.5ex] 0 & 1 \end{matrix} \, \bigg)\,, \nn\\[1ex]
		P_{12}  = T^{1} + i T^{2} = \bigg(\, \begin{matrix}0 & 1 \\[.5ex] 0 & 0 \end{matrix} \,\bigg)\,, &\qquad\qquad   P_{21}  = T^{1} - i T^{2} = \bigg( \,\begin{matrix}0 & 0 \\[.5ex] 1 & 0 \end{matrix}\,\bigg) \, ,
		\end{align}
		being $T^i$ the $SU(2)$ generators. The equivalent building blocks to the EWET terms $u_\mu$ and $\mT$ are, respectively
		\begin{equation}\label{eq:dic3}
		L_{\mu}\, \equiv \, i U D_{\mu}U^{\dagger}\,=\,u\,u_\mu\, u^\dagger\,, \qquad\qquad \tau_{L}\, \equiv\, U\, T^{3}\,U^{\dagger}\,=\, \Frac{1}{g'}\,u\, \mT \,u^\dagger\,,
		\end{equation}
		satisfying the following identities:
		\begin{align} \label{eq:dic4}
		D_{\mu}\tau_{L} &\,=\, i [L_{\mu},\tau_{L}], \nn\\
		D_{\mu}L_{\nu}-D_{\nu}L_{\mu} &\,=\, g W_{\mu\nu} \,-\, g' B_{\mu\nu}\tau_{L}\,+\, i[L_{\mu},L_{\nu}], \nn\\
		\bra U^{\dagger} L_{[\mu}L_{\nu]}U P_{12}\ket_2 & \,=\, -2 \bra \tau_{L}L_{[\mu}\ket_2 \bra U P_{12}U^{\dagger} L_{\nu]}\ket_2,  \nn\\
		\bra U^{\dagger} L_{[\mu}L_{\nu]}U P_{21}\ket_2 & \,=\, 2 \bra \tau_{L}L_{[\mu}\ket_2 \bra U P_{21}U^{\dagger} L_{\nu]}\ket_2.
		\end{align}
		Hence, we are able to relate the bosonic sector of both the EWET and the MUC basis, making use of these relations \cite{Fingerprints}:
		\begin{align} \label{eq:dic8}
		u^{2}  \,=\, U, \qquad &\qquad (u^{\dagger})^{2} \,=\, U^{\dagger}, \nn\\
		\hat W^{\mu}  \,=\, -g\, W^{\mu,a} T^{a},\qquad\quad  \hat B^{\mu} &\,=\, - g' B^{\mu} T^{3}, \qquad\quad  \hat X^{\mu} \,=\, - g' B^{\mu} \nn\\
		\mathcal{T} \,=\, - u\, g' &T^{3}\, u^{\dagger} \,=\, -g' u^{\dagger} \tau_{L}\, u\,, \nn\\
		f^{\pm}_{\mu\nu}   = -(g\, u^{\dagger}W_{\mu\nu}^{a} &T^{a}u\,\pm\, g' u B_{\mu\nu}T^{3}u^{\dagger}) \,.
		\end{align}
		Another relevant equation for the basis translation regarding the derivatives of the gauge strength tensors can be found in eq.~(\ref{eq:dic5}).
		
		Fermions in the MUC basis are also organized in a different way. If we consider a right-hand fermion field, $\psi$ ($\psi_R$ or $U^\dagger \psi_L$), and  a tensor $X$ transforming as a right-handed object, $X\to g_R X\,g_R^\dagger\,$, a general fermion bilinear can be decomposed into the projectors of eq.~(\ref{eq:dic1}) as follows:
		\begin{align}\label{eq:dic2}
		\bar\psi \Gamma X \psi &\,=\, \bar\psi \Gamma (P_{+}-P_{-}) \psi \;\bra X T^{3}\ket_2 \,+\, \bar\psi \Gamma (P_{+}+P_{-}) \psi \;\bra \frac{X}{2} \ket_2 \nn\\
		&\,+\, \bar\psi \Gamma P_{12} \psi \;\bra X P_{21}\ket_2 \,+\, \bar\psi \Gamma P_{21} \psi \;\bra X P_{12}\ket_2\,,
		\end{align}
		where $\Gamma$ is a Dirac matrix written in the same convention than the EWET. This relation maybe useful for dealing with right-handed structures like $U^{\dagger}W_{\mu\nu}U$, $iD_{\mu}U^{\dagger}U=U^{\dagger}L_{\mu}U$ or $T^{3}_{R}$. Analogous relations apply to left-handed field combinations. The equivalence between light fermion bilinears is set in table \ref{tab:dicbilinears}
		\renewcommand{\arraystretch}{1.6}
		\begin{table}[!ht]
			\centering
			\begin{tabular}[t]{|c|c|}
				\hline
				EWET & MUC \\
				\hline
				$(J_S)_{mn} = \bar \xi_n \xi_m$ & $(\bar\psi_{L}u)_{n}(u\psi_{R})_{m}+(\bar\psi_{R}u^{\dagger})_{n}(u^{\dagger}\psi_{L})_{m}$\\
				\hline
				$(J_P)_{mn} = \bar \xi_n\,i\gamma^{5}\, \xi_m$ & $i(\bar\psi_{L}u)_{n}(u\psi_{R})_{m}-i(\bar\psi_{R}u^{\dagger})_{n}(u^{\dagger}\psi_{L})_{m}$\\
				\hline
				$(J^\mu_V)_{mn} = \bar \xi_n\,\gamma^{\mu}\, \xi_m$ & $(\bar\psi_{L}u)_{n}\gamma^{\mu}(u^{\dagger}\psi_{L})_{m}+(\bar\psi_{R}u^{\dagger})_{n}\gamma^{\mu}(u\psi_{R})_{m}$\\
				\hline
				$(J^{\mu}_A)_{mn} = \bar \xi_n\,\gamma^{\mu}\gamma^{5}\, \xi_m$ & $-(\bar\psi_{L}u)_{n}\gamma^{\mu}(u^{\dagger}\psi_{L})_{m}+(\bar\psi_{R}u^{\dagger})_{n}\gamma^{\mu}(u\psi_{R})_{m}$\\
				\hline
				$(J^{\mu\nu}_T)_{mn} = \bar \xi_n\,\sigma^{\mu\nu}\, \xi_m$ & $(\bar\psi_{L}u)_{n}\sigma^{\mu\nu}(u\psi_{R})_{m}+(\bar\psi_{R}u^{\dagger})_{n}\sigma^{\mu\nu}(u^{\dagger}\psi_{L})_{m}$\\
				\hline
			\end{tabular}
			\caption{Fermion bilinear conversion for the EWET and the MUC bases. This conversion applies exactly for leptons fields,\ie $J_\Gamma^l$ and $\psi = l$ for the EWET and MUC bases, respectively. Quark singlet bilinears, $J_\Gamma^q$ and $\psi=q$, are identical but quarks acquire a color index, $\xi^a$ and $q^a$, respectively. For quark octet bilinears, $J_\Gamma^8$, both EWET and MUC structures carry a $SU(3)_C$ generator, $T^a$, within the bilinear which is contracted with one another from outside the bilinear.}
			\label{tab:dicbilinears}
		\end{table}

		\section{Purely bosonic sector}
		
		In the first place, we analyze the operator conversion for the purely bosonic operators (table \ref{tab:bosonic-Op4}). The conversion table from the EWET basis to the MUC and the Longhitano bases is placed in table \ref{tab:dicbosonic}.

		\renewcommand{\arraystretch}{1.8}
		\begin{table}[!b]
			\centering
			\begin{tabular}[t]{|c|c|c|}
				\multicolumn{3}{c}{} \\[-3ex]
				\hline
				\multicolumn{3}{|c|}{Bosonic $P$-even operators}
				\\
				\hline
				\multicolumn{3}{c}{}
				\\[-3ex]
				\hline
				EWET & MUC & Longhitano \\
				\hline
				$\mathcal{O}_{1}$ & 
				$\mathcal{O}_{XU1}$ & 
				$\mathcal{O}^L_{1}$	%
				\\ \hline 
				$\mathcal{O}_{2}$ & 
				$\displaystyle \mathcal{O}_{Xh2} + \frac{1}{2}\mathcal{O}_{Xh1}$ & 
				---	
				\\ \hline	
				$\mathcal{O}_{3}$ & 
				$\displaystyle -\frac{1}{2} \left( \mathcal{O}_{XU7} + \mathcal{O}_{XU8} \right)$ & 
				$\displaystyle \frac{1}{2} \left( \mathcal{O}^L_{2} - \mathcal{O}^L_{3} \right)$
				\\ \hline	
				$\mathcal{O}_{4}$ & 
				$\mathcal{O}_{D2}$ & 
				$\mathcal{O}^L_{4}$	%
				\\ \hline
				$\mathcal{O}_{5}$ & 
				$\mathcal{O}_{D1}$ & 
				$\mathcal{O}^L_{5}$	%
				\\ \hline
				$\mathcal{O}_{6}$ & 
				$\mathcal{O}_{D7}$ & 
				---	
				\\ \hline
				$\mathcal{O}_{7}$ & 
				$\mathcal{O}_{D8}$ &
				---	
				\\ \hline
				$\mathcal{O}_{8}$ & 
				$\mathcal{O}_{D11}$ &
				---	
				\\ \hline
				$\mathcal{O}_{9}$ & 
				explained below &
				---	
				\\ \hline
				$\mathcal{O}_{10}$ & 
				$\displaystyle \frac{1}{v^{2}}\mathcal{O}_{\beta}$ &
				$\mathcal{O}^L_{0}$	%
				\\ \hline
				$\mathcal{O}_{11}$ & 
				$\begin{matrix} \mbox{redundant when} \\[-1ex] \hat X^{\mu} \rightarrow - g' B^{\mu} \end{matrix}$ &
				$\begin{matrix} \mbox{redundant when} \\[-1ex] \hat X^{\mu} \rightarrow - g' B^{\mu} \end{matrix}$
				\\ \hline
				$\mathcal{O}_{12}$ & 
				$\mathcal{O}_{Xh3}$ &
				---
				\\ \hline
				\multicolumn{3}{c}{} \\[-3ex]
			\end{tabular}
		\end{table}
		\renewcommand{\arraystretch}{1.8}
		\begin{table}[!ht]
			\centering
			\begin{tabular}[t]{|c|c|c|}
				\hline
				\multicolumn{3}{|c|}{Bosonic $P$-odd operators}
				\\
				\hline
				\multicolumn{3}{c}{}
				\\[-3ex]
				\hline
				EWET & MUC & $\quad\;$Longhitano$\quad\;$ \\
				\hline
				$\mathcal{\widetilde O}_{1}$ & 
				$\displaystyle \frac{1}{2}\left(\mathcal{O}_{XU7} - \mathcal{O}_{XU8}\right)$ & 
				$\displaystyle -\frac{1}{2}\left(\mathcal{O}^L_{2} + \mathcal{O}^L_{3}\right)$
				\\ \hline		
				$\mathcal{\widetilde O}_{2}$ &
				$\displaystyle \mathcal{O}_{Xh2} - \frac{1}{2}\mathcal{O}_{Xh1}$ &
				---
				\\ \hline
				$\mathcal{\widetilde O}_{3}$ & 
				explained below &
				---
				\\ \hline
			\end{tabular}
			\caption{\small{Basis conversion from EWET to MUC and Longhitano for the purely bosonic operators. MUC basis has trivial conversions except for $\mO_9$ and $\widetilde \mO_3$. The Longhitano basis is only adressed for the EWET Higgsless operators.  P-even (P-odd) operators are shown in the upper (lower) block.}
			}
			\label{tab:dicbosonic}
		\end{table}
		
		In order to establish a conversion for the MUC basis it is necessary to identify all the operators that respect the symmetries of the EWET,\ie $CP$ even transformations and custodial symmetry. In addition, an explicit chiral suppression will be assumed for the MUC operators containing the spurion $\tau_L$, following the same criteria that in the EWET spurion field $\mT$. The same thing applies for any operator bringing non-SM weak couplings. The list of structures satisfying these conditions is shortened to 12 operators: $\mathcal{O}_{\beta}$, $\mathcal{O}_{D1}$, $\mathcal{O}_{D2}$, $\mathcal{O}_{D7}$, $\mathcal{O}_{D8}$, $\mathcal{O}_{D11}$, $\mathcal{O}_{Xh1}$, $\mathcal{O}_{Xh2}$, $\mathcal{O}_{Xh3}$, $\mathcal{O}_{XU1}$, $\mathcal{O}_{XU7}$ and  $\mathcal{O}_{XU8}$; denoted as in \cite{Buchalla:2013rka}. The EWET set of purely bosonic operator, however, is formed by 15 different structures (see table \ref{tab:bosonic-Op4}). A consistent basis conversion requires the same number of operators for the matching. Nonetheless, this does not mean that neither the EWET is reduntant nor the MUC basis is incomplete, due to the following considerations.
		
		In the first place, $\mO_{11} = \hat X_{\mu\nu} \hat X^{\mu\nu}$ is a redundant operator when we set the SM value for this field, $\hat X_\mu = -g' B_\mu$, according to eq.~(\ref{eq:SMgauge}). Indeed, when taken the SM prescription, the EoM of this field impose a constraint over the EoMs of the right-handed vector current of $\hat B_\mu^3$ and the hypercharge current of $B_\mu$, being 
		\be
		 0 \, =\, \bar\psi_{L} \gamma_{\mu} \left(\frac{B-L}{2}\right) \psi_{L} + \bar\psi_{R} \gamma_{\mu}  \left(\frac{B-L}{2}\right) \psi_{R}\,.
		\ee
		In the second place, the operators $\mO_9$ and $\widetilde \mO_3$ are not present in table \ref{tab:dicbosonic} because they are rewritten as two-fermion operators (and many other bosonic operators already present) in the MUC basis, by means of partial integration. In order to establish the relation, we make use of the eq.~(\ref{eq:dic14}). Therefore, we can redefine this operators in terms of $\bra u_\mu J_V^{q\,\mu} \ket_2$ and $\bra u_\mu J_A^{q\,\mu} \ket_2$, introduced in eq.~(\ref{eq:L3F}).\footnote{The partial integration also brings the lepton chiral dimension $\hat d = 3$ operators $\bra u_\mu J_V^{l\,\mu} \ket_2$ and $\bra u_\mu J_A^{l\,\mu} \ket_2$. Notwithstanding, recall that these structures were already reabsorbed in the gauge sources in the appendix \ref{app:simpOp3}.} As a consequence, we are able to relate these missing operators to the two-fermion MUC structures $\tilde{\mathcal{O}}_{q V1}^{L}$ and $\tilde{\mathcal{O}}_{q V1}^{R}$, using the following relations 
		\begin{align} \label{eq:dic15}
		\bra u_{\mu} J^{q\,\mu}_{V} \ket_2 &\,=\,\tilde{\mathcal{O}}_{q V1}^{R}+\tilde{\mathcal{O}}_{q V1}^{L} \,, \nn\\
		\bra u_{\mu} J^{q\,\mu}_{A} \ket_2 &\,=\,\tilde{\mathcal{O}}_{q V1}^{R}-\tilde{\mathcal{O}}_{q V1}^{L} \,.
		\end{align}

		The conversion from the EWET to the Longhitano basis (table \ref{tab:dicbosonic}) only applies to the Higgsless and color singlet bosonic operators,\ie $\mO_{1,\ldots,5, 10,11}$. Actually, operator $\mO_{10}$ is considered in \cite{Longhitano:1980iz, Longhitano:1980tm} to be a LO contribution, since no further chiral suppression is included in the spurion $\mT$ in the Longhitano set. Nevertheless, phenomenology has rejected this contribution to be a dominant one. In regard to $\mO_{11}$, the same conclusions as in the MUC basis apply.
		
		\section{Two-fermion operators}
		
		NLO operators containing both bosonic fields and one single two-fermion bilinear are built differently as well, following distinct criteria. While the EWET organizes their operators according to custodial symmetry invariance building blocks, the two-fermion MUC operators are set with the premise that they become single $W^\pm$ or $Z$ operators in the unitary gauge. Notice that in this section, and also in the next one referring to four-fermion operators, the bases comparison in only established between the EWET and MUC ones, since the Longhitano basis only includes purely bosonic objects.
		
		The different design of the EWET and the MUC bases leads us to introduce some intermediate operator set for the second one, so that its elements, made out of combinations of the original basis operators, are custodially invariant too. We denote them with a small\footnote{Do not confuse them with $P$-odd EWET operators, carrying a wide tilde.} tilde $\sim$. As a consequence, we find 25 (12+13) independent operators classified in four subsets: vector current operators, denoted as $\tilde \mO_{\psi V}$; scalar current operators, referred as $\tilde \mO_{\psi S}$; tensor ones, $\tilde \mO_{\psi T}$; and also dipole operators, tagged as $\tilde \mO_{\psi X}$; and in two fermion families. Therefore, for each subset $\psi$ can be either a lepton, $l$, or a quark, $q$. The correspondence to the MUC basis is found to be \cite{Colorful}
		\begin{align} \label{eq:dic10}
		\tilde{\mathcal{O}}_{\psi V1}^{L} &\equiv \bar\psi_{L}\gamma_{\mu}L^{\mu}\psi_{L} \,\xrightarrow{\,\psi\,=\,q\,}\, -(2\mathcal{O}_{\psi V2}+\mathcal{O}_{\psi V3}+\mathcal{O}_{\psi V3}^{\dagger} )\,, \nn\\
		\tilde{\mathcal{O}}_{\psi V2}^{L} &\equiv \bar\psi_{L}\gamma_{\mu}\{\tau_{L},L^{\mu}\}\psi_{L}\,  \,\xrightarrow{\,\psi\,=\,q\,}\,\,-\mathcal{O}_{\psi V1}\,,\nn\\
		\tilde{\mathcal{O}}_{\psi V1}^{R} &\equiv \bar\psi_{R}\gamma_{\mu}U^{\dagger}L^{\mu}U\psi_{R} \, \,\xrightarrow{\,\psi\,=\,q\,}\,\, -(\mathcal{O}_{\psi V4}-\mathcal{O}_{\psi V5}+\mathcal{O}_{\psi V6}+\mathcal{O}_{\psi V6}^{\dagger} )\,,\nn\\
		\tilde{\mathcal{O}}_{\psi V2}^{R} &\equiv \bar\psi_{R}\gamma_{\mu}U^{\dagger}\{\tau_{L},L^{\mu}\}U\psi_{R} \, \,\xrightarrow{\,\psi\,=\,q\,}\,\,-\frac{1}{2}(\mathcal{O}_{\psi V4}+\mathcal{O}_{\psi V5})\,, \nn\\
		\tilde{\mathcal{O}}_{\psi S1} &\equiv \bar\psi_{L}U\psi_{R} \left(\partial_{\mu}\frac{h}{v}\right) \left(\partial^{\mu}\frac{h}{v}\right)  \, \,\xrightarrow{\,\psi\,=\,q\,}\,\, \mathcal{O}_{\psi S14}+\mathcal{O}_{\psi S15}\,,\nn\\
		\tilde{\mathcal{O}}_{\psi S2} &\equiv \bar\psi_{L}L_{\mu}U\psi_{R} \left(\partial^{\mu}\frac{h}{v}\right)  \, \,\xrightarrow{\,\psi\,=\,q\,}\,\, \mathcal{O}_{\psi S10}-\mathcal{O}_{\psi S11}+\mathcal{O}_{\psi S12}+\mathcal{O}_{\psi S13}\,,\nn\\
		\tilde{\mathcal{O}}_{\psi S3} &\equiv \bar\psi_{L}L_{\mu}L^{\mu}U\psi_{R}  \, \,\xrightarrow{\,\psi\,=\,q\,}\,\,\frac{1}{2}(\mathcal{O}_{\psi S1}+\mathcal{O}_{\psi S2}) \,, \nn\\
		\tilde{\mathcal{O}}_{\psi T1} &\equiv \bar\psi_{L}\sigma_{\mu\nu}L^{\mu}L^{\nu}U\psi_{R}  \, \,\xrightarrow{\,\psi\,=\,q\,}\,\, \mathcal{O}_{\psi T1}-\mathcal{O}_{\psi T2}+2\mathcal{O}_{\psi T3}-2\mathcal{O}_{\psi T4}\,, \nn\\
		\tilde{\mathcal{O}}_{\psi T2} &\equiv \bar\psi_{L}\sigma_{\mu\nu}L^{\mu}U\psi_{R} \left(\partial^{\nu}\frac{h}{v}\right)  \, \,\xrightarrow{\,\psi\,=\,q\,}\,\, \mathcal{O}_{\psi T7}-\mathcal{O}_{\psi T8}+\mathcal{O}_{\psi T9}+\mathcal{O}_{\psi T10} \,, \nn\\
		\tilde{\mathcal{O}}_{\psi X1} &\equiv g' \bar\psi_{L}\sigma_{\mu\nu}U\psi_{R} B^{\mu\nu} \, \,\xrightarrow{\,\psi\,=\,q\,}\,\,\mathcal{O}_{\psi X1}+\mathcal{O}_{\psi X2}\,,\nn\\
		\tilde{\mathcal{O}}_{\psi X2} &\equiv g' \bar\psi_{L}\sigma_{\mu\nu}\tau_{L}U\psi_{R} B^{\mu\nu} \, \,\xrightarrow{\,\psi\,=\,q\,}\,\,\mathcal{O}_{\psi X1}-\mathcal{O}_{\psi X2}\,, \nn\\
		\tilde{\mathcal{O}}_{\psi X3} &\equiv g \bar\psi_{L}\sigma_{\mu\nu}W^{\mu\nu}U\psi_{R} \, \,\xrightarrow{\,\psi\,=\,q\,}\,\,\mathcal{O}_{\psi X3}-\mathcal{O}_{\psi X4}+\mathcal{O}_{\psi X5}+\mathcal{O}_{\psi X6} \,, \nn\\
		\tilde{\mathcal{O}}_{\psi X4} &\equiv g_{s} \bar \psi_{L} \sigma_{\mu\nu}G^{\mu\nu}U\psi_{R} \, \,\xrightarrow{\,\psi\,=\,q\,}\,\, \mathcal{O}_{\psi X7}+\mathcal{O}_{\psi X8} \,.
		\end{align}
		being the last intermediate operator, $\tilde{\mathcal{O}}_{\psi X4}$, not present in the lepton set. Notice that eq.~(\ref{eq:dic10}) reflects the general form of this auxiliary basis (left-hand side of the arrow), valid for both leptons and quarks, although the equation is only specialized for the quark set of MUC operators (right-hand side of the arrow) and not for the lepton copy.
		
		Once identified the custodial building blocks, we show in table \ref{tab:dic2fermion} the actual equivalence with the EWET operators. It should be mentioned that there are just 21 (10+11) two-fermion structures in the EWET basis, 4 operators less than in MUC. Indeed, two of the remaining operators compensate precisely the two operator excess present in the purely bosonic case, related through eq.~(\ref{eq:dic14}) and eq.~(\ref{eq:dic15}); while the other two operators correspond to the $\mO_{\hat d}(p^3)$ leptonic structures in eq.~(\ref{eq:L3F}) that can be reabsorbed in the gauge sources attending to eq.~(\ref{eq:WBtransfP3}), as shown in the appendix \ref{app:simpreabsorb}.

		%
		\begin{table} [!t]
			\begin{center}
				\renewcommand{\arraystretch}{1.8}
				\begin{tabular}{|c|c|c|c|c|}
					\hline
					\multicolumn{5}{|c|}{Two-fermion $P$-even operators --- leptons or quarks} \\
					\hline
					\multicolumn{5}{c}{} \\	[-3ex] \cline{1-2}\cline{4-5}
					EWET & MUC & & EWET & MUC
					\\ \cline{1-2}\cline{4-5}
					$\mO_1^{\psi^2_f}$  & 
					$2\tilde{\mathcal{O}}_{\psi S3}+\text{h.c.}$  & &
					$\mO_5^{\psi^2_f}$ & 
					$i\tilde{\mathcal{O}}_{\psi S2}+\text{h.c.} $						
					\\ \cline{1-2}\cline{4-5}
					$\mO_2^{\psi^2_f}$ & 
					$2 i\tilde{\mathcal{O}}_{\psi T1}+\text{h.c.} $ & &
					$\mO_6^{\psi^2_f}$ & 
					$\frac{1}{2}(-\tilde{\mathcal{O}}_{\psi V2}^{L}+\tilde{\mathcal{O}}_{\psi V2}^{R}) $
					\\  \cline{1-2}\cline{4-5}
					$\mO_3^{\psi^2_f}$ &
					$-\tilde{\mathcal{O}}_{\psi X2}-\tilde{\mathcal{O}}_{\psi X3}+\text{h.c.} $ & &
					$\mO_7^{\psi^2_f}$ &
					$\tilde{\mathcal{O}}_{\psi S1}+\text{h.c.} $ 
					\\ \cline{1-2}\cline{4-5}
					$\mO_4^{\psi^2_f}$ &
					$-\tilde{\mathcal{O}}_{\psi X1}+\text{h.c.} $ & &
					$\mO_8^{\psi^2_q}$ &
					$\frac{1}{2}\tilde{\mathcal{O}}_{\psi X4} \quad(\dagger)$
					\\ \cline{1-2}\cline{4-5}
					%
					\multicolumn{5}{c}{} \\	 
					\hline
					\multicolumn{5}{|c|}{Two-fermion $P$-even operators --- leptons or quarks}	\\
					\hline
					\multicolumn{5}{c}{} 
					\\[-3ex] \cline{1-2}\cline{4-5}
					EWET & MUC & & EWET & MUC
					\\ \cline{1-2}\cline{4-5}
					$\widetilde\mO_1^{\psi^2_f}$ &
					$\tilde{\mathcal{O}}_{\psi X2}-\tilde{\mathcal{O}}_{\psi X3}+\text{h.c.}$ & &
					$\widetilde\mO_3^{\psi^2_f}$ &
					$\frac{1}{2}(\tilde{\mathcal{O}}_{\psi V2}^{L}+\tilde{\mathcal{O}}_{\psi V2}^{R}) $
					\\ \cline{1-2}\cline{4-5}
					$\widetilde\mO_2^{\psi^2_f}$ &
					$-\tilde{\mathcal{O}}_{\psi T2}+\text{h.c.} $
					\\ \cline{1-2}	
					\multicolumn{5}{c}{} 
					\\[-3ex]		
				\end{tabular}
				\caption{\small
					Conversion for the EWET and MUC bases for the two-fermion operators. The identification is done through the intermediate operators written in eq.~(\ref{eq:dic10}). This table has two copies: one for leptons, $f=\psi=l$, and another one for quarks, $f=\psi=q$; having both of them the same formal structure, but $(\dagger)$ the operator $\mO_8^{\psi^2_q}$, which only is present in the quark set. P-even (P-odd) operators are shown in the upper (lower) block.}
				\label{tab:dic2fermion}
			\end{center}
		\end{table}
		%

		\section{Four-fermion operators}
		The procedure to make the equivalence between four-fermions EWET and MUC operators is analogous to the previous section. First, it is required to identify the $CP$-even custodial invariant structures in an intermediate basis and, second, establish the relation between both bases.
		
		A general four-fermion operator in the MUC basis has the form 
		\begin{equation}\label{eq:dic16}
		\left(\bar\psi_{\alpha,a}^{i,p}\Gamma_{\alpha\beta}\psi_{\beta,b}^{j,q}\right)\left(\bar\psi_{\gamma,c}^{k,r}\Gamma_{\gamma\delta}\psi_{\delta,d}^{l,s}\right),
		\end{equation}		
		where Greek letters indicate spinorial indices, which remain usually omitted since spinor contraction within the brackets is always assumed. Otherwise, $a,b,c,d$ and $i,j,k,l$ correspond to $SU(3)$ and $SU(2)$ indices; while $p,q,r,s$ stand for a quark or a lepton. 
		As well as in the EWET, Fierz identities, detailed in the appendix \ref{app:fierzidentities}, reduce significantly the number of four-fermions, forcing to select which operators are considered as independent and which redundant. In the MUC convention, $SU(2)$ and $SU(3)$ indices always contract the same pair of fermions,\ie if the $SU(2)$ indices $i$ and $j$ are contracted in eq.~(\ref{eq:dic16}), then the $SU(3)$ ones $a$ and $b$ must be contracted too. Analogously, the contraction of $i$ with $l$ implies that $a$ and $d$ are also connected. However, the possible redundancies and combinations among all the index contractions depend heavily on the particle content of the four-fermion operators. In any case, we disregard all those structures carrying different particles within the same bracket, which are expressed through other operators of the basis.   
		
		Nonetheless, the formal structure of the operators in the intermediate MUC basis for four-fermion structures, denoted with a small tilde $\sim$, is similar for the distinct fermion classes and they only differ in the number of them being independent. According to the bilinear particle content, we find 6 distinct structures in the lepton-lepton case, 12 operators in the quark-quark set and 14 in the lepton-quark mixed one, as well as it happens in the EWET case (see table \ref{tab:4fermion}). In the following, we show the possible custodial invariants to be constructed for these operators, divided in three subsets: chiral current structures, $\tilde \mO_{4L/R}^{\psi\psi'}$, scalar current operators, $\tilde \mO_{4S}^{\psi\psi'}$ and tensor current terms, $\tilde \mO_{4T}^{\psi\psi'}$; being $\psi, \psi'=l,q$ the fermion type of the first and second bilinear, respectively. We find
		\begin{align*} 
		\tilde{\mathcal{O}}_{4LL1}^{\psi\psi'} &\equiv (\bar\psi_{L}\gamma_{\mu}\psi_{L})(\bar\psi'_{L}\gamma^{\mu}\psi'_{L})  \,\xrightarrow{\,\psi,\psi'=\,q\,}\, \mathcal{O}_{LL1}\, , \hspace{5.5cm} \nn\\
		\tilde{\mathcal{O}}_{4LL2}^{\psi\psi'} &\equiv (\bar\psi_{L}\gamma_{\mu}T^{a}\psi_{L})(\bar\psi'_{L}\gamma^{\mu}T^{a}\psi'_{L})  \,\xrightarrow{\,\psi,\psi'=\,q\,}\, \mathcal{O}_{LL2}+\frac{1}{12}\mathcal{O}_{LL1}\,, \nn\\
		\tilde{\mathcal{O}}_{4LR1}^{\psi\psi'} &\equiv (\bar\psi_{L}\gamma_{\mu}\psi_{L})(\bar\psi'_{R}\gamma^{\mu}\psi'_{R})  \,\xrightarrow{\,\psi,\psi'=\,q\,}\, \mathcal{O}_{LR1}+\mathcal{O}_{LR3}\,, \nn\\
		\tilde{\mathcal{O}}_{4LR2}^{\psi\psi'} &\equiv (\bar\psi_{L}\gamma_{\mu}T^{a}\psi_{L})(\bar\psi'_{R}\gamma^{\mu}T^{a}\psi'_{R})  \,\xrightarrow{\,\psi,\psi'=\,q\,}\, \mathcal{O}_{LR2}+\mathcal{O}_{LR4}\,, \nn\\
		\end{align*}\begin{align} \label{eq:19}
		\tilde{\mathcal{O}}_{4RR1}^{\psi\psi'} &\equiv (\bar\psi_{R}\gamma_{\mu}\psi_{R})(\bar\psi'_{R}\gamma^{\mu}\psi'_{R})  \,\xrightarrow{\,\psi,\psi'=\,q\,}\, \mathcal{O}_{RR1}+\mathcal{O}_{RR2}+2\mathcal{O}_{RR3}\,, \nn\\
		\tilde{\mathcal{O}}_{4RR2}^{\psi\psi'} &\equiv (\bar\psi_{R}\gamma_{\mu}T^{a}\psi_{R})(\bar\psi'_{R}\gamma^{\mu}T^{a}\psi'_{R})  \,\xrightarrow{\,\psi,\psi'=\,q\,}\, \frac{1}{3}\mathcal{O}_{RR1}+\frac{1}{3}\mathcal{O}_{RR2}+2\mathcal{O}_{RR4}\,, \nn\\
		\tilde{\mathcal{O}}_{4S1}^{\psi\psi'} &\equiv (\bar\psi_{L}U\psi_{R})(\bar\psi'_{L}U\psi'_{R})+\text{h.c. }  \,\xrightarrow{\,\psi,\psi'=\,q\,}\, \mathcal{O}_{FY1}+\mathcal{O}_{FY3}+2\mathcal{O}_{ST5}+\text{h.c.}\,, \nn\\
		\tilde{\mathcal{O}}_{4S2}^{\psi\psi'} &\equiv (\bar\psi_{L}U\psi_{R})(\bar\psi'_{R}U^{\dagger}\psi'_{L}) \,\xrightarrow{\,\psi,\psi'=\,q\,}\, \mathcal{O}_{FY5}+\mathcal{O}_{FY5}^{\dagger} -\frac{1}{12}(\mathcal{O}_{LR1}+\mathcal{O}_{LR3}) \nn\\
		& -\frac{1}{2}(\mathcal{O}_{LR2}+\mathcal{O}_{LR4})+\frac{1}{6}(\mathcal{O}_{LR12}-\mathcal{O}_{LR10})-\mathcal{O}_{LR11}+\mathcal{O}_{LR13}\,, \nn\\
		\tilde{\mathcal{O}}_{4S3}^{\psi\psi'} &\equiv (\bar\psi_{L}U T^{a}\psi_{R})(\bar\psi'_{L}U T^{a}\psi'_{R})+\text{h.c. }  \,\xrightarrow{\,\psi,\psi'=\,q\,}\, \mathcal{O}_{FY2}+\mathcal{O}_{FY4}\nn\\
		& +2\mathcal{O}_{ST7}+\text{h.c.}\,, \nn\\
		\tilde{\mathcal{O}}_{4S4}^{\psi\psi'} &\equiv (\bar\psi_{L}U T^{a}\psi_{R})(\bar\psi'_{R}U^{\dagger} T^{a}\psi'_{L}) \,\xrightarrow{\,\psi,\psi'=\,q\,}\, \mathcal{O}_{FY6}+\mathcal{O}_{FY6}^{\dagger} -\frac{1}{9}(\mathcal{O}_{LR1}+\mathcal{O}_{LR3})\,\nn\\
		& +\frac{1}{12}(\mathcal{O}_{LR2}+\mathcal{O}_{LR4})-\frac{5}{72}(\mathcal{O}_{LR12}-\mathcal{O}_{LR10})+\frac{1}{6}(\mathcal{O}_{LR11}-\mathcal{O}_{LR13})\,, \nn\\
		\tilde{\mathcal{O}}_{4T1}^{\psi\psi'} &\equiv (\bar\psi_{L}U\sigma_{\mu\nu}\psi_{R})(\bar\psi'_{L}U\sigma^{\mu\nu}\psi'_{R})+\text{h.c. }  \,\xrightarrow{\,\psi,\psi'=\,q\,}\, -\frac{20}{3}(\mathcal{O}_{FY1}+\mathcal{O}_{FY3})\nn\\
		& -16(\mathcal{O}_{FY2}+\mathcal{O}_{FY4})-8\mathcal{O}_{ST5}-\frac{16}{3}\mathcal{O}_{ST6}-32\mathcal{O}_{ST8}+\text{h.c.}\,, \nn\\
		\tilde{\mathcal{O}}_{4T2}^{\psi\psi'} &\equiv (\bar\psi_{L}U\sigma_{\mu\nu}T^{a}\psi_{R})(\bar\psi'_{L}U\sigma^{\mu\nu}T^{a}\psi'_{R})+\text{h.c. }  \,\xrightarrow{\,\psi,\psi'=\,q\,}\, -\frac{32}{9}(\mathcal{O}_{FY1}+\mathcal{O}_{FY3}) \nn\\
		& -\frac{4}{3}(\mathcal{O}_{FY2}+\mathcal{O}_{FY4})-\frac{64}{9}\mathcal{O}_{ST6}-8\mathcal{O}_{ST7}+\frac{16}{3}\mathcal{O}_{ST8}+\text{h.c.} \,.
		\end{align}
		As in the two-fermion intermediate operators, the last equation definitions stand for all the fermion types, although their exact correspondence with the MUC basis is only set for the quark-quark bilinear operators, as indicated with the arrows. The four-lepton and the two-lepton two-quark case are slightly different. Four-lepton operators, however, are not sensitive to operators including explicit $SU(3)_C$ generators, $T^a$, so they must be removed. In the mixed two-lepton and two-quark case, all the above 12 relations from eq.~(\ref{eq:19}) apply, but the number of operators increases since the bilinears can be lepton-quark or quark-lepton. Indeed, the number of operators in this case is 14 because all the relations are symmetric under $\psi \leftrightarrow \psi'$ but $\tilde \mO^{\psi\psi'}_{4LR1}$ and  $\tilde \mO^{\psi\psi'}_{4LR2}$.
				
		The conversion between the EWET and the MUC basis for four-fermion operators is set in table \ref{tab:dic4fermion}.

		%
		%
		\begin{table}[b] 
			\renewcommand{\arraystretch}{1.8}
			\begin{center}
				\begin{tabular}{|c|c|}
					\hline
					\multicolumn{2}{|c|}{\hspace{1cm} Four-fermion $P$-even operators --- lepton-lepton bilinears \hspace{1cm}} \\
					\hline
					\multicolumn{2}{c}{} \\	[-3ex] 
					\hline
					EWET & MUC 
					\\   \hline			
					$\mO^{\psi^2_l\psi^2_l}_1$ &
					$\tilde{\mathcal{O}}_{4S1}^{ll}+2\tilde{\mathcal{O}}_{4S2}^{ll}$ 
					\\ \hline
					%
					%
					$\mO^{\psi^2_l\psi^2_l}_2$ & 
					$- \tilde{\mathcal{O}}_{4S1}^{ll}+2\tilde{\mathcal{O}}_{4S2}^{ll} $ 
					\\ \hline 
					$\mO^{\psi^2_l\psi^2_l}_3$ &
					$\tilde{\mathcal{O}}_{4LL1}^{ll}+\tilde{\mathcal{O}}_{4RR1}^{ll}+2\tilde{\mathcal{O}}_{4LR1}^{ll}  $ 
					\\   \hline
					$\mO^{\psi^2_l\psi^2_l}_4$ &
					$\tilde{\mathcal{O}}_{4LL1}^{ll}+\tilde{\mathcal{O}}_{4RR1}^{ll}-2\tilde{\mathcal{O}}_{4LR1}^{ll}  $ 
					\\ \hline
					$\mO^{\psi^2_l\psi^2_l}_5$ &
					$\tilde{\mathcal{O}}_{4T1}^{ll}  $ 	
					\\  \hline
					\multicolumn{2}{c}{} 
					\\ \hline
					\multicolumn{2}{|c|}{Four-fermion $P$-odd operators --- lepton-lepton bilinears} \\
					\hline
					\multicolumn{2}{c}{} \\	[-3ex] \hline 
					EWET & MUC 
					\\ \hline
					$\widetilde \mO^{\psi^2_l\psi^2_l}_1$ &
					$\tilde{\mathcal{O}}_{4RR1}^{ll}-\tilde{\mathcal{O}}_{4LL1}^{ll}  $ 
					\\   \hline
					\multicolumn{2}{c}{} \\	[-3ex]
				\end{tabular}
			\end{center}
		\end{table}
		%

		\renewcommand{\arraystretch}{1.8}
		\begin{table}[!ht]
			\centering
			\begin{tabular}[t]{|c|c|}
				\hline
				\multicolumn{2}{|c|}{Four-fermion $P$-even operators --- lepton-quark bilinears} \\
				\hline
				\multicolumn{2}{c}{}
				\\[-3ex] \hline
				EWET & MUC \\
				\hline
				$\mathcal{O}^{\psi^2_l\psi^2_q}_{1}$ & $-\frac{1}{6}\tilde{\mO}^{lq}_{4S1}-\tilde{\mO}^{lq}_{4S3}-\frac{1}{24}\tilde{\mO}^{lq}_{4T1}-\frac{1}{4}\tilde{\mO}^{lq}_{4T2}-\frac{1}{3}\tilde{\mO}^{lq}_{4LR1}-2\tilde{\mO}^{lq}_{4LR2} $ \\
				\hline
				$\mathcal{O}^{\psi^2_l\psi^2_q}_{2}$ & $\frac{1}{6}\tilde{\mO}^{lq}_{4S1}+\tilde{\mO}^{lq}_{4S3}+\frac{1}{24}\tilde{\mO}^{lq}_{4T1}+\frac{1}{4}\tilde{\mO}^{lq}_{4T2}-\frac{1}{3}\tilde{\mO}^{lq}_{4LR1}-2\tilde{\mO}^{lq}_{4LR2} $ \\
				\hline
				$\mathcal{O}^{\psi^2_l\psi^2_q}_{3}$ & $\tilde{\mO}^{lq}_{4S1}+2\tilde{\mO}^{lq}_{4S2}  $ \\
				\hline
				$\mathcal{O}^{\psi^2_l\psi^2_q}_{4}$ & $- \tilde{\mO}^{lq}_{4S1}+2\tilde{\mO}^{lq}_{4S2} $ \\
				\hline
				$\mathcal{O}^{\psi^2_l\psi^2_q}_{5}$ & $\frac{1}{3}\tilde{\mO}^{lq}_{4LL1}+\frac{1}{3}\tilde{\mO}^{lq}_{4RR1}+2\tilde{\mO}^{lq}_{4LL2}+2\tilde{\mO}^{lq}_{4RR2}-\frac{4}{3}\tilde{\mO}^{lq}_{4S2}-8\tilde{\mO}^{lq}_{4S4}  $ \\
				\hline
				$\mathcal{O}^{\psi^2_l\psi^2_q}_{6}$ & $\frac{1}{3}\tilde{\mO}^{lq}_{4LL1}+\frac{1}{3}\tilde{\mO}^{lq}_{4RR1}+2\tilde{\mO}^{lq}_{4LL2}+2\tilde{\mO}^{lq}_{4RR2}+\frac{4}{3}\tilde{\mO}^{lq}_{4S2}+8\tilde{\mO}^{lq}_{4S4}  $ \\
				\hline
				$\mathcal{O}^{\psi^2_l\psi^2_q}_{7}$ & $\tilde{\mO}^{lq}_{4LL1}+\tilde{\mO}^{lq}_{4RR1}+2\tilde{\mO}^{lq}_{4LR1}  $ \\
				\hline
				$\mathcal{O}^{\psi^2_l\psi^2_q}_{8}$ & $\tilde{\mO}^{lq}_{4LL1}+\tilde{\mO}^{lq}_{4RR1}-2\tilde{\mO}^{lq}_{4LR1}  $ \\
				\hline
				$\mathcal{O}^{\psi^2_l\psi^2_q}_{9}$ & $-2\tilde{\mO}^{lq}_{4S1}-12\tilde{\mO}^{lq}_{4S3}+\frac{1}{6}\tilde{\mO}^{lq}_{4T1}+\tilde{\mO}^{lq}_{4T2}  $ \\
				\hline
				$\mathcal{O}^{\psi^2_l\psi^2_q}_{10}$ & $\tilde{\mO}^{lq}_{4T1}  $ \\
				\hline
				\multicolumn{2}{c}{}
				\\	\hline
				\multicolumn{2}{|c|}{Four-fermion $P$-odd operators --- lepton-quark bilinears} \\
				\hline
				\multicolumn{2}{c}{}
				\\[-3ex] \hline
				EWET & MUC \\
				\hline
				$\mathcal{\widetilde O}^{\psi^2_l\psi^2_q}_{1}$ & $-\frac{1}{3}\tilde{\mO}^{lq}_{4LL1}-2\tilde{\mO}^{lq}_{4LL2}+\frac{1}{3}\tilde{\mO}^{lq}_{4RR1}+2\tilde{\mO}^{lq}_{4RR2} $ \\
				\hline
				$\mathcal{\widetilde O}^{\psi^2_l\psi^2_q}_{2}$ & $-\frac{1}{3}\tilde{\mO}^{ql}_{4LL1}-2\tilde{\mO}^{ql}_{4LL2}+\frac{1}{3}\tilde{\mO}^{ql}_{4RR1}+2\tilde{\mO}^{ql}_{4RR2} $ \\
				\hline
				$\mathcal{\widetilde O}^{\psi^2_l\psi^2_q}_{3}$ &$\tilde{\mO}^{lq}_{4RR1}-\tilde{\mO}^{lq}_{4LL1}  $ \\
				\hline
				$\mathcal{\widetilde O}^{\psi^2_l\psi^2_q}_{4}$ &$\tilde{\mO}^{ql}_{4RR1}-\tilde{\mO}^{ql}_{4LL1}  $ \\
				\hline
			\end{tabular}
			\label{tab:dic:fourfermi}
		\end{table}

		\renewcommand{\arraystretch}{1.8}
		\begin{table}[!ht]
			\centering
			\begin{tabular}[t]{|c|c|}
				\hline
				\multicolumn{2}{|c|}{Four-fermion $P$-even operators --- quark-quark bilinears} \\
				\hline
				\multicolumn{2}{c}{}
				\\[-3ex] \hline
				EWET & MUC \\
				\hline
				$\mathcal{O}^{\psi^2_q\psi^2_q}_{1}$ & $-\frac{1}{6}\tilde{\mO}^{qq}_{4S1}-\tilde{\mO}^{qq}_{4S3}-\frac{1}{24}\tilde{\mO}^{qq}_{4T1}-\frac{1}{4}\tilde{\mO}^{qq}_{4T2}-\frac{1}{3}\tilde{\mO}^{qq}_{4LR1}-2\tilde{\mO}^{qq}_{4LR2} $ \\
				\hline
				$\mathcal{O}^{\psi^2_q\psi^2_q}_{2}$ & $\frac{1}{6}\tilde{\mO}^{qq}_{4S1}+\tilde{\mO}^{qq}_{4S3}+\frac{1}{24}\tilde{\mO}^{qq}_{4T1}+\frac{1}{4}\tilde{\mO}^{qq}_{4T2}-\frac{1}{3}\tilde{\mO}^{qq}_{4LR1}-2\tilde{\mO}^{qq}_{4LR2} $ \\
				\hline
				$\mathcal{O}^{\psi^2_q\psi^2_q}_{3}$ & $\tilde{\mO}^{qq}_{4S1}+2\tilde{\mO}^{qq}_{4S2}  $ \\
				\hline
				$\mathcal{O}^{\psi^2_q\psi^2_q}_{4}$ & $- \tilde{\mO}^{qq}_{4S1}+2\tilde{\mO}^{qq}_{4S2} $ \\
				\hline
				$\mathcal{O}^{\psi^2_q\psi^2_q}_{5}$ & $\frac{1}{3}\tilde{\mO}^{qq}_{4LL1}+\frac{1}{3}\tilde{\mO}^{qq}_{4RR1}+2\tilde{\mO}^{qq}_{4LL2}+2\tilde{\mO}^{qq}_{4RR2}-\frac{4}{3}\tilde{\mO}^{qq}_{4S2}-8\tilde{\mO}^{qq}_{4S4}  $ \\
				\hline
				$\mathcal{O}^{\psi^2_q\psi^2_q}_{6}$ & $\frac{1}{3}\tilde{\mO}^{qq}_{4LL1}+\frac{1}{3}\tilde{\mO}^{qq}_{4RR1}+2\tilde{\mO}^{qq}_{4LL2}+2\tilde{\mO}^{qq}_{4RR2}+\frac{4}{3}\tilde{\mO}^{qq}_{4S2}+8\tilde{\mO}^{qq}_{4S4}  $ \\
				\hline
				$\mathcal{O}^{\psi^2_q\psi^2_q}_{7}$ & $\tilde{\mO}^{qq}_{4LL1}+\tilde{\mO}^{qq}_{4RR1}+2\tilde{\mO}^{qq}_{4LR1}  $ \\
				\hline
				$\mathcal{O}^{\psi^2_q\psi^2_q}_{8}$ & $\tilde{\mO}^{qq}_{4LL1}+\tilde{\mO}^{qq}_{4RR1}-2\tilde{\mO}^{qq}_{4LR1}  $ \\
				\hline
				$\mathcal{O}^{\psi^2_q\psi^2_q}_{9}$ & $-2\tilde{\mO}^{qq}_{4S1}-12\tilde{\mO}^{qq}_{4S3}+\frac{1}{6}\tilde{\mO}^{qq}_{4T1}+\tilde{\mO}^{qq}_{4T2}  $ \\
				\hline
				$\mathcal{O}^{\psi^2_q\psi^2_q}_{10}$ & $\tilde{\mO}^{qq}_{4T1}  $ \\
				\hline
				\multicolumn{2}{c}{}
				\\	\hline
				\multicolumn{2}{|c|}{Four-fermion $P$-odd operators --- quark-quark bilinears} \\
				\hline
				\multicolumn{2}{c}{}
				\\[-3ex] \hline
				EWET & MUC \\
				\hline
				$\mathcal{\widetilde O}^{\psi^2_q\psi^2_q}_{1}$ & $-\frac{1}{3}\tilde{\mO}^{qq}_{4LL1}-2\tilde{\mO}^{qq}_{4LL2}+\frac{1}{3}\tilde{\mO}^{qq}_{4RR1}+2\tilde{\mO}^{qq}_{4RR2} $ \\
				\hline
				$\mathcal{\widetilde O}^{\psi^2_q\psi^2_q}_{2}$ &$\tilde{\mO}^{qq}_{4RR1}-\tilde{\mO}^{qq}_{4LL1}  $ \\
				\hline
			\end{tabular}
			\caption{\small
				Conversion for the EWET and MUC bases for the four-fermion operators. The identification is done through the intermediate operators written in eq.~(\ref{eq:19}). There are three different particle sets: lepton-lepton, lepton-quark and quark-quark bilinears, organized in separated tables. For each set, P-even (P-odd) operators are shown in the upper (lower) block.}
			\label{tab:dic4fermion}
		\end{table}

		\chapter{Spin-1 field representations}
		
		Spin-1 quantum fields differ from other kinds of fields in the singular fact that there is not a unique form to describe them. The theory itself leaves some freedom in the representation of this sort of fields, which lead to different spin-1 formalisms, being the Proca vector field and the rank-2 antisymmetric tensor formalisms the most renowned ones. Their phenomenology and predictions, however, are expected to be exactly the same, since the selection of any of them as a field representation does not respond to any particular physical reason.
		
		In this appendix we prove, indeed, this last statement. Nevertheless, we first introduce some concepts and notation about the antisymmetric field representation, prior to justify the equivalence of both formalisms mentioned above.
		
		\section{Comparison between Proca and Antisymmetric formalisms} \label{app:spin1comparison}
		\markboth{APPENDIX E.\hspace{1mm} SPIN-1 FIELD REPRESENTATIONS}{E.1.\hspace{1mm} THE PROCA AND THE ANTISYMMETRIC FORMALISMS}
		
		While the Proca vector formalism describes spin-1 particles as vector fields, $R_\mu$, e.g.,the SM gauge fields; the Antisymmetric representation is a spin-1 field formalism formulated through spin-1 rank-2 tensor fields, antisymmetric under any Lorentz index permutation,\ie $R_{\mu\nu} = -R_{\nu\mu}$. The general kinetic Lagrangians of the Proca and the Antisymmetric \cite{Ecker:1988te,Gasser:1983yg} fields are\footnote{For simplicity, we compare both formalisms for a a general object $R$ which is a singlet under $SU(2)$ and $SU(3)$.} 
		\begin{align}
		\mL^{\text{Proca}}_{\hat R}  & =   -\Frac{1}{4}  \hat R_{\mu\nu}\,  \hat R^{\mu\nu}  \,-\, \Frac{1}{2} M_{R}^2\,\hat R_{\mu}\, \hat R^{\mu}\,, \nn\\
		\mL^{\text{Antisym.}}_R &= -\Frac{1}{2}\, \partial^\mu R_{\mu\nu}\,\partial_\lambda
		R^{\lambda\nu} \, + \, \Frac{1}{4}\, M_R^2\, R_{\mu\nu} R^{\mu\nu} \,
		\, ,
		\end{align}
		where $\hat R_{\mu\nu}\equiv \partial_\mu \hat R_\nu - \partial_\nu \hat R_\mu$  is the usual Proca field strength tensor. They yield the following EoM, respectively:
		\begin{align} \label{eq.EOMantisym}
		\partial_\mu \hat R^{\mu\nu} \,+\, M_R^2 \hat R^\nu & \,=\, 0\,, \nn\\
		\partial^\mu \partial_\lambda \, R^{\lambda\nu}\,
		- \, \partial^\nu \partial_\lambda \, R^{\lambda\mu}\,
		+ \, M_R^2 \, R^{\mu\nu}\, &= \, 0 \, .
		\end{align}
		In addition, taking the derivative in the last equation for the antisymmetric case displays also an interesting relation
		\be
		\partial_\mu \, (\partial^2+M_R^2)\, R^{\mu\nu}\, = \, 0\,  .
		\ee

		Prior to analyzing how the Proca and the Antisymmetric propagators are built, it is convenient to define the transverse and longitudinal Lorentz projectors, being respectively,
		\be \label{eq:propagators}
		P_T^{\mu\nu}(q)\,=\,g^{\mu\nu}\,-\,\frac{q^\mu q^\nu}{q^2}\,, \qquad\qquad
		P_L^{\mu\nu}(q)\,=\, \frac{q^\mu q^\nu}{q^2}\,.
		\ee
		On the one hand, we can use these elements in order to construct the Proca propagator, which can be written in an arbitrary gauge, $R_\xi$, like
		\bear
		i \Delta_R^{\mu\nu}(q) & =&
		\Frac{i}{M_R^2-q^2}\; P_T^{\mu\nu}(q) \,  + \,
		\Frac{i\, \xi}{\xi\, M_R^2 - q^2} \; P_L^{\mu\nu}(q)
		\nn\\[10pt]
		& = & \Frac{i}{M_R^2-q^2} \,
		\left\{ g^{\mu\nu} \, - \, \Frac{(\xi-1)\, q^2}{\xi\, M_R^2-q^2}\; P_L^{\mu\nu}(q)
		\right\} \, ,
		\eear
		just as the SM propagator of the electroweak gauge sources $\hat W_\mu$ and $\hat B_\mu$. In particular, taking $\xi\rightarrow \infty$ one recovers the unitary gauge usual formulation for the propagator for the Proca fields
		\bear
		i  \Delta_R^{\mu\nu}(q) & =&
		\Frac{i}{M_R^2-q^2}\; P_T^{\mu\nu}(q) \,  + \,
		\Frac{i }{M_R^2 } \; P_L^{\mu\nu}(q)
		\nn\\[5pt]
		& = & \Frac{i}{M_R^2-q^2} \,
		\left\{ g^{\mu\nu} \, - \, \Frac{q^2}{M_R^2}\; P_L^{\mu\nu}(q)  \right\} \, .
		\eear

		On the other hand, the Antisymmetric propagator requires the construction of some auxiliary four-index tensors, expressed in terms of the projectors in eq.~(\ref{eq:propagators}), and defined as
		\bear
		\label{eq.defA}
		\mA_{\mu\nu,\rho\sigma}(q) &\equiv &\Frac{1}{2q^2} \,
		\left[\, g_{\mu\rho}q_\nu q_\sigma -
		g_{\rho\nu}q_\mu q_\sigma - (\rho \leftrightarrow \sigma) \,\right]
		\,=\, \Frac{1}{2}\, P_T^{\mu\rho}(q)\,  P_L^{\nu\sigma}(q) \nn\\
		&&\hspace{-1.5cm}
		-\, \Frac{1}{2}\, P_T^{\mu\sigma}(q)\,  P_L^{\nu\rho}(q)
		- \Frac{1}{2}\, P_T^{\nu\rho}(q)\,  P_L^{\mu\sigma}(q)
		+ \Frac{1}{2}\, P_T^{\nu\sigma}(q)\,  P_L^{\mu\rho}(q)
		\nn\\
		&&\hspace{-1.5cm} =
		\, \Frac{1}{2}\, g^{\mu\rho}\,  P_L^{\nu\sigma}(q)\,
		\, -\, \Frac{1}{2}\, g^{\mu\sigma}\,  P_L^{\nu\rho}(q)\,
		\, -\, \Frac{1}{2}\, g^{\nu\rho}\,  P_L^{\mu\sigma}(q)\,
		\, +\, \Frac{1}{2}\, g^{\nu\sigma}\,  P_L^{\mu\rho}(q)\,
		\, ,
		\nn\\[8pt]
		\Omega_{\mu\nu,\rho\sigma}(q) &\equiv & -\,\Frac{1}{2q^2}\,
		\left[ g_{\mu\rho}\, q_\nu q_\sigma -
		g_{\rho\nu}\, q_\mu q_\sigma  - q^2\, g_{\mu\rho}g_{\nu\sigma} -
		(\rho\leftrightarrow\sigma) \,\right]
		\nn\\
		&&\hspace{-1.5cm} =
		\, \Frac{1}{2} \, P_T^{\mu\rho}(q)\,  P_T^{\nu\sigma}(q)\,
		\, -\, \Frac{1}{2} \, P_T^{\mu\sigma}(q)\,  P_T^{\nu\rho}(q)\,
		\, ,
		\nn\\[8pt]
		\mI_{\mu\nu,\rho\sigma} &\equiv & \Frac{1}{2}\, \left(
		g_{\mu\rho}\, g_{\nu\sigma} - g_{\mu\sigma}\, g_{\nu\rho}\right) \, .
		\eear
		Notice that the three tensors are antisymmetric under the exchanges $\mu \leftrightarrow \nu$ and $\rho \leftrightarrow \sigma$. Besides, these elements play an analogous role in the Antisymmetric formalism as the conventional projectors do in the Proca case. They satisfy, indeed, the projector algebra:
		\be
		\Omega\cdot \mA\, =\, \mA\cdot \Omega\, =\, 0 \,, \qquad
		\mA\cdot \mA\, = \,\mA\,,\qquad
		\Omega \cdot \Omega\, = \,\Omega \, , \qquad
		\mA \,+\,\Omega\, = \,\mI \, ,
		\ee
		and also
		\be
		q^\mu\,\Omega_{\mu\nu,\rho\sigma}(q) \, = \,
		q^\nu\,\Omega_{\mu\nu,\rho\sigma}(q) \, = \,
		q^\rho\,\Omega_{\mu\nu,\rho\sigma}(q) \, = \,
		q^\sigma\,\Omega_{\mu\nu,\rho\sigma}(q) \, = \, 0\, .
		\ee
		Hence, we are able to characterize properly the Antisymmetric propagator in terms of these quantities:
		\bear \label{eq:Apropagator}
		\bra R^{\mu\nu}R^{\rho\sigma}\ket_F \, &=& \,i\,
		\Delta^{\mu\nu,\rho\sigma}(q)  \,= \,
		\, \Frac{2i}{M_R^2-q^2} \,
		\mA^{\mu\nu,\rho\sigma}(q)
		\, + \, \Frac{2i}{M_R^2} \, \Omega^{\mu\nu,\rho\sigma}(q)
		\nn\\[1ex]
		& =& \, \Frac{2i}{M_R^2-q^2}\,\left\{ \mI^{\mu\nu,\rho\sigma}
		-\Frac{q^2}{M_R^2}\;\Omega(q)^{\mu\nu,\rho\sigma}\right\}
		\nn\\[1ex]
		& =& \, \Frac{2i}{M_R^2} \; \mI^{\mu\nu,\rho\sigma} \, +\,
		\Frac{2i}{M_R^2-q^2}\, \Frac{q^2}{M_R^2}\,\mA(q)^{\mu\nu,\rho\sigma} \, .
		\eear

		Finally, polarization vectors in the Antisymmetric formalism, $\epsilon_{(i)}^{\mu\nu}(p)$, can be related to the standard Proca polarization vectors, $\epsilon_{(i)}^\mu(p)$. Therefore, given a momentum $p$ and a polarization $i$, the outgoing spin-1 matrix element in the Antisymmetric representation reads
		\be
		\bra 0\, |\, R^{\mu\nu}\, |\, R(p,\epsilon_{_{(i)}})\ket \,
		= \, \epsilon^{\mu\nu}_{_{(i)}}(p)\; = \;
		\Frac{i}{M_R} \, \left[ p^\mu\epsilon_{_{(i)}}^\nu(p)
		-p^\nu\epsilon_{_{(i)}}^\mu(p)\right] \, .
		\ee
		The products of two polarization vectors in the Proca and the Antisymmetric representations, summed over all the physical polarizations\footnote{Recall that $\epsilon_{(i)}(p) \cdot p = 0$ and $p^2=M_R^2$.} are found to be, respectively,
		\begin{align}
		\sum_{i=1,2,3}\epsilon^{\alpha}_{_{(i)}}(p)\,\epsilon^{\beta}_{_{(i)}}(p)^*\, &=\,\left(-g^{\alpha\beta}+\Frac{p^\alpha p^\beta}{M_V^2} \right) \nn\\[1ex]
		\sum_{i=1,2,3} \, \epsilon^{\mu\nu}_{_{(i)}}(p)\, \epsilon^{\rho\sigma}_{_{(i)}}(p)^* \,
		&= \, - \,
		2 \, \mA(p)^{\mu\nu,\rho\sigma}\, .
		\end{align}

		\section{Spin-1 formalism equivalence} \label{app:spin1equivalence}
		
		For some time, it has been controversial which field representations (if any) should be used for computing a given physical process involving spin-1 particles. Naively, calculations show quite different results and lead to an incompatible phenomenology. Nevertheless, it is clear that an unphysical selection for spin-1 field representation cannot generate a distinct outcome for any physical observable. 
		
		We prove the equivalence of the Proca and the Antisymmetric formalism at the path integral level, showing that it is possible to relate them with a change of variables in the generating functional. Without losing any generality, we consider a massive resonance state in the Proca representation, $\hat R^8_3$, being a triplet under $SU(2)_{L+R}$ and an octet under the color $SU(3)_C$ group. It will be easily generalized to the rest of resonance states afterwards. As indicated in eq.~(\ref{eq:formalRLagrangian}), a given high-energy Lagrangian can be split in a resonance and a non-resonance Lagrangian. In particular, 
		\begin{align} \label{eq:fullProcaAntisymL}
		\mL^{\rm (P)}[\hat{R}^8_3,\phi_j] \; =\;
		\mL^{\rm (P)}_{\hat{R}}[\hat{R}^8_3,\phi_j] \; +\; \mL^{\rm (P)}_{\text{non-R}}[\phi_j] \, , \nn\\
		\mL^{\rm (A)}[R^8_3,\phi_j] \; =\;
		\mL^{\rm (A)}_{R}[R^8_3,\phi_j] \; +\; \mL^{\rm (A)}_{\text{non-R}}[\phi_j] \, ,
		\end{align}
		where $\phi_j$ include all the light fields considered in the EWET and being
		\begin{align} \label{eq:LagP-R}
		\mL^{\rm (P)}_{\hat{R}}[\hat{R}^8_3,\phi_j]\, &=\, -\Frac{1}{4}\,\bra \hat{R}^8_{3\,\mu\nu}\, \hat{R}^{8\,\mu\nu}_3 \ket_{2,3}
		\,+\, \Frac{1}{2}\, M_{R^8_3}^2\, \bra \hat{R}^8_{3\,\mu} \hat{R}^{8\,\mu}_3\ket_{2,3}
		\nn\\
		& +\,  \bra \hat{R}^8_{3\,\mu}\, \hat{\chi}_{\hat{R}^8_3}^\mu\,+\, \hat{R}^8_{3\,\mu\nu}\, \hat{\chi}_{\hat{R}^8_3}^{\mu\nu} \ket_{2,3}
		\, , \nn\\[1ex]
		\mL^{\rm (A)}_{R}[R^8_3,\phi_j]\, &=\, -\Frac{1}{2}\, \nabla^\mu R^8_{3\,\mu\nu}\,\nabla_\lambda
		R^{8\,\lambda\nu}_3 \, + \, \Frac{1}{4}\, M_{R^8_3}^2\, R^8_{3\,\mu\nu} R^{8\,\mu\nu}_3 \nn\\
		& +\,  \bra  R^8_{3\,\mu\nu}\, {\chi'}_{R^8_3}^{\,\mu\nu} \ket_{2,3}
		\, ,
		\end{align}
		with $\hat R^8_{3\,\mu\nu} = \nabla_\mu \hat{R}^8_{3\,\nu} -\nabla_\nu \hat{R}^8_{3\,\mu}$ the resonance Proca strength field tensor and 
		\be
		{\chi'}_{R^8_3}^{\,\mu\nu}= \chi_{R^8_3}^{\mu\nu} \, + \, \Frac{1}{2} (\nabla^\mu \chi_{R^8_3}^\nu \,-\, \nabla^\nu \chi_{R^8_3}^\mu )
		\ee
		a compact structure for all Antisymmetric resonance interactions.
		
		In order analyze the equivalence between this Lagrangian and the Antisymmetric one, we write the generating functional of the light fields where the Proca $\hat R^8_3$ fields are explicitly integrated,
		\begin{align} \label{eq:genfun1}
		Z[\phi_j] \,&=\, \mN \, \Int [d\hat{R}^8_3]\;
		\exp\left\{
		i\Int{\rm d^d x}\,\,  \mL^{\rm (P)}[\hat{R}^8_3,\phi_j]
		\right\}
		\nn\\[5pt]
		&=\,  \mN' \, \Int [dR^8_3] \,[d\hat{R}^8_3]\;
		\exp\left\{
		i\Int{\rm d^d x}\,
		\left( \mL^{\rm (P)}[\hat{R}^8_3,\phi_j]  \,+\, \Frac{1}{4} \bra R^8_{3\,\mu\nu} R^{8\,\mu\nu}_3 \ket_{2,3}
		\right)\right\} \, . \nn\\
		\end{align}
		In the second line, there has been added the Antisymmetric term $\bra R^8_{3\,\mu\nu} R^{8\,\mu\nu}_3 \ket_{2,3}$, which only alter the identity by incorporating a global factor submitted in $\mN'$. After this step, the next change of variables, analogous to \cite{Bijnens:1995,Kampf:2006}, is required
		\be
		R^{\mu\nu} \longrightarrow M_{R^8_3} R^{8\,\mu\nu}_3 -\hat{R}^{8\,\mu\nu}_3
		+  2 \left(\hat{\chi}_{\hat{R}^8_3}^{\mu\nu}-\frac{1}{2} \bra \hat{\chi}_{\hat{R}^8_3}^{\mu\nu}\ket_2 -\frac{1}{3}\bra \hat{\chi}_{\hat{R}^8_3}^{\mu\nu}\ket_3 + \frac{1}{6} \bra \hat{\chi}_{\hat{R}^8_3}^{\mu\nu}\ket_{2,3}\right)\,. \quad
		\ee
		When applied to eq.~(\ref{eq:genfun1}) it yields
		\begin{align}
		Z[\phi_j] \,&=\,  \mN'' \, \Int [dR^8_3] \,[d\hat{R}^8_3]\,
		\exp\Bigg\{
		i\Int{\rm d^d x}\,
		\bigg( \Delta \mL^{\rm (A)}[R^8_3,\phi_j]  \,+\, \Frac{M_{R^8_3}^2}{2}\, \bra \hat{R}^8_{3\,\mu} \hat{R}^{8\,\mu}_3\ket_{2,3} \nn\\
		& +\, \bra \hat{R}^8_{3\,\mu} \mJ^\mu \ket_{2,3}\bigg)\Bigg\}\,,
		\end{align}
		provided one introduces the source $\mJ^\mu$ \cite{Kampf:2006} and identifies the remaining piece as a contribution to the kinetic Antisymmetric Lagrangian plus some non-resonance terms, being
		\begin{align}
		\mJ^\mu &= \hat{\chi}_{\hat{R}^8_3}^\mu\,  +\, M_{R^8_3}\, \nabla_{\nu} R^{8\,\nu\mu}_3 \, , \nn\\[5pt]
		\Delta \mL^{\rm (A)}[R^8_3,\phi_j] &=\,
		\Frac{1}{4}\, M_{R^8_3}^2\, \bra R^8_{3\,\mu\nu} R^{8\,\mu\nu}_3\ket_{2,3} \, +\, M_{R^8_3}\, \bra R^8_{3\,\mu\nu} \hat{\chi}^{\mu\nu}_{\hat{R}^8_3}\ket_{2,3}
		\,  \nn\\ 
		& \hspace{-1.5cm}+\, \bigg( \bra \hat{\chi}_{\hat{R^8_3}\,\mu\nu}\, \hat{\chi}_{\hat{R^8_3}}^{\mu\nu}\ket_{2,3}\,-\,\Frac{1}{2}\,\bra \bra  \hat{\chi}_{\hat{R^8_3}\,\mu\nu}\ket_2
		\bra  \hat{\chi}^{\mu\nu}_{\hat{R^8_3}}\ket_2 \ket_3  -\,\Frac{1}{3}\,\bra\bra  \hat{\chi}_{\hat{R^8_3}\,\mu\nu}\ket_3
		\bra  \hat{\chi}^{\mu\nu}_{\hat{R^8_3}}\ket_3 \ket_2 \nn\\
		& \hspace{-1.5cm} +\,\Frac{1}{6}\,\bra  \hat{\chi}_{\hat{R^8_3}\,\mu\nu}\ket_{2,3}\bra  \hat{\chi}_{\hat{R^8_3}}^{\mu\nu}\ket_{2,3}\bigg) \, .
		\end{align}
		At this point, it is possible to perform a Gaussian integration in the path integral over $\hat R^8_3$ and recover the full Antisymmetric Lagrangian from eq.~(\ref{eq:fullProcaAntisymL}) (resonance and non-resonance contributions),
		\begin{align}
		Z[\phi_j] \,&=\, \tilde{\mN} \Int [dR^8_3]\,
		\exp\Bigg\{
		i\Int{\rm d^d x}
		\Bigg( \Delta  \mL^{\rm (A)}[R^8_3,\phi_j] - \Frac{1}{2 M_R^2}\,
		\bigg[\bra \mJ_\mu \mJ^\mu \ket_{2,3}  - \Frac{1}{2}\bra \mJ_\mu \ket^2_{2,3} \nn\\
		& - \Frac{1}{2}\bra \mJ_\mu \ket^2_{2,3}\bigg]
		\Bigg)\Bigg\} \, =\, \widetilde{\mN} \; \Int [dR^8_3]\;
		\exp\left\{
		i\Int{\rm d^d x} \,\, \mL^{\rm (A)}[R^8_3,\phi_j]
		\right\} \, ,
		\end{align}
		where 
		\begin{align} \label{eq:equivalencenonresonantap}
		\displaystyle \mL^{\rm (A)}_{\text{non-R}} & \,=\,
		\left( \bra \hat{\chi}_{\hat{R}^8_3\, \mu\nu}\,
		\hat{\chi}_{\hat{R}^8_3}^{\mu\nu}\ket_{2,3}
		\, -\, \Frac{1}{2}\, \bra \bra  \hat{\chi}_{\hat{R}^8_3}^{\mu\nu}\ket_2
		\bra  \hat{\chi}_{\hat{R}^8_3 \,\mu\nu}^{\phantom{\mu}}\ket_2 \ket_3 \right. 
		\nn\\
		\displaystyle 
		& \qquad \,-\, \left. \Frac{1}{3}\, \bra \bra  \hat{\chi}_{\hat{R}^8_3}^{\mu\nu}\ket_3 
		\bra  \hat{\chi}_{\hat{R}^8_3 \,\mu\nu}^{\phantom{\mu}}\ket_3 \ket_2 \,+\, \Frac{1}{6}\,\bra \hat{\chi}_{\hat{R}^8_3\, \mu\nu}\ket_{2,3}^2 \right)
		\nn\\
		\displaystyle 
		& \,-\, 
		\Frac{1}{2M_{R^8_3}^2}\, \left(  \bra  \hat{\chi}_{\hat{R}^8_3}^{\mu}\,
		\hat{\chi}_{\hat{R}^8_3\,\mu}^{\phantom{\mu}} \ket_{2,3}
		-\Frac{1}{2}\, \bra \bra \hat{\chi}_{\hat{R}^8_3}^{\mu}\ket_2
		\bra \hat{\chi}_{\hat{R}^8_3\, \mu}^{\phantom{\mu}}\ket_2 \ket_3 \right. 
		\nn\\
		\displaystyle 
		& \qquad \,-\, \left.  \Frac{1}{3}\, \bra \bra  \hat{\chi}_{\hat{R}^8_3}^{\mu}\ket_3 
		\bra  \hat{\chi}_{\hat{R}^8_3 \,\mu}^{\phantom{\mu}}\ket_3 \ket_2 \,+\, \Frac{1}{6}\,\bra \hat{\chi}_{\hat{R}^8_3\, \mu}\ket_{2,3}^2 \right) 
		\,\, +\,\, \mL^{\rm (P)}_{\text{non-R}}
		\, .
		\end{align}
		Finally, in order to properly recover the resonance contribution for the Antisymmetric Lagrangian, the following identification is made
		\begin{align} \label{eq:equivalencetensorap}
		\displaystyle {\chi'}^{\,\mu\nu}_{R^8_3} \,& = \,  \chi_{R^8_3}^{\mu\nu} \, + \, \Frac{1}{2} \left(\nabla^\mu \chi_{R^8_3}^\nu \,-\, \nabla^\nu \chi_{R^8_3}^\mu\right) \nn\\
		& = \, 
		\Frac{1}{2 M_{R^8_3}} \left(\nabla^\mu \hat{\chi}_{\hat{R}^8_3}^\nu
		\,-\, \nabla^\nu \hat{\chi}_{\hat{R}^8_3}^\mu\right)
		\, +\, M_{R^8_3}\, \hat{\chi}_{\hat{R}^8_3}^{\mu\nu}  \,.
		\end{align}

		Hence, the Proca and the Antisymmetric formalisms are proved to be equivalent. Nonetheless, the conversion is not only requested for the high-energy resonance Lagrangian, but the non-resonant contribution too. 
		
		The results of eq.~(\ref{eq:equivalencetensorap}) and eq.~(\ref{eq:equivalencenonresonantap}) can be easily generalized to all the types of spin-1 heavy states, being vector or axial-vector in whichever n-plet representation under the custodial symmetry and the color group. The general resonance interaction tensor and the non-resonant Lagrangian equivalences between the Proca and the Antisymmetric formalisms are found to be, respectively,
		\begin{align}
		& \displaystyle \chi_{R^m_n}^{\mu\nu} \, + \, \Frac{1}{2} \left(\nabla^\mu \chi_{R^m_n}^\nu \,-\, \nabla^\nu \chi_{R^m_n}^\mu\right) &  & &\nn\\
		& \qquad = \, 
		\Frac{1}{2 M_{R^m_n}} \left(\nabla^\mu \hat{\chi}_{\hat{R}^m_n}^\nu
		\,-\, \nabla^\nu \hat{\chi}_{\hat{R}^m_n}^\mu\right)
		\, +\, M_{R^m_n}\, \hat{\chi}_{\hat{R}^m_n}^{\mu\nu} & \qquad
		& (R=V,\, A)\, , &
		\end{align}
		and
		\begin{align}
		\displaystyle \mL^{\rm (A)}_{\text{non-R}} & \,=\,
		\sum_{R=V,A} \left\{ \sum_{\hat R^1_1,\hat R^1_3,\hat R^8_1,\hat R^8_3} \left[ \left( \bra \hat{\chi}_{\hat{R}^m_n\, \mu\nu}\,
		\hat{\chi}_{\hat{R}^m_n}^{\mu\nu}\ket_{2,3}
		\, -\, \Frac{1}{2}\, \bra \bra  \hat{\chi}_{\hat{R}^m_n}^{\mu\nu}\ket_2
		\bra  \hat{\chi}_{\hat{R}^m_n \,\mu\nu}^{\phantom{\mu}}\ket_2 \ket_3 \right. \right. \right.
		\nn\\
		\displaystyle 
		& \qquad \,-\, \left. \Frac{1}{3}\, \bra \bra  \hat{\chi}_{\hat{R}^m_n}^{\mu\nu}\ket_3 
		\bra  \hat{\chi}_{\hat{R}^m_n \,\mu\nu}^{\phantom{\mu}}\ket_3 \ket_2 \,+\, \Frac{1}{6}\,\bra \hat{\chi}_{\hat{R}^m_n\, \mu\nu}\ket_{2,3}^2 \right)
		\nn\\
		\displaystyle 
		& \,-\, 
		\Frac{1}{2M_{R^m_n}^2}\, \left(  \bra  \hat{\chi}_{\hat{R}^m_n}^{\mu}\,
		\hat{\chi}_{\hat{R}^m_n\,\mu}^{\phantom{\mu}} \ket_{2,3}
		-\Frac{1}{2}\, \bra \bra \hat{\chi}_{\hat{R}^m_n}^{\mu}\ket_2
		\bra \hat{\chi}_{\hat{R}^m_n\, \mu}^{\phantom{\mu}}\ket_2 \ket_3 \right. 
		\nn\\
		\displaystyle 
		& \qquad \,-\, \left.  \Frac{1}{3}\, \bra \bra  \hat{\chi}_{\hat{R}^m_n}^{\mu}\ket_3 
		\bra  \hat{\chi}_{\hat{R}^m_n \,\mu}^{\phantom{\mu}}\ket_3 \ket_2 \,+\, \Frac{1}{6}\,\bra \hat{\chi}_{\hat{R}^m_n\, \mu}\ket_{2,3}^2 \right) \bigg]
		\Bigg\}
		\,\, +\,\, \mL^{\rm (P)}_{\text{non-R}}
		\, .
		\end{align}

		\chapter{Short-distance constraints computation} \label{app:SD}
		
		The determination of the Wilson coefficients associated to the short-distance local Lagrangian, formally identical to the EWET LECs, can only be fixed by the requirement of the high-energy theory to behave properly in the short-distance regime. It implies the assumption of some properties that prevent the EFT to break at large energies, imposed over a set of Green functions. These particular calculations are both sensitive to the resonant and non-resonant high-energy Lagrangians in such a way that once the high-energy constraints are incorporated the Wilson coefficients of the local Lagrangian are fixed.
		
		This procedure is performed for every high-energy coupling associated to a non-local operator. However, we only focus in those ones related to spin-1 heavy states due to the non-unique representation for these fields. Indeed, local contribution values depend strongly in the chosen representation, either Proca or Antisymmetric. Otherwise, spin-0 and fermion resonances do not bring the same ambiguity since they are unequivocally set.
		
		In this appendix, we analyze most of the Green-functions not computed explicitly in chapter \ref{ch:spin1}. There, we already studied the vector and axial-vector form functions, $\mmF^\mJ_{\phi\phi}(s)$, the fermion-Goldstone forward scattering, $\mM_{\psi\phi\to\psi\phi}$ and the generic Green function in eq.~(\ref{eq:4F-GreenP}) involving all single Lorentz index interactions.
		
		\section{Two-Goldstone scattering amplitudes}
		
		We consider a general scattering process involving four Goldstone bosons with polarizations $a,b,c,d$, respectively,
		\begin{align}
		T[\varphi^a (p_1)\, \varphi^b(p_2) \to \varphi^c(p_3)\, \varphi^d(p_4)]
		\, &=\,
		A(s,t,u)\,\delta_{ab}\delta_{cd} \,+\, A(t,s,u)\,\delta_{ac}\delta_{bd}
		\,\nn\\
		&+\, A(u,s,t)\,\delta_{ad}\delta_{bc} \, ,
		\end{align}
		where $s,t,u$ are the so-called Mandelstam variables. The underlying $SU(2)_{L+R}$ and the crossing symmetry, actually, allow us to write this computation in terms of the generic amplitude $A$. Depending on the selection of the Proca or the Antisymmetric formalism to describe the spin-1 resonance fields, the following results are obtained, respectively:
		\begin{align}
		A (s,t,u)^{\rm SDP}\, &= \,
		\Frac{g_{\hat V}^2}{ v^4 } \, \left[\, \Frac{ t\, (s^2-u^2) }{t-M_{V^1_3}^2}
		\, + \, \Frac{u\, (s^2-t^2)}{u-M_{V^1_3}^2}\,\right]
		\nn\\
		&+ \,
		\Frac{\widetilde{g}_{\hat A}^2}{ v^4 } \, \left[\, \Frac{ t\, (s^2-u^2) }{t-M_{A^1_3}^2}
		\, + \, \Frac{u\, (s^2-t^2)}{u-M_{A^1_3}^2}\,\right] \, + \,
		\Frac{2c_d^2}{v^4}\,\Frac{s^2}{M_{S_1^1}^2-s}  \, +\, \Frac{s}{v^2} 
		\nn\\  
		& +\,
		\Frac{4}{v^4}\,\left[ 2\,\mF_5^{\rm SDP}\, s^2 \, +\, \mF_4^{\rm SDP}\,  (t^2+u^2)\,\right] \, ,
		\nn\\[8pt]
		A (s,t,u)^{\rm SDA}\, &=\,
		\Frac{G_V^2}{v^4}\,\left[\, \Frac{s^2-u^2}{t-M_{V^1_3}^2} \,
		+\, \Frac{s^2-t^2}{u-M_{V^1_3}^2} \, \right]\, +\,
		\Frac{\widetilde{G}_A^2}{v^4}\,\left[\, \Frac{s^2-u^2}{t-M_{A^1_3}^2} \,
		+\, \Frac{s^2-t^2}{u-M_{A^1_3}^2}\,\right]
		\nn\\ 
		&+ \,
		\Frac{2c_d^2}{v^4}\,\Frac{s^2}{M_{S_1^1}^2-s} \, +\,
		\Frac{s}{v^2}  \, +\,   \Frac{4}{v^4}\,\left[ 2\,\mF_5^{\rm SDA}\, s^2 \, +\, \mF_4^{\rm SDA}\,  (t^2+u^2)\,\right]\, .
		\end{align}
		Notice that the resonance exchange contributions (proportional to $g_{\hat V}^2$, $g_{\hat A}^2$ and $G_V^2$, $G_A^2$ in each case, respectively) present a different high-energy behavior. As in the previous cases in section \ref{sec:SD}, this discrepancy is originated by the different internal structure of the resonance formalisms. While the Antisymmetric resonance contribution (and also the scalar contribution mediated by $c_d$) grows with energy as $E^2$ and it does not break the Froissart bound for the cross section, the Proca one exhibits an unacceptable growing proportional to $E^4$ which violates unitarity. Therefore, non-trivial short-distance local terms arise in this case in order to fix the short-distance regime. Likewise, the Antisymmetric values for their analogous local terms are forced to be zero, as well. They yield
		\begin{align} \label{eq:F45SD}
		\mF_4^{\rm SDP}\,&=\,
		\Frac{g_{\hat V}^2}{4} \, + \, \Frac{\widetilde g_{\hat A}^2}{4}\, , &
		\qquad\qquad
		\mF_5^{\rm SDP} \,&=\, -\, \Frac{g_{\hat V}^2}{4} \, - \, \Frac{\widetilde g_{\hat A}^2}{4}\,, & \nn\\
		\mF_{4}^{\rm SDA} \,&=\, 0\,, & \qquad  \mF_{5}^{\rm SDA}\,&=\,  0 \, .&
		\end{align}

		One can perform similar calculations for scattering processes involving the Higgs field, like $\varphi \varphi \to h h$ or $hh\to hh$. These amplitudes receive contributions from both the vector and axial-vector resonance exchanges and the short-distance local operators $\mO_{6,7}$ for the Goldstone-Higgs scattering and $\mO_8$ for the four-Higgs forward scattering, as well as other non-concerning contributions for this study. Once analyzed these amplitudes, the Proca and Antisymmetric short-distance coupling terms acquire the following values:
		\begin{align} \label{eq:F678SD}
		\mF_6^{\rm SDP}\,&=\,  -v^2( \widetilde{\lambda}_1^{hV\,\, 2} + \lambda_1^{hA\,\, 2})\, , &
		\quad\;
		\mF_7^{\rm SDP} \,&=\, v^2(\lambda_1^{hA\,\, 2} +  \widetilde{\lambda}_1^{hV\,\, 2}) \, &
		\quad\;
		\mF_8^{\rm SDP} \,&=\, 0 \,,& \nn\\
		\mF_{6}^{\rm SDA} \,&=\, 0\,, & \quad  \mF_{7}^{\rm SDA}\,&=\,  0 \, ,& \quad \mF_8^{\rm SDA} \,&=\, 0 \,.&
		\end{align}
		Notice that $\mF_8$ does not receive any contributions from the SD Lagrangian. Indeed, this is found to be the only EWET NLO operator not sensitive to the resonance interactions.
		
		\section{Two-point vector and axial-vector correlators}
		
		We construct the two-point Green functions for the current correlators, both the vector and the axial-vector ones, defined as
		\be
		i \int {\rm d}^4x\; {\rm e}^{iq(x-y)}\, \bra 0\,|\, T\left[ \mJ^{\mu}_{a} (x) \, \mJ'^{\,\nu}_{b}(y)^\dagger \right]  |\,0\ket =
		\delta_{ab}\, ( -g^{\mu\nu} q^2 + q^\mu q^\nu)\,\Pi_{\mJ \mJ'} (q^2)  \, , \vspace{2mm}
		\ee
		where $\mJ,\mJ'\in\{\mV,\mA\}$ stand for the chiral currents already introduced in eq.~(\ref{eq:VAcurrent}). The computation at LO for the vector, axial-vector and mixed correlators show
		\begin{align*} 
		\Pi_{\mV \mV}(q^2) &= \left\{ \bat
		\Frac{f_{\hat V}^2\, q^2}{M_{V^1_3}^2-q^2}
		\, + \, \Frac{\widetilde f_{\hat A}^2\, q^2}{M_{A^1_3}^2-q^2}
		\, - \, 2\,\left(\mF_1^{\rm SDP} + 2\, \mF_2^{\rm SDP}\right)\,,
		& \quad\quad\,\mbox{\small  (SDET-P)}\, ,
		\\[12pt]
		\Frac{F_V^2}{M_{V^1_3}^2-q^2} \, + \, \Frac{\widetilde F_A^2}{M_{A^1_3}^2-q^2}
		\, - \, 2\,\left(\mF_1^{\rm SDA} + 2\, \mF_2^{\rm SDA}\right)\,,
		& \quad\quad\,\mbox{\small  (SDET-A)}\, ,
		\ea\right.
		\end{align*}
		\begin{align*}
		\Pi_{\mA \mA}(q^2) &= \left\{ \bat 
		\Frac{f_{\hat A}^2\, q^2}{M_{A^1_3}^2-q^2}
		 +  \Frac{\widetilde f_{\hat V}^2\, q^2}{M_{V^1_3}^2-q^2}
		\,  - \, \Frac{v^2}{q^2}
		\, +\,  2\,\left(\mF_1^{\rm SDP} - 2\, \mF_2^{\rm SDP}\right)\,,
		& \,\mbox{\small  (SDET-P)}\, ,
		\\[12pt]
		\Frac{F_A^2}{M_{A^1_3}^2-q^2}  +  \Frac{\widetilde F_V^2}{M_{V^1_3}^2-q^2} \,
		- \, \Frac{v^2}{q^2}
		\,+ \, 2\,\left(\mF_1^{\rm SDA} - 2\, \mF_2^{\rm SDA}\right)\,,
		& \,\mbox{\small  (SDET-A)}\, ,
		\ea\right.
		\end{align*}
		\begin{align}\label{eq:PiVVAAVA}
		\Pi_{\mV \mA}(q^2) &= \left\{ \bat
		-\,\Frac{f_{\hat V}\,\widetilde f_{\hat V}\, q^2}{M_{V^1_3}^2-q^2}
		\, - \, \Frac{f_{\hat A}\, \widetilde f_{\hat A}\, q^2}{M_{A^1_3}^2 - q^2}
		\, + \, 4\,\widetilde \mF_2^{\rm SDP}\,,
		& \quad\qquad\qquad\qquad\,\mbox{\small  (SDET-P)}\, .
		\\[12pt]
		-\,\Frac{F_V\, \widetilde F_V}{M_{V^1_3}^2-q^2} \, -\, \Frac{F_A\, \widetilde F_A}{M_{A^1_3}^2 - q^2}
		\, + \, 4\,\widetilde \mF_2^{\rm SDA}\,,
		& \quad\qquad\qquad\qquad\,\mbox{\small  (SDET-A)}\, .
		\ea\right.
		\end{align}
		The quantity $\Pi_{\mV \mV}(q^2)-\Pi_{\mA \mA}(q^2)$ is an order parameter of EWSB. Since it can only be proportional to symmetry breaking parameters, this correlator difference is required to fulfill some soft conditions in order to keep an stable  high-energy regime. In particular, it must satisfy the unsubtracted dispersion relations introduced in eq.~(\ref{eq:dispersionrels}).
		As well, the correlator $\Pi_{\mV\mA}(q^2)$ is demanded to be zero at large energies, for similar reasons. These two SD constraints yield
		\begin{align}\label{eq:F12SD}
		\mF_1^{\rm SDP} \, &=\, -\frac{1}{4}\,\left(
		f_{\hat V}^2 -\widetilde f_{\hat V}^2 - f_{\hat A}^2 + \widetilde f_{\hat A}^{  2   } \right) \, ,&
		\widetilde\mF_2^{\rm SDP}  \, &=
		-\frac{1}{4}\,\left( f_{\hat V}\,\widetilde f_{\hat V} + f_{\hat A}\,\widetilde f_{\hat A}\right)\, , &
		\nn\\
		\mF_1^{\rm SDA} \,&=\, 0 \, ,& \qquad
		\widetilde\mF_2^{\rm SDA} \, &= 0 \, . &
		\end{align}

		Additionally, the vector and axial-vector correlators (alone) still present a different high-energy behavior. In principle, one could assume these functions to vanish in the UV, although this is not a necessary and model-independent requirement of the resonance theory.\footnote{Indeed, this condition is not fulfilled in QCD.} It would lead to 
		\begin{align} \label{eq:F2SD}
		\mF_2^{\rm SDP}  \,&=\,
		-\frac{1}{8}\,\left( f_{\hat V}^2 +\widetilde f_{\hat V}^2 + f_{\hat A}^2 +\widetilde f_{\hat A}^2\right)\, ,
		\nn\\
		\mF_2^{\rm SDA} \,&=\, 0 \, .
		\end{align}
		Otherwise, one should include a non-trivial contribution $\Delta\mF_2^{\rm SDP} = \mF_2^{\rm SDA}$ in order to stay as general as possible, not doing any assumption in regard to the background UV theory.
		
		\section{Fermion form factors}
		We analyze the following matrix elements 
		\be \label{eq:appFF}
		\bra \psi_f (p_1) \,|\, \mJ^\mu \,| \, \psi_f (p_2)\,  \ket\; =\;
		\bar{u}_f(p_1) \left[ \Gamma_\mJ^\mu\, \mathbb{F}^{\mJ\,f}_{1}(q^2)
		+  \frac{i}{2}\, q_\nu\, \sigma^{\mu\nu}\, \mathbb{F}^{\mJ\,f}_{2}(q^2)
		\right] u_f(p_2)\, ,
		\ee
		with $f=l,q$ and $\Gamma^\mu _{\mV_3,\,\mV_{(0)}}=\gamma^\mu$
		and $\Gamma^\mu_{\mA_3}=\gamma^\mu \gamma_5$ the Dirac structure associated to the corresponding fermion currents, $\mJ = \mV_3,\,\mA_3,\, \mV_{(0)}$, being 
		\be
		\mV_{(0)}^\mu \;\equiv\; \Frac{\partial S}{\partial \hat v^{(0)}_\mu}\, ,
		\qquad\qquad\qquad\qquad
		\hat v^{(0)}_\mu\; =\; \hat{X}_\mu  \, ,
		\ee
		whereas the vector and axial-vector ones are already defined in eq.~(\ref{eq:VAcurrent}). In particular, we will establish some short-distance constraints over the second form factor in eq.~(\ref{eq:appFF}), also called magnetic\footnote{It is also common to define this form factor with a different normalization $\mathbb{F}^\mJ_{2'} (q^2)= m_\psi \,\mathbb{F}^\mJ_{2} (q^2)$.} form factor. Taking $\mJ=\mV_3$ and assuming the massless limit, the resonance and non-resonance Lagrangians contribute to this element with
		\begin{align}
		\mathbb{F}^{\mV_3\,f}_{2}(s)  & = \left\{ \bat
		-4\sqrt{2}\; T^3_f \,
		\left( \Frac{f_{\hat{V}} c_{0\,f}^{\hat{V}} s }{M_{V^1_3}^2-s} +\Frac{\widetilde{f}_{\hat{A}} \widetilde{c}_{0\,f}^{\hat{A}} s}{M_{A^1_3}^2-s} \,-\, \sqrt{2}\, \mF_3^{\psi^2_f ,\, {\rm SDP}} \right)\,,
		& \,\quad\mbox{\small (SDET-P)} \, ,
		\\[10pt]
		-4\sqrt{2}\; T^3_f \,  \left( \Frac{F_V C_{0\,f}^V}{M_{V^1_3}^2-s} +\Frac{\widetilde{F}_A \widetilde{C}_{0\,f}^A}{M_{A^1_3}^2-s} \,-\, \sqrt{2} \,\mF_3^{\psi^2_f ,\, {\rm SDA}} \right)\,,
		& \,\quad\mbox{\small  (SDET-A)}\, .
		\ea\right.\quad
		\end{align}
		Well-behaved theories require the magnetic form factor to vanish at large momenta. Hence, the values of the local Lagrangian couplings are set to
		\begin{align}
		\mF_3^{\psi^2_f,\,\mathrm{SDP}} \,& = \, -\frac{1}{\sqrt{2}}\, \left( f_{\hat V}\,
		c_{0\,f}^{{\hat{V}}} \, + \, \widetilde f_{\hat A}\,
		\widetilde{c}_{0\,f}^{{\hat{A}}} \right) \, ,
		\nn \\
		\mF_3^{\psi^2_f,\,\mathrm{SDA}} \,& = \,  0\, ,
		\end{align}
		for either leptons and quarks, $f=l,q$.
		
		Analogously, a similar outcome is computed when changing the external fermion current $\mJ = \mA_3, \mV_{(0)}$. As in the previous case, their magnetic form factors are also demanded to vanish in the UV. Hence, it yields for the following two-fermion SD couplings 
		\begin{align}
		& \mF_3^{\psi^2_f} \,=\, \mF_3^{\psi^2_f,\,\mathrm{SDP}} \,,& \qquad 
		& \mF_4^{\psi^2_f} \,=\, \mF_4^{\psi^2_f,\,\mathrm{SDP}} \,,& \qquad
		& \widetilde \mF_1^{\psi^2_f} \,=\, \widetilde \mF_1^{\psi^2_f,\,\mathrm{SDP}} \,,& 
		\nn\\
		& \mF_3^{\psi^2_f,\,\mathrm{SDA}} \,=\, 0\,, & \qquad 
		& \mF_4^{\psi^2_f,\,\mathrm{SDA}} \,=\, 0\,, & \qquad
		& \widetilde \mF_1^{\psi^2_f,\,\mathrm{SDA}} \,=\, 0\,. &
		\end{align}

	\end{appendices}
	
	
	\chapter*{Resumen de la tesis}
	\addcontentsline{toc}{chapter}{Resumen de la tesis}
	\markboth{RESUMEN DE LA TESIS}{}
	
	El Modelo Est\'andar ({\it{Standard Model}}, SM) de f\'isica de part\'iculas representa el mejor marco te\'orico para describir el mundo subat\'omico. Su importancia radica en que nos ha permitido adentrarnos en un nuevo nivel m\'as profundo de la rea\-lidad f\'isica y del mundo que nos rodea. En el a\~no 2012, fue confirmada la \'ultima part\'icula fundamental que completaba el modelo, el bos\'on de Higgs, en el {\it{Large Hadron Collider}} (LHC) y en la actualidad no hay ninguna evidencia de ninguna nueva part\'icula ex\'otica ni de ninguna desviaci\'on significativa de alg\'un par\'ametro de la teor\'ia (las masas de los neutrinos pueden ser incluidas f\'acilmente). Adem\'as, son ya numerosos los modelos de altas energ\'ias que han sido descartados o que han sido pospuestos a energ\'ias superiores. Por otra parte, es bien sabido que el SM no es la teor\'ia definitiva de la f\'isica de part\'iculas. Hay todav\'ia algunas cuestiones esenciales a las cuales este modelo no puede dar respuesta.
	
	En lugar de buscar nueva f\'isica mediante la elecci\'on de alg\'un modelo en concreto o la detecci\'on directa de nuevos estados, esta tesis est\'a ideada desde un punto de vista independiente del modelo y trata de identificar cualquier fuente de nueva f\'isica de una manera indirecta, es decir, a trav\'es de discrepancias con respecto a las predicciones del SM. Es por ello que las teor\'ias efectivas de campos ({\it{effective field theories}}, EFTs) son la mejor estrategia para llevar a cabo este prop\'osito y su uso constituye uno de los pilares de este trabajo. Al no tener que realizar ninguna hip\'otesis sobre c\'omo se rige la f\'isica a m\'as altas energ\'ias ni qu\'e teor\'ia fundamental subyace, las conclusiones de esta tesis son v\'alidas para cualquier modelo espec\'ifico que satisfaga las premisas de la teor\'ia efectiva. Por lo tanto, no asumir ninguna condici\'on salvo que sea imprescindible y tener planteamientos generales es una prioridad.
	
	Otra de las motivaciones de esta tesis la encontramos en la analog\'ia que se establece con respecto a la teor\'ia quiral de perturbaciones ({\it{chiral perturbation theory}}, \chpt) y la teor\'ia quiral de resonancias ({\it{resonance chiral theory}}, R$\chi$T). Mientras que las interacciones fuertes est\'an regidas por la cromodin\'amica cu\'antica ({\it{quantum chromodynamics}}, QCD), su computaci\'on es t\'ecnicamente muy compleja y en algunos casos impracticable, especialmente en el r\'egimen de m\'as bajas energ\'ias. Ciertamente, es mucho m\'as conveniente emplear \chpt\ y R$\chi$T para describir las interacciones fuertes en este rango del espectro. En este escenario, los mesones pasan a ser los elementos primarios y las variables efectivas de la teor\'ia, en lugar de los quarks y gluones, y, de hecho, no es siquiera necesario asumir QCD para su formulaci\'on. En esta tesis nos planteamos si un escenario similar podr\'ia ser posible en el contexto de las interacciones electrod\'ebiles. En esta situaci\'on, el SM desempe\~na el papel de \chpt\ como teor\'ia de bajas energ\'ias, mientras que las resonancias electrod\'ebiles ser\'ian an\'alogas a R$\chi$T. Seguramente existe una teor\'ia ultravioleta (UV) que explica la f\'isica a\'un inexplorada, como hace QCD en la comparaci\'on; no obstante, este hecho no es relevante para el prop\'osito de este trabajo, ya que ni el an\'alisis de las resonancias ni las conclusiones obtenidas dependen de ella.
	
	Teniendo como sustrato las teor\'ias efectivas, nuestro punto de partida es la construcci\'on de la teor\'ia efectiva eletrod\'ebil ({\it{electroweak effective theory}}, EWET). Est\'a concebida como una EFT de car\'acter ampliamente general en la que est\'an incluidos todos los grados de libertad (ligeros) del SM, as\'i como las simetr\'ias que lo gobiernan y que dan lugar a las interacciones fuertes y electrod\'ebiles. La EWET est\'a constituida, en primer lugar, por quarks y leptones, aunque solo consideramos una \'unica generaci\'on. Los gluones, invariantes bajo el grupo de color $SU(3)_C$, median las interacciones fuertes; mientras que los bosones gauge electr\'odebiles tambi\'en est\'an incluidos y se transforman bajo el grupo de simetr\'ia extendido $SU(2)_L\otimes SU(2)_R \otimes U(1)_X$, donde $L$ y $R$ hacen referencia a las componentes quirales y $X\equiv (B-L)/2$, a la semidiferencia entre los n\'umeros bari\'onico y lept\'onico. Consideramos, adem\'as, un bos\'on escalar tipo Higgs de masa $m_h=125$ GeV, singlete bajo los grupos de simetr\'ia anteriores. A diferencia del bos\'on de Higgs del SM, este campo en la EWET no forma parte de un doblete de campos escalares complejos. De esta forma, no realizamos ninguna suposici\'on acerca de c\'omo est\'a acoplado esta part\'icula al modelo. La \'unica hip\'otesis que efectuamos es el patr\'on de ruptura de la simetr\'ia, consistente con la fenomenolog\'ia existente en la actualidad. El grupo de simetr\'ia electrod\'ebil se rompe espont\'aneamente tal que, $SU(2)_L\otimes SU(2)_R \rightarrow SU(2)_{L+R}$, dando lugar a tres bosones de Goldstone que, a su vez, dan lugar a las polarizaciones longitudinales y justifican la masa de los bosones gauge.
	
	La EWET est\'a organizada de acuerdo a lo que denominamos contaje quiral ({\it chiral power counting}), fundamentado en el comportamiento de los distintos campos cuánticos a bajas energ\'ias. Es por tanto posible asignar a cada operador que conforma la teor\'ia una dimensi\'on quiral, as\'i como tambi\'en a los denominados {\it building blocks}, es decir, a los elementos con los que se construyen los operadores. Este contaje permite establecer una jerarqu\'ia entre todas las interacciones a nivel de Lagrangiano y la renormalizaci\'on est\'a garantizada orden a orden. Un rasgo caracter\'istico de este contaje quiral es que ni el Higgs ni los bosones de Goldstone incrementan la dimensi\'on quiral. Este hecho implica, en primer lugar, que el bos\'on de Higgs no est\'a necesariamente d\'ebilmente acoplado al resto del SM y, en segundo lugar, que la EWET es no lineal con respecto a los Goldstones. 
	
	Para el prop\'osito de esta tesis, solo requerimos de aquellos operadores de m\'as baja dimension quiral, o {\it leading order} (LO), y los que inmediatamente les preceden en cuanto a dimension quiral, o {\it next-to-leading order} (NLO); que co\-rresponden a los Lagrangianos de orden $\hat d=2$ y $\hat d=4$, respectivamente. Adem\'as, solo se analizan operadores invariantes bajo la simetr\'ia discreta $CP$, ya que los que no satisfacen esta propiedad se encuentran considerablemente supri\-midos fenomenol\'ogicamente. Las interacciones LO, descritas en las ec.~(\ref{eq:LscalarEWET}) y ec.~(\ref{eq:L_SMinL_EWET}), constituyen en esencia el Lagrangiano del SM con un sector escalar extendido. Por otra parte, el Lagrangiano NLO, mostrado en la ec.~(\ref{eq:L4EWET}), est\'a formado por 12 (3) operadores puramente bos\'onicos con paridad par (impar), 15 (6) t\'erminos que incluyen 2 fermiones, y 25 (7) con 4 fermiones. El listado detallado lo podemos encontrar en las tablas \ref{tab:bosonic-Op4}, \ref{tab:2fermion} y \ref{tab:4fermion}. Cabe destacar que el n\'umero de elementos de estas listas no es trivial, ya que se ha realizado un importante esfuerzo para minimizar esta base. Para ello han sido precisas numerosas t\'ecnicas de simplificaci\'on y redefiniciones. En particular, hemos probado que todos los operadores con dimensi\'on quiral $\hat d=3$ pueden ser reabsorbidos o reexpresados en t\'erminos de los anteriores. Adicionalmente, mostramos una conversi\'on expl\'icita de los operadores de la EWET a otras bases ya existentes en la literatura \cite{Buchalla:2012qq,Buchalla:2013rka,Krause:2016uhw, Longhitano:1980iz, Longhitano:1980tm}.
	Un ejemplo de un operador de la EWET, concretamente de tipo puramente bos\'onico es el siguiente:
	\be \label{eq:example1conclusions}
	\mO_1 \,=\, \Frac{1}{4}\bra {f}_+^{\mu\nu} {f}_{+\, \mu\nu}
	- {f}_-^{\mu\nu} {f}_{-\, \mu\nu}\ket_2  \qquad \text{con}\qquad \mF_1\,\mO_1 \subset \mL^{\hat d=4}_{\text{NLO}}\,.
	\ee
	Los t\'erminos $f^{\mu\nu}_{\pm}$ son uno de los {\it building blocks} mencionados anteriormente. Este en particular tiene dimension quiral $\hat d=2$, por lo que es f\'acil deducir que el operador completo tiene dimensi\'on quiral $\hat d=4$ y, por lo tanto, pertenece al Lagrangiano NLO, $\mL^{\hat d=4}_{\text{NLO}}$.
	
	Todos los operadores est\'an parametrizados en t\'erminos de sus respectivos coeficientes de Wilson, tambi\'en denominados en este contexto constantes de acoplamiento de baja energ\'ia ({\it low energy coupling constants}, LECs), como es el caso de $\mF_1$ en el ejemplo previo de la ec.~\ref{eq:example1conclusions}. Estos par\'ametros se caracte\-rizan por ser los mejores indicadores indirectos en el rango de bajas energ\'ias de la presencia de estados masivos. De hecho, un evento es considerado de nueva f\'isica si alguna LEC exhibe una variaci\'on significativa (m\'as de cinco desviaciones est\'andar) con respecto a los valores predichos por el SM (cero en el caso de las LECs de orden $\mO_{\hat d}(p^4)$). En la literatura existente, son numerosos los trabajos en los que se analiza el patr\'on de las LECs para modelos espec\'ificos, especialmente aquellos donde el bos\'on de Higgs est\'a introducido linealmente a trav\'es de un doblete de campos escalares complejos y se encuentra d\'ebilmente acoplado. Entre ellos, los podemos encontrar tanto a nivel \'arbol \cite{Brehmer:2015rna, deBlas:2014mba, delAguila:2010mx, delAguila:2008pw, delAguila:2000rc, Corbett:2015lfa, Bar-Shalom:2014taa, Pappadopulo:2014qza}, como a un {\it loop} expandido en serie de potencias de alg\'un par\'ametro de peque\~no valor \cite{HLM:16, FRW:15, DEQ:15, Drozd:2015rsp, Huo:15, Huo:15b, dAKS:16, Boggia:2016asg, Fuentes-Martin:2016uol, BS:94}. En cambio, en esta tesis analizamos las contribuciones de las LECs en un entorno no lineal, ya que de esta manera no es necesario hacer ninguna hip\'otesis sobre c\'omo el bos\'on de Higgs y la nueva f\'isica en general est\'an acoplados al resto del SM.
	
	Con la mirada puesta en \chpt\ y en R$\chi$T como su extensi\'on con resonancias mes\'onicas en el contexto de las interacciones fuertes, construimos la teor\'ia de resonancias electrod\'ebiles en un marco de trabajo de car\'acter general, independiente de modelo y, a su vez, para ser una herramienta \'util para explorar nuevos estados masivos en el rango del TeV en adelante. En general, denominamos re\-sonancia a cualquier objeto de alta energ\'ia que est\'e vinculado de alguna forma a las interacciones electrod\'ebiles y que no est\'e predicho por el SM. Esto se manifiesta formalmente en la teor\'ia de resonancias a nivel de Lagrangiano con la inclusi\'on de los siguientes estados masivos: resonancias bos\'onicas de spin-0 y spin-0 con n\'umeros cu\'anticos $J^{PC}=0^{++},\,0^{-+},\,1^{--},\,1^{++}$ y resonancias fermi\'onicas de spin-(1/2). Adem\'as, todos estos objetos est\'an presentes en la teor\'ia formando tanto singletes como n-pletes bajo los grupos de simetr\'ia custodial, $SU(2)_{L+R}$, y de color, $SU(3)$. Cabe destacar que, a pesar de que el t\'ermino resonancia nos puede recordar a teor\'ias fuertemente acopladas, este trabajo tambi\'en es v\'alido para modelos d\'ebilmente acoplados, si bien es cierto que el enfoque quiz\'a es m\'as adecuado para los primeros. De hecho, las conclusiones de esta tesis son ge\-neralizables a cualquier teor\'ia que incluya resonancias. 
	
	El Lagrangiano de resonancias est\'a constituido por un conjunto extenso de operadores de alta energ\'ia en los que interact\'uan una \'unica resonancia con otros campos ligeros ya presentes en la EWET. Se incluyen, por lo tanto, todos aquellos operadores cuyas contribuciones a baja energ\'ia implican a los Lagrangianos LO y NLO de la EWET. Asimismo, las interacciones entre dos o m\'as estados pesados son procesos de mayor orden quiral al analizado en este trabajo y, por lo tanto, no son estudiadas. De manera similar a la teor\'ia de bajas energ\'ias, se ha realizado un esfuerzo notable para eliminar todas las redundancias y todas las dependencias que existen entre las distintas estructuras, con el prop\'osito de simplificar y reducir el Lagrangiano de resonancias en la medida de lo posible.
	
	El procedimiento para obtener las contribuciones de las resonancias en t\'ermi\-nos de la EWET es equivalente al empleado en \chpt\ \cite{Ecker:1988te, Ecker:1989yg, Cirigliano:2006hb}, por lo que utilizaci\'on es ya bien conocida. Formalmente, el m\'etodo se basa en integrar las resonancias de la acci\'on que rige la teor\'ia a altas energ\'ias. Una vez obtenidas las soluciones en primera aproximaci\'on de las ecuaciones de movimiento (EdM) de las resonancias, son sustituidas en el Lagrangiano de forma que el resultado queda expresado en funci\'on de campos ligeros \'unicamente. Es posible entonces conocer el orden quiral que deriva de las interacciones con estados pesados y proyectar sus efectos en los Lagrangianos LO y NLO de bajas energ\'ias.
	
	La forma gen\'erica de estas contribuciones a las LECs queda reflejada en la siguiente expresi\'on:
	\be
	\Delta \mF_i \, \sim \Frac{1}{M_R^2} \times g_1 \times g_2 \,,
	\ee
	donde $M_R$ es la masa de la resonancia en cuesti\'on y $g_1$ y $g_2$ son dos acoplamientos cualesquiera del Lagrangiano de altas energ\'ias que parametrizan sendas interacciones. As\'i pues, el patr\'on completo de estas estructuras configura la totalidad de las trazas de las resonancias en las LECs, es decir, la firma que los estados pesados rubrican en la teor\'ia de bajas energ\'ias y que advierten de su existencia en el espectro. El conjunto de todas las contribuciones a las LECs, tanto puramente bos\'onicas como aquellas con uno o dos pares fermi\'onicos se encuentran en las tablas \ref{tab:LECbosonic}, \ref{tab:LEC2fermion} y \ref{tab:LEC4fermionlepton}--\ref{tab:LEC4fermionquark}, respectivamente.
	
	La siguiente interacci\'on con una resonancia vectorial (spin-1) singlete de color y triplete bajo el grupo de simetr\'ia custodial, es uno de los ejemplos que se analizan en esta tesis: 
	\be \label{eq:example2conclusions}
	\Frac{1}{2\sqrt{2}}\,\bra ( F_{ V}\, f_+^{\mu\nu} + \widetilde{F}_{ V} \, f_-^{\mu\nu} )\,V^1_{3\,\mu\nu} \ket_2\,.
	\ee
	En particular, el campo pesado est\'a descrito en el formalismo Antisim\'etrico, mientras que 
	$F_V$ y $\widetilde F_V$ son los acoplamientos asociados a estas interacciones con\-cretas con este objeto masivo (con paridad par e impar, respectivamente). Una vez integrada la resonancia de la ecuaci\'on previa haciendo uso de su EdM, queda reemplazada por un conjunto de campos ligeros con dimensi\'on quiral $\hat d=2$. De esta manera, la ec.~(\ref{eq:example2conclusions}) queda reflejada en una estructura de dimensi\'on quiral $\hat d=4$, a su vez proporcional a combinaciones de los acoplamientos $F_V$ y $\widetilde F_V$. Al proyectar este resultado en la base de operadores NLO de la EWET, se obtiene que tres de las LECs son sensibles a estas interacciones:
	\bear \label{eq:example3conclusions}
	\Delta \mF_1\,=\, - \Frac{F_V^2-\widetilde{F}_V^2}{4M_{V^1_3}^2}\,, &&\qquad \qquad
	\Delta \mF_2\,=\,- \Frac{F_V^2+\widetilde{F}_V^2}{8M_{V^1_3}^2}\,, \nn\\
	\Delta \widetilde \mF_2 &=& - \Frac{F_V\,\widetilde{F}_V^2}{4M_{V^1_3}^2}\,.
	\eear
	El total de todas las contribuciones de alta energ\'ia a las LECs es, de hecho, la suma de todas las contribuciones individuales, obtenidas de manera an\'aloga a este ejemplo ilustrativo.
	
	Una de las partes centrales de esta tesis es el an\'alisis de estados pesados de spin-1, tanto vectores como vectores axiales (o axiales, simplemente). Al contrario que el resto de las resonancias, estos objetos se pueden implementar (y se implementan) en el modelo con distintas representaciones de campos cu\'anticos. Concretamente en este trabajo, nos centramos en los formalismos de vectores de Proca (Proca) y en el formalismo de tensores antisim\'etricos de rango 2 (Antisim\'etrico), que son a su vez los m\'as extendidos en la literatura. La utilizaci\'on de una u otra representaci\'on no es una decisi\'on trivial ya que aparentemente dan lugar a distintas predicciones sobre las LECs. De hecho, si consideramos la interacci\'on de la ec.~(\ref{eq:example2conclusions}) pero descrita en el formalismo de Proca, obtenemos que no contribuye al Lagrangiano de la EWET a NLO. Indudablemente, este suceso esconde un problema de fondo, pues la selecci\'on de la representaci\'on no es f\'isica sino una elecci\'on. Al igual que con este operador en particular, el pro\-blema persiste para todos los procesos que involucran resonancias de spin-1 a nivel \'arbol: las predicciones a baja energ\'ia de las resonancias dependen fuertemente del formalismo utilizado.
	
	En este trabajo mostramos que ambas representaciones de campos de spin-1 son equivalentes, como cab\'ia esperar, mediante un cambio de variables en la acci\'on a nivel de integral de camino. La confusi\'on generada es debida principalmente a la falsa premisa de que los Lagrangianos de interacci\'on de resonancias deben producir los mismos resultados, cuando esta es solo una parte de las involucradas en la resoluci\'on. En cambio, la equivalencia tiene lugar una vez considerada toda la estructura de la EFT: 
	\be \label{eq:equivalenceconclusions}
	\mL_R^{(P)} \, + \, \mL_{{\rm non}-R}^{(P)} \, = \, \mL_R^{(A)} \, + \, \mL_{{\rm non}-R}^{(A)} \,,
	\ee
	donde $P$ y $A$ hacen referencia a Proca y Antisim\'etrico; y $\mL_R$ y $\mL_{\text{non}-R}$
	a los Lagrangianos de interacci\'on con y sin resonancias de la teor\'ia de altas energ\'ias, respectivamente. Cabe destacar que, al igual que los Lagrangianos con resonancias son diferentes en ambas representaciones, tambi\'en lo son los Lagrangianos sin resonancias, ya que tienen distintos coeficientes de Wilson, como tambi\'en son distintos los acoplamientos de bajas energ\'ias, pues estamos tratando con distintas teor\'ias efectivas. Uno de los hitos de este trabajo es, en consecuencia, la determinaci\'on de manera expl\'icita de las identidades algebraicas que relacionan tanto los Lagrangianos de la ec.~(\ref{eq:equivalenceconclusions}) como sus constantes de acoplamiento. Una de las identidades que obtenemos es, por ejemplo, la establecida entre los acoplamientos $F_V$ and $\widetilde F_V$ del operador de la ec.~(\ref{eq:example2conclusions}) en formalismo Antisim\'etrico, con sus versiones duales en el Lagrangiano de Proca, $f_{\hat V}$ y $\widetilde f_{\hat V}$,
	\be \label{eq:equivalenceconclusions2}
	f_{\hat V} \,=\, \Frac{F_V}{M_{V^1_3}}\,, \qquad\qquad \widetilde f_{\hat V} \,=\, \Frac{\widetilde F_V}{M_{V^1_3}}\,,
	\ee
	donde en esta representaci\'on el operador contiene los mismos campos ligeros que en la ec.~(\ref{eq:equivalenceconclusions}) pero, en lugar de con $V^1_{3\,\mu\nu}$, interaccionan con el tensor $\nabla_\mu \hat V^1_{3\,\nu} - \nabla_\nu \hat V^1_{3\,\mu}$.
	
	Sin embargo, pese a que somos capaces de relacionar ambos formalismos, la teor\'ia de altas energ\'ias no est\'a un\'ivocamente fijada, ya que no podemos conocer a priori qu\'e contribuciones del Lagrangiano de altas energ\'ias sin resonancias intervienen en un determinado proceso que involucre a estos estados pesados. Para poder eliminar esta incertidumbre es necesaria la inclusi\'on de ciertas condiciones de distancias cortas ({\it short-distance constraints}). Se trata de un conjunto de requerimientos que asumimos con el fin de que la teor\'ia de altas energ\'ias sea consistente y se comporte correctamente en su l\'imite ultravioleta. Entre ellos destacan la conservaci\'on de la unitariedad o el comportamiento de los factores de forma y los correladores en el l\'imite de altas energ\'ias. Una vez incorporadas estas condiciones en nuestra teor\'ia de resonancias, concretamente sobre un conjunto determinado de funciones de Green, los coeficientes de Wilson de los Lagrangianos no resonantes de altas energ\'ias adquieren un valor exacto y, de esta manera, quedan fijados.
	
	Una de las condiciones de distancias cortas que computamos se impone directamente sobre la diferencia entre correladores (funciones de Green de dos puntos) cuando las fuentes externas son corrientes vectoriales o axiales, $\Pi_{\mV \mV}-\Pi_{\mA \mA}$. Para simplificar este ejemplo, vamos a restringir las contribuciones de estados de alta energ\'ia a la resonancia vectorial singlete bajo el grupo de color y triplete bajo simetr\'ia custodial, $V^1_3$, como en los casos anteriores. De esta manera, solo el operador $\mO_1$, definido en la ec.~(\ref{eq:example1conclusions}), interviene en el proceso, aunque en este caso en la teor\'ia de altas energ\'ias. La peculiaridad de la funci\'on de Green que estamos estudiando es justamente que es un par\'ametro de orden de la ruptura espont\'anea de la simetr\'ia electrod\'ebil ({\it electroweak symmetry breaking}, EWSB). Esto implica que el resultado del c\'alculo de dicha diferencia solo puede ser proporcional a par\'ametros de la EWSB, como el valor de expectaci\'on del vac\'io, $v$. En cambio, al incorporar las resonancias obtenemos
	\begin{align} \label{eq:SDconclusions1}
	(\Pi_{\mV \mV} - \Pi_{\mA \mA})(q^2) \,&=\, \Frac{v^2}{q^2}\, +\, \Frac{(F_V^2 \,-\,\widetilde F_V^2)}{M_{V^1_3}^2-q^2}\, -\,4\,\mF_1^{\rm SDA}\,,  &\qquad\quad \mbox{\small  (SDET-A)} \nn\\
	(\Pi_{\mV \mV} - \Pi_{\mA \mA})(q^2) \,&=\, \Frac{v^2}{q^2}\,+\, \Frac{(f_{\hat V}^2\,-\, \widetilde f_{\hat V}^2) \, q^2}{M_{V^1_3}^2-q^2} \,-\, 4\, \mF_1^{\text{SDP}}\,, & \qquad\quad \mbox{\small  (SDET-P)}	
	\end{align}
	en los formalismos Antisim\'etrico (SDET-A) y Proca (SDET-P), respectivamente, siendo $\Delta \mF^{\text{SDA}}_1$ y $\Delta \mF^{\text{SDP}}_1$ los coeficientes de Wilson del Lagrangiano sin resonancias (de altas energ\'ias). Si incorporamos el requisito de cortas distancias, estos dos par\'ametros se ven inmediatamente forzados a tomar los valores
	\be \label{eq:SDconclusions2}
	\Delta \mF^{\text{SDA}}_1\,=\,0\,, \qquad\qquad \Delta \mF^{\text{SDP}}_1\,=\, - \Frac{1}{4}\,(f_{\hat V}^2-\widetilde{f}_{\hat V}^2)\,=\, - \Frac{F_V^2-\widetilde{F}_V^2}{4M_{V^1_3}^2} \,,
	\ee
	donde, adicionalmente, hemos hecho uso de la ec.~(\ref{eq:equivalenceconclusions2}) en el t\'ermino de la derecha. Si comparamos precisamente esta expresi\'on con la ec.~(\ref{eq:example3conclusions}), concluimos que tanto el formalismo Antisim\'etrico como el de Proca dan lugar a las mismas predicciones para las LECs.
	
	A lo largo de este trabajo, demostramos que este \'ultimo resultado es extensible al resto de las LECs, de forma que las trazas de las resonancias a bajas energ\'ias son consistentes con independencia del formalismo empleado. Adem\'as, explicamos cu\'al de las dos representaciones es la m\'as conveniente a la hora de estudiar un proceso f\'isico cualquiera que involucre estados pesados de spin-1. Uno de los rasgos m\'as significativos de la dualidad Proca-Antisim\'etrico es su complementariedad. Esto significa que si un proceso con resonancias en una de las dos representaciones da lugar a una contribuci\'on no nula a cierto orden de la EWET, la otra es nula, y viceversa. Por supuesto, siempre existe una contribuci\'on local sin resonancias que corrige esta discrepancia, pero su c\'alculo no es trivial. Concluimos que el formalismo Antisim\'etrico es m\'as adecuado siempre que la interacci\'on con resonancias se pueda expresar con operadores cuyos campos ligeros tengan dos \'indices de Lorentz abiertos; mientras que el formalismo de Proca es m\'as adecuado cuando estos campos ligeros tengan un \'unico \'indice de Lorentz. En definitiva, la justificaci\'on de una u otra representaci\'on es puramente cinem\'atica.

	Finalmente, realizamos un an\'alisis fenomenol\'ogico de la teor\'ia de resonancias electrod\'ebiles, restringido al sector puramente bos\'onico, con paridad par y sin incluir operadores que violen la simetr\'ia custodial. El Lagrangiano resultante, que denominamos Lagrangiano reducido, queda descrito en la ec.~(\ref{eq:Lpheno}). Contiene 8 constantes de acoplamiento de resonancias ($\kappa_W$, $c_d$, $\lambda_{hS_1}$, $d_P$, $F_V$, $G_V$, $F_A$ y $\Lambda_1^{hA}$) relacionadas con 4 estados pesados distintos ($S^1_1$, $P^1_3$, $V^1_{3\,\mu\nu}$ y $A^1_{3\,\mu\nu}$), de forma que estudiamos un espacio de 12 par\'ametros independientes (que es todav\'ia mode\-radamente amplio). Para tratar de reducir el n\'umero de grados de libertad del sistema, incorporamos tambi\'en una serie de condiciones de distancias cortas tales como condicionar el comportamiento asint\'otico de los factores de forma vecto\-ria\-les y axiales, y dos reglas de super convergencia denominadas {\it Weinberg sum rules} (WSRs), siendo la segunda m\'as restrictiva. De hecho, esta \'ultima regla de suma, a diferencia del resto de condiciones asumidas anteriormente, no es un requisito indispensable para tener teor\'ias bien comportadas a muy altas energ\'ias. No obstante, se satisface en la gran mayor\'ia de modelos presentes en la literatura. Una vez incorporadas las condiciones de distancias cortas, las predicciones de las LECs del Lagrangiano reducido quedan caracterizadas por 7 par\'ametros independientes, como se muestra en la tabla \ref{tab:pheno}. Adicionalmente, si restringimos este marco fenomenol\'ogico a los estados vectoriales y axiales, todos los efectos de estos estados masivos en la EWET pueden ser descritos \'unicamente en t\'erminos de las masas de estas resonancias, $M_{V^1_3}$ y $M_{A^1_3}$, respectivamente.
	
	A la hora de contrastar las predicciones de la teor\'ia de resonancias con los datos experimentales disponibles nos fijamos en los par\'ametros {\it oblique}, concretamente $S$ y $T$, y en los l\'imites actuales de las LECs. Los par\'ametros {\it oblique} son un conjunto de observables que miden la desviaci\'on de la nueva f\'isica respecto de las predicciones del SM en magnitudes relacionadas con las correcciones radiativas de la teor\'ia electrod\'ebil. En este trabajo, se analizan estos observables para las resonancias vectoriales y axiales (estados de spin-1) del Lagrangiano reducido hasta NLO. Para ello, se incorporan en el c\'alculo todas las condiciones de distancias cortas, incluidas las dos WSRs \cite{Pich:2013fea}, haciendo menci\'on expl\'icita sobre si solo se incluye la primera regla de suma o ambas. As\'i pues, considerando un escenario en el que se satisfacen las dos WSRs, se obtiene con un nivel de confianza del 68\% (95\%)
	\be \label{eq:phenomassesconclusions}
	M_{A^1_3} \gsim M_{V^1_3} \,>\, 5\, (4)\,\mbox{TeV}\,, \qquad\qquad \kappa_W = \Frac{ M_{V^1_3}}{M_{A^1_3}} \,\in\, [0.97\, (0.94)\,,\, 1]\,.
	\ee
	donde $\kappa_W$, que parametriza el acoplamiento Higgs-Goldstone, queda fijado como el cociente entre las masas de las resonancias al asumir la segunda WSR. 
	Adem\'as de establecer cotas para estas masas, tambi\'en se puede extraer del estudio de los par\'ametros $S$ y $T$ informaci\'on relevante acerca de los l\'imites que estas resonancias imponen sobre las LECs del Lagrangiano reducido (esencialmente las LECs puramente bos\'onicas). Se concluye que las resonancias imponen fuertes l\'imites sobre estos acoplamientos, del orden de $|\mF_i| \lsim 10^{-3}$ (95\% C.L.; con $i=1,3,4,6,9$), como se muestra en la ec.~(\ref{eq:LECsbounds}) y en la figura \ref{fig:LECs1}.
	
	En cambio, los l\'imites actuales que proporcionan los datos de los colisio\-nadores LHC (run-I) y LEP todav\'ia no son lo suficientemente acotados como para intuir la presencia de alg\'un tipo de resonancia a trav\'es de las LECs. Este hecho es incluso m\'as acusado en lo que se refiere a acoplamientos que involucran al bos\'on de Higgs, cuya f\'isica de precisi\'on todav\'ia se encuentra en una fase muy inicial. En las ecs.~(\ref{eq:phenoF1b}, \ref{eq:LECspheno31}, \ref{eq:LECspheno32}) se muestran las incertidumbres experimentales actuales para las LECs analizadas. Todas ellas presentan unos l\'imites de exclusi\'on (95\%) de orden $\mF_i \sim 10^{-1}$, excepto el caso de $\mF_1\sim 10^{-3}$, por lo que no es posible confirmar ni excluir ninguno de los estados de alta energ\'ia. Actualmente, se est\'an analizando nuevos datos experimentales del LHC (run-II) y hay previsiones de mejoras significativas para algunas de las constantes de acoplamiento, particularmente para los acoplamientos an\'omalos triples y cu\'articos. En los pr\'oximos a\~nos tambi\'en se espera una mejora de la precisi\'on de $\mF_1$ con la construcci\'on de nuevos colisionadores $e^+e^-$, que permitir\'a determinar la existencia de resonancias en el rango del TeV. 
	
	La teor\'ia de resonancias electrod\'ebiles conforma un marco de trabajo id\'oneo para la identificaci\'on de forma indirecta de estados de alta energ\'ia. No obs\-tante, todav\'ia quedan algunas actualizaciones pendientes para futuros trabajos. Principalmente, es necesario promocionar tanto la EWET como la misma teor\'ia de resonancias a una teor\'ia que incluya las tres generaciones de fermiones simult\'aneamente. Esta ampliaci\'on de la teor\'ia requiere, por otra parte, un enorme esfuerzo algebraico para construir de forma consistente una base de operadores tanto en la teor\'ia de altas como de bajas energ\'ias. A su vez, ser\'ia posible analizar la fenomenolog\'ia en el sector fermi\'onico, en la que tambi\'en habr\'ia que considerar nuevas condiciones fermi\'onicas de distancias cortas. En definitiva, es importante hacer un estudio fenomenol\'ogico global en futuros trabajos. Adicionalmente, se han de a\~nadir al an\'alisis realizado los datos experimentales de LHC (run-II) re\-ferentes a las LECs bos\'onicas cuando est\'en disponibles. Por \'ultimo, ser\'ia tambi\'en de inter\'es traducir algunos de los modelos específicos con estados pesados similares a los estudiados al lenguaje de la teor\'ia de resonancias electrod\'ebiles.


\end{document}